\documentclass[12pt,english]{article}
\usepackage{times,epsfig,amsmath,longtable}
\usepackage[T1]{fontenc}
\usepackage[latin1]{inputenc}
\usepackage{geometry,amssymb,graphicx,babel}
\title{CUORE: A CRYOGENIC UNDERGROUND OBSERVATORY FOR RARE EVENTS}

\def\tect{$^{130}$Te }
\def\tectn{$^{130}$Te}
\def\xect{$^{130}$Xe }
\def\xectn{$^{130}$Xe}
\def\tecv{$^{128}$Te }
\def\tecvn{$^{128}$Te}

\def\pbdd{$^{210}$Pb }
\def\pbddn{$^{210}$Pb}
\def\podd{$^{210}$Po }
\def\udt{$^{238}$U }

\def\thdt{$^{232}$Th }
\def\thdtn{$^{232}$Th}
\def\tld{$^{208}$Tl }
\def\tldn{$^{208}$Tl}
\def\bidq{$^{214}$Bi }

\def\kq{$^{40}$K }

\def\coss{$^{60}$Co }
\def\cosn{$^{60}$Co}

\def\amnu{$\vert\langle m_{\nu} \rangle\vert$~}
\def\amnun{$\vert\langle m_{\nu} \rangle\vert$}
\def\nmamnu{\vert\langle m_{\nu} \rangle\vert}

\def\BBz{$\beta\beta(0\nu)$~}
\def\BBm{$\beta\beta(0\nu,\chi)$~}
\def\BBd{$\beta\beta(2\nu)$~}
\def\BB{$\beta\beta$~}

\def\ca{$\sim$}

\def\dot{$\cdot$}
\def\pom{$\pm$ }

\def\teod{TeO$_2$~}
\def\teodn{TeO$_2$}
\def\be{\begin{equation}}
\def\ee{\end{equation}}
\def\gohm{G$\Omega$}

\def\per{$\times$}
\def\gagamma{g_{a\gamma\gamma}}
\def\ciccio{5$\times$5$\times$5 cm$^3$ }
\def\magro{3$\times$3$\times$6 cm$^3$ }

\renewcommand{\bf}{\textbf}

\newcommand{\lsim}{\mathrel{\mathop{\kern 0pt \rlap
  {\raise.2ex\hbox{$<$}}}
  \lower.9ex\hbox{\kern-.190em $\sim$}}}
\newcommand{\gsim}{\mathrel{\mathop{\kern 0pt \rlap
  {\raise.2ex\hbox{$>$}}}
  \lower.9ex\hbox{\kern-.190em $\sim$}}}

\begin{document}
\ifx\href\undefined\else\hypersetup{linktocpage=true}\fi 
\maketitle

\begin{center}


Dipartimento di Fisica dell'Universit\`{a} di Milano-Bicocca \\
e Sezione di Milano dell'INFN,Milano I-20126, Italy\\
\vskip 1.8 truemm
Dipartimento di Ingegneria Strutturale del Politecnico di
Milano, \\
Milano I-20133, Italy\\
\vskip 1.8 truemm
Department of Physics and Astronomy, University of South Carolina, \\
Columbia, South Carolina 29208 USA\\
\vskip 1.8 truemm
Laboratori Nazionali del Gran Sasso, \\
I-67010, Assergi (L'Aquila), Italy\\
\vskip 1.8 truemm
Dipartimento di Fisica dell'Universit\`{a} di Firenze \\
e Sezione di Firenze dell'INFN, Firenze I-50125, Italy\\
\vskip 1.8 truemm
Lawrence Berkeley National Laboratory, \\
Berkeley, California 94720, USA\\
\vskip 1.8 truemm
Lawrence Livermore National Laboratory, \\
Livermore, California, 94551, USA\\
\vskip 1.8 truemm
Laboratorio de Fisica Nuclear y Altas Energias, \\
Universidad de Zaragoza, 50009 Zaragoza, Spain\\
\vskip 1.8 truemm
Kamerling Onnes Laboratory, Leiden University, \\
2300 RAQ, Leiden, The Netherlands\\
\vskip 1.8 truemm
Dipartimento di Scienze Chimiche, Fisiche e Matematiche dell'Universit\`a dell'Insubria \\
e Sezione di Milano dell'INFN, Como I-22100, Italy\\
\vskip 1.8 truemm
Dipartimento di Fisica dell'Universit\`{a} di Genova \\
e Sezione di Genova dell'INFN, Genova I-16146, Italy\\
\vskip 1.8 truemm
University of California, Berkeley, California 94720 USA\\
\vskip 1.8 truemm
Laboratori Nazionali di Legnaro, \\
I-35020 Legnaro (Padova), Italy\\
\vskip 1.8 truemm
Dipartimento di Fisica dell'Universit\`{a} di Roma\\
e Sezione di Roma 1 dell'INFN, Roma  I-16146, Italy\\
\vskip 3.0 truemm
{\bf ( The CUORE Collaboration )}\\
\end{center}

Professor Ettore Fiorini, Universit\`{a} di Milano Bicocca, Spokesman.

\newpage

\vskip 1.5 truecm

\begin{center}
SCIENTIFIC PERSONAL MEMBERS OF THE CUORE COLLABORATION
\end{center}
\vskip 1.5 truecm

R.~Ardito$^{1,2}$,
C.~Arnaboldi$^{1}$,
D.~R.~Artusa$^{3}$,
F.~T.~Avignone~III$^{3}$,
M.~Balata$^{4}$,
I.~Bandac$^{3}$,
M.~Barucci$^{5}$,
J.W.~Beeman$^{6}$,
F.~Bellini$^{14}$,
C.~Brofferio$^{1}$,
C.~Bucci$^{4}$,
S.~Capelli$^{1}$,
F.~Capozzi$^{1}$,
L.~Carbone$^{1}$,
S.~Cebrian$^{7}$,
C.~Cosmelli$^{14}$,
O.~Cremonesi$^{1}$,
R.~J.~Creswick$^{3}$,
I.~Dafinei$^{14}$,
A.~de~Waard$^{8}$,
M.~Diemoz$^{14}$,
M.~Dolinski$^{6,11}$,
H.~A.~Farach$^{3}$,
F.~Ferroni$^{14}$,
E.~Fiorini$^{1}$,
G.~Frossati$^{8}$,
C.~Gargiulo$^{14}$,
E.~Guardincerri$^{10}$,
A.~Giuliani$^{9}$,
P.~Gorla$^{7}$,
T.D. Gutierrez$^{6}$,
E.~E.~Haller$^{6,11}$,
I.~G.~Irastorza$^{7}$,
E.~Longo$^{14}$,
G.~Maier$^{2}$,
R. Maruyama$^{6,11}$,
S.~Morganti$^{14}$,
S.~Nisi$^{4}$,
E.~B.~Norman$^{13}$,
A.~Nucciotti$^{1}$,
E.~Olivieri$^{5}$,
P.~Ottonello$^{10}$,
M.~Pallavicini$^{10}$,
V.~Palmieri$^{12}$,
E.~Pasca$^{5}$,
M.~Pavan$^{1}$,
M.~Pedretti$^{9}$,
G.~Pessina$^{1}$,
S.~Pirro$^{1}$,
E.~Previtali$^{1}$,
B. Quiter$^{6,11}$,
L.~Risegari$^{5}$,
C.~Rosenfeld$^{3}$,
S.~Sangiorgio$^{9}$,
M.~Sisti$^{1}$,
A.~R.~Smith$^{6}$,
S.~Toffanin$^{12}$,
L.~Torres$^{1}$,
G.~Ventura$^{5},$
N.~Xu$^{6},$
~and~L.~Zanotti$^{1}$\\

\begin{enumerate}
\item{Dipartimento di Fisica dell'Universit\`{a} di 
Milano-Bicocca e Sezione di Milano dell'INFN, Milano I-20126,
Italy} 

\item{Dipartimento di Ingegneria Strutturale del Politecnico di
Milano, Milano I-20133, Italy}

\item{Dept.of Physics and Astronomy,
University of South Carolina, Columbia, South Carolina, USA 29208}

\item{Laboratori Nazionali del Gran Sasso, I-67010,
Assergi (L'Aquila), Italy}

\item{Dipartimento di Fisica dell'Universit\`{a} di
Firenze e Sezione di Firenze dell'INFN, Firenze I-50125, Italy}

\item{Lawrence Berkeley National Laboratory, Berkeley,
California, 94720, USA}

\item{Laboratorio de Fisica Nuclear y Altas Energias,
Universid\`{a}d de Zaragoza, 50009 Zaragoza, Spain}

\item{Kamerling Onnes Laboratory, Leiden University,
2300 RAQ, Leiden, The Netherlands}

\item{Dipartimento di Fisica e
Matematica dell'Universit\`{a} dell'Insubria e Sezione di Milano
dell'INFN, Como I-22100, Italy}

\item{Dipartimento di Fisica dell'Universit\`{a} di
Genova e Sezione di Genova dell'INFN, Genova I-16146, Italy}

\item{University of California, Berkeley, California
94720, USA}

\item{Laboratori Nazionali di Legnaro, 
I-35020 Legnaro ( Padova ), Italy}

\item{Lawrence Livermore National Laboratory, 
Livermore, California, 94550, USA}

\item{Dipartimento di Fisica dell'Universit\`{a} di Roma e Sezione di Roma 1 dell'INFN, Roma  I-16146, Italy}
\end{enumerate}

\vskip 1.5 truecm

\clearpage

\newcounter{dic}
\setcounter{page}{1}
\tableofcontents
\newpage
\listoffigures{}
\newpage
\listoftables
\cleardoublepage

\section{Executive summary}

Recently, neutrino oscillation experiments have unequivocally demonstrated that neutrinos have mass and that the neutrino mass eigenstates mix. These experiments have yielded valuable information on the mixing angles and on the mass differences of the three eigenstates.  They cannot, however, determine the scale of the neutrino mass, which is fixed by the lightest neutrino mass eigenvalue. This can only be directly determined by beta decay end point spectral shape measurements, or in the case of Majorana neutrinos, by the observation and measurement of the neutrinoless double-beta decay half-life. The oscillation experiments yield values for the mixing angles and mass differences accurate enough to allow the prediction of a range of values of the effective mass of the Majorana electron neutrino.
The CUORE experiment is designed with a sensitivity capable of probing all but a small portion of this range.

CUORE is a proposed array of 988, 750 g, \teod bolometers operating at 8 to 10 milli-degrees Kelvin, hanging on the bottom of the mixing chamber of a dilution refrigerator.
Natural tellurium has a 33.87\% abundance of \tect, which is an even-even nucleus that can only decay by double-beta decay. This large natural abundance eliminates the requirement for the very expensive isotopic enrichment required in all of the other proposed next generation experiments. The signal of neutrinoless double beta decay would be a sharp peak at 2528.8 keV in the bolometric energy spectrum. The proposed array is a cylindrical geometry of 19 towers of 52 detectors each. Each tower consists of thirteen, four-detector modules. One such tower has been successfully constructed and is being operated in the Gran Sasso Laboratory as a test experiment, and also as an independent, first-step next generation double-beta decay experiment called CUORICINO.

Thus far the stability, the energy resolution, as well as the background counting rates, imply that the expanded array, CUORE, will be a very competitive next generation experiment. In fact, operating CUORICINO for 3 years with the present background level will yield a half-life sensitivity for neutrinoless double-beta decay of 6.1\per 10$^{24}$ years, corresponding to an effective mass of the electron neutrino of the order of 0.3 eV. This will be superior to the present upper bound on the effective electron-neutrino mass set by the $^{76}$Ge experiments.
The early CUORICINO data have already shown us what and where the sources of background are and how to reduce them. We conclude that the required reduction in the background is possible. This would allow the full CUORE array to achieve a 5-year sensitivity of the order of 30~meV to the effective neutrino mass. Another order of magnitude reduction will present a real challenge, but is possible. In that case, a 5-year sensitivity in the effective mass would be around 15 meV. In the least optimistic case above, and in the case of inverted hierarchy of the neutrino mass spectrum, a null experiment would strongly disfavour the Majorana character of neutrinos, whereas a definite observation would confirm that neutrinos are Majorana particles. Only double-beta experiments can achieve this goal. 
The technology is now proven, and with adequate funding, there are no show-stoppers. It is proposed to build the CUORE experiment in the Laboratori Nazionali del Gran Sasso (LNGS) in Assergi, Italy, the present site of CUORICINO. CUORE is by far the most cost effective next generation double-beta decay experiment proposed.

\clearpage

\section{CUORE Science Motivation}
\label{sec:physics}

The neutrino mass range of interest favoured by the results of neutrino oscillation experiments is now within the grasp of well-designed \BBz decay experiments. A well established technique has been developed by the MIBETA and CUORICINO experiments that made dramatic improvements in cryogenic spectroscopy. The realization that the technology is available to achieve such a fundamental physics goal provides the basic motivation for the CUORE experiment.

To convey the importance of neutrino mass and of the CUORE experiment, we present a theoretical motivation and a brief recapitulation of past double-beta decay experiments, drawing heavily on our own completed work. We present our new technological capabilities to show how the CUORE Collaboration is positioned to make rapid strides toward neutrino mass discovery, as well as measuring the Majorana mass scale. The CUORE apparatus also has application in searches for Cold Dark Matter and Solar Axions and more in general for rare nuclear events.\\

\begin{table} [h]
\begin{center}    
    \begin{tabular}{lcc}
        \hline\hline
        Parameter & Best fit & 3$\sigma$  \\
        \hline
        $\delta m^2_{21}\: [10^{-5} eV^2]$\ \ \ \ \ & 6.9 & 5.4--9.5 \\
        $\delta m^2_{32}\: [10^{-3} eV^2]$\ \ \ \ \  & 2.6 & 1.4--3.7 \\
        $\sin^2\theta_{1}$ & 0.52 0.31--0.72  \\ 
        $\sin^2\theta_{2}$ & 0.006 & $\leq$ 0.054  \\
        $\sin^2\theta_{3}$ & 0.30 0.23--0.39 \\ 
        \hline\hline
    \end{tabular}
    \caption{ \label{tab:nures} Best-fit values and 3$\sigma$ intervals for the three-flavour neutrino oscillation parameters from global data including solar (SK and SNO), atmospheric (SK and MACRO), reactor (KamLAND and CHOOZ) and accelerator (K2K) experiments~\cite{maltoni03} ($\delta m^2_{ij}=m^2_i-m^2_j$).}
\end{center}
\end{table}

\subsection{Theoretical Motivation of Neutrinoless Double-Beta
Decay Experiments}

Neutrinoless double-beta decay is by now an old subject \cite{elliott02,tretyak02,zdesenko02,ejiri02,morales03,cremonesi03,giuliani03}. What is new is the fact that positive observation of neutrino oscillations in atmospheric \cite{fukuda00,toshito01,hayato03,ambrosio03}, solar \cite{ahmad01,ahmad02,ahmed03,fukuda02,smy03}, reactor \cite{eguchi03,apollonio99} and accelerator \cite{ahn03} neutrinos gives new motivation for more sensitive searches. 
In fact, recently published constraints on the mixing angles of the neutrino-mixing matrix \cite{pascoli02,maltoni03,feruglio02,bilenky01,klapdor01} make a strong case that if neutrinos are Majorana particles, there are many scenarios in which next generation double-beta decay experiments should be able to observe the phenomenon and measure the effective Majorana mass of the electron neutrino, \amnu, which would provide a measure of the neutrino mass scale $m$. One fact is clear: neutrino oscillation experiments can only provide data on the mass differences of the neutrino mass-eigenstates. The absolute scale can only be obtained from direct mass measurements, $^{3}H$ end point measurements for example \cite{osipowicz01}, or in the case of
Majorana neutrinos, more sensitively by neutrinoless double-beta decay. The time for large, next generation \BB-decay experiments has arrived, for if the mass scale is below $\sim $ 0.35 {\it eV}, \BB-decay may be the only hope for measuring it.

\subsection{Neutrino Mixing Matrix}

Using the Chau and Keung parametrization of the Pontecorvo-Maki-Nakagawa-Sakata (PMNS) neutrino mixing matrix~\cite{chau84}:
\begin{eqnarray} 
\left( {\begin{array}{c}
 \nu _e \\ \nu _\mu \\ \nu _\tau
\end{array} } \right) = UV\left( {\begin{array}{c}
 \nu _1 \\ \nu _2 \\ \nu _3
\end{array} } \right) & = &
\left( {\begin{array}{ccc}
 {c_3 c_2 } & {s_3 c_2 } & {s_2 e^{ - i\delta }}\\
 { - s_3 c_1 - c_3 s_1 s_2 e^{i\delta }}& {c_3 c_1 - s_3 s_1 s_2
e^{i\delta }}& {s_1 c_2 }\\
 {s_3 s_1 - c_3 c_1 s_2 e^{i\delta }}& { - c_3 s_1 - s_3 c_1 s_2
e^{i\delta }}& {c_1 c_2 }\\
\end{array} } \right) \times \nonumber \\
&& \times \left( {\begin{array}{ccc}
 1 & 0 & 0 \\
 0  & e^{i{\phi _2 }/2} & 0 \\
 0 & 0 & e^{i (\phi _3/2 + \delta) }
\end{array} } \right)
\left( {\begin{array}{c}
 \nu _1 \\ \nu _2 \\ \nu _3
\end{array} } \right)
\label{eq:CKMM}
\end{eqnarray}

\noindent where $c_i \equiv \cos \theta _i $, $s_i \equiv \sin \theta _i $, V is a diagonal matrix containing Majorana CP phases that do not appear in neutrino oscillations. While this looks very complicated and populated with many unknowns, neutrino oscillation data \cite{smy03,ahmed03,hayato03,ambrosio03,eguchi03,apollonio99,ahn03} have constrained all three of the angles and squared neutrino mass differences as shown in Table \ref{tab:nures}.

Considering the values found in Table \ref{tab:nures}, we find it is possible to make the approximation that $\theta _2 \equiv 0$ and assume maximal mixing for the atmospheric neutrino oscillation data, or $\theta _1 \cong 45^ \circ $. Accordingly,
\begin{equation}\label{eq:noa} 
U \cong \left( {\begin{array}{ccc}
 {c_3} & {s_3} & 0\\
 { - s_3 c_1}& {c_3 c_1}& {s_1}\\
 {s_3 s_1 }& { - c_3 s_1}& {c_1}\\
\end{array} } \right) 
\cong \left( {\begin{array}{ccc}
 \frac{\sqrt 3} {2} & \frac{1}{2} & 0\\
 { -{\frac{1} {2 \sqrt 2}}} & {\frac{\sqrt 3}{2 \sqrt 2}}& {\frac {1}{\sqrt 2}}\\
 {\frac{1}{2 \sqrt 2}}& { -{\frac{\sqrt 3}{2 \sqrt 2}}}& {\frac{1}{\sqrt 2}}\\
\end{array} } \right) 
\end{equation}

\noindent where $c_3 = {\frac{\sqrt 3 }{2}}$, $s_3 = {\frac{1}{2}}$ and $c_1 = s_1 = {\frac{1}{\sqrt 2}}$ were used in the second matrix.

\subsection{Neutrinoless Double-Beta Decay}\label{sec:bb0n}

\begin{figure}[htbp]
 \begin{center}
 \includegraphics[width=0.8\textwidth]{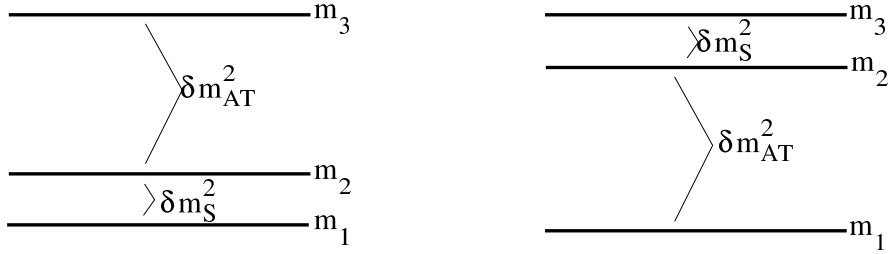}
 \end{center}
 \caption{Normal and inverted mass hierarchies schemes. Quasi-degenerate hierarchy corresponds to the case $m_1 >> \delta m_AT >> \delta m_S$. }
 \label{fig:mhyer}
\end{figure}

The decay rate for the process involving the exchange of a Majorana neutrino can be expressed as follows:
\begin{equation}\label{eq:t0n} 
\left[ T _{1/2}^{0\nu } \right]^{-1} = {\frac{\nmamnu^2}{m_e^2}} F_N .
\end{equation}

\noindent where 
\begin{equation}\label{eq:fn}
F_N \equiv G^{0\nu}\vert M^{0\nu} \vert ^2
\end{equation}

\noindent with $G^{0\nu }$ being the two-body phase-space factor including coupling constants and  $M^{0\nu} $ the \BBz nuclear matrix element. The quantity \amnu is the effective Majorana electron neutrino mass given by:
\begin{equation}\label{eq:meff} 
\nmamnu \equiv \vert \vert
U_{e1}^L \vert ^2m_1 + \vert U_{e2}^L \vert ^2m_2 e^{i\phi _2 } +
\vert U_{e3}^L \vert ^2m_3 e^{i\phi _3 }\vert ,
\end{equation}

\noindent where $e^{i\phi _2 }$ and $e^{i\phi _3 }$ are the Majorana CP phases ($\pm $1 for CP conservation) and $m_{1,2,3} $ are the mass eigenvalues.
The measured values of $\delta m_{21}^2$ ($\delta m_S^2 $ solar) and $\delta m_{32}^2$ ($\delta m_{AT}^2 $ atmospheric) given in table~\ref{tab:nures} motivate the pattern of masses in two possible hierarchy schemes shown in Figure \ref{fig:mhyer}.

In the case of inverted hierarchy the correct expression for \amnu can be obtained by interchanging the first and third columns of UV, i.e. by interchanging the roles of m$_1$ and m$_3$.\par
In general, prior to the approximation $\theta _2 = 0 = s_2 $, in the normal hierarchy case we have:
\begin{equation}\label{eq:meffnh} 
\nmamnu_{NH} = \vert c_3^2 c_2^2
m_1 + s_3^2 c_2^2 e^{i\phi _2 }m_2 + s_2^2 e^{i\phi _3 }m_3 \vert.
\end{equation}

\noindent while in the inverted hierarchy we have:
\begin{equation}\label{eq:meffih} 
\nmamnu _{IH}= \vert s_2^2 e^{i\phi _3 }m_1 + s_3^2 c_2^2 e^{i\phi _2 }m_2 + c_3^2 c_2^2 m_3 \vert.
\end{equation}

With the values and errors (3$\sigma )$ from Table \ref{tab:nures}, these becomes
\begin{equation}\label{eq:meffrisnh} 
\nmamnu_{NH} = \vert (0.69_{ -0.11}^{ +0.04} ) m_1 + (0.30_{ -0.08}^{ +0.07} ) e^{i\phi _2
}m_2 + ( < 0.054) e^{i\phi _3 }m_3 \vert .
\end{equation}

\begin{equation}\label{eq:meffrisih} 
\nmamnu_{IH} = \vert ( < 0.054) e^{i\phi _3 }m_1 + (0.30_{ -0.08}^{ +0.07} ) e^{i\phi _2
}m_2 + (0.69_{ -0.11}^{ +0.04} ) m_3 \vert .
\end{equation}

From the relations in Figure \ref{fig:mhyer}, we can write $m_2 = \sqrt {\delta m_S^2 + m_1^2 } $ and $m_3 = \sqrt {\delta m_{AT}^2 + \delta m_{S}^2 + m_1^2 } $ in the case of normal hierarchy and 
$m_2 = \sqrt {\delta m_{AT}^2 + m_1^2 } $ and $m_3 = \sqrt {\delta m_S^2 + \delta m_{AT}^2 + m_1^2}$ in the case of inverted hierarchy. Equations (\ref{eq:meffnh}) and (\ref{eq:meffih}) can be therefore expressed in terms of mixing angles, $\delta m_S^2 $, $\delta m_{AT}^2 $, and CP phases as follows \cite{barger02,pascoli02}:
\begin{equation}\label{eq:meffnh1} 
\nmamnu_{NH} = \left| {
c_2^2 c_3^2 m_1 + c_2^2 s_3^2 e^{i\phi _2 }\sqrt {\delta m_S^2 +
m_1^2 } + s_2^2 e^{i\phi _3 }\sqrt {\delta m_{AT}^2 + \delta m_S^2 + m_1^2 } }
\right|
\end{equation}

\begin{equation}\label{eq:meffih1} 
\nmamnu _{IH}= \vert s_2^2 e^{i\phi _3 }m_1 + 
s_3^2 c_2^2 e^{i\phi _2 }\sqrt {\delta m_{AT}^2 + m_1^2 } + 
c_3^2 c_2^2 \sqrt {\delta m_S^2 + \delta m_{AT}^2 + m_1^2} \vert.
\end{equation}

With the approximation $\theta _2 \equiv 0$ and the further approximation of $\delta m_S^2 < < \delta m_{AT}^2 $, equations (\ref{eq:meffnh1}) and (\ref{eq:meffih1}) are rewritten as follows:
\begin{equation}\label{eq:meffnh2} 
\nmamnu = m_1 \left|
{ c_3^2 + s_3^2 e^{i\phi _2 }\sqrt {1 + \frac{\delta m_s^2
}{m_1^2 }}  } \right|
\end{equation}

\begin{equation}\label{eq:meffih2} 
\nmamnu _{IH}=  \sqrt {\delta m_{AT}^2 + m_1^2 }
\vert s_3^2 e^{i\phi _2 } + c_3^2  \vert.
\end{equation}

Actual values of \amnu for different hierarchies have been recently evaluated by many authors on the base of the latest results of neutrino oscillation experiments. The results of the analysis of Pascoli et al. are summarized for reference in table~\ref{tab:amn90}~\cite{pascoli02}.
\begin{table}[ht]
\begin{center}
\begin{tabular}{ccccc} 
\hline
\rule{0pt}{0.5cm} $\sin^2 \theta$
& $\nmamnu^{\rm max}_{\rm NH}$ & $\nmamnu^{\rm min}_{\rm IH}$ &  $\nmamnu^{\rm max}_{\rm IH}$ &
$\nmamnu^{\rm min}_{\rm QD}$ \\ 
\hline \hline
0.0  & 3.7 & 8.7 & 50.6 & 47.9 \\
0.02 & 4.6 & 8.6 & 49.6 & 42.8 \\
0.04 & 5.3 & 9.9 & 48.6 & 45.4 \\
\hline
\end{tabular}
\caption{\label{tab:amn90} 
Present constraints on \amnu.
The maximal values of  \amnu (in units of $10^{-3}$ eV) for the normal (NH)  and inverted (IH) hierarchy cases, and the minimal values of  \amnu (in units of $10^{-3}$ eV) for the IH and QD (quasi-degenerate) cases, for the 90\% C.L. allowed values of $\delta m_S$, $\theta_S$ and $\delta m_{AT}$.
}
\end{center}
\end{table}

\subsection{Experimental Prospects} \label{bwmw}

It is interesting to compare the values given in Table \ref{tab:amn90} with the projected sensitivity of the CUORE experiment for \tect \BBz.
The projected 5 years sensitivity, discussed in detail later in this proposal, is 2.1\per 10$^{26}$ y. Using eq.~\ref{eq:t0n} and \ref{eq:fn} we can obtain the corresponding sensitivity on \amnu.
A list of values of F$_N$ calculated in the framework of different 
nuclear models are summarized in table \ref{tab:nmev}.

\begin{table}[h]
\begin{center}
\begin{tabular}{lllccc} \hline\hline
&Authors/Ref.&Method&T$_{1/2}$(\tect)&F$_N$(\tect)&F$_N$($^{76}$Ge)\\
&&&(10$^{23}$~y)&(10$^{-13}$ y$^{-1}$)&(10$^{-13}$ y$^{-1}$)\\
\hline
QRPA&Staudt et al., 1992~\cite{staudt92}
&pairing (Paris)&0.77-0.88&29-34&5.9-10\\
&&pairing (Bonn)&0.9-1.1&24-29&4.5-8.9\\

QRPA&Staudt et al., 1990~\cite{staudt90}
&&5&5.22&1.12\\

&Pantis et al., 1996~\cite{pantis96}
&no p-n pairing&8.64&3.0&0.73\\
&&p-n pairing&21.1&1.24&0.14\\

&Vogel et al., 1986~\cite{vogel86}
&&6.6&3.96&0.19\\

&Tomoda et al., 1987~\cite{civitarese87}
&&5.2&5.0&1.2\\

&Civitarese et al., 2003\cite{civitarese03}
&&5.2&5.0&0.7\\


&Barbero et al., 1999~\cite{barbero99}
&&3.36&7.77&0.84\\

&Simkovic et al, 1997~\cite{simkovic97}
&Full RQRPA&4.66&5.6&0.27\\

&Simkovic, 1999~\cite{simkovic99}
&pn-RQRPA&14.5&1.79&0.62\\

&Suhonen et al., 1992~\cite{suhonen92}
&&8.34&3.13&0.72\\

&Muto et al., 1989~\cite{muto89}
&&4.89&5.34&1.1\\

&Stoica et al., 2001~\cite{stoica01}
& large basis&10.7&2.44&0.65\\
&& short basis&9.83&2.66&0.9\\

&Faessler et al., 1998~\cite{faessler98}
&&9.4&2.78&0.83\\

&Engel et al., 1989~\cite{engel89}
&generalized &&&\\
&&seniority&2.4&10.9&1.14\\

&Aunola et al., 1998~\cite{aunola98}
& WS&4.56&5.72&0.9\\
&& AWS&5.16&5.06&1.33\\

&Rodin et al., 2003~\cite{rodin03}
&QRPA &34.6&0.75&0.45\\

&&RQRPA &37.9&0.69&0.36\\

SM&Haxton et al., 1984~\cite{haxton84}
&weak coupling&1.6&16.3&1.54\\

&Caurier et al., 1996~\cite{caurier96}
&large basis&58&0.45&0.15\\

OEM&Hirsh et al., 1995~\cite{hirsh95}
&&7.3&3.6&0.95\\

\hline\hline
\end{tabular}
\caption{\BBz nuclear factors of merit F$_N$ for \tect and $^{76}$Ge according to different evaluation methods (QRPA: Quasi Random Phase Approximation, SM: Shell Model and OEM: Operator Expansion Method) and authors. The foreseen \BBz half-lifetime for \tect (\amnu=1~eV) is also reported.} 
\label{tab:nmev}
\end{center}
\end{table}

Using the average value of these values, we find that the CUORE experiment predicts a sensitivity of \ca 30 meV. This value probes a significant portion of the range of values of $m_{1}$ given in the table \ref{tab:amn90}. 
We note that some of the proposed experiments already consider a possible extension to even larger masses. This is for instance the case for the full ten tons proposal of GENIUS. The cost of this experiment would amount to two billion dollars and would require a very long delay for the enrichment of such a large mass of $^{76}${\it Ge}. A similar extension to ten tons of CUORE would cost at least an order of magnitude less, and would avoid the long delay required by enrichment. The option for an "enriched CUORE" will be considered later.

Since $\delta m_S^2 << \delta m_{AT}^2$ and $\theta_2$ is small (Tab.~\ref{tab:nures} and Fig.~\ref{fig:mhyer}), equations (\ref{eq:meffnh2}) and (\ref{eq:meffih2}) can be used to constrain the sum of the masses of all the neutrino eigenstates ($\Sigma\equiv\Sigma_j m_j$) \cite{barger02}
\begin{equation}\label{eq:sumlimit} 
2 \nmamnu + \sqrt { \nmamnu ^2\pm \delta m_{AT}^2 } \le
\Sigma \le \frac{2\nmamnu + \sqrt { \nmamnu ^2\pm \delta
m_{AT}^2 \cos^2 (2\theta _3 )} }{\vert \cos (2\theta _3 ) \vert}
\end{equation}

\noindent where the plus is for normal hierarchy and the minus for inverted hierarchy. Equation (\ref{eq:sumlimit}) can be simplified significantly for values of \amnu achievable in next generation experiments. 


When $\delta m_{AT}^2 << \Sigma ^2$ eq.~(\ref{eq:sumlimit}) transforms into
\begin{equation}\label{eq:sumlimit1}
\nmamnu \le
\frac{\Sigma }{3} \le 2 \nmamnu
\end{equation}

\noindent which, using the sensitivity projection given above, gives 90 meV $ \le \Sigma  \le $ 180 meV. CUORE will therefore improve the WMAP result ($\Sigma  \le $ 700 meV~\cite{spergel03}) further constraining the Majorana neutrinos contribution to the Cosmic Microwave Background fluctuations.

In a similar manner, CUORE would be very complementary to the proposed KATRIN $^{3}$H neutrino mass experiment. KATRIN has a projected sensitivity of 200 {\it meV} \cite{osipowicz01}, therefore should it observe a positive effect, and CUORE place a bound \amnu  << $m_{KATRIN}$, it would be clear evidence that neutrinos are Dirac particles.

In conclusion, according to most theoretical analyses of of present neutrino experiment results, next-generation DBD experiments with mass sensitivities of the order of 10 meV may find the Majorana neutrino with a non-zero effective electron neutrino mass, if the neutrino is self-conjugate and the neutrino mass spectrum is of the quasi-degenerate type or it has inverted hierarchy. Majorana massive neutrinos are common predictions in most theoretical models, and the value of a few $10^{-2}$ eV predicted for its effective mass, if reached experimentally will test its Majorana nature. DBD experiments with even better sensitivities (of the order of meV) will be essential to fix the absolute neutrino mass scale and possibly to provide information on CP violation. It is therefore evident that next generation neutrinoless double-beta decay experiments are the next important step necessary for a more complete understanding of the physics of neutrinos. 
In this proposal we describe the CUORE experiment, show how it could reach the required sensitivity, and give details of the work breakdown, cost estimate, and main project milestones.

\clearpage
\section{Status of Neutrinoless Double-Beta Decay Experiments}\label{sec:DBD}

In the Standard Model, neutrinos and their antiparticles, antineutrinos, are different particles, but no experimental proof has been provided so far. Nuclear double beta decay addresses both the question of the neutrino nature (Dirac or Majorana) and of its possible Majorana mass. 
The standard nuclear Double Beta Decay is a process where a nucleus (A,Z) transforms into its isobar (A,Z\pom 2) with the emission of two electrons (or positrons) and their corresponding neutrinos ($\nu_{e}$ or $\overline{\nu_{e}}$), say: (A,Z)$\rightarrow$(A,Z+2)+2e$^{-}$+2$\overline{\nu_{e}}$.  The most direct way to determine if neutrinos are Majorana particles is to explore, in potential nuclear double beta emitters (A,Z), if they decay without emitting neutrinos (A,Z) $\rightarrow$ (A,Z+2) + $2e^{-}$, violating the lepton number conservation. For this non-standard \BBz process to happen, the emitted neutrino in the first neutron decay must be equal to its antineutrino and match the helicity of the neutrino absorbed by the second neutron.

Phenomenologically that implies the presence of a mass term or a right-handed coupling. Well-known arguments of Kayser, Petcov, Rosen and Schechter and Valle \cite{schechter82} show that in the context of any gauge theory, whatever mechanism be responsible for the neutrinoless decay, a Majorana neutrino mass is required. In fact \cite{kayser87}, the observation of a \BBz decay implies a lower bound for the neutrino mass, i.e. at least one neutrino eigenstate has a non-zero mass.

Another form of neutrinoless decay, (A, Z) $\rightarrow$ (A, Z+2) + $2e^{-}+\chi$ may reveal also the existence of the Majoron ($\chi$), the Goldstone boson emerging from the spontaneous symmetry breaking of B--L (Baryon minus Lepton number), of most relevance in the generation of Majorana neutrino masses and of far-reaching implications in Astrophysics and Cosmology.

Double Beta Decay (DBD) is a classical research topic, which current neutrino physics results have put on the front-line as an unique probe to explore and elucidate, in the foreseeable future, important open questions left unanswered by the oscillation experiments, namely the determination of the absolute neutrino mass scale.



The DBD are extremely rare processes. In the two neutrino decay mode their half-lives range from $T_{1/2}\sim10^{18}$ y to $10^{25}$ y. On the other hand, the non-standard neutrinoless decay has not yet been found. The experimental investigation of these phenomena requires a large amount of DBD emitter, in low-background detectors with the capability for selecting reliably the signal from the background. In the case of the neutrinoless decay searches, the detectors should have a sharp energy resolution, or good tracking of particles, or other discriminating mechanisms. There are several natural and enriched isotopes that have been used in experiments with tens of kilograms. Some of them could be produced in amounts large enough to be good candidates for next generation experiments. The choice of the emitters should be made also according to its two-neutrino half-life (which could limit the ultimate sensitivity of the neutrinoless decay), according also to its nuclear factor-of-merit and according to the experimental sensitivity that the detector can achieve in the case that the emitter is, at the same time, the detector, as is the case of CUORE. As explained at length later on, the CUORE experiment has chosen as its \BB emitter natural tellurium oxide, as the natural isotopic abundance of\tect is about 34\%. Finally the extremely low background required for high sensitivity double beta decay searches requires the development of techniques for identifying, reducing and suppressing the background of all types and origins in the detectors and their environments . All these conditions set the strategies to search for the neutrinoless double beta decay.

The expected signal rate depends on the nuclear matrix element, but the dispersion of results of the current calculations (tab.~\ref{tab:nmev}) makes uncertain the interpretation of the experimental output.

Improvements in the precision of the theoretical evaluations of the matrix elements are essential. Experimental studies of nuclear structures relevant to DBD will help calculations of the matrix elements. The exploration of the conventional two-neutrino double beta decay of several potential double beta emitters and its comparison with theory must serve to help in determining the most suitable nuclear model. These improvements will hopefully provide accurate nuclear matrix elements for the neutrinoless DBD which are crucial for extracting the effective Majorana mass parameter. Table \ref{teor2nhl} gives a summary of theoretical predictions for \BBd half-lives of several emitters in some representative nuclear models.

\begin{table*}[t] 
\begin{center}
\begin{tabular*}{\textwidth}{l@{\extracolsep{\fill}}llllllll}
\hline

                 Model
		 & \multicolumn{3}{c}{SM}
                 & \multicolumn{2}{c}{QRPA}
                 & \multicolumn{1}{c}{$1^+D$}
                 & \multicolumn{1}{c}{OEM}
                 & \multicolumn{1}{c}{MCM}
                 \\ \hline

                 Reference
		 & \cite{haxton82}
                 & \cite{haxton93}
                 & \cite{caurier96,caurier99}
                 & \cite{vogel86}
                 & \cite{staudt90}
                 & \cite{abad84}
                 & \cite{hirsh95}
                 & \cite{suhonen98}
                 \\ \hline
$^{48}$Ca($10^{19}$y) & 2.9 & 7.2 & 3.7 &  &  &  &  &
\\
 $^{76}$Ge($10^{21}$y) & 0.42 & 1.16 & 2.2 & 1.3 & 3.0 & &
0.28 &1.9
\\
 $^{82}$Se($10^{20}$y) & 0.26 & 0.84 & 0.5 & 1.2 & 1.1 & 2.0
& 0.88 & 1.1

\\
 $^{96}$Zr($10^{19}$y) &  &  &  & 0.85 & 1.1 &  &  & .14-.96
\\
 $^{100}$Mo($10^{19}$y) &  &  &  & 0.6 & 0.11 & 1.05 & 3.4 &
0.72
\\
 $^{116}$Cd($10^{19}$y) &  &  &  &  & 6.3 & 0.52 &  & 0.76
\\
 $^{128}$Te($10^{24}$y) & 0.09 & 0.25 &  & 55 & 2.6 & 1.4 &  &
\\
 $^{130}$Te($10^{21}$y) & 0.017 & 0.051 & 2.0 & 0.22 & 1.8 & 2.4 & 0.1 &
\\
 $^{136}$Xe($10^{21}$y) &  &  & 2.0 & 0.85 & 4.6 &  &  &
\\
 $^{150}$Nd($10^{19}$y) &  &  &  &  & 0.74 &  &  &
\\ \hline
\end{tabular*}
\caption{Predicted theoretical half-lives $T_{1/2}^{2\nu}$ in some representative nuclear models.} \label{teor2nhl}
\end{center}
\end{table*}

Due to the extraordinary importance of the neutrinoless double beta decay in extracting information on the neutrino mass scale, the nuclear matrix element issue has become of paramount importance both through direct calculation and through information obtained from nuclear reactions (of type (p,n), for instance), or through "virtual" $\beta^{-}$ and $\beta^{+}$ processes which may give informations on the two branches connecting the relevant nuclear states in the double beta transition through the intermediate states.

In the case of the two-neutrino double beta decay, the half-lives are customarily expressed as $[T_{1/2}^{2\nu}\ (0^+\rightarrow 0^+ )^{-1}=G_{2\nu} \mid M_{GT}^{2\nu} \mid^{2}$, where $G_{2\nu}$ is an integrated kinematical factor \cite{doi85} and $M_{GT}^{2\nu}$ the nuclear \BB Gamow Teller matrix element. 
Analogously, the neutrinoless decay half-life (as far as the mass term contribution is concerned) is expressed by means of equations (\ref{eq:t0n}) and (\ref{eq:fn}) through the nuclear factor-of-merit F$_N$, the nuclear matrix-element $M^{0\nu}$, the effective neutrino mass parameter \amnu and the integrated kinematic factor $G_{0\nu}$~\cite{doi85}.

Concerning the neutrino mass question, the discovery of a \BBz decay will show that at least one of the Majorana neutrino eigenstates has a mass equal or larger than $\nmamnu = m_e/(F_NT_{1/2}^{0\nu})^{1/2}$ eV, where $T_{1/2}^{0\nu}$ is the neutrinoless half life. On the contrary, when only a lower limit of the half-life is obtained, one gets only an upper bound on \amnu, but not an upper bound on the masses of any neutrino. In fact, \amnu can be much smaller than the actual neutrino masses. The \amnu bounds depend on the nuclear model used to compute the \BBz matrix element. The \BBd decay half-lives measured till now constitute bench-tests to verify the reliability of the nuclear matrix element calculations which, obviously, are of paramount importance to derive the Majorana neutrino mass upper limit. As stated above the wide range of the nuclear matrix elements calculations for a given emitter is the main source of uncertainty in the derivation of the neutrino mass bound.

\subsection{Main strategies for Double Beta Decay searches}

The experimental signatures of the nuclear double beta decays are in principle very clear: in the case of the neutrinoless decay, one should expect a peak (at the Q$_{\beta\beta}$ value) in the two-electron summed energy spectrum, whereas two continuous spectra (each one of well-defined shape) will feature the two-neutrino and the Majoron-neutrinoless decay modes (the first having a maximum at about one third of the Q value, and the latter shifted towards higher energies). In spite of such characteristic imprints, the rarity of the processes under consideration make their identification very difficult. 
Such remotely probable signals have to be disentangled from a background due to natural radioactive decay chains, cosmogenic-induced activity, and man-made radioactivity, which deposit energy on the same region where the \BB decays do it but at a faster rate. Consequently, the main task in \BB-decay searches is to diminish the background by using the state-of-the-art ultra-low background techniques and, hopefully, identifying the signal.

To measure double beta decays three general approaches have been followed: geochemical, radiochemical, and direct counting measurements. In the geochemical experiments, isotopic anomalies in daughters of \BB decaying nuclei over geological time scales are investigated. Determination of the gas retention age of the ore is important. They are inclusive $2\nu+0\nu$ measurements, not distinguishing $2\nu$ from $0\nu$ modes. However, when $T_{1/2,exp.meas.}^{2\nu+0\nu(geoch.)} \ll T_{1/2,exp.bound}^{0\nu(direct)}$, most of the decay is through $2\nu$ mode. The finite half-lives measured geochemically in the cases of $^{82}$Se, $^{96}$Zr, $^{128,130}$Te and radiochemically in the case of $^{238}$U can be regarded as \BBd half-life values. Also the $T_{1/2}$ values measured geochemically can be taken as a bound for $T_{1/2}^{0\nu}$, because $T_{1/2}^{0\nu}$ (or the half-life of whatever decay mode) cannot be shorter than $T_{1/2,exp.}$. Consequently, bounds on \amnu can be derived from
geochemical half-life measurements.

Another way to look for double beta decay are the radiochemical experiments, by noticing that when the daughter nuclei of a double beta emitter are themselves radioactive, they can be accumulated, extracted and counted. If the daughter has a half-life much smaller than $10^{9}$y and has no other long-lived parents, its presence can be only due to \BB. Noticeable examples are that of $^{238}U \to \, ^{238}Pu$ (88 y, $\alpha$ decay) and $^{244}Pu \to \, ^{244}Cm$ (18 y, $\alpha$ decay).

Most of the recent activity, however, refers to direct counting experiments, which measure the energy of the \BB emitted electrons and so the spectral shapes of the $2\nu$, $0\nu$, and $0\nu \chi$ modes of double beta decay. Some experimental devices track also the electrons (and other charged particles), measuring the energy, angular distribution, and topology of events. The tracking capabilities are useful to discriminate the \BB signal from the background.

The types of detectors currently used are~\cite{avignone04}:
\begin{itemize}
\item{Calorimeters} where the detector is also the \BB source~\cite{antonio60} (Ge diodes, scintillators -- CaF$_2$, CdWO$_4$ --, thermal detectors  --TeO$_{2}$ --, Xe ionization chambers, \dots). They are calorimeters which measure the two-electron sum energy. Notable examples are IGEX, Heidelberg-Moscow, Milano-Gran Sasso (MiDBD), CUORICINO and the proposed CUORE and Majorana.
\item{Tracking detectors} of the inhomogeneous (source$\neq$detector) type (Time Projection Chambers TPC, drift chambers, electronic detectors). In this case, the \BB source plane(s) is placed within the detector tracking volume, defining two --or more-- detector sectors. Leading examples of tracking devices are the Irvine TPC's and the NEMO and ELEGANTS series.
\item{Tracking calorimeters} They are tracking devices where the tracking volume is also the \BB source, for example a Xenon TPC (CALTECH/PSI/Neuchatel) and the future EXO.
\end{itemize}

Well-known examples of \BB emitters measured in direct counting experiments are $^{48}$Ca, $^{76}$Ge, $^{96}$Zr, $^{82}$Se, $^{100}$Mo, $^{116}$Cd, $^{130}$Te, $^{136}$Xe, $^{150}$Nd.

The strategies followed in the \BB searches are different. Calorimeters of good energy resolution and almost 100\% efficiency (e.g. Ge-detectors and tellurite bolometers) are well suited for \BBz searches. They lack, obviously, the tracking capabilities to identify the background on an event-by-event basis but they have, in favour, that their sharp energy resolutions do not allow the leakage of too many counts from ordinary double beta decay into the neutrinoless region, and so the sensitivity intrinsic limitation is not too severe. On the contrary, the identification capabilities of the various types of chambers make them very well suited for \BBd and \BBm searches. However, their energy resolution is, at least presently, rather modest and the efficiency is only of a few percent. Furthermore, the ultimate irreducible background source in these devices when looking for \BBz decay will be that due to the standard \BBd decay. The rejection of background provided by the tracking compensates, nevertheless, the figure of merit in \BBz searches.

Modular calorimeters can have reasonable amounts of \BB emitters (Heidelberg/Moscow, IGEX, MiDBD and CUORICINO experiments) or large quantities (like CUORE, Majorana and GENIUS). Tracking detectors, instead, cannot accommodate large amounts of \BB emitters in the source plate. Recent versions of tracking devices have 10 kg and more (NEMO3). On the other hand, TPC devices are planning to reach one ton and more of Xenon (EXO).

The general strategy followed to perform a neutrinoless double beta decay experiment is simply dictated by the expression of the half-life
\begin{equation}\label{eq:bbhl}
T_{1/2}^{0\nu}\simeq ln2 \times \frac{N\cdot t}{S}
\end{equation}

\noindent where N is the number of \BB emitter nuclei and S the number of recorded counts during time t (or the upper limit of double beta counts consistent with the observed background). 
In the case of taking for S the background fluctuation (when that might be possible) one has the so-called detector factor-of-merit or \emph{neutrinoless sensitivity} (1$\sigma$) which for homogeneous (source$=$detector) devices - for which the background rate scales with the detector mass - reads
\begin{equation}\label{eq:bbs}
F_D=4.17 \times 10^{26}(f/A)(Mt/B\Gamma)^{1/2}\varepsilon_{\Gamma}\;\; \rm{years} 
\end{equation}

\noindent where B is the background rate (c/(keV kg y)), M the mass of \BB emitter (kg), $\varepsilon_{\Gamma}$ the detector efficiency in the energy interval $\Gamma$ around Q$_{\beta\beta}$ ($\Gamma =$ FWHM) and t the running time measurement in years (f is the isotopic abundance and A the mass number). 
A slightly different behaviour is expected when the background rate is very low ("zero background" or $B \Gamma M t << 1$ hypothesis) 
\begin{equation}\label{eq:bbs0}
F_D \sim f M t \varepsilon_{\Gamma}/A\;\; \rm{years}
\end{equation}

The other guideline of the experimental strategy is to choose, according to equation (\ref{eq:t0n}), a \BB emitter of large nuclear factor of merit $F_N=G_{0\nu} \mid M^{0\nu} \mid^2$, where the kinematical factor qualifies the goodness of the $\rm Q_{\beta\beta}$ value and $M^{0\nu}$ the likeliness of the transition. 
Notice that the upper limit on \amnu is given by $\nmamnu \le m_e/(F_D F_N)^{1/2}$, or in terms of its experimental and theoretical components 
\begin{equation}\label{eq:bbml}
\nmamnu \le 2.5\times 10^{-8} \times (A/f)^{1/2} \times (B\Gamma/Mt)^{1/4} 
\times \epsilon_{\Gamma}^{-1/2} \times G_{0\nu}^{-1/2} 
\times\mid M^{0\nu}\mid^{-1} eV
\end{equation}

\subsection{Overview of main experiments and synopsis of results}

It is out of the scope of this Proposal to fully review the experimental status of the field. We refer the reader to various recent or completed reviews, also including comparison with theory \cite{elliott02,tretyak02,zdesenko02,ejiri02,morales03,cremonesi03,giuliani03,barabash04,avignone04,elliott04}. We will sketch in the following a briefing of the main techniques and results and summarize the current experimental situation in Tables \ref{tab:limits} and \ref{tab:constraintsresults}.
We will obviously devote more attention to the bolometric experiments with \teodn, MiDBD and CUORICINO as the antecedents of CUORE.


The Germanium case, as prototype of the calorimetric approach where the \BB source is at the same time detector medium, has a special meaning for this Proposal. The search for the nuclear double beta decay of the isotope $^{76}$Ge started as earlier as in 1967 by the Milano group \cite{fiorini67,fiorini70} which is now leading the CUORE Collaboration. Other groups both from USA (USSB-LBL-UCB, Caltech, PNL-USC) and Europe (Zaragoza, PSI-Neuchatel, H/M) continued this search with germanium detectors. 
The mastering of the techniques in the fabrication of Ge detectors and the continuous reduction of radioactive background achieved steadily since the first prototypes (PNL-USC and USSB-LBL) made the Ge detectors a reference in the field of double beta decay and other rare event physics phenomena. The  effective neutrino mass bound obtained from the neutrinoless half life limits achieved with Ge detectors has been steadily improving along the last decades and now, the current most stringent bounds are led by the results obtained with Germanium experiments (IGEX, H/M). As they are now classical reference results for neutrinos mass bounds, we will describe them even briefly.

Two recent experiments have looked for the double beta decay of $^{76}$Ge. They both employ several kilograms of enriched $^{76}$Ge (86\%) in sets of detectors: the IGEX Collaboration experiment \cite{aalseth99} (a set of three detectors of total mass 6.0 kg) in the Canfranc Underground Laboratory (Spain) and the Heidelberg/Moscow Collaboration experiment \cite{gunther97} (a set of five detectors amounting to 11 kg) in Gran Sasso and 2.1 kg in the Baksan Neutrino Observatory (Russia). Both experiments were designed to get the highest possible sensitivity for neutrinoless double beta decay of $^{76}$Ge: a large amount of the $^{76}$Ge isotope, in detectors of good energy resolution and very low radioactive background. Very efficient shieldings, both active and passive, were used in the two experiments. As a result, in the case of IGEX the background recorded in the energy region between 2.0 and 2.5 MeV is about 0.2 c/keV kg y prior to Pulse Shape Discrimination and most of the background in the relevant Q$_{\beta\beta}$ region of 2039 keV is accounted for by cosmogenic activated nuclei ($^{68}$Ge and $^{60}$Co). Background reduction through Pulse Shape Discrimination successfully eliminates multisite events, characteristic of non-\BB events, leading to less than $\sim0.07$c/keV.kg.y. IGEX data corresponding to 8.87 kg y in $^{76}$Ge provided (using the statistical estimator recommended by the Particle Data Group) a half-life lower bound of $T_{1/2}^{0\nu }\geq 1.57\times 10^{25} y$ at 90\% C.L. for the complete data set with PSD \cite{aalseth00,gonzales00,aalseth02}. Accordingly, the limit on the neutrino mass parameter is 0.33--1.31~eV. The uncertainties originate from the spread in the calculated nuclear structure parameters. Data from one of the IGEX detectors, RG--3 --which went underground in Canfranc several years ago-- corresponding to 291 days, were used to set a value for the $2\nu$-decay mode half-life by simply subtracting the MC-simulated background. The best fit to the stripped data corresponds to a half-life $T^{2\nu}_{1/2}=(1.45 \pm 0.20) \times 10^{21}$ y.
The Heidelberg/Moscow experiment operates five p-type HPGe detectors of enriched $^{76}$Ge (86\%) with a total active mass of 10.96 kg, corresponding to 125.5 mol of $^{76}$Ge. The recorded background spectrum of the detectors of the H/M experiment corresponding to an exposure of 47.4 kg.y substracted by the MonteCarlo simulation of the background, provides a half-life for the \BBd-decay at 68\% C.L. $T^{2\nu}_{1/2}=\{1.55\pm0.01(stat)^{+0.19}_{-0.15}(syst)\}\times10^{21}$y in agreement with the IGEX result. To derive the neutrinoless decay half-life limit from the H/M experiment, the raw data of all five detectors as well as data with pulse shape analysis are considered. No indication for a peak at the Q-value\BBz-decay is seen in none of the two data sets (the first 200 d of measurement of each detector were suppressed, because of possible interference with the cosmogenic $^{56}$Co). Data prior to PSD show a background around the Q-value of ($0.19\pm0.1$) c/keV.kg.y similar to that of IGEX. Both background figures, in IGEX and H/M, together with nthe recent results of CUORICINO to be reported later are the lowest ever obtained without using discrimination mechanisms. The set of H/M data analyzed with PSD corresponds to 35.5 kg.y of exposure and its background in the energy region between 2000 and 2080 keV is ($0.06\pm0.01$)counts/(keV.kg.y). Following the method proposed by PDG, the limit on the half-life is \cite{klapdorB01} $T^{0\nu}_{1/2}\ge 1.9\times10^{25}$ y at 90\% C.L. No evidence for neutrinoless DBD has been reported so far \cite{tretyak02,barabash04,elliott02,elliott04} with the exception of the claimed discovery of the decay of $^{76}$Ge reported by a subset of the Heidelberg-Moscow collaboration \cite{klapdorC01}. This claim has been contested by various authors \cite{aalsethB02,zdesenkoB02} and also by other members of the same Heidelberg-Moscow Collaboration \cite{bakalyarov03}. A new analysis in favour of the previous claim has however been  published recently \cite{klapdor04,klapdor04a}.
After a new statistical analysis of the same set of data, some of the authors of the H/M collaboration conclude \cite{klapdorC01,klapdor02,klapdor03,klapdorB03} that there exists evidence of a neutrinoless double beta peak. This result has been very widely contested \cite{feruglio02,aalsethB02,zdesenkoB02,bakalyarov03}.

Another technique, this time of the detector-different-from-source type, employs Time Projection Chambers. An example of a TPC, successfully operated along many years, was the Time Projection Chamber of the UC Irvine group which played an historical role in the discovery of the conventional double beta decay. It was a rectangular box filled with helium and located underground at 290 m.w.e. (Hoover Dam). A central \BB source plane divides the volume into two halves, whereas a magnetic field is placed perpendicular to the source plane. Electrons emitted from the source follow helical trajectories from where the momentum and the angles of the $\beta$-particles are determined.
The \BB signal is recognized as two electrons emitted from a common point in the source, with no other associated activity, during some time before and after the event. The \BB source is thin enough (few mg/cm$^2$) to allow $\alpha$-particles to escape and be detected for tagging the background. The UCI TPC series of experiments, finished more than five years ago, had the merit of having obtained the first direct observation of the double beta decay ($^{82}$Se). They also successfully measured the two-neutrino double beta decay of $^{100}$Mo, $^{150}$Nd and $^{48}$Ca.

The NEMO series of detectors are \cite{piquemal99,sarazin00,lalanne02} electron tracking devices (with open Geiger cells) filled with helium gas. An external calorimeter (plastic scintillator) covers the tracking volume and measures the $\beta$ energies and time of flight. The \BB source is placed in a central vertical plane and is divided in two halves, one enriched and another of natural abundance (of about 150 grams each), to monitor and subtract the background. To identify a \BB signal, one should have a 2e-track with a common vertex ($cos\alpha <0.6$) in the source plus two fired plastic scintillators (E deposition$>$200 keV each). The two-electron events are selected by time of flight
analysis (in the energy range of \BB). For instance, NEMO 2 has been operating for several years at the Modane Underground Laboratory (Frejus Tunnel) at 4800 m.w.e and has measured the \BBd decays of $^{100}$Mo, $^{116}$Cd, $^{82}$Se and $^{96}$Zr.
A new, bigger detector of the NEMO series, NEMO 3, has recently started its data
taking with sources of up to 10 kg of $^{100}$Mo and other emitters \cite{vasiliev03}.

The ELEGANTS V detector of the University of Osaka (placed successively in Kamioka and Oto) is an electron tracking detector which consists of two drift chambers for $\beta$-trajectories, sixteen modules of plastic scintillators for $\beta$ energies and timing measurement, and twenty modules of NaI for X- and $\gamma$-rays identification. The \BB signals should appear as two tracks in the drift chamber with the vertex in the source plus two signals from two plastic scintillator segments. Both enriched and natural sources (of about 100 grams) are employed in the detector for background monitoring and subtraction. This detector has measured the \BBd decay of $^{116}$Cd, $^{100}$Mo, and several other emitters \cite{ejiri99}.
A new variant of ELEGANTS is searching for the double beta decay of $^{48}$Ca.

The Caltech/PSI/Neuchatel Collaboration \cite{vuilleumier93,farine96} has investigated the double beta decay of $^{136}$Xe in the Gotthard Tunnel (3000 m.w.e.) by using a time projection chamber where the Xenon, enriched up to 62.5\% in $^{136}$Xe, with a total mass of
m$=$3.3 kg, was both the source and the detector medium, i.e. a calorimeter plus a tracking device. It had a cylindrical drift volume of 180 fiducial litres at a pressure of 5 atm. The \BB signal appears as a continuous trajectory with
distinctive end features: a large angle multiple scattering and increase charge deposition (charge ``blobs'') at both ends. As usual, the \BB topology gives powerful background rejection, leading to a figure of $\rm B \sim 10^{-1}-10^{-2}$ c/keV kg y (at 2480 keV).

Other results worth mentioning are those of the ITEP and Florence-Kiev groups. The ITEP group measured \cite{artemiev95} the double beta decay of $^{150}$Nd (40 g) with a TPC of $\sim 300$ litres filled with CH$_4$ at atmospheric pressure, in a 700 gauss magnetic field. On the other hand, the group of INR at Kiev \cite{bizzeti02} is investigating the double beta decay of $^{116}$Cd with cadmium tungstate ($^{116}$CdWO$_4$) scintillator crystals. 

To conclude this synopsis, Table \ref{tab:limits} summarizes the neutrinoless half-life limits derived in the various experiments.
From such neutrinoless half-life limits, one can derive Majorana neutrino mass bounds according to the nuclear model chosen. Table \ref{tab:constraintsresults} shows the range of bounds derived from the most sensitive experiments.

\begin{table}[h] 
\begin{center}
\begin{tabular}{llcc}
\hline
Emitter & Experiment & $T_{1/2}^{0\nu} >$ &
\sl{C.L.\%} \\ \hline
$^{48}$Ca & HEP Beijing & $1.8 \times 10^{22}$ y & 68 \\
$^{76}$Ge & MPIH/KIAE & $1.9 \times 10^{25}$ y & 90 \\
          & IGEX  & $1.6 \times 10^{25}$ y & 90 \\
$^{82}$Se & UCI & $2.7 \times 10^{22}$ y & 68 \\
          & NEMO 3 & $4.7 \times 10^{22}$ y & 90 \\
$^{96}$Zr & NEMO 2 & $1.3 \times 10^{21}$ y & 90 \\
$^{100}$Mo & LBL/MHC/UNM & $2.2 \times 10^{22}$ y & 68 \\
           & UCI & $2.6 \times 10^{21}$ y & 90 \\
           & Osaka & $5.5 \times 10^{22}$ y & 90 \\
           & NEMO 3 & $6 \times 10^{22}$ y & 90 \\
$^{116}$Cd & Kiev & $1.7 \times 10^{23}$ y & 90 \\
           & Osaka & $2.9 \times 10^{21}$ y & 90 \\
           & NEMO 3 & $1.6 \times 10^{22}$ y & 90 \\
$^{130}$Te & Milano & $2.1 \times 10^{23}$ y & 90 \\
& CUORICINO & $5.5 \times 10^{23}$ y & 90 \\
$^{136}$Xe &
Caltech/UN/PSI & $4.4 \times 10^{23}$ y & 90 \\
 $^{136}$Xe & Rome
& $1.2 \times 10^{24}$ y & 90 \\
$^{150}$Nd & UCI & $1.2 \times 10^{21}$ y & 90 \\ 
           & NEMO 3 & $1.4 \times 10^{21}$ y & 90 \\ \hline
\end{tabular}
\caption{Limits on Neutrinoless Decay Modes} \label{tab:limits}
\end{center}
\end{table}

\begin{table}[h] \centering
\begin{center}
\begin{tabular}{ll}
\hline Experiment & \amnu $<$~~~(eV)\\
\hline
IGEX enrich. $^{76}$Ge&($0.33\sim1.35$)~(6.0kg)\\
H/M enrich. $^{76}$Ge&($0.35\sim1.05$)~(11kg)\\
MiDBD nat. $^{130}$Te&($0.62\sim3.1$)~(6.8kg)\\
CUORICINO nat. $^{130}$Te&($0.38\sim2.2$)~(40.7kg)\\
\hline
\end{tabular}
\caption{Current best constraints (upper limits) on \amnu.} \label{tab:constraintsresults}
\end{center}
\end{table}

\subsection{Thermal detector approach to double beta decay}

A series of bolometer experiments has been carried out by the Milan group since 1989 in the Gran Sasso Laboratory, searching for the double beta decay of $^{130}$Te (sec. \ref{sec:MiCUORICINO}). The working principle of these thermal detectors is explained in section \ref{sec:bolometers} of this Proposal. This saga of experiments, their development and achievements, are the origin of the present CUORE proposal. The increase in temperature produced by the energy released in the crystal due to a nuclear event (i.e. \BB), is measured by means of a sensor in thermal contact with the absorber. The Milano group has been using Tellurium oxide crystals as absorbers, and glued NTD Ge thermistors as sensors. Natural Tellurium contains 34\% of $^{130}$Te. After using successively TeO$_2$ crystals of 73 g and 334 g, as well as a set of four of these large crystals, a tower-like array of 20 crystals of 340 g in a copper frame at $\sim 10$ mK was installed and operated: that was the so-called MiDBD experiment which successfully got the second best neutrino mass bound (after the Ge results). An enlarged version of MiDBD, called CUORICINO (meaning small CUORE) is now taking data in Gran Sasso (see \ref{sec:MiDBD} and \ref{sec:CUORICINO})
and the limits on the effective neutrino mass obtained in two months only are already the best except from those of the Ge experiments (Table~\ref{tab:constraintsresults}).

\subsection{Next generation experiments}
The requirements for a next-generation experiment can easily  be deduced by reference to equations (\ref{eq:bbhl}) and (\ref{eq:bbs}). 
To improve the sensitivity of \amnu by a factor of 100, the quantity $Nt/S$ must be increased by a factor of $10^{4}$. The quantity $N$ can easily be increased by a factor of $\sim 10^{2}$ over present experiments, so that $t/S$ must also be improved by that amount. Since practical counting times can only be increased by a factor 2-4, the background should be reduced by a factor 25-50 below the present levels. These are approximately the target parameters of the next-generation neutrinoless double-beta decay experiments.\\
\\
The CAMEO proposal involves placing isotopically enriched parent isotopes at the center of BOREXINO. One example given involves 65 kg of $^{116}$CdWO$_{4}$ scintillation crystals. The collaboration predicts a sensitivity of $\nmamnu \sim$60 meV, and with 1000 kg the prediction  is $\nmamnu \sim$20 meV \cite{bellini01}.

As already stated in sec. \ref{sec:physics}, CUORE (sec.~\ref{sec:CUORE}) is a proposed cryogenic experiment with 19 towers of 52 detectors, each one being a 750 g \teod bolometer. It will be the only future detector not requiring isotopic enrichment. This means that it would utilize natural abundance Te, containing 33.8\% $^{130}$Te. A detailed description of CUORE, its performances, sensitivity and physics prospects is given in the present Proposal. A pilot experiment CUORICINO comprising one CUORE tower, is  now taking data at LNGS (section \ref{sec:CUORICINO}). With equivalent background, CUORE would be as sensitive as 400-950 kg of Ge enriched to 86\% $^{76}$Ge, depending on the nuclear matrix elements used to derive \amnu. It will be performed in the \emph{Laboratori Nazionali del Gran Sasso} (LNGS). It has been already approved by the Scientific Committee of this laboratory and its location in Hall A of the same Laboratory has been established.

EXO is a large proposed TPC, either high-pressure gas or liquid, of enriched $^{136}$Xe. This novel technique involves schemes for locating, isolating, and identifying the daughter $^{136}$Ba$^{+}$ ion by laser resonance ionization spectroscopy. EXO is planned as a zero-background experiment and it is expected to be one of the relevant experiments of the future. Its plans include the use of 1 to 10 tons of Xenon. A program of research and development is underway at the Stanford Linear Accelerator (SLAC)~\cite{danilov00}.

GENIUS is a proposal to use between 1.0 and 10 tons of "naked" germanium detectors, isotopically enriched to 86\% in $^{76}$Ge, directly submerged in a large tank of liquid nitrogen functioning both as a cooling medium and a clean shield. Extensive studies were made based on certain assumptions, measurements, and Monte Carlo simulations. In Ref. \cite{klapdor98,klapdorB98}, the authors claim a sensitivity range of 1-10 meV for \amnu, using $10^{3}$-$10^{4}$ kg of enriched Ge. A research and development program is underway in the Gran Sasso Laboratory to develop the techniques for cooling and operating "naked" Ge detectors in liquid nitrogen for extended periods \cite{klapdorB02,klapdorC03}.

The Majorana Project is a proposed significant expansion of the IGEX experiment, utilizing newly developed segmented detectors along with pulse-shape discrimination techniques. It proposes 500 fiducial kg of Ge isotopically enriched to 86\% in $^{76}$Ge in the form of 200-250 detectors. Each detector will be segmented into electrically independent volumes, each of which will be instrumented with the new PSD system. 

The MOON experiment is a proposed major extension of the ELEGANTS experiment. It will utilize between 1 and 3 tons of Mo foils, isotopically enriched to 85\% in $^{100}$Mo, inserted between plastic scintillators. It will have coincidence and tracking capabilities to search for \BBz decay as well as solar neutrinos. This novel technique for detecting solar neutrinos depends on the special properties of the nuclear decay schemes of $^{100}$Mo and its daughters, allowing both event and background identification \cite{ejiri00}.

This list of proposals should produce several experiments with sensitivities sufficient to actually observe \BBz decay or obtain stringent upper bounds on \amnu reaching the sensitivity range implied by recent neutrino oscillation results. The attainability of such a goal strongly depends on the true capability of these projects to reach the required background levels in the \BBz region. An experimental confirmation of the (often optimistic) background predictions of the various projects (even if extrapolated from the results of lower scale successful experiments) is therefore worthwhile and the construction of preliminary test setups is absolutely needed.
\\


Frequently appearing publications on the subject almost always refer to the importance of conducting next-generation neutrinoless double-beta decay experiments. A complete understanding of the neutrino mass matrix depends on three types of data: neutrino oscillations, tritium beta-decay measurements, and neutrinoless double-beta decay. Each is analogous to one leg of a three-legged stool, and each is necessary for the complete picture. The case for a significant investment in next-generation experiments of all three types is compelling.

\clearpage
\section{CUORE experimental prospects}

As discussed in the previous sections, a number of recent theoretical interpretations of atmospheric, solar and accelerator neutrino experiments imply
that the effective Majorana mass of the electron neutrino, \amnu, expressed in Eq.~\ref{eq:meff}, could be in the range 0.01 $eV$ to the present bounds. 

Considering this range, could a next generation \BBz - decay experiment detect it? If so, what technique would be the best for a possible discovery experiment? We will address these questions in an effort to demonstrate that CUORE, an array of 988, 750 g \teod bolometers, is one of the best approaches presently available. It can be launched without isotopic enrichment nor extensive Research and Development (R \& D) , and it can achieve next generation sensitivity.

The CUORE project originates as a natural extension of the succesfull MiDBD and CUORICINO \tect experiments, described before, where for the first time large arrays of bolometers were used to search for \BBz - decay. The good results obtained so far prove that the \emph{bolometric} technique although novel is competitive and alternative to the traditional \emph{calorimetric $Ge$} technique. The pertinent details of the detectors of the CUORE array and its operation at temperature of $\sim 10\;mK$, as well as the background issues, electronics, DAQ and data-analysis are discussed in other parts of this proposal. In the following we will discuss the Physics potential of CUORE regarding Double Beta Decay, Dark Matter, Solar Axions searches, and other rare processes.


Rare event hysics is playing a significant role in Particle Astrophysics and Cosmology. Examples of such rare phenomena could be the detection of non-baryonic particle dark matter (axions or WIMPs), supposedly filling a substantial part of the galactic halos, nucleon decay or neutrinoless double beta decay. These rare signals, if detected, would be important evidence of new physics beyond the Standard Model of Particle Physics, and would have far-reaching consequences in Cosmology. The experimental achievements accomplished during the last decade in the field of ultra-low background detectors have led to sensitivities capable of searching for such rare events. Dark matter detection experiments have largely benefited from the techniques developed for double beta decay searches. We expect a similar symbiosis also in the case of CUORE and CUORICINO for which Dark Matter detection will be therefore another important scientific objective.

Due to the low rates of the proposed processes, the essential requirement of these experiments is to achieve an extremely low radioactive background and large mass. Accordingly the use of radiopure detector components and shieldings, the instrumentation of mechanisms of background identification, the operation in an ultra-low background environment, in summary, the use of the state-of-the-art of low background techniques is mandatory. In some of these phenomena, as in the case of the interaction of a particle dark matter with ordinary matter, a very small amount of energy is deposited, and the sensitivity needed to detect events within such a range of energies, relies on how low the energy threshold of detection is. In addition, to increase the chances of observing such rare events, a large amount of detector mass is in general needed, and so most of the experiments devoted to this type of searches are planned to have large detector mass, while keeping the other experimental parameters (background, energy thresholds and resolutions) optimized. 
On the other hand, the recent development of cryogenic particle detection \cite{booth96} has led to the extended use of thermal detectors \cite{fiorini84,ltd9} to take advantage of the low energy threshold and good energy resolution required for the theoretically expected thermal signals. This detection technology has also the advantage of enlarging the choice of materials which can be used, either as DM targets or \BB decay emitters. After a long period of R\&D to master the techniques used in cryogenic particle detectors, low temperature devices of various types are now applied to the detection of double beta decay or particle dark matter \cite{morales99a}. 
A major example of this development were the MiDBD (\emph{Milano Double Beta Decay}) experiment (sec.~\ref{sec:MiDBD}), which successfully operated a 20 \teod crystal array of thermal detectors of a total mass of 6.8 kg and CUORICINO discussed below. With the objective of going to larger detector masses and of improving the sensitivity achieved in the smaller arrays, the CUORE (\emph{Cryogenic Underground Observatory for Rare Events}) project \cite{fiorini98,arnaboldi03} was born some years ago as a substantial extension of MiDBD. The objective of CUORE is to construct an array of 988 bolometric detectors with cubic crystal absorbers of tellurite of 5 cm side and of about 750 g of mass each. The crystals will be arranged in a cubic compact structure and the experiment will be installed in the Gran Sasso Underground Laboratory (LNGS). 

The material to be employed first is \teodn, because the main goal of CUORE is to investigate the double beta decay of \tectn, although other absorbers could also be used to selectively study several types of rare event phenomenology as it will be discussed later. Apart from the MiDBD experiment, already completed, a smaller and intermediate stage of CUORE, CUORICINO is presently taking data at LNGS. It consists of 44 cubic crystals slighly larger with respect to those foreseen for CUORE (790~g) and 18 smaller crystals (330~g) for a  total mass of $\sim\,$40.7~kg. CUORICINO is, by far, the largest cryogenic detector on the stage. Its early results are excellent (see section \ref{sec:CUORICINO}) and represent the most eloquent demonstration of the CUORE feasibility.

In the following, the Physics Potential of the CUORE experiment with respect to its double beta decay discovery potential and its detection capability of WIMP and axions is presented. For this purpose, we will rely on the CUORE background expected sensitivity described in other sections of this proposal (\ref{sec:CUORICINO} and \ref{sec:simulation}) and tested with the results already obtained in CUORICINO. 

In CUORICINO a background improved by a factor 1.5 with respect to the one measured in MiDBD-II has been  observed in the \BBz region.  The analysis of the low energy region, relevant for dark matter search, has not been completed but we expect a similar improvement there also. For what concerns CUORE on the other hand, MiDBD and CUORICINO have shed light on how the background contamination dominating the two regions of interest is mostly the intrinsic background due to bulk and surface contaminations of the constructing materials. With the presently achieved quality of low  contamination materials and considering the worst possible condition for bulk contaminations (i.e. all the contamination equal to the present 90\%/,C.L. measured upper limits) we proved that the corresponding contribution to the CUORE background will be \ca 0.004~counts/(keV~kg~y) at the \BBz transition energy and \ca 0.025~counts/(keV~kg~d) near threshold (section \ref{sec:simulation}). 
Even at these conservative values therefore, bulk contaminations will not represent a problem for CUORE sensitivity. A larger background contribution is however expected when extrapolating to CUORE the surface contributions measured in CUORICINO. A dedicated R\&D aiming at reducing the radioactive surface contaminations of the detector structure (mainly copper) is therefore required. A reduction by an order of  magnitude of the surface contributions seems attainable in a reasonable time (less than 2 years) by simply improving the surface cleaning procedures used for CUORICINO (e.g. by using cleaner solutions). This would be enough to reduce the \BBz background contributions to the level of few counts/keV/ton/y. Further improvements are still possible (e.g. active surface bolometers) and are considered by the CUORE collaboration as part of a second R\&D phase aiming at an even better sensitivity both for \BBz and dark matter searches (sec.~\ref{sec:INNO}).
In particular, the goal of this CUORE technical R\&D will be the reduction of background to the level of $\sim$0.001~counts/(keV~kg~y) at the \BBz transition energy and to $\sim$0.01~counts/(keV~kg~d) at threshold.  The potential of the experiment in this background configuration will also be discussed.

Regarding the expected threshold and resolution, energy thresholds of \ca 5 keV and energy resolutions lower than 1 keV at the 46 keV line of \pbdd were obtained in some test measurements carried out at LNGS ("Hall C" setup) for the early CUORE R\&D program~\cite{pirro00,giulani02}. We will assume therefore conservative values of 10 keV for the energy threshold and of 1 keV for the energy resolutions at threshold, as observed in MiDBD and CUORICINO. As far as the energy resolutions obtained in the double beta decay region, values of 3 keV at 2615 keV were achieved in some of the above mentioned test measurements. On the other hand, a 9 keV average value has been measured in the CUORICINO calibration spectra. A possible energy resolution interval of 5-10 keV (taking into account possible improvements in the foreseen CUORE R\&D, sec.~\ref{sec:INNO}) will be therefore assumed for CUORE. 

Taking into account these expectations, we discuss in the following the prospects of CUORE for double beta decay searches, for WIMP detection, for solar axion exploration and for searches on rare nuclear events.

\subsection{Double beta decay prospects}\label{sec:Qprospects}

The main scientific objective of the CUORE detector is the search for the neutrinoless double-beta decay of the \tect isotope contained in the (natural) \teod crystals.

In fact, cryogenic thermal detectors provide several double-beta nuclei to be explored in "active" source/detector calorimeters as it will be discussed later. Some of them have been tested and others are already in running detectors, like $^{48}$Ca in CaF$_2$, \tect in \teodn, and $^{116}$Cd in CdWO$_4$. As far as the Tellurium Oxide is concerned, the \tect isotope is a good candidate for double beta decay searches: its isotopic content in natural Tellurium is 33.87\%, and its \BB Q-value ($Q_{\beta\beta}=2528.8\pm1.3$ keV) is reasonably above the main radioimpurity lines when looking for a neutrinoless signal. Moreover, this Q-value happens to be between the peak and the Compton edge of the 2615 keV line of $^{208}$Tl, which leaves a clean window to look for the signal. Finally, it has a fairly good neutrinoless nuclear factor-of-merit F$_N$ (sec.~\ref{sec:bb0n}). It is in fact apparent (tab.~\ref{tab:nmev}) that, no matter which nuclear model is used to compute the neutrinoless double-beta decay matrix elements, the nuclear factors of merit of \tect are a factor 5--10 more favorable than those of $^{76}$Ge (the emitter where the best neutrinoless double beta decay half-life limits have been achieved so far). This translates into a factor 2 to 3 better as far as the \amnu (Majorana neutrino mass parameter) bounds are concerned.

The detector factor-of-merit F$_D^{0\nu}$ (eq.~(\ref{eq:bbs})), or detection sensitivity, introduced earlier by Fiorini, provides an approximate estimate of the neutrinoless half-life limit (1$\sigma$) achievable with a given detector. In the case of a \ciccio \teod bolometers (for which the \BBz detector efficiency is 0.86) we have F$_D^{0\nu}$ \ca 7.59\per 10$^{23} \sqrt{\frac{Mt}{B\Gamma}}$, with M the crystal mass in kg and B the background in counts per keV per year and per kg of detector mass.

To demonstrate the full CUORE physics potential, let us evaluate F$_D$ according to our  conservative projection for the CUORE array (750 kg of \teod). As described above, a background of B=0.01 c/keV/kg/y would be achievable with a slight improvement of the current available material selection and cleaning techniques (sections \ref{sec:MATE}, \ref{sec:SURF} and \ref{sec:simulation}) and exploting the possibility to build a dedicated cryostat with low activity materials and effective shields (sec.~\ref{sec:cryogenic}). Assuming finally an energy resolution $\Gamma$(2.5 MeV)=5 keV, we would get F$_D^{0\nu}$ of 9.4\per 10$^{25}\sqrt{t}$ years (6.5\per 10$^{25} \sqrt{t}$ years for $\Gamma$=10 keV), which in t years of statistics would provide \amnu bounds in the range 0.036--0.2~t$^{-1/4}$ eV (according to QRPA models F$_N$ predictions, tab.~\ref{tab:nmev}).
However, the R\&D to be carried out in CUORE, if successful, would provide a value of B\ca 0.001 c/keV/kg/y, i.e. a detection sensitivity of F$_D$\ca 2.96\per 10$^{26} \sqrt{t}$ years (2.1\per 10$^{26} \sqrt{t}$ years for $\Gamma$=10 keV), or \amnu bounds in the range \ca 0.02--0.11~t$^{-1/4}$ eV. \teod crystals made with \tect enriched material have been already operated in MiDBD and CUORICINO, making an enriched CUORE a feasible option (sec.~\ref{sec:enrich}). Assuming a 95\% enrichment in \tect and a background level of b=0.001 c/keV/kg/y, the sensitivity would become F$_D$\ca 8.32\per10$^{26} \sqrt{t}$ years. For an exposure of 5 years, the corresponding \amnu bounds would range from 8 meV to 45 meV depending on the nuclear matrix element calculations.

\subsection{Other physics with CUORE}\label{Others}
Even if double beta decay, particularly in its neutrinoless channel, is the main scientific aim of CUORE, other interesting searches can be tackled with this massive detector. For instance, as stressed above, dark matter detection experiments have largely benefitted in the past from the techniques developed for double beta decay searches and we have valid motivations to expect such a symbiosis also in the case of CUORE and CUORICINO. A number of other searches for rare processes (rare nuclear $\alpha$ and $\beta$ decays, electron and nucleon stability, etc.) could however take advantage of the large mass and low background of CUORE. We will discuss in the following, the potential of CUORE for such searches.

\subsubsection{WIMP detection} \label{wimpdetection}

Recent cosmological observations \cite{bachcall99} provide compelling evidence for the existence of an important component of non-baryonic cold dark matter in the Universe. Among the candidates to compose this matter, Weakly Interacting Massive Particles (WIMPs) and axions are the front runners. The lightest stable particles of supersymmetric theories, like the neutralino \cite{jungman96}, constitute a particular class of WIMPs.

Under the hypothesis that WIMPS are the main component of the dark matter, these particles should fill the galactic halos and explain the flat rotation curves which are observed in spiral galaxies. The detection of such particles could be attempted both by means of direct and indirect methods. The direct detection of WIMPs relies on the measurement of their elastic scattering off the target nuclei of a suitable detector\cite{morales99a}. The non relativistic and heavy (GeV -- TeV) WIMPs could hit a detector nucleus producing a nuclear recoil of a few keV. Because of the small WIMP-matter interaction cross sections the rate is extremely low. In the case of SUSY WIMPs, most of the cross section predictions \cite{ellis01,bottino01,bergstrom00} (derived using MSSM as the basic frame implemented with different unification hypothesis) encompass a range of values several orders of magnitude (the so-called scatter plots) providing rates rangingfrom 1 c/kg/day down to 10$^{-5}$ c/kg/day according to the particular SUSY model.

It is well known that the predicted signal for the WIMP elastic scattering has an exponentially decaying energy dependence, hardly distinguishable from the background recorded in the detector. The simple comparison of the theoretical WIMP spectrum with the one experimentally obtained, provides an exclusion curve (at a given confidence level), as dark matter component of the halo, of those WIMPs with masses ($m$) and cross sections on nucleons ($\sigma$) which yield spectra above the measured experimental rate. To claim a positive identification of the WIMP, however, a distinctive signature is needed. The only identification signals of the WIMP explored up to now are provided by the features of the Earth's motion with respect to the dark matter halo. In particular, the annual modulation \cite{drukier86} is originated by the combination of the motion of the solar system in the galactic rest frame and the rotation of the Earth around the Sun. Due to this effect, the incoming WIMP velocities in the detector rest frame change continuously during the year, having a maximum in summer and a minimum in winter in the northern hemisphere. Therefore the total WIMP rate changes in time with an oscillating frequency which corresponds to an annual period and a maximum around the beginning of June.

The relative annual variation of the signal is small (a few percent) so in order to detect it one needs large detector masses to increase statistics and several periods of exposure to minimize systematics. Several experiments have already searched for this effect \cite{sarsa96,abriola99,belli96} and since 1997 the DAMA group has reported a positive signal \cite{bernabei00} which has been appearing throughout four yearly periods.  Recently, the DAMA experiment has extended up to seven yearly periods the annual modulation effect, which stands now at 6.3 $\sigma$ level \cite{bernabei03}. The present situation is no doubt exciting. On one hand that result has triggered an intense activity in the field; on the other, the experimental sensitivities of various types of underground detectors are entering the supersymmetric parameter space \cite{bottino01} and in particular several of them \cite{abusaidi00,benoit01,morales00,smith02} are excluding, to a larger or lesser degree (within standard framework models), the region of mass and cross-section where the reported WIMP is supposed to exist. This is particularly true after the recent results of the CDMS experiment~\cite{akerib04} whose disagreement with DAMA is statistically more significant. It is then compulsory to explore this effect with other nuclear targets, and that will be one of the purposes of CUORE. We will discuss in the following the capabilities of CUORE to exclude or detect WIMPs using the total time-integrated experimental rate and comparing it with the predicted nuclear recoil rate. To look for the annual modulation signal in CUORE, which in principle has enough mass to be sensitive to it, one needs to know its stability performance. The analysis of the CUORE potential for annual modulation searches will be performed -following statistical consideration- (see Ref. \cite{cebrian01}), with the proviso that systematic uncertainties are under control. Data on the stability of CUORICINO will be crucial to assess such hypotheses.

To calculate the theoretical WIMP rate, standard hypothesis and astrophysical parameters are assumed, i.e., that the WIMPs form an isotropic, isothermal, non-rotating halo (the isothermal sphere model) of density $\rho=0.3 $ GeV/cm$^3$, which has a maxwellian velocity distribution with $v_{rms}=270$ km/s (with an upper cut corresponding to an escape velocity of 650 km/s), and a relative Earth-halo velocity of $v_r=230$ km/s). Other, more elaborated halo models, which have been considered recently \cite{halomodels} would lead to different results. The same applies when other astrophysical parameters are employed or when uncertainties in the halo WIMPs velocity distribution are included \cite{astroparticle}. The theoretical predicted rate is expressed in terms of the mass and cross-section of the WIMP-matter interaction. The cross sections are normalized per nucleon assuming a dominant scalar interaction, as is expected, for instance, for one of the most popular dark matter candidates, the neutralino:

\begin{center}
\begin{equation}\label{norm}
    \sigma_{N\chi} = \sigma_{n\chi}A^2 \frac{\mu^2_{W,N}}{\mu^2_{W,n}}
\end{equation}
\end{center}

\noindent where $A$ is the target (oxygen and tellurium) mass number, $\mu^2_{W,N}$, is the WIMP-nucleus reduced mass, and $\mu^2_{W,n}$ the WIMP-nucleon reduced mass. The Helm parameterization \cite{engel91} is used for the scalar nucleon form factor. The $(m,\sigma)$ exclusion plot is then derived by requiring the theoretically predicted signal for each $m$ and $\sigma$ in each energy bin to be less than or equal to the (90\% C.L.) the Poisson upper limit of the recorded counts. The bin width is assumed to be equal to the detector resolution.

\begin{figure}
 \begin{center}
 \includegraphics[width=0.75\textwidth]{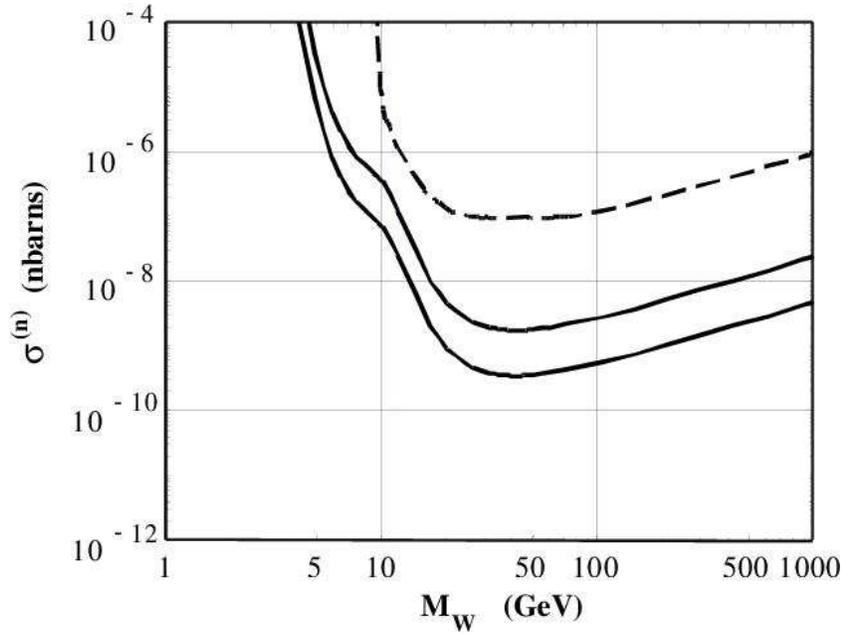}
 \end{center}
 \caption{Exclusion projected for 1 year of CUORE assuming a threshold of 10 keV, a low energy resolution of 1 keV, and low energy background levels of 0.05 and 0.01 c/keV/kg/day respectively. The dashed line corresponds to the MiDBD result.}
 \label{fig:cuore_exclusion}
\end{figure}

In Fig.~\ref{fig:cuore_exclusion}, the exclusion plots for coherent spin-independent WIMP-matter interaction are shown for two possible values of the background of CUORE: 0.05 and 0.01 c/keV/kg/day1 and 0.1 which we expect to be attainable after a dedicated R\&D directed to improve the detector performance in the low energy region.

Notice, moreover, that values of a few 0.01 c/keV/kg/day have been obtained above 10 keV in the raw spectra of Germanium experiments (e.g. IGEX \cite{morales00}) without using mechanisms of background  rejection, and so it does not seem impossible to achieve such equivalent small values in crystal thermal detectors of tellurium (with only phonons, and no discrimination mechanism). To draw the two exclusion contours of Fig. \ref{fig:cuore_exclusion}, a low energy resolution of 1 keV, an energy threshold of 10 keV, the two quoted values of the background of CUORE (0.05 and 0.01 c/keV/kg/day) and an exposure of 1 year (750 kg$\cdot$year) have been assumed.

As previously noted, CUORE is characterized by a detector mass large enough to search for the annual modulation signal. As it is well known, an essential requirement to estimate the prospects of any detector to search for annual modulation is to have a superb control of systematic errors and to assure that the stability of the various experimental parameters, which might mimic periodic variations of the signals, are kept within a small fraction of the (already small) expected signal. The various changes of the set-up, crystals and shielding of the MiDBD experiment have not provided a definitive estimation of the long-term stability parameters of MiDBD; it will be however measured in CUORICINO. Possible instabilities are that of the electronic gain and the ensuing time fluctuation of the energy scale (both in energy thresholds and energy resolutions), the temperature variations, the possible fluctuation in time of the efficiency with which the triggered noise is rejected.

The fact that we are dealing with a signal depending on time, which typically amounts to a fraction between 1\% and 7\% of the average count rates, reinforces the need for a control of the stability of the experiment well below that range over long periods of time.


When assuming that all these fluctuations are controlled well below the levels needed ($<$1\%), then one can proceed to analyze the sensitivity of CUORE to the annual modulation signal on purely statistical grounds. This has been first attempted in \cite{ramachers98} and \cite{hasenbalg98}, but a more extensive and rigorous approach is followed in ref. \cite{cebrian01} where sensitivity plots for several types of detectors (and experimental parameters) are presented.


Following the guidelines of that reference, the sensitivity of a given experimental device to the annual modulation signal can be precisely quantified by means of the $\delta$ parameter, defined from the likelihood function or, equivalently, from the $\chi^2$ function of the cosine projections of the data (for further details see ref. \cite{cebrian01}):

\begin{equation}\label{eq:asimptotic}
  \delta^2=y(\sigma=0)-y_{min}\simeq\chi^2(\sigma=0)-\chi^2_{min}.
\end{equation}

This parameter measures the statistical significance of the modulation signal detected in an experimental set of data. However, for a given ($m,\sigma$) and a given experiment, the expected value $\langle \delta^2 \rangle$ can be estimated using the expression derived in ref. \cite{cebrian01}:

\begin{equation}\label{eq:magic}
  \langle \delta^2 \rangle=\frac{1}{2}\sum_{k}
  \frac{S_{m,k}(\sigma,m_W)^2\Delta
  E_k}{b_{k}+S_{0,k}}MT\alpha+2
\end{equation}

\noindent where $S_{m,k}$ and $S_{0,k}$ are the modulated and non-modulated parts of the WIMP signal in the $k$th energy bin of $\Delta E_k$ width, $b_k$ is the background in that energy bin and $MT\alpha$ the effective exposure, being $\alpha$ a coefficient accounting for the temporal distribution of the exposure time around modulation maxima and minima ($\alpha = 1/n \sum_{i=1}^n \cos^2 \omega (t_i -t_0)$ for $n$ temporal bins).

\begin{figure}
 \begin{center}
 \includegraphics[width=0.75\textwidth]{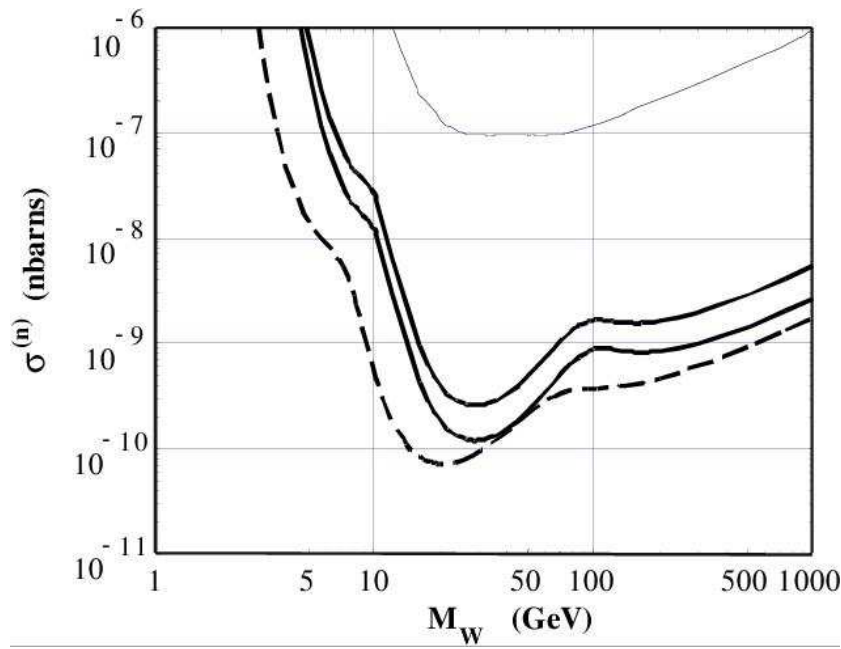}
 \end{center}
 \caption{The solid lines represent the sensitivity plot in the ($m,\sigma$) plane for CUORE, assuming a threshold of 10 keV, two years of exposure (1500 kg year) and flat backgrounds of 0.05 and 0.01 c/keV/kg/day. It has been calculated for $\langle\delta^2\rangle=$ 5.6 (see the text). The sensitivity curve has been also calculated for a possible threshold of 5 keV with a background of 0.01 c/keV/kg/day (dashed line).}
 \label{cuore_mod}
\end{figure}

Using this equation we have estimated the region that could be within reach for CUORE with the above mentioned assumptions on the background levels. We have fixed a value of 5.6 for $\langle \delta^2 \rangle$ that corresponds to 50\% probability of obtaining a positive result at 90\% C.L. 
In figure \ref{cuore_mod} curves are shown obtained for a threshold of 10 keV, two years of exposure and two assumed flat backgrounds of 0.05 and 0.01 c/keV/kg/day (solid lines). The possibility of a lower threshold of 5 keV with a background of 0.01 c/keV/kg/day is also shown (dashed line).

In conclusion, CUORE will be able to explore (and/or exclude) WIMPs lying in large regions of their parameter space. 

The capability of CUORE to investigate the DAMA region through the exclusion plot (time integrated method) relies on getting a background of 0.1 c/keV/kg/day from 10 keV onwards, independently of more elaborated time modulation methods which require an exhaustive control of the stability of the experiment.
However, CUORE could also attempt to look for annual modulation of WIMP signals provided that the stability of the experiment is sufficient.

We would like to note in addition that the nuclei of Tellurium are totally different from those of  Germanium and close in atomic number to Iodine. The ground state of 93\% of the Te nuclei is 0$^{+}$. Comparison among the results from these nuclei seems therefore very interesting.

\subsubsection{Solar axion detection} \label{sec:axions}

Axions are light pseudoscalar particles which arise in theories in which the Peccei-Quinn U(1) symmetry has been introduced to solve the strong CP problem \cite{peccei77}. They could have been produced in early stages of the Universe being attractive candidates for the cold dark matter (and in some particular scenarios for the hot dark matter) responsible to 1/3 of the ingredients of a flat universe. Dark matter axions can exist in the mass window $10^{-2(3)}$ eV $<m_{a}\leq 10^{-6}$ eV, but hadronic axions could exist with masses around the eV.

Axions could also be copiously produced in the core of the stars by means of the Primakoff conversion of the plasma photons. In particular, a nearby and powerful source of stellar axions would be the Sun. The solar axion flux can be easily estimated \cite{vanbibber89,creswick98} within the standard solar model, resulting in an axion flux of an average energy of about 4 keV that can produce detectable X-rays when converted in an electromagnetic field. We would like to stress that, although we focus on the axion because its special theoretical motivations, all this scenario is also valid for any generic pseudoscalar (or scalar) particle coupled to photons \cite{masso95}. Needless to say that the discovery of any type of pseudoscalar or scalar particle would be extremely interesting in Particle Physics. We will keep our discussion, however, restricted to the case of solar axions.

Single crystal detectors provide a simple mechanism for solar axion detection \cite{creswick98,paschos94}. Axions can pass in the proximity of the atomic nuclei of the crystal where the intense electric field can trigger their conversion into photons. The detection rate is enhanced if axions from the Sun coherently convert into photons when their incident angle with a given crystalline plane fulfills the Bragg condition. This induces a correlation of the signal with the position of the Sun which can be searched for in the data and allows for background subtraction. 

The potentiality of Primakoff conversion in crystals relies in the fact that it can explore a range of axion masses ($m_a\gsim 0.1$ keV) not accessible to other direct searches. Moreover it is a relatively simple technique that can be directly applied to detectors searching for WIMPs.

Primakoff conversion using a crystal lattice has already been employed in two germanium experiments: SOLAX \cite{avignone98} and COSME-II \cite{irastorza00} with the ensuing limits for axion-photon coupling $\gagamma \lsim 2.7\times 10^{-9}$ GeV$^{-1}$ and $\gagamma \lsim 2.8\times 10^{-9}$ GeV$^{-1}$ respectively. Also the DAMA collaboration has analyzed 53437 kg-day of data of their NaI set up \cite{bernabei01}, in a search for solar axions, following the techniques developed in ref. \cite{avignone99}, where a calculation of the perspectives of various crystals detectors (including NaI) for solar axion searches has been made. The DAMA result $\gagamma \lsim 1.7\times 10^{-9}$ GeV$^{-1}$ improves slightly the limits obtained with other crystal detectors \cite{avignone98,irastorza00} and agrees with the result predicted in ref.~\cite{avignone99}. These "crystal helioscopes" constraints are stronger than that of the Tokyo axion helioscope \cite{inoue02} for $m_a \gsim 0.26$ eV and do not rely on astrophysical considerations (i.e. on Red Giants or HB stars dynamics \cite{raffelt99}). The orientation of the crystal was not known so that the data were analyzed taking the angle corresponding to the most conservative limit.

It has been noted that the model that yields the solar axion fluxes used to calculate the expected signals is not compatible with the constraints coming from helioseismology if $\gagamma \gsim 10^{-9}$ GeV$^{-1}$ \cite{schlattl99}. This would imply a possible inconsistency for solar axion limits above that value, and sets a minimal goal for the sensitivity of future experiments.

The use of CUORE to search for solar axions via Bragg scattering should have some advantages with respect to germanium detectors, because of the higher Z, larger mass and known orientation of the crystals. Since the cross-section for Primakoff conversion depends on the square of the atomic number, \teod will be a better candidate than Germanium. Needless to say that a low energy threshold is mandatory because the expected signal lies in the energy region 2 keV $\lsim E \lsim$ 10 keV and is peaked at $E\simeq 4$ keV. As for dark matter searches, this will of course require specific R\&D directed to a better performance in the low energy region.

A detailed analysis has been performed \cite{avignone99} for a \teod crystal (which has a tetragonal structure \cite{ewbank83}) assuming different values for the experimental parameters. As it is shown in Ref. \cite{avignone99}, the bound on axion-photon coupling which a given experiment can achieve can be estimated through the expression:

\begin{equation} \label{eq:gagg}
g_{a\gamma \gamma }<g_{a\gamma \gamma }^{\lim }\simeq k\left(
\frac b{\rm c/keV/kg/day}\frac{\rm kg}M\frac{\rm years}T\right)
^{1/8}\times 10^{-9}{\rm  GeV}^{-1}
\end{equation}

\noindent where $k$ depends on the crystal structure and material, as well as on the experimental threshold and resolution. For the case of \teod and a threshold of 5 keV, $k$ has been calculated to be $k=2.9$ assuming an energy resolution of 1 keV. The computation of this expression for some assumed values of the experimental parameters is shown in table \ref{tab:cuoreax} for CUORE. Flat backgrounds and 2 years of exposure are assumed.

\begin{figure}
 \begin{center}
 \includegraphics[width=0.8\textwidth]{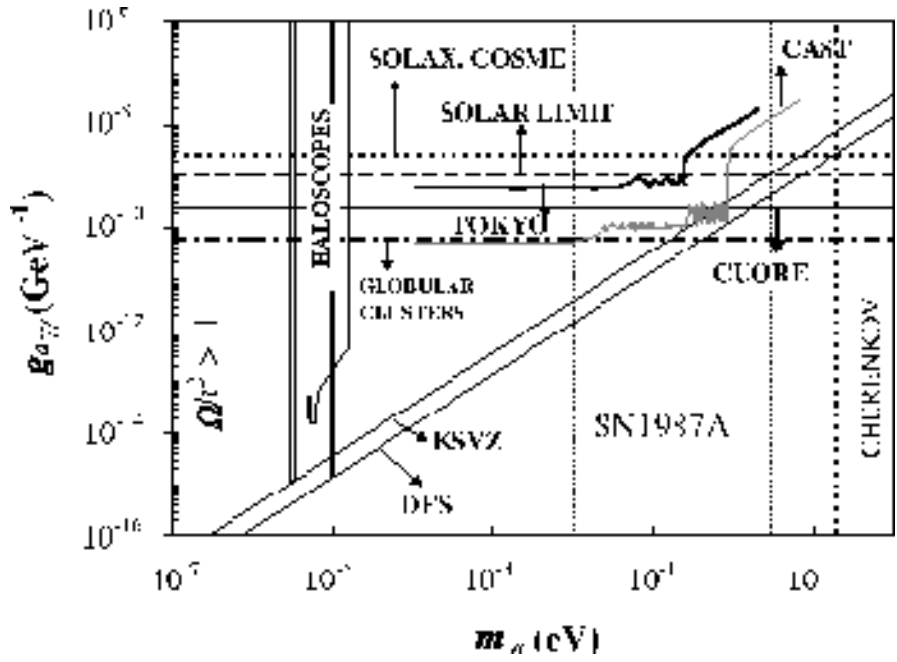}
 \end{center}
 \caption{Best bound attainable with CUORE (straight line labelled "CUORE") compared with others limits.}
 \label{fig:axions}
\end{figure}

It is worth noticing the faible dependence of the ultimate achievable axion-photon coupling bound on the experimental parameters, background and exposure MT: the 1/8 power dependence of $g_{a\gamma\gamma}$ on such parameters softens their impact in the final result. The best limit shown in table \ref{tab:cuoreax} is in fact only one order of magnitude better than the present limits of SOLAX and COSME-II. The $\gagamma$ bound that CUORE could provide is depicted comparatively to other limits in figure \ref{fig:axions}.

\begin{table}[h] \centering
  \begin{tabular}{ccccc} \hline\hline
    {\bf Mass}  &  {\bf Resolution}&  {\bf Threshold}   & {\bf Background } &  $g_{a\gamma \gamma }^{\lim }$ {\bf (2 years)}\\
     {\bf (kg)}  &  {\bf (keV)}  &  {\bf (keV)} & {\bf (c/kg/keV/day)} &  {\bf (GeV$^{-1}$)}\\ \hline

     750       &  1   & 5    &  0.01  &  6.5$\times 10^{-10}$\\
     750       &  1   & 5    &  0.05  &  8.0$\times 10^{-10}$\\
     \hline\hline
  \end{tabular}
    \caption{Expected limits on the photon-axion coupling for 2 years of exposure of CUORE assuming the quoted values for the experimental parameters}\label{tab:cuoreax}
\end{table}

The limit which can be expected from CUORE is better than the helioseismological bound mentioned before (see Figure \ref{fig:axions} and Table \ref{tab:cuoreax}). Notice that in both cases an energy threshold of $E_{thr}\sim5$ keV (and resolution of $\sim$1 keV) has been assumed. 
As was described at length in Ref. \cite{avignone99}, the crucial parameters for estimating the perspectives on solar axion detectors with crystals rely on the energy threshold and resolution (appearing in $k$), and the level of background achieved (although the influence of this parameter is damped by a factor 1/8). In particular, a threshold of 5-8 keV would loose most of the axion signal. Other crystal detectors with, say, Ge or NaI (GENIUS, MAJORANA, GEDEON, DAMA, LIBRA, ANAIS, \dots) could surpass CUORE as axion detectors because the energy thresholds of these projects are supposed to be significantly lower. Also the background is expected to be better.

Nevertheless, it should be stressed that the bounds on $g_{a\gamma \gamma}$ obtained with this technique in the various proposed crystal detector arrays stagnate at a few $\times 10^{-10}$ GeV$^{-1}$, not too far from the goal expected for CUORE, as has been demonstrated in \cite{avignone99}. There are no realistic chances to challenge the limit inferred from HB stars counting in globular clusters \cite{raffelt99} and a discovery of the axion by CUORE would presumably imply either a systematic effect in the stellar-count observations in globular clusters or a substantial change in the theoretical models that describe the late-stage evolution of low-metallicity stars. To obtain lower values of $\gagamma$ one should go to the magnet helioscopes like that of Tokyo \cite{inoue02} and that of CERN (CAST experiment \cite{zioutas99} currently taking data). 

In particular, until recently the best experimental bound of $\gagamma$ published came from the Tokyo helioscope: $\gagamma \leq 6 \times 10^{-10}$ GeV$^{-1}$ for $m_{a} \lesssim 0.03$ eV and $\gagamma \leq 6.8-10.9 \times 10^{-10}$ GeV$^{-1}$ for $m_{a} \sim 0.05-0.27$ eV. The sensitivity of CAST is supposed to provide a bound $\gagamma \leq 5 \times 10^{-11}$ GeV$^{-1}$ or even lower. A recent run of CAST has improved the Tokyo limit with a bound $\gagamma \leq 3 \times 10^{-10}$ for $m_{a} \leq 0.05$ eV.

\subsubsection{Searches for other rare processes}
Searches for rare atomic or nuclear events have provided and are going to provide a very rich harvest of important results, some of them probably beyond our imagination. We believe that the technique of massive thermal detectors is particularly suitable to search for many  of  them, but we present here only a few examples just to stimulate our theoretical friends to suggest others.

\paragraph{Electron stability}
Searches on this decay \cite{back02}, can represent an important by-product for CUORE. An inclusive experiment on this process can be carried out by detecting the disappearance of an atomic electron followed by the consequent decay (or  series of decays) of the outer-shell electrons to fill the vacancy  left by  this disappearance. These searches, which have relevance also in connection with charge conservation and Pauli principle validity, can be carried out by looking for sharp lines corresponding to decays in the atomic tellurium structure. The exclusive charge violating decay of an electron in a neutrino and a photon can be revealed by the presence of  peak at 255 keV, spread however by the Doppler broadening of the atomic electron motion. Even if not fully exploiting the excellent energy resolution of CUORE, this decay can be efficiently searched, due to the large mass and low background expected for this detector. 

\paragraph{Nucleon stability}
CUORE cannot obviously compete with the various experiments specifically designed to investigate specific channels of this decay~\cite{tretyak01}. It can however provide competitive results  on the inclusive decays of a proton or neutron through the search for delayed coincidences between the nucleon decay signature (e.g. high energy releases by nucleon decay products or prompt $\gamma$'s from nuclear re-arrangement processes) and possible unstable produced nuclei. Possible nucleon decays leading to the production of unstable nuclei emitting gamma rays are listed in Table~\ref{tab:Ndecay} . Delayed gamma rays emitted in these secondary decays could be detected in nearby crystals. If the half-lifetime of the unstable produced nuclei would be reasonably short an effective coincidence selection could be applied.\\
Unusual decay modes (e.g. $n\to 3\nu$) have been recently suggested as possible dominant modes~\cite{mohapatra03}. The CUORE unique capabilities for the detection of prompt low energy $\gamma$--rays following such "invisible" decay channels could provide competitive sensitivities for the search of these decay modes which may have been missed by previous dedicated experiments. In particular, results from CUORE can be complementary to those recently obtained by the SNO collaboration.

\begin{table}[ht] \centering
\caption{Possible nucleon decays leading to unstable nuclei characterized by the emission of gamma rays.}\label{tab:Ndecay}

\begin{tabular}{lcclcc}
\hline\hline
Nucleus & Abundance \% & Decay & Product & Lifetime & Gamma Rays (keV)\\
\hline
$^{120}$Te & 0.096 & p & $^{119}$Sn & 38.1 h & 24\\
&& n & $^{119}$Te & 16 h & 644.\\
&&& $^{119m}$Te & 4.69 d & 154, 1213, 271 \\
$^{122}$Te & 2.6 & n & $^{121}$Te & 16.8 d & 573, 508\\
&&& $^{121m}$Te & 154 d & 1102, 37\\
$^{123}$Te & 0.908 & p & $^{122}$Sb & 2.72 d & 564\\
&&& $^{122m}$Sb & 4.191 m & 61, 76\\
$^{124}$Te & 4.816 & n & $^{123m}$Te & 119.7 d & 159\\
$^{125}$Te & 7.14 & p & $^{124}$Sb & 60.2 d & 603, 1691, 773 \\
&&& $^{124m}$Sb & 20.2 m & 603, 946, 498\\
$^{126}$Te & 18.95 & n & $^{125m}$Te & 58 d & 36\\
$^{128}$Te & 31.69 & p & $^{127}$Sb & 3.84 d & 686, 474, 784\\
&& n & $^{127}$Te & 9.4 h & 418, 380\\
&&& $^{127m}$Te & 109 d & 58\\
$^{130}$Te & 33.8 & p & $^{129}$Sb & 4.4 h & 813, 915, 545\\
&&& $^{129m}$Sb & 17.7 m &  760, 658, 434\\
&& n & $^{129}$Te & 1.16 h & 460, 28\\
&&& $^{129m}$Te & 33.6 d & 696\\
\hline\hline
\end{tabular}
\end{table}

\paragraph{Rare beta processes}
\begin{description}
\item{\bf{$^{113}$Cd}}
An experiment carried out with a large CdWO$_4$ crystal~\cite{alessandrello94} has determined the transition energy and lifetime of this nucleus and measured for the first time the shape of the beta spectrum. The study of this transition was also repeated later by exploiting the scintillating properties of CdWO$_4$~\cite{danevich96}.  These results, can be further improved if a large CdWO$_4$ would be inserted into the CUORE detector.

\item{\bf{$^{123}$Te}}
Evidence for the existence of the second forbidden electron capture of this nucleus has been reported by various authors.   An experiment recently carried out by the Milano group~\cite{alessandrello03} has shown that these evidences could be simply mimicked  by other background effects and that this decay has actually never been found.  The reported values of the half-lifetime for this decay have therefore  to be still considered only as lower limits. The discovery of this decay in CUORE would have relevant implications in nuclear structure theory~\cite{civitarese01}.
\end{description}

\paragraph{Rare alpha decays}

\begin{description}
\item{\bf{ $^{209}$Bi}}
A search for the possible decay of this naturally occurring isotope of Bismuth, commonly regarded as the heavier stable nucleus, was proposed in our previous letter of intent. Recently, however a bolometric experiment~\cite{pdemarcillac03} has detected the $\alpha$--decay of this nucleus with a lifetime of  $(1.9 \pm  0.2 ) \times 10^{19}$ years. Better measurements and a better study of this decay could be carried out in CUORE.
\item{\bf{ $^{180}$W}}
This nucleus can energetically $\alpha$--decay into $^{176}$Hf. Indication of this rare decay was obtained with conventional detectors~\cite{danevich03} but unambiguously detected for the first time thanks to the simultaneous measurement of phonon and light signals with the CRESST cryogenic detectors~\cite{cozzini04} (T$_{1/2}$ = (1.8 \pom 0.2) \per 10$^{18}$ y, Q = (2516.4 \pom 1.1 (stat.) \pom 1.2 (sys.)) keV ). This and other rare nuclear alpha decays could be investigated with great accuracy in CUORE.
\end{description}

\clearpage
\section{Principles of operation of TeO$_2$ bolometers} \label{sec:bolometers}

In a very naive approach  a thermal detector is a sensitive calorimeter which measures the energy deposited by a single interacting particle through the corresponding temperature rise.

This is accomplished by using suitable materials (dielectric crystals, superconductors below the phase transition, etc.) and by running the detector at very low temperatures (usually below 100 mK) in a suitable cryostat (e.g. dilution refrigerators). According to the Debye law, in fact, the heat capacity of a single dielectric and diamagnetic crystal at low temperature is proportional to the ratio ($T/\Theta_D)^3$ ($\Theta_D$ is the Debye temperature) so that, providing that the temperature is extremely low, a small energy release in the crystal results in a measurable temperature rise. 

This temperature change can be measured by means of a proper thermal sensor.

A low temperature detector (LTD or ``bolometer'') consists therefore of three main components (Fig.\ref{fig:model} (a)).

\begin{description}
\item [\textbf{Particle absorber}]: it is the sensitive mass of the device where the particles deposit their energy. The absorber material can be chosen quite freely, the only requirements being, in fact, a low heat capacity and the capability to stand the cooling in vacuum. The absorber can therefore be easily realised with materials containing any kind of unstable isotopes and many interesting searches are therefore possible (e.g. beta decay spectroscopy, neutrinoless double beta decay and dark matter). The absorber mass can range from few micrograms to almost one kilogram.

\item [\textbf{Temperature sensor}]: it represents a crucial element of the thermal detector needed to measure the temperature rise induced by the particle interaction. Among the variety of thermometers so far developed, doped semiconductor thermistors are preferred because of their ease of use and the reproducibility with which they can be produced in large quantities.

\item [\textbf{Thermal link}]:  it mainly depends on the way the massive crystal is held to the structure (heat sink). It influences the time response of the device.

\end{description}

Absorber masses usually characterize LTD's applications and can range from few hundred micrograms (X-ray spectroscopy or neutrino mass direct measurements) to several hectograms (Dark Matter or DBD searches).

\subsection{Basic principles of thermal particle detectors}\label{subsec:basic}

\begin{figure}
 \begin{center}
 \includegraphics[width=1\textwidth]{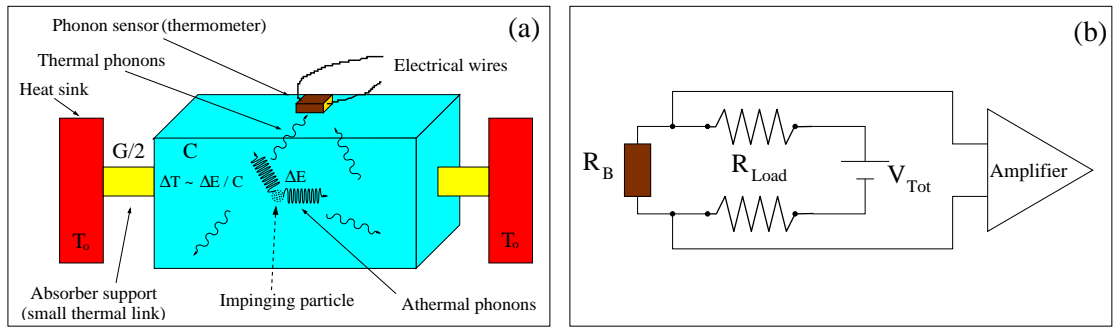}
 \end{center}
 \caption{Simplified model of a thermal detector (a) and schematic read-out of a resistive thermometer (b)}
 \label{fig:model}
\end{figure}

The basic idea behind the thermal detector is very simple \cite{moseley94}. Let us assume that the detector has a heat capacity C (absorber) in thermal connection with a heat sink (kept at temperature $T_0$) through a thermal conductance G (fig.~\ref{fig:model}(a)). 

Let $T(t)$ be the temperature of the absorber as a function of time and let us assume that $\Delta T \equiv \vert T(t) - T_0 \vert \ll T_0 $ for all times t, so that $C$ and $G$ can be treated as constants. Then an instantaneous deposition of an energy $\Delta E$ in the absorber gives rise to a temperature pulse whose time development is described by:
\begin{equation}
\Delta T(t) =  \frac{\Delta E}{C}\exp \left(- \frac{t}{\tau} \right) \qquad  {\rm with} \quad  
\tau = 
\frac{C}{G}
\end{equation}

As $T_{max}$ is given by $\Delta E/C$, one immediately sees that, in order to have a significant percentage increase of the temperature (for an energy release in the keV-MeV range) it is necessary to have a very small heat capacity. As already stated above, this can be achieved only operating at very low temperatures. 

Very small specific heats can be attained by exploiting the strong decrease of the specific heat with temperature. The specific heat of a dielectric and diamagnetic crystal is in fact due only to lattice contributions~\cite{fiorini84}, with a temperature dependance following the Debye law
\begin{equation}\label{eq:Debye}
C(T) =  \beta \left(\frac{T}{\Theta_D} \right)^3 \qquad T<  \Theta_D
\end{equation}

\noindent where $\beta = 1994 \; J \; K^{-1}$mol$^{-1}$ and $\Theta_D$ is the Debye temperature of the crystal.

In a very simplified model \cite{mccammon93} in which all primary phonons undergo the thermalization process, an evaluation of the energy resolution can be obtained. In this case in fact the energy resolution of the detector is only limited by the thermodynamic fluctuations in the number of thermal phonons exchanged with the heat sink,  which produces random fluctuations in the energy content of the absorber. Since the mean energy of thermal phonons is $\sim$kT (k is the Boltzman constant), the average total number of thermal phonons N is proportional to C(T)/k and, assuming a Poisson statistics, this imply  a contribution to the energy resolution given by

\begin{equation}\label{eq:E_res}
N=\frac{C(T)}{k}\;, \; \Delta N = \sqrt{N}\Longrightarrow\sigma_E = \Delta N \cdot kT = \sqrt{\frac{c(T)}{k}}kT  = \xi\sqrt{kC(T)T^2}
\end{equation}

\noindent where $\xi$ is a dimensionless factor (generally of the order of unity) whose actual value depends on the details of the temperature sensor, of the thermal link and of the temperature dependance  of $C(T)$. 
It should  be stressed that, according to eq.~\ref{eq:E_res}, $\sigma_E$ is \emph{independent} of $E$, with impressive ``theoretical'' expectations even in the case of a massive crystal: a \teod  crystal with a mass of 1 kg operating at 10 mK could energy depositions of the order of few MeV with a resolution of about 20 eV. It should be also stressed, however, that this ``thermodynamic limit'' doesn't take into account other sources of energy fluctuations (e.g. metastable electron-hole states or long-lived non thermal phonons) which participate to the energy deposition process and whose contributions to the energy resolution worsening should be therefore minimized. One should in addition take into account fluctuations in the gain due to temperature instabilities which could normally worsen the resolution.

\subsection{Semiconductor thermistors}\label{subsec:thermistors}
Thermistors are heavily doped high resistance semiconductors, with an impurity concentration slightly below the metal-insulator transition \cite{mott56}. 
\emph{Good} thermistors require a higly homogeneous doping concentration (see sec\ref{sec:modular}) which is achieved, in the case of NTD (Neutron Transmutation Doped) thermistors, by means of thermal neutron irradiation throughout the entire volume \cite{haller95} (sec.~\ref{sec:ntd}). 
The electrical conductivity of these heavily doped semiconductors strongly depends on the temperature according to the \emph{hopping}  and \emph{variable range hopping} (VRH) mechanisms \cite{miller61}. In general the resistivity varies with  temperature as  
\begin{equation}\label{eq:resistivity}
\rho = \rho_{0} e^{{\left(\frac{T_0}{T} \right)}^\gamma} 
\end{equation}

\noindent where $\rho_0$, $T_0$ and $\gamma$ depends on the doping concentration. In the case of VRH, the value of $\gamma$ is 1/2.

Thermistors are usually parameterized by their sensitivity $A$, defined as
\begin{equation}\label{eq:sens}
A(T) = \left|\frac{d\ln R}{d \ln T} \right|  = \gamma \left( \frac{T_0}{T}   \right)^\gamma \;\;\;
{\rm with} \;R(T) = \rho(T)\frac{l}{s} = R_0\exp \left( \frac{T_0}{T}   \right)^\gamma
\end{equation}

\noindent where \emph{l} and \emph{s} are the distance between the contacts and the cross section of the thermistor, respectively. The value of the sensitivity is usually in the range  2$\div$10.

\begin{figure}
 \begin{center}
 \includegraphics[width=1\textwidth]{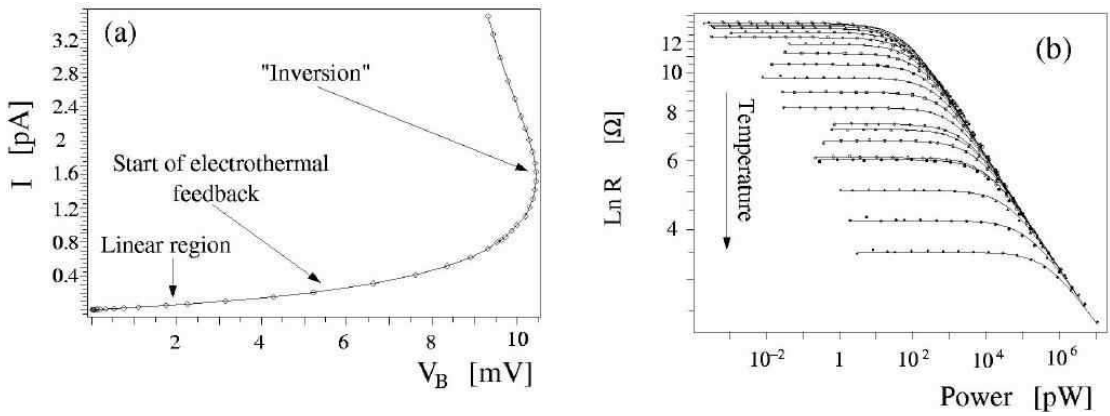}
 \end{center}
 \caption{Typical load curve for a thermistor at T= 8 mK) (a) and resistance-power curves for a thermistor at different base temperatures (b).}
 \label{fig:characterization}
\end{figure}

NTD germanium thermistors developed and built at LBNL have been used in all the large mass LTD's developed by the Milano group. They are made by exposing a ultrapure germanium single crystal to a nuclear reactor neutron flux. They are characterized by an excellent reproducibility even if they are in part made by hand. The values of R$_0$, T$_0$ and $\gamma$ must be experimentally measured for each thermistor (characterization process). 
This measurement is performed by coupling the thermistors to a low temperature heat sink using a high conductivity epoxy. The temperature $T_B$ (base temperature) of the heat sink is then varied (15$\div$ 50 mK) while a steady current $I$ (bias current) flows trough the thermistors and a voltage $V = IR$ appears across them. This produces a power dissipation which raises the temperature and acts back on the resistance, until an equilibrium is reached. This phenomenon makes the $V-I$ relation deviate from linearity. This characteristic behaviour of bolometers is often referred to as ``electrothermal feedback''. The static resistance is simply the ratio V/I while the  dynamic resistance (the only one actually involved in the preamplifier matching requirements) is the tangent at  the $V-I$ curve. By further increasing  the bias current the dynamic resistance crosses the so called \emph{inversion point} (where it vanishes) and becomes then negative.
For semiconductor thermistors, a typical $V-I$ curve, usually referred to as \emph{load curve}, is represented in Fig.\ref{fig:characterization} (a). 
By a combined  fit to a set of load curves at different base temperatures, all the 
thermistor parameters are evaluated (see Fig.\ref{fig:characterization} (b))

\subsection{Detector operation}\label{subsec:detect_op}
\begin{figure}
 \begin{center}
 \includegraphics[width=1\textwidth]{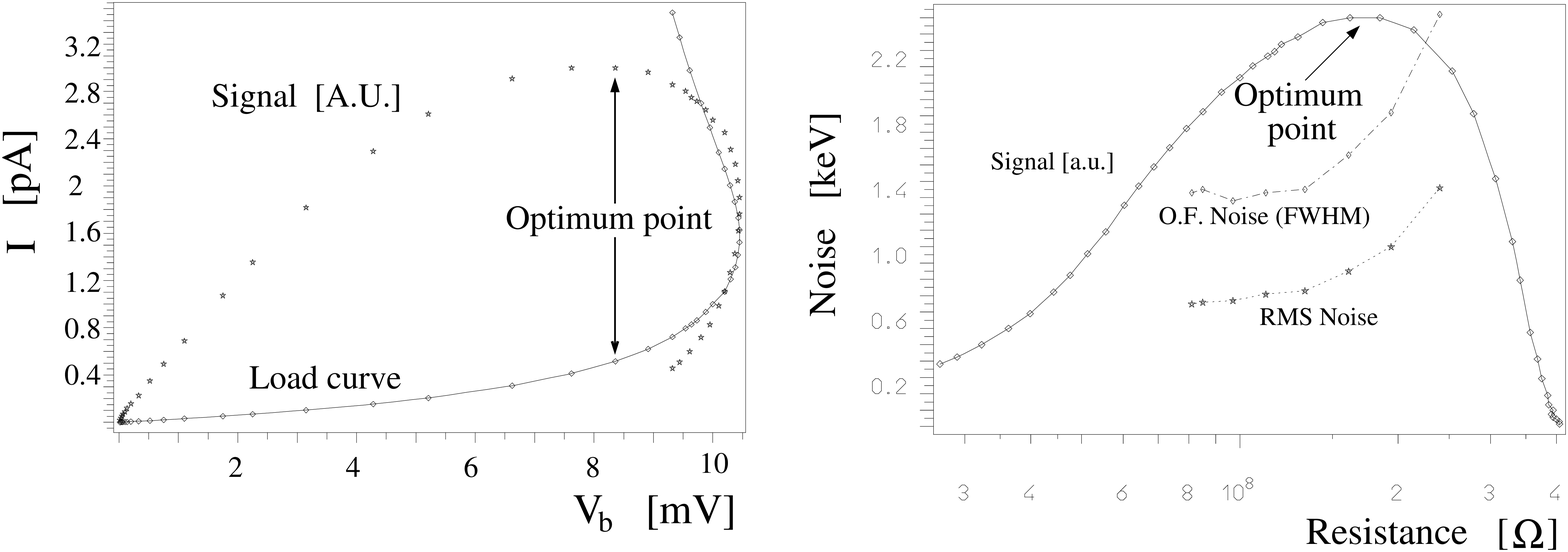}
 \end{center}
 \caption{\emph{left panel:} evaluation of the \emph{optimum point} of a bolometer; \emph{right panel}: evaluation of operation point which maximizes the signal/noise ratio.}
 \label{fig:op_point}
\end{figure}

As mentioned in section \ref{subsec:basic}, in first approximation, the thermal pulse produced by an energy release in the absorber is characterized by a very fast rise time and an exponential decay time determined by the physical characteristics of the individual detector. Such a thermal pulse is then converted into an electrical one by means of the thermistor bias circuit and detector optimization is equivalent to determining the best configuration for a maximum pulse amplitude.
Considering the basic circuit shown in Fig. \ref{fig:model}(b), it's straightforward to obtain the relation between the maximum potential drop across the thermistor (V$_b$), the thermistor parameters and the deposited energy:
\begin{equation}\label{eq:maximum}
\Delta V_b = \frac{R_{Load}R_B}{(R_{Load}+R_B)^2}\;V_{Tot} A \frac{\Delta T_b}{T_b} \; \approx \; { \frac {E}{CT_b}A\sqrt{P}\sqrt{R_B}, 
} 
\end{equation}

\noindent where E is the deposited energy, A is the sensitivity given above in equation (\ref{eq:sens}), $T_b$ and $\Delta T_b$ are the operating temperature and the temperature increase of the bolometer, C is the heat capacity of the absorber and P is the electrical power dissipated on the thermistor by the Joule effect.
This expression vanishes both in the limit $P\rightarrow 0$  and $P \rightarrow \infty$. When the thermistor is operated as a detector sensor, a steady current $I$ flows through it and, in a steady condition, its electrical and thermal parameters are described by a point on the detector $V-I$ curve. This point is usually named \emph{working point}, or \emph{optimum point} when corresponding to the maximum detector response (maximum amplitude of the voltage pulse).
The \emph{optimum point} of each detector must be determined experimentally. This is accomplished by means of a fixed resistance (\emph{heater}) directly glued onto the crystal through which it is possible to dissipate a known joule power into the crystal in a manner perfectly equivalent to a particle energy deposit. The \emph{optimum point} correponds to the bias voltage for which the amplitude of the \emph{heater} pulses is maximum. A typical bolometer response as a function of detector bias is shown in Fig.\ref{fig:op_point}.\par

In fact, the actual \emph{optimum point} is chosen by investigating the signal/noise ratio. Since the electronic sources of noise decrease with the resistance of the bolometer the best \emph{working point} corresponds to a bias voltage slightly higher than the one corresponding to the \emph{optimum point} (see Fig.\ref{fig:op_point}).
The use of the \emph{heater} also plays a fundamental role in the stabilization of the  bolometer response which changes with the temperature of the heat sink~\cite{alessandrello98s}.

\subsection{Bolometer applications}
Since much information on the initial particle interaction is washed out by the thermalization process (position, momentum and interaction type), true cryogenic detectors are sensitive only to the total deposited energy which, however, can be measured very accurately (sec. \ref{subsec:basic}). In fact, the intrinsic energy resolution of such devices can be as low as a few eV.
Main advantages of thermal detectors with respect to conventional ones are however represented by their sensitivity to low or non-ionizing events (Dark Matter (DM) searches) and material choice flexibility. Time response is sometimes poor (large mass bolometer pulse rise/decay times are of the order of few tens/hundreds msec or more) but, due to the very low counting rates, this is not a limitation at all for DBD (or DM) searches. It should be stressed that the flexibility in the choice of materials, besides allowing an experimental investigation on any DBD active isotope, means also the possibility of selecting materials characterized by a large natural isotopic abundance (a crucial parameter showing a linear dependence in the DBD lifetime sensitivity).
Besides prohibitive costs, in fact, isotopic enrichment is not always easily attainable and the possibility to choose the DBD candidate element characterized by the highest natural isotopic abundance represents a big advantage. Because of the calorimetric approach, the thermal detectors detection efficiency is generally of the order of unity. 
Concerning energy resolution, up to now the actual detectors are still far from reaching the limits imposed by the thermodynamic fluctuations of the system internal energy (eq.~\ref{eq:E_res}). Actual values are in the 1-10 keV range (FWHM) and seem to scale little with the detector mass (similar values were obtained in detectors whose masses scaled from few grams to few hectograms) while showing a quasi-linear dependence on the energy. Energy resolution is therefore still limited by other contributions (e.g. electronic and microphonic noise, vibrations of the system, etc.) and a good detector design (thermistor choice, mounting system, etc.), a low noise read-out electronics and an effective data acquisition and data analysis systems are of crucial importance for a well performing detector. So far, most of the R~\&~D has been devoted to optimizing the detector performance.
Although energy resolution doesn't seem to scale with detector mass, there is a limit to the single detector mass (imposed not only by the increase of the heat capacity but also by the technical feasibility of large perfect crystals). In the case of the \teod , such a limit has been set so far to about one kilogram but larger masses could be possible in the near future (section \ref{sec:INNO}). 
The only possibility for a detector mass in the ton range is therefore a large array of LTD's and the main issue is therefore reproducibility. This item has been faced and solved few years ago by the Milano group with the realization of a \teod 20 bolometer array (340 g each) and, more recently, by the successful operation of CUORICINO (section \ref{sec:CUORICINO}).

\subsection{Detector model}\label{sec:model}

A rough approximation of the NTD thermistor voltage pulse amplitude $\Delta V_b$ produced by an energy deposition E into the \teod absorber can be obtained by assuming $R_{Load}>>R_B$ in eq. (\ref{eq:maximum}):
\begin{equation}
 \Delta V_b = V_b \cdot A \cdot \frac{\Delta T_b}{T_b}
   = V_b \cdot A \cdot \frac{E}{C\;T_b}
 \label{eq:pulse}
\end{equation}
Using the heat capacity discussed earlier, a one $MeV$ energy deposition into the \teod absorber results in $\Delta T_b = 7\times 10^{-5}K$. Using equation (\ref{eq:pulse}), this results in $\Delta V_b \sim 600 \; \mu V$, assuming typical electrical and thermal paramaters at the detector operating point. However, the observed pulses are lower by a factor $2$ to $3$, therefore a more quantitative model is needed.

\begin{figure}
 \begin{center}
 \includegraphics[width=0.9\textwidth]{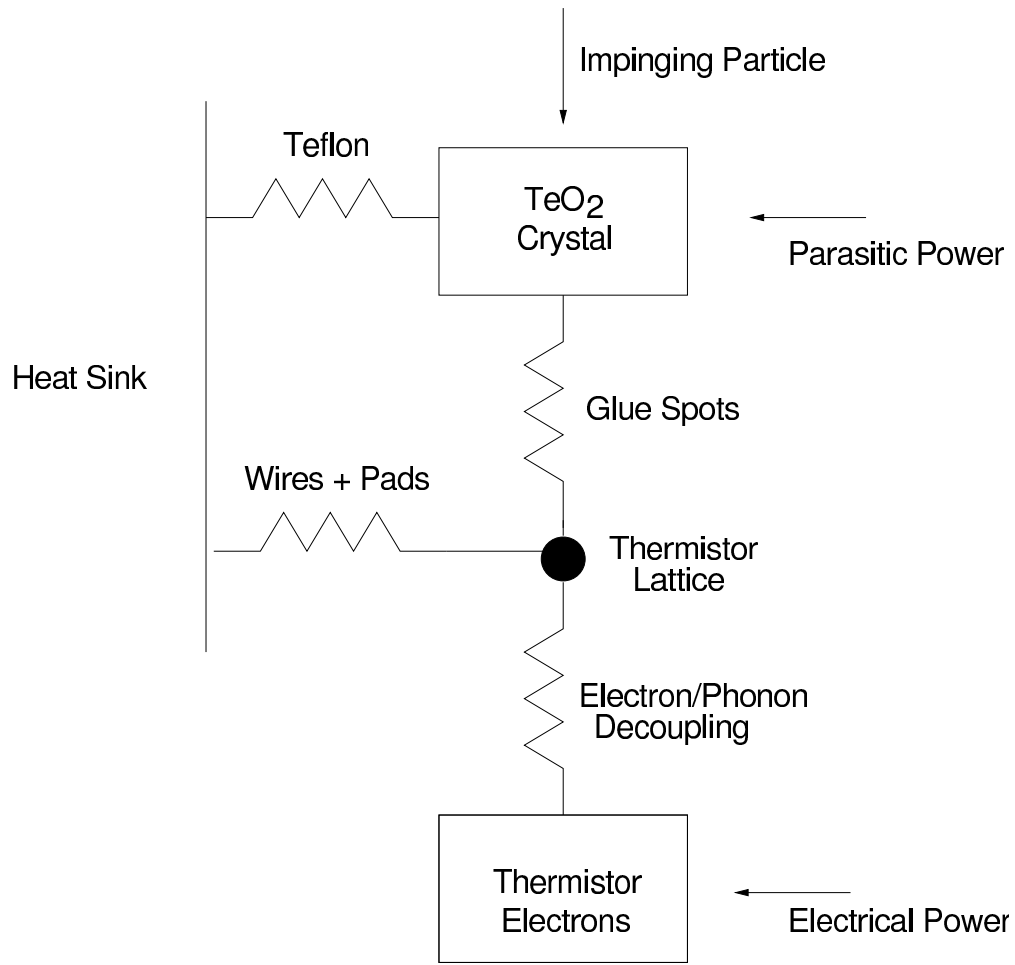}
 \end{center}
 \caption{Network representing the detector model.}
 \label{fig:network}
\end{figure}

A semi-quantitative Thermal Model (TM) has been developed which describes the detector in terms of three thermal nodes with finite heat capacities: the \teod crystal absorber, $Ge$  thermistor lattice, and $Ge$ thermistor electrons. For simplification, the lattice component of heat capacity is assumed to be negligible. The network describing the detector is shown in Fig.~\ref{fig:network}. The two heat capacities, and four thermal conductances and the parasitic power from the vibrations and power dissipated in the thermistor electrons, are the model parameters.
A typical energy deposition of 1~$MeV$ in the crystal is used and it is assumed that it is instantaneously thermalized.

A computer code was developed in Milan to determine the static behaviour. The required input data are: the heat sink temperature, the temperature behaviour of the thermal conductances shown in Fig.~\ref{fig:network} (partially determined with {\em ad hoc} measurements), the parasitic power in the crystal, and a set of values of electrical Joule power dissipated in the thermistor electrons. The code determines a complete solution to the static problem, for each power level, and yields the temperature of the three thermal nodes, the resistance of the thermistor, and the current through it for each voltage-current (or power resistance) point of a load curve.
When the static operation point is determined, the time evolution of the pulse is computed.
The main aim of the TM is to have a reliable prediction of the detector figure of merit on the basis of the known thermal parameters. To address this question it is necessary to solve the non trivial problem of how to compare the performance of two detectors.
A more sensitive thermistor would yield larger amplitude pulses, but they are also more sensitive to spurious noise, for example microphonics, crosstalk, and slow heat pulses from vibrations that excite the 1-5 $Hz$ portion of the noise spectrum. A lower heat sink temperature yields larger pulses at the price of higher resistance and more spurious noise.

Recent experience shows that the best way to define a detector figure of merit is by fixing the heat sink temperature between $ 7\; mK $ and $ 11\; mK $ and by measuring
a load curve. Using particle or heater pulses, the operation point on the load curve,
corresponding to the highest signal is determined. For each heat sink temperature,
one determines a pair of points, the thermistor resistance, $R$, and the signal amplitude, $V$, usually expressed in microvolts per $MeV$, corresponding to the maximum pulse amplitude for a given heat sink temperature. One obtains a characteristic curve for that detector by plotting voltage vs resistance, which to a good approximation is a straight line on a log-log plot. This is called the Detector Merit Curve (DMC). Experience shows that a detector has constant signal-to-noise ratio along a DMC.  No large differences in performance are noted for heat sink temperatures of $8\; mK$ or $12\; mK$, because the signal level and the spurious noise offset one another. It is better to avoid the extremities of the DMC that lead to higher spurious noise or low signals.

The best way to compare two different detectors is by using their DMCs. If the DMC
of detector A lies systematically above that of detector B, the performance of detector
A is superior. It provides the highest signals for a fixed thermistor resistance value.
The main value of the TM is to predict the DMC given the temperature behaviour of the thermal parameters discussed above.

\begin{figure}
 \begin{center}
 \includegraphics[width=0.7\textwidth]{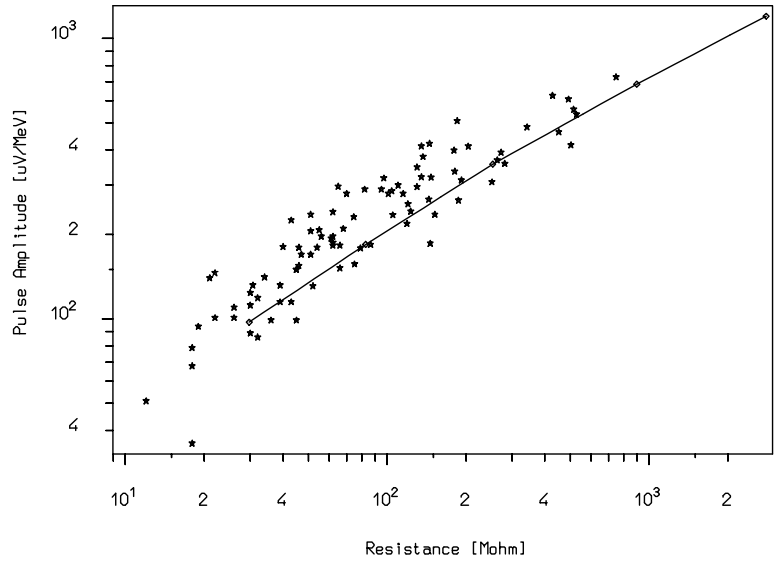}
 \end{center}
 \caption{Detector Merit Curve: simulation (solid curve) and experimental
 points for the 20 detector array.}
 \label{DMC}
\end{figure}

As a test of the model, a typical detector of the MiDBD 20-element array was simulated. These have different masses and geometries than the CUORE crystals. A DMC was then constructed using the model for this detector. It was then compared to the real DMC measured for the 20-detector array. A significant spread among the experimental points, for all 20-detectors was found. There were two reasons for this; intrinsic detector differences and the presence of points not corresponding exactly to the maximum signal. Nevertheless the simulated DMC lies inside of the experimental points and has the correct slope (see Fig.~\ref{DMC}). The detector simulations are not distinguishable from a typical detector of the array.

\begin{figure}
 \begin{center}
 \includegraphics[width=0.7\textwidth]{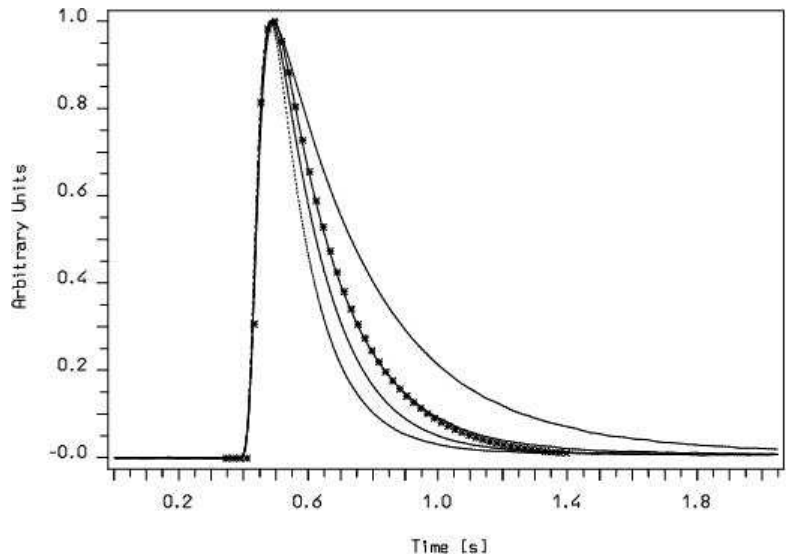}
 \end{center}
 \caption{Real pulses (full line) and simulated pulses (line+points).}
 \label{pulse}
\end{figure}

The TM must correctly predict not only pulse amplitude, but also time constants. At a typical operating resistance ($\sim 100\, M \Omega$) the predicted rise time (10\%-90\%) and decay time, are respectively 40~$ms$ and 430~$ms$, and lie inside the distribution of these parameters for the 20-detector array. In Fig.~\ref{pulse}, the simulated pulse is compared with real pulses collected with four detectors, randomly chosen among the 20 elements. It is evident that the pulse-shape of the simulated detector is very similar to that of a typical detector of the array. From this it was concluded that the TM is capable of explaining semi-quantitative performance of the 20 - element array.

The TM can be used in parallel with specific low temperature experimental tests, to define the optimum properties of the CUORE element, and the results thus obtained imply the following  optimum parameters for the CUORE elements:
\begin{enumerate}
\item{$TeO_2$ absorber dimensions: $50 \times 50 \times 50\;mm$, with a mass of 750~$g$;}

\item{NTD Thermistor \# 31, $T_0=3.0\,K$, $R_0=1.5\,\Omega$, dimensions $3\times 3\times 1$~$mm$,
     with a contact distance of $3\,mm$, and with a resistance of $50\,M\Omega$
     at $T=10\,mK$; }
\item{absorber-thermistor coupling made with 9 epoxy spots, 0.5 to 0.8 $mm$ in
diameter, and 50 $\mu m$ high;}
\item{gold connecting wires 50 $\mu m$ in diameter, and 15 $mm$ in length;}
\item{crystals mounted on PTFE blocks with a total
    PTFE-to-crystal contact area equal to about a few square $mm$.}
\end{enumerate}

In spite of the reasonable agreement obtained between the TM predictions and the performance of the 340~g \teod detectors, the comprehension of the detector operation principles is far to be complete. The proponents would like to stress in fact that this agreement was obtained with the {\em average} detector behaviour, while the TM seems to fail when one attempts to fully simulate an individual device. In particular, one observes a general trend according to which the simulated pulse amplitudes are smaller than the experimental ones. Inconsistency appears also in the pulse shape. This is true in particular for the CUORICINO 790~g detectors, as described below.

In the CUORICINO case, the proponents have used an improved version of the code implementing the thermal model, capable not only of simulating but also of fitting to experimental data of the load curve of individual detectors.  After fitting, the agreement between the simulated and the real curve is excellent. Of course the fit procedure changes the initial values of the thermal conductances, as measured in the specific experiments mentioned above. However, this modification is not drastic and can be justified with the inevitably not complete reproducibility of delicate manual operations like thermistor gluing and crystal insertion in the Teflon blocks. This proves that the {\em static} part of the model describes the detector satisfactorily. Problems show up when dealing with the {\em dynamic} behaviour. The simulated pulse, reconstructed using the thermal parameters extracted from the static behaviour of a specific detector and, in addition, the lattice heat capacity and the electron heat capacity (taken as a free parameter), is too small, typically a fraction 3 to 5, with respect to the experimental pulses collected for that specific detector. The decay time is also in disagreement, being typically 3 to 5 times too fast. Note that this disagreement cannot be naively solved by postulating a crystal heat capacity lower than expected: in fact, this can solve the discrepancy between the pulse amplitudes, but at the price of making the disagreement between decay times even more severe. In addition, while it would not be surprising to find higher than expected specific heats for the \teod crystals at low temperatures, it is really difficult to justify lower values. The impression is that a model based on heat flows, which assumes thermal distribution of the involved excitations (phonons in the absorber, electrons in the thermistor), is not totally adequate for the description of the detector performance. Some attempts to modify the model by introducing athermal phonons directly absorbed in the thermistor seem to improve our ability to describe the pulse shape and amplitude. However, work is in progress and for the moment no conclusive statement can be made about the mechanism of pulse formation.

\clearpage
\section{Previous results of MiDBD and CUORICINO} \label{sec:MiCUORICINO}

The MiDBD experiment, carried out in Hall A of LNGS by the Milano group, was only the last of a series of experiments on neutrinoless DBD of \tectn, based on the bolometric technique. In fact, the improvements in the detector performance obtained over the course of years of measurements with detectors of constantly increasing mass, led to a large mass, low background experiment whose results finally demonstrated the competitiveness of the low temperature calorimeters. The idea of CUORE proceeds just from the successful results of MiDBD which can be therefore considered as a first step towards CUORE. The possibility to use the available MiDBD cryogenic system to test a single module of CUORE, thus checking the feasibility of this second generation experiment, finally lead to CUORICINO. Approved and funded in 2001, CUORICINO is actually the most sensitive experiment on neutrinoless DBD of \tectn, and is presently running in Hall A of LNGS, in the same cryostat used for MiDBD.

\subsection{LNGS cryogenic setups}\label{sec:lngssetup}

Two cryogenic setups were installed in the eighties at LNGS by the Milano group. The first, used in the past for the MiDBD experiment (Hall A) is now hosting the CUORICINO detector. The second (Hall C) has been dedicated to the research and development activities for CUORICINO and CUORE. Both cryogenic setups consist of  dilution refrigerators having powers of 1000 $\mu W$ and 200 $\mu W$ at 100 mK respectively. They are both equipped with a dedicated Helium liquefier, providing a substantial recovery of the Helium and preventing Helium contamination of the tunnel atmosphere, and are housed inside Faraday cages to suppress electromagnetic interference. A much larger experimental volume is  available in the Hall A cryogenic setup. \\

\begin{figure}[h]
    \begin{center}
      \includegraphics[width=0.7\textwidth]{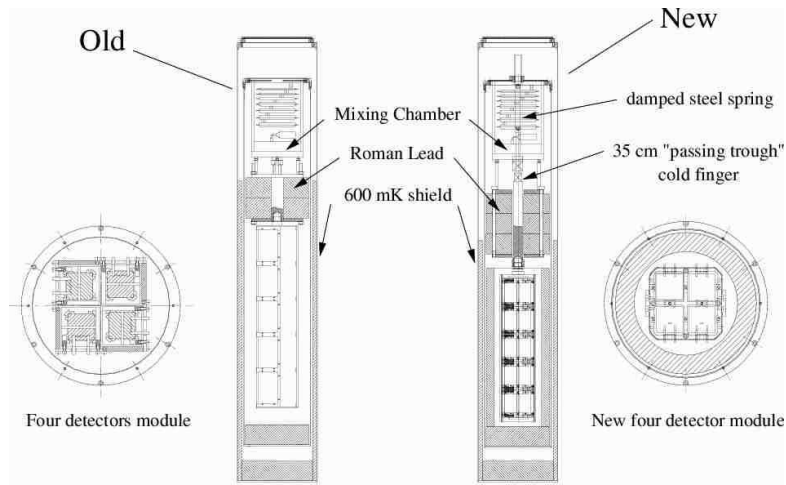}
    \end{center}
\caption{Scheme of the MiDBD detectors.}
\label{fig:setup}
\end{figure}

All the materials used for construction of MiDBD and CUORICINO were analysed to determine their radioactive contamination levels. These measurements were carried out by means of two large Ge detectors installed in the Gran Sasso underground Low Radioactivity Laboratory. The level of the radon contamination in the air of the Laboratory is continuously monitored. Both dilution refrigerators are equipped with heavy shields against environmental radioactivity. In particular, the Hall A dilution refrigerator is shielded with two layers of lead of 10 cm minimum thickness each. The outer layer is of commercial low radioactivity lead, while the internal one is made with special lead with a \pbdd contamination of 16 $\pm$ 4 Bq/kg. The external lead shields are surrounded by an air-tight box flushed with fresh nitrogen from a dedicated evaporator to avoid radon contamination of the gas close to the cryostat.  

\begin{figure}
    \begin{center}
      \centering\includegraphics[width=0.8\textwidth]{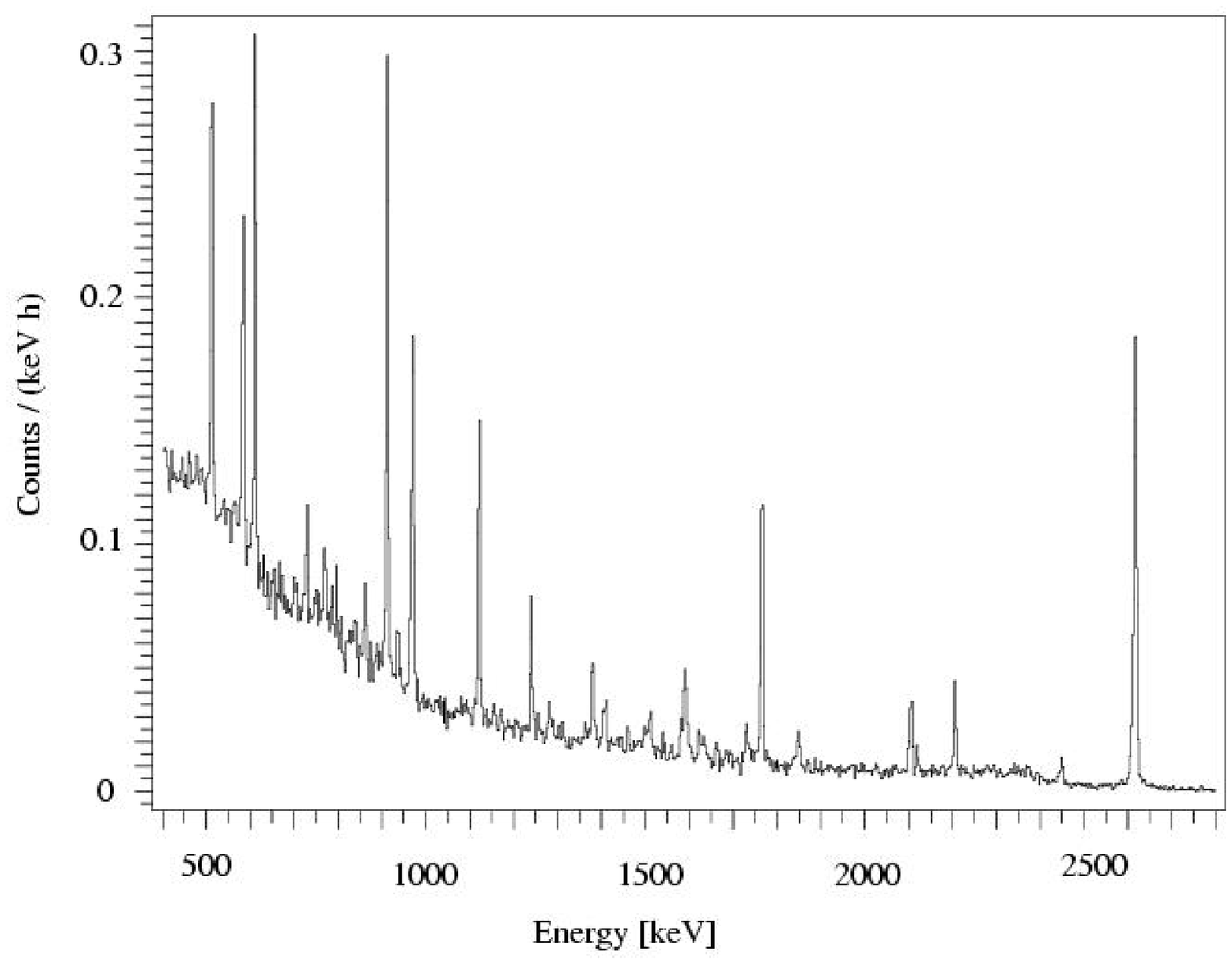}%
    \end{center}%
\caption{\label{fig:MiDBDUThcal} An U+Th calibration, sum spectrum of the 20 detectors} 
\end{figure}

In order to shield the detectors against the unavoidable radioactive contamination of some fundamental components of the dilution refrigerator thick layers of Roman lead are placed inside the cryostat just around the detectors. A borated polyethylene neutron shield (10 cm) was added in 2001 to the hall A cryostat. 

\subsection{MiDBD experiment}\label{sec:MiDBD}
The main goal of MiDBD was the study of neutrinoless and two neutrino double beta decay  of \tect to  \xect using the thermal method. In addition, its results on the detector performance, and on the background, were demonstrated to be particularly useful for the CUORE and CUORICINO development. 

\begin{figure}
 \begin{center}
 \includegraphics[width=0.7\textwidth]{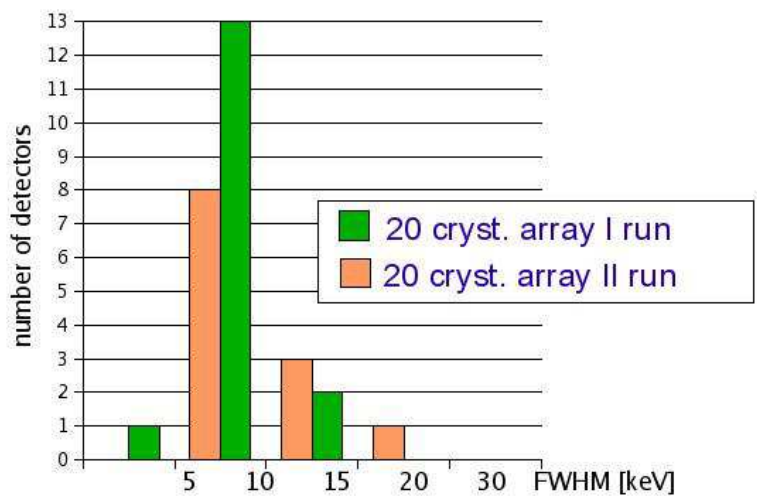}
 \end{center}
 \caption{Distribution of the single MiDBD detector energy resolutions (FWHM) on the \tld \ 2615 keV line.}
 \label{fig:MiDBDdE}
\end{figure}

MiDBD consisted of an array of 20 crystals of \teod of \magro each (340 g), for a total mass of 6.8 kg. It was operated in the Hall A dilution refrigerator until December 2001. Sixteen MiDBD crystals were made of natural \teodn. Of the remaining four, two were isotopically enriched at 82.3~\% in \tecv and other two at 75.0~\% in \tectn. Mass spectrometer measurements determined that, due to the crystallization process, these enrichment factors were lower than the original ones (94.7~\% and 92.8~\%, respectively). The array, held in a copper tower-like frame mounted inside the inner vacuum chamber (IVC) of the dilution refrigerator, was maintained at a constant temperature of about 10 mK. 

\begin{figure}
    \begin{center}
      \centering\includegraphics[width=0.8\textwidth]{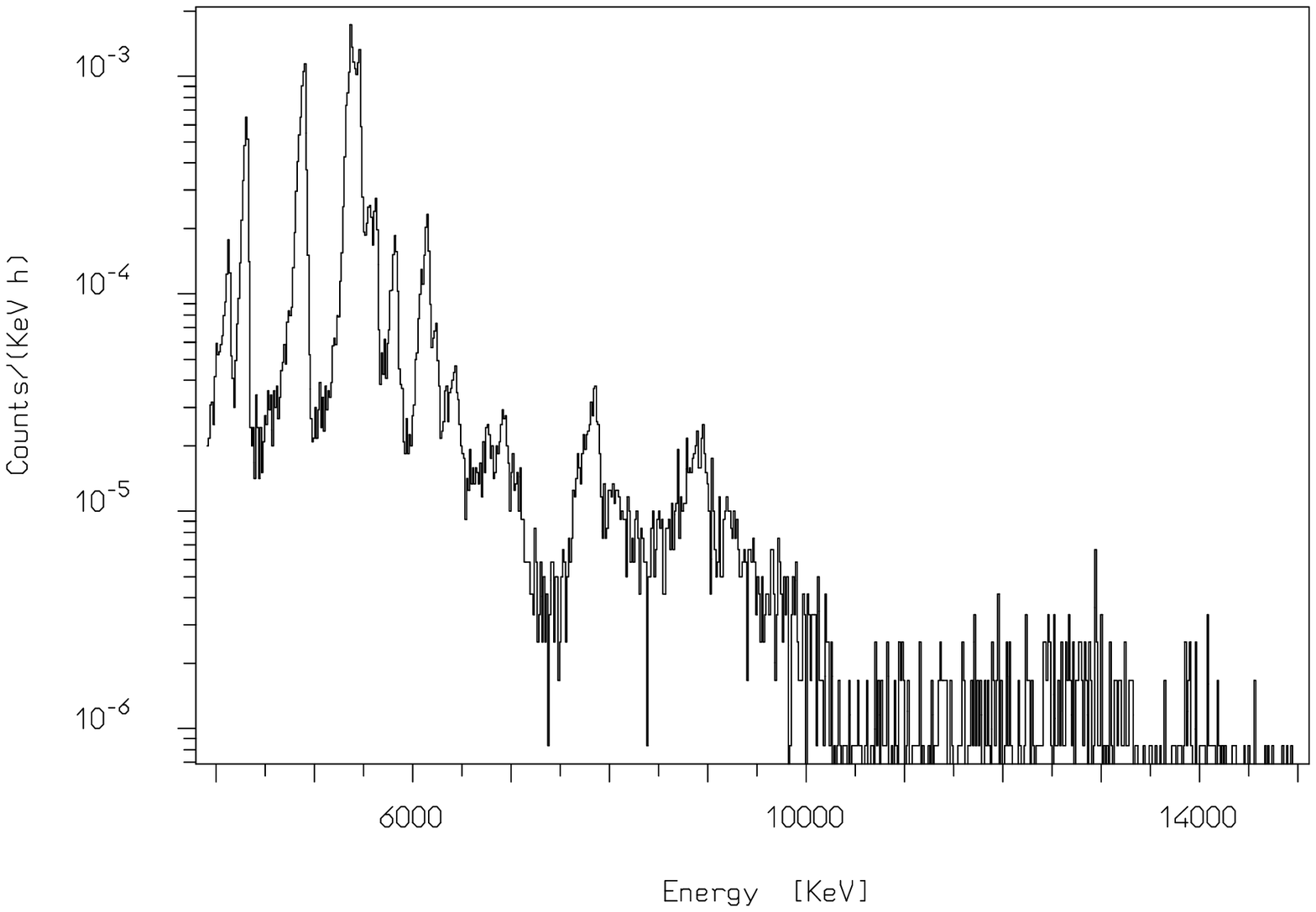}%
    \end{center}%
\begin{center}
\caption{\label{fig:MiDBDabkg} Background spectrum: the alpha particle region} 
\end{center}
\end{figure}

A Neutron Transmutation Doped (NTD) thermistor and a Si doped resistor were glued on the surface of each crystal. The former was used to read-out the thermal signal while the latter acted as a heater to generate a reference heat pulse in the crystal in order to monitor continuously the gain of each bolometer\cite{alessandrello98s}. The electronic readout was accomplished with a room temperature low noise differential voltage-sensitive preamplifier, followed by an amplifier and an antialiasing Bessel filter. The MiDBD detector was shielded with an internal layer of Roman lead of 1 cm minimum thickness (Fig. \ref{fig:setup}). In order to shield the detectors against the unavoidable radioactive contamination due to some fundamental components of the dilution refrigerator a layer of 10 cm Roman lead was placed above the tower of the array inside the cryostat. A similar layer, also inside the cryostat, was placed below the tower.
The significant effort was carried out to check both the knowledge of the background sources and the performance of the new detector mounting system proposed for CUORICINO and CUORE. This led to a second run (MiDBD-II) in 2001 in which the detector structure was heavily modified (Fig. \ref{fig:setup}). The MiDBD tower was completely rebuilt with a structure very similar to the one designed for CUORE/CUORICINO in which all the surfaces of the \teod crystals and of the mounting structure were subject to a dedicated etching process aiming to reduce radioactive contaminants. On this occasion, due to the more compact structure of the tower, an enlargement of the internal Roman lead shields was also realized. An external neutron shield (10 cm of borated polyethylene) was added in June 2001. 
MiDBD data collection was divided in runs of about 15-20 days. At the beginning and at the end of each run the detector was calibrated with an exposure, lasting about 3 days, to a mixed U and Th source placed at the two opposite side of the 20 crystal tower, just in contact with the external cryostat thermal shields.\\

The excellent reproducibility of the MiDBD detectors is shown in Fig. \ref{fig:MiDBDUThcal} where the total calibration spectrum obtained by summing the spectra of the 20 detectors is presented. The overall energy resolution in the total calibration spectrum and in the total background spectrum are reported in Table \ref{tab:FWHMs}.The single detector energy resolution distributions at the \tld 2615 keV $\gamma$-line are shown in Fig. \ref{fig:MiDBDdE}.

\begin{table}[b]
\begin{center}
\caption{\label{tab:FWHMs} FWHM resolutions (keV) in a calibration spectrum and in the 84000 hours $\times$ crystal background spectrum.}
\vspace{2mm}
\begin{tabular}{ccccc}
\hline\hline
\multicolumn{5}{l}{Calibration}\\
\hline
E$_\gamma $ [keV] & 583 & 911 & 1764 & 2615 \\
FWHM [keV] & 3.6 & 4.3 & 5.9 & 8.1\\
\hline\hline
\multicolumn{5}{l}{Background}\\
\hline
E$_\gamma $ [keV] & 583 & 911 & 1460 & 2615 \\
FWHM [keV]& 4.2& 5.1& 6.7& 9.3\\
\hline\hline
\end{tabular}
\end{center}
\end{table}

\begin{figure}
    \begin{center}
      \centering\includegraphics[width=0.9\textwidth]{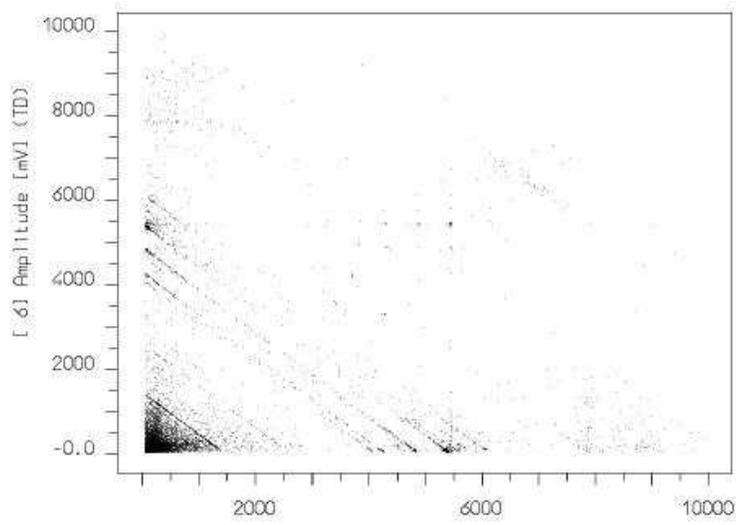}%
    \end{center}%
\caption{\label{fig:MiDBDSplot} Scatter plot of coincident events between all possible MiDBD detector pairs.} 
\end{figure}

The two MiDBD runs totalled \ca 31,508 (MiDBD-I) and \ca 5,690 (MiDBD-II) kg\per hours of effective running time, respectively. Sum spectra were obtained both with no anticoincidence cut and  by operating each detector in anticoincidence with all the others (``anticoincidence cut''). In all these spectra the main lines due to the natural activity of the \thdt and \udt chains, of \kq and the lines at 1173 and at 1332 keV  due to cosmogenic \coss were observed. 

\begin{figure}
     \begin{center}
      \centering\includegraphics[width=0.95\textwidth]{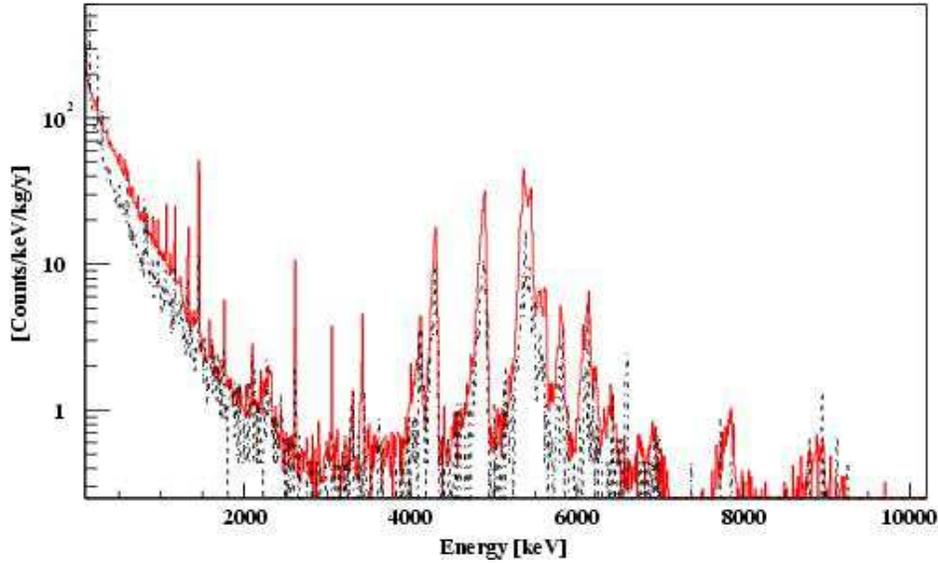}
    \end{center}%
\caption{MiDBD-I (continuous line) and MiDBD-II (dashed line)background spectra after anticoincidence cut.}
\label{fig:bkg_MiDBD}
\end{figure}

The sum spectra of the two runs obtained after applying the \emph{anticoincidence cut} are shown in Fig.~\ref{fig:bkg_MiDBD}. A clear reduction in both continuum and $\gamma$/$\alpha$ peaks intensities is observed. This proves the effectiveness of the changes applied in MiDBD-II and provide information on the localization of the main background sources (see Table \ref{tab:red_fact}).  

\begin{table}[bth]
\begin{center}
\vspace{2mm}
\begin{tabular}{cccc}
\hline\hline
\multicolumn{4}{l}{Reduction factors - peaks}\\
\hline
U and Th $\gamma$'s & $^{60}$Co & $^{40}$K & U and Th $\alpha$'s\\
$\ge$ 2 & $\sim$ 2 & $\sim$ 6 & $\sim$ 2\\
\hline\hline
\multicolumn{4}{l}{Reduction factors - continuum}\\
\hline
[1 - 2]  MeV & [2 - 3]  MeV & DBD region & [3 - 4]  MeV\\
$\sim$ 1.3& $\sim$ 1.6 & 1.5 $\pm$ 0.3 & $\sim$ 2.4\\
\hline\hline
\end{tabular}
\end{center}
\caption{\label{tab:red_fact} Background reduction factors measured between the two runs of MiDBD.}
\end{table}

A background analysis, supported by the Monte Carlo simulation of the detector,  allows us to disentangle the different sources responsible of the MiDBD background in the two runs and to understand which of the changes applied in MiDBD-II are responsible of the observed reductions. The conclusions thus obtained are the following:
\begin{itemize}
\item{the $\gamma$ peaks above \ca 500 keV of the U and Th chains and the two \coss $\gamma$ peaks at 1.173 and 1.332 MeV are mainly due to sources outside the detector, these peaks are substantially reduced by the increased shield of Roman lead;}
\item{the \kq $\gamma$ peak is probably due to multiple sources, its reduction being due both to the surface treatements and to the increased shield of Roman lead;}
\item{the $\alpha$ peaks of the \udt and \thdt chains, clearly visible in the background spectrum above 4 MeV, are mainly due to surface contamination of the crystals, their reduction has to be accounted for by the surface treatment of the crystals;}
\item{\udt and \thdt $\gamma$ peaks below $\sim$ 500 keV are attributed to a surface contamination of the copper mounting structure;}
\item{low intensities $\gamma$ lines of rather short living isotopes are explained by the cosmogenic activation of both copper and tellurium.}
\end{itemize}

According to our Monte Carlo simulation of the MiDBD detectors, the foregoing sources can completely account for the observed background (peaks and continuum) measured in the two runs.

A semi-quantitative analysis was carried out considering only MiDBD-I data for which the available statistics were more significant. In particular, the sources responsible of the background measured in the \BBz region were identified as follows:
\begin{enumerate}
\item{\thdt contaminations external to the detector structure (only the $\gamma$ emission from this chain can contribute to the background, mainly throught the 2615 keV line of \tldn)}
\item{\udt and \thdt surface contaminations of the crystals  ($\alpha$, $\beta$ and  $\gamma$ emissions from these chains contribute to the background)}
\item{\udt and \thdt surface contamination of the copper mounting structure ($\alpha$, $\beta$ and  $\gamma$ emissions from these chains contribute to the background)}
\end{enumerate}

The first source is identified from the clear 2615 keV \tld line seen in the spectra of both MiDBD runs. Its localization is based on the analysis of the relative intensities of the $\gamma$ lines (of different energies) belonging to the \thdt chain. Environmental $\gamma$ flux in the underground laboratory (measured with a Ge detector just close to our experimental set-up) and the measured contamination of the external lead shields lead us to rule out these two possible sources for the \tld line. The contamination seems therefore to be in the cryostat, and it could reasonably be the sum of the contamination of the various thermal shields that surround our detector.
In order to extrapolate the contribution of this source to the background in the \BBz region, we analyzed the calibration spectra obtained with an external \thdt source located just outside the cryostat. The shape of the calibration spectrum is in fact in good agreement with the shape of the simulated spectra for radioactive \thdt contaminations distributed over the cryostat thermal shields. The observed ratio between the area of \tld line at 2615 keV and the integral in the \BBz region is 14\pom 2. When compared with the 2615 keV rate observed in the background spectrum (129~counts/kg/y), this translates into a contribution of 0.11~counts/keV/kg/y which is \ca 20\% of the total \BBz background rate.

\begin{figure}
    \begin{center}
      \centering\includegraphics[width=0.9\textwidth]{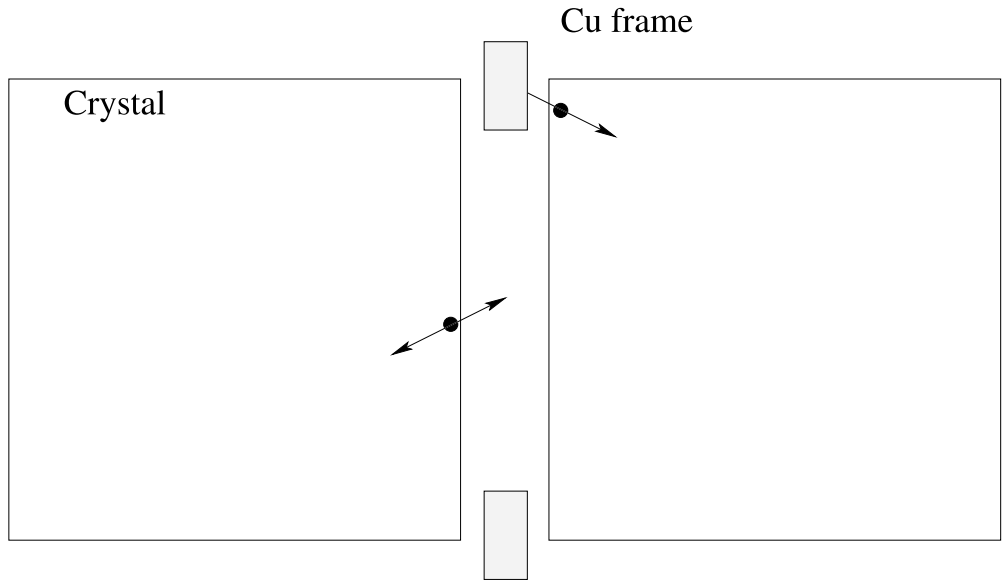}%
    \end{center}%
\caption{\label{fig:coinc} Scheme of a surface $\alpha$ decay.} 
\end{figure}

The second source can be easily recognized on the basis of the $\alpha$ peaks visible in the background spectra. Its location on the surface is deduced from the low energy tails of the peaks and by the coincidence scatter plots. These features exclude in fact the possibility of a bulk origin. Indeed the peak energies correspond to the transition energies of the decays implying that both the $\alpha$ particle and the nuclear recoil contribute to the signal. The decay can only be ``contained'' in the crystal and the contamination producing  the peaks is localized at a sufficient depth of the crystal surface. The low energy tail can be explained in terms of the fraction of the decays occurring very near to the surface, where part of the $\alpha$ or of the recoil energy is lost outside the crystal. Since the ``lost'' energy can be however observed by a second detector faced to the one in which the decay occurred (fig.~\ref{fig:coinc}), a straight line with a negative slope equal to -1 is expected when plotting coincident events in different faced detectors as is shown in Fig. \ref{fig:MiDBDSplot}. 
This kind of analysis not only yields a further proof that $\alpha$ peaks are due to a surface contaminations of the crystals, but also proves that this contamination produces a continuum background extending far below the $\alpha$ decay transition energy. 
In particular this source can partially - but not completely - account for the rather flat continuous background measured between 3 and 4 MeV, a region where no known relevant $\gamma$ nor $\alpha$ lines from natural and cosmogenic radioactivity are present. 
By the analysis of the coincident events from $\alpha$ particles we obtained a rate R$_D$=0.13 counts/keV/kg/y, in the 3--4 Mev interval, for the events with multiplicity 2 (i.e. events in which 2 crystals are simultaneously hit). The only possible origin for these events, is radioactive decay produced in the fraction of crystal surface facing other crystals (S$_{Active}$). On the other hand, when the decay occurs in the fraction of crystal surface (S$_{Inert}$) which faces inert materials (e.g. copper), this contributes to the  background remaining in the anticoincidence spectrum. By simply considering the geometric evaluation of the ratio S$_{Inert}$/S$_{Active}$ (average Monte Carlo evaluation) we can obtain from R$_D$ the expected contribution to the anticoincidence spectrum in the 3--4 MeV region: 0.16\pom 0.05~counts/keV/kg/y. In order to extrapolate this rate back to the \BBz region we need to know the shape of the background spectrum. In particular, various estimates of the ratio R$^s_{3-4}$/R$^s_{\beta\beta}$ (where R$^s$ are the counting rates in the energy regions of interest after the anticoincidence cut) were obtained on the basis of Monte Carlo simulations of crystal surface contaminations of different depths and density profiles. 
Unfortunately the ratio depends critically on these parameters and the average Monte Carlo evaluation of the ratio has therefore a large error.
The extrapolated contribution to the \BBz region background is 0.37\pom 0.1~counts/keV/kg/y corresponding to \ca 43\pom10\% of the measured rate.

Concerning the third source, the prove of its relevance is less direct and is mainly based on the observation that the continuous background measured between 3 and 4 MeV cannot be completely explained by the surface contamination of the crystals.

Other possible sources for this background were taken into account and excluded as relevant for MiDBD: neutrons (ruled out by comparing the background rates measured with and without the neutron shield), muon induced neutrons (ruled out by simulation of the production rate of neutrons in the experimental set up), contamination of other components close to the detectors (such as the Teflon spacers placed between the copper frame and the crystals, the NTD thermistors, the Si heaters, for which the measured  contamination is too low to account for the observed background). Finally, some tests done with dedicated detectors in the hall C set-up proved that the etching of the copper structure in a clean environment yields a reduction of the background in the 3-4 MeV region. 
This fact, and the agreement of Monte Carlo simulation with the measured spectra, indicate that \udt or \thdt surface contamination of the copper should be the missing background source.

As in the case of the second background source mentioned above, the contributions of the third source to the background rate can be obtained by extrapolating the rate in the 3--4 MeV to the \BBz region.
The 3--4 MeV region rate (after the anticoincidence cut) can be evaluated by subtracting the crystal surface contribution from the total one: 0.33\pom 0.1~counts/keV/kg/y. Once again the extrapolation to the \BBz region is model dependent (even if, according to our Monte Carlo evaluations, less so in this case) and the extrapolated contribution the \BBz region background is  0.2\pom 0.1~counts/keV/kg/y (\ca 23\pom 11\%).
The results of this analysis are summarized in table~\ref{tab:MiDBD_contrib}

\begin{table}[bth]
\begin{center}
\caption{\label{tab:MiDBD_contrib} Estimate of the weight of the different sources responsible for the background measured in the MiDBD-I run.}
\vspace{2mm}
\begin{tabular}{cccc}
\hline\hline
Source & \tld & \BBz region & 3-4 MeV region\\
\hline
\teod \udt and \thdt surface contamination & - &43\pom 13\% &33\pom 10\% \\
Cu \udt and \thdt surface contamination & \ca 10\% &23\pom 11\% &67\pom 20\% \\
\thdt contamination of cryostat Cu shields & \ca 90\% &20\pom 5\% & -  \\
\hline\hline
\end{tabular}
\end{center}
\end{table}

\begin{figure}
    \begin{minipage}[c]{0.95\textwidth}%
      \centering\includegraphics[width=1\textwidth]{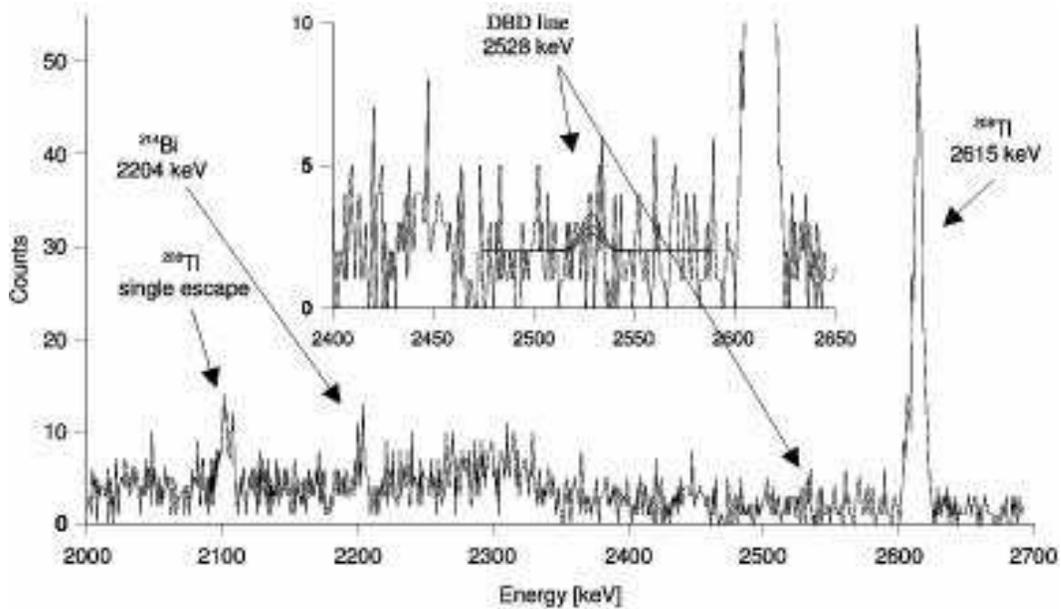}%
    \end{minipage}%
\caption{Total spectrum (in anticoincidence) in the region of neutrinoless DBD obtained with the twenty  crystal array. The solid curves represent the best fit (lowest curve) and the 68~\% and 90~\% C.L. excluded signals.}
\label{fig:sbb0n}
\end{figure}

The sum of the spectra obtained in anticoincidence in the two runs  in the  region above 2000 keV is shown in Fig. \ref{fig:sbb0n}. It corresponds  to \ca 3.56 kg \per year of \teod  and to \ca 0.98  kg \per year of \tect. The clear peaks corresponding to the lines at 2104 keV (single escape of the 2615 keV \tld line), at 2204 keV ($^{214}$Bi) and at 2615 keV (\tld), confirm the reproducibility of the array during both runs. 

No peak appeared in the region of \BBz of  \tectn, where the rates were, respectively, of 0.59 \pom 0.06 and 0.33 \pom 0.11 counts/keV/kg/y, when operated in anticoincidence. 
Also, no peak appeared at the energies  corresponding to neutrinoless DBD of \tect  to excited levels of \xectn, nor at the energy of 867 keV corresponding to \BBz of \tecvn. Fit parameters and 90~\% C.L. limits for the various decay processes (Tab. \ref{tab:milimits}) were evaluated with a maximum likelihood procedure\cite{baker84}. Similar results were however obtained following the approach proposed by G.J.~Feldman and R.D.~Cousins~\cite{feldman98}.
 
\begin{table}
\caption{Half lifetime limits (90~\% C.L.) on lepton violating and conserving channels deduced from the MiDBD data analysis. E$_0$ is the energy analyzed to obtain T$_{1/2}$, while \emph{a.c.} indicates that the anticoincidence spectrum was used.}
\label{tab:milimits}
\begin{center}
\begin{tabular}{@{}cccccc}
\hline\hline
Isotope & Transition & Cuts & E$_0$ & Efficiency & T$_{1/2}$\\
& & & (keV) & (\%) & (years)\\
\hline
\tect & $0\nu: 0^+ \to 0^+$ & a.c. & 2528.8 & 84.5 & $> 2.1 \times 10^{23}$\\
\tect & $0\nu: 0^{+*} \to 0^+$ & none & 1992.8 & 7.9 & $> 3.1 \times 10^{22}$\\
\tect & $0\nu: 0^{+*} \to 2^+$ & none & 1992.8 & 37.5 & $> 1.4 \times 10^{23}$\\
\tect & $2\nu: 0^+ \to 0^+$ & a.c. &  &   & $(6.1\pm 1.4^{+2.9}_{-3.5}) \times 10^{20}$\\
\tect & $1\chi: 0^+ \to 0^+$ & a.c. &  &  & $> 2.2 \times 10^{21}$\\
\tect & $2\chi: 0^+ \to 0^+$ & a.c. &  &  & $> 0.9 \times 10^{21}$\\
\tecv & $0\nu: 0^+ \to 0^+$ & a.c. & 867.2 & 97.9 & $> 1.1 \times 10^{23}$\\
\hline\hline
\end{tabular}
\end{center}
\end{table}

\subsection{CUORICINO experiment}\label{sec:CUORICINO}
\begin{figure}

 \begin{center}
 \includegraphics[height=16cm]{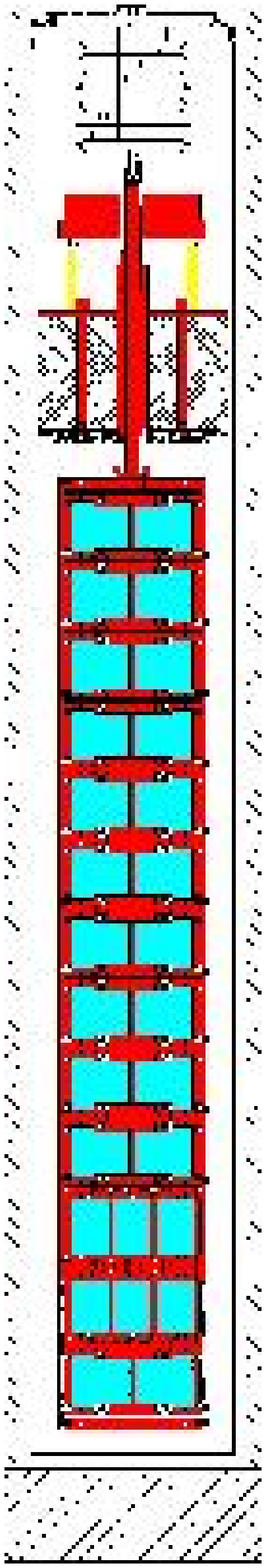}
 \hskip 1.1cm
 \includegraphics[height=16cm]{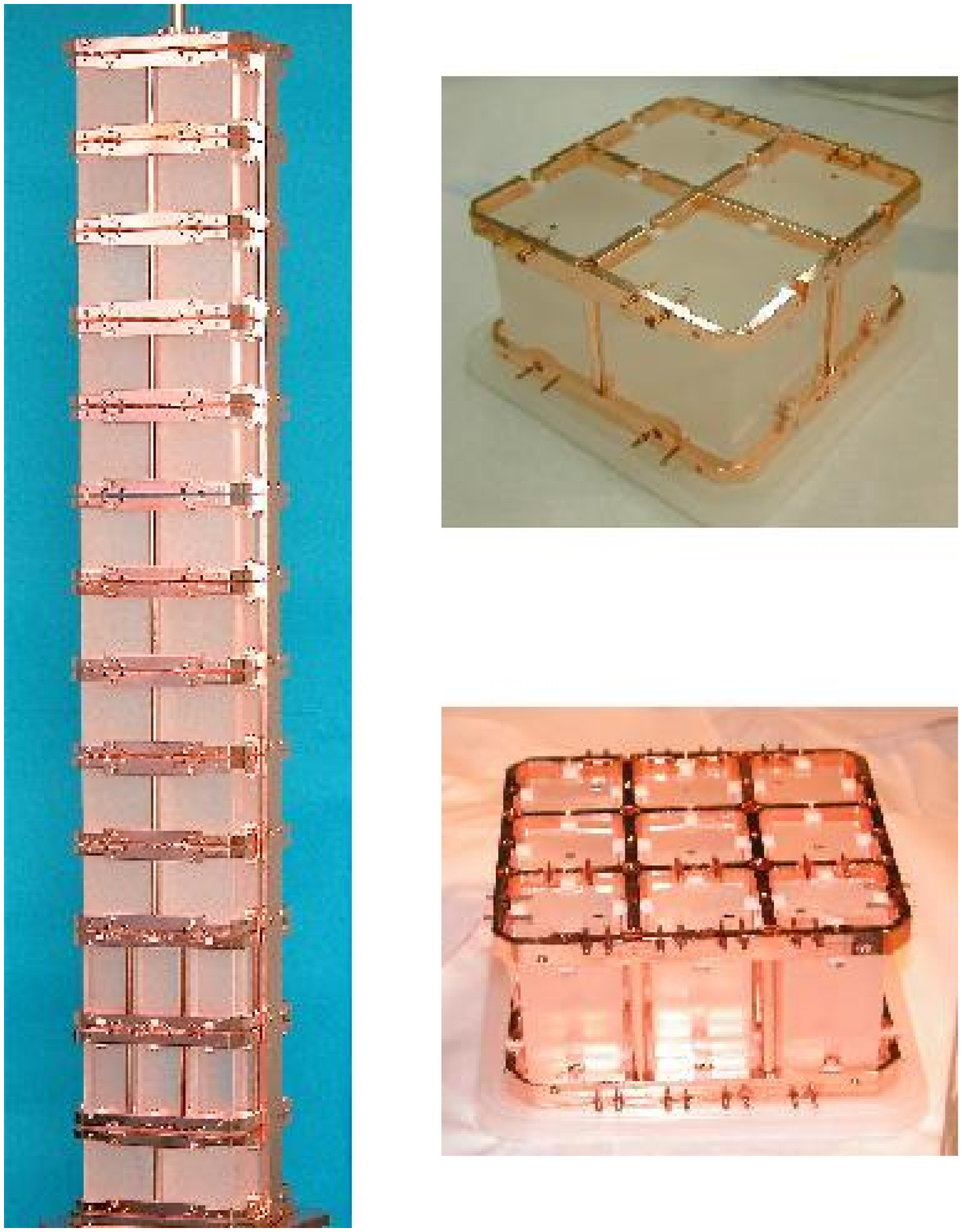}
 \end{center}
 \caption{The CUORICINO detector: scheme of the tower and internal roman lead shields (left), the 13 planes tower (centre), the 4 crystal module (top right) and the 9 crystal module (bottom right).}
  \label{fig:cuoricino}
\end{figure}

CUORICINO (Fig.~\ref{fig:cuoricino}) is a tower made by 13 planes containing 62 crystals of \teod. 44 of them are cubes of 5 cm side while the dimension of the others is \magro. All crystals are made with natural tellurite, apart two
\magro crystals which are enriched in \tecv and two others enriched in \tect with isotopic abundance of 82.3 \% and 75 \%, respectively. The total mass of CUORICINO is 40.7 kg, the largest by more than an order of magnitude with respect to any cryogenic detector. More details on the preparation of the crystals and on the mechanical structure of the array is reported elsewhere \cite{arnaboldi04}. In order to shield against the radioactive contaminants from the materials of the refrigerator, a 10 cm layer of Roman lead with \pbdd activity of <4 mBq/kg \cite{alessandrello98s} is inserted inside the cryostat immediately above the CUORICINO tower. A 1.2 cm lateral layer of the same lead is framed around the array to reduce the activity of the thermal shields. The cryostat is externally shielded by two layers of Lead of 10 cm minimal thickness each. While the outer is made by common Lead, the inner one has a \pbdd activity of 16 \pom 4 Bq/kg. An additional layer of 2 cm of electrolytic Copper is provided by the cryostat thermal shields. The background due to environmental
neutrons is reduced by a layer of Borated Polyethylene of 10 cm minimum thickness. The refrigerator operates inside a Plexiglass anti-radon box flushed with clean N$_2$ and inside a Faraday cage to reduce electromagnetic interferences.
Thermal pulses are recorded by means of Neutron Transmutation Doped (NTD) Ge thermistors thermally coupled to each crystal. Calibration and stabilization of each bolometer is performed by means of a resistor attached to each absorber and acting as an heater. The front-end electronics for all the \magro and for 20 of
the \ciccio detectors are located at room temperature. 
In the so called cold electronics, applied to the remaining 24 detectors, the preamplifier is located in a box at \ca 100 Kelvin near the detector to reduce the noise due to microphonics, which would be very dangerous when searching for WIMPS. More details on the read-out electronics and DAQ are reported in~\cite{arnaboldi04}.
CUORICINO is operated at a temperature of \ca 8 mK with a spread of \ca one mK. A routine energy calibration is performed before and after each subset of runs, which lasts about two weeks, by exposing the array to two thoriated tungsten wires inserted in immediate contact with the refrigerator. All runs where the average difference between the initial and final calibration is larger than the experimental error in the evaluation of the peak position were discarded.
During the first cool down 12 of the \ciccio and one of the \magro crystals were lost, due to disconnection of a few of the thermalizers which allow the transmission of the electric signals from the detectors to room temperature~\cite{arnaboldi04}. Since the active mass was of \ca 30 kg and the energy resolution excellent, we decided to continue running for a few months before warming up the array and attack the problem, which has been now fully solved.
The data presented here refers to this first run and to the new data obtained in about two months with a second run where the contacts of only two bolometers were lost. The sum of the spectra of the \ciccio and \magro crystals in the region of the neutrinoless DBD is shown in Fig.~\ref{fig:fondoBbb}. The background at the energy of neutrinoless DBD (i.e. 2510-2580 keV) is of 0.21 \pom 0.02 and 0.19 \pom 0.06 counts/kg/keV/y for the \ciccio and \magro crystals respectively.
No evidence is found at the energy expected for neutrinoless DBD of \tect. By appling a maximum likelihood procedure \cite{baker84,barnett96} we obtain a 90\% C.L. lower limit of 1.0 \per 10$^{24}$ years on the half-lifetime for \BBz of this nucleus. The unified approach of G.I.Feldam and R.D.Cousins \cite{feldman98,groom00} leads to a similar result. The upper bounds on the effective neutrino mass that can be extracted from our result depend strongly on the values adopted for the nuclear matrix elements. As in our previous paper \cite{arnaboldi04} we consider all vailable theoretical calculations \cite{suhonen98,tretyak02,elliott02,elliott04} apart those based on the shell model which are not considered as valid for heavy nuclei \cite{faessler98}. We have also not considered for the moment the calculation by Rodin et al. \cite{rodin03} since in the case of \tect they are based on the not yet established value for the two neutrino DBD lifetime. In particular the adopted value of 2.7 \per 10$^{21}$ years \cite{bernatowicz03} is the largest among all geochemical ones \cite{tretyak02,elliott02,elliott04}.
Taking into account the above mentioned uncertainties, our lower limit leads to a constraint on the effective neutrino mass ranging from 0.26 to 1.4 eV and partially covers the mass span of 0.1 to 0.9 eV indicated by H.V.~Klapdor-Kleingrothaus et.al.~\cite{klapdor04}.

\begin{figure}
 \begin{center}
 \includegraphics[width=0.99\textwidth]{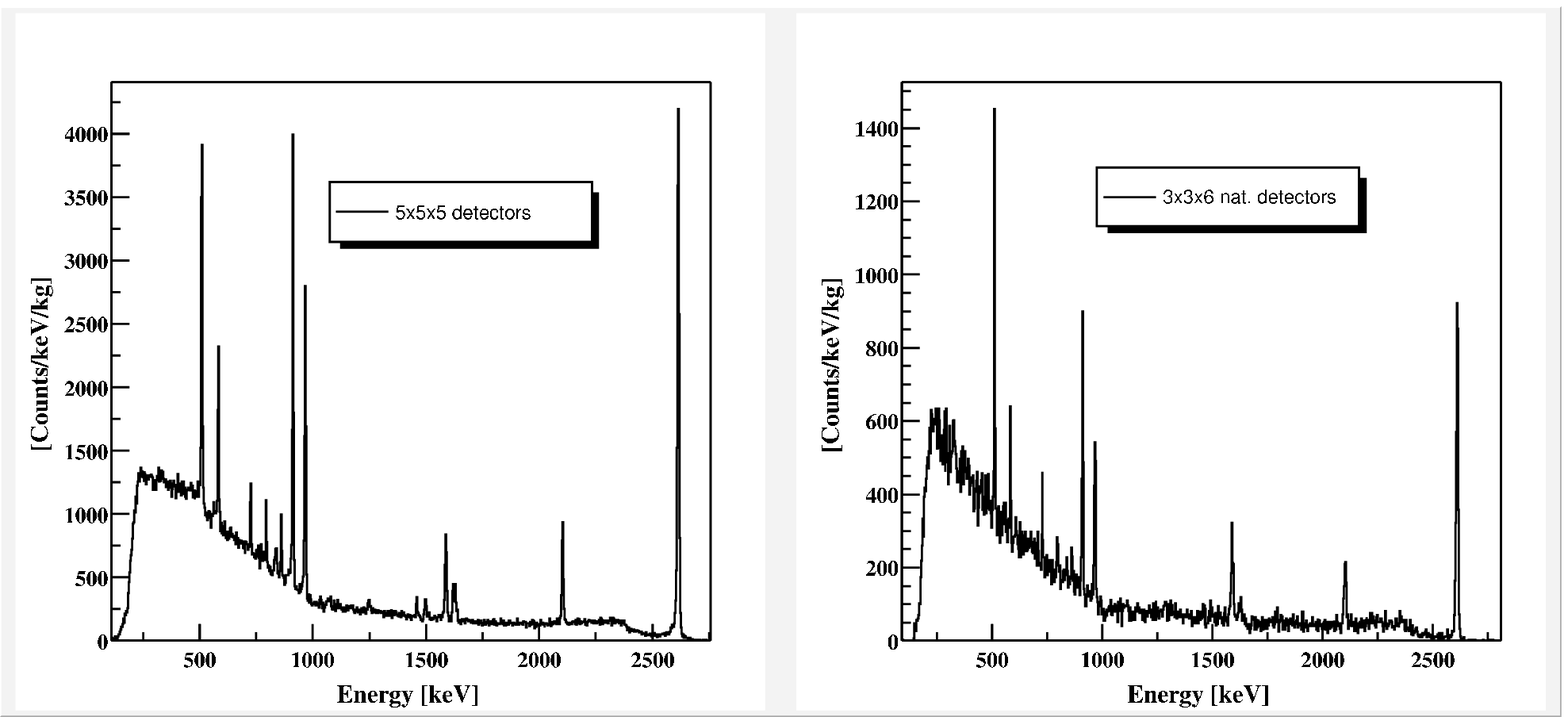} 
 \end{center} 
 \caption{Summed calibration spectrum (\thdt source just outside the crysotat) from all the operating \ciccio and \magro crystals.} 
\label{fig:QcalibrazioneBX} 
\end{figure}

\begin{figure}
 \begin{center}
 \includegraphics[width=0.7\textwidth]{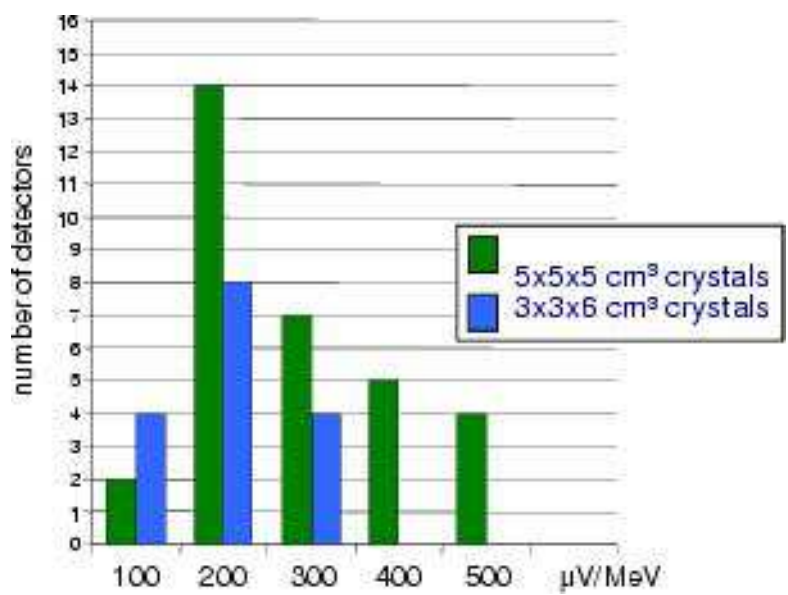}
 \end{center}
 \caption{Distribution of the single CUORICINO detector energy responses normalized to 1 kg of \teodn.}
 \label{fig:Qdr}
\end{figure}

\begin{figure}
 \begin{center}
 \includegraphics[width=0.7\textwidth]{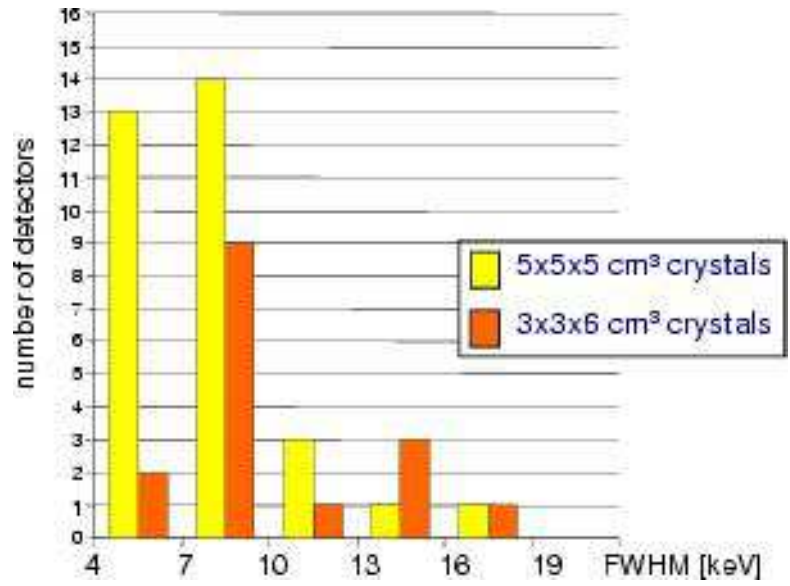}
 \end{center}
 \caption{Distribution of the single CUORICINO detector energy resolutions (FWHM) at the \tld 2615 keV line.}
 \label{fig:Qde}
\end{figure}

\subsubsection{CUORICINO background measurements}
\begin{figure}
 \begin{center}
 \includegraphics[width=0.9\textwidth]{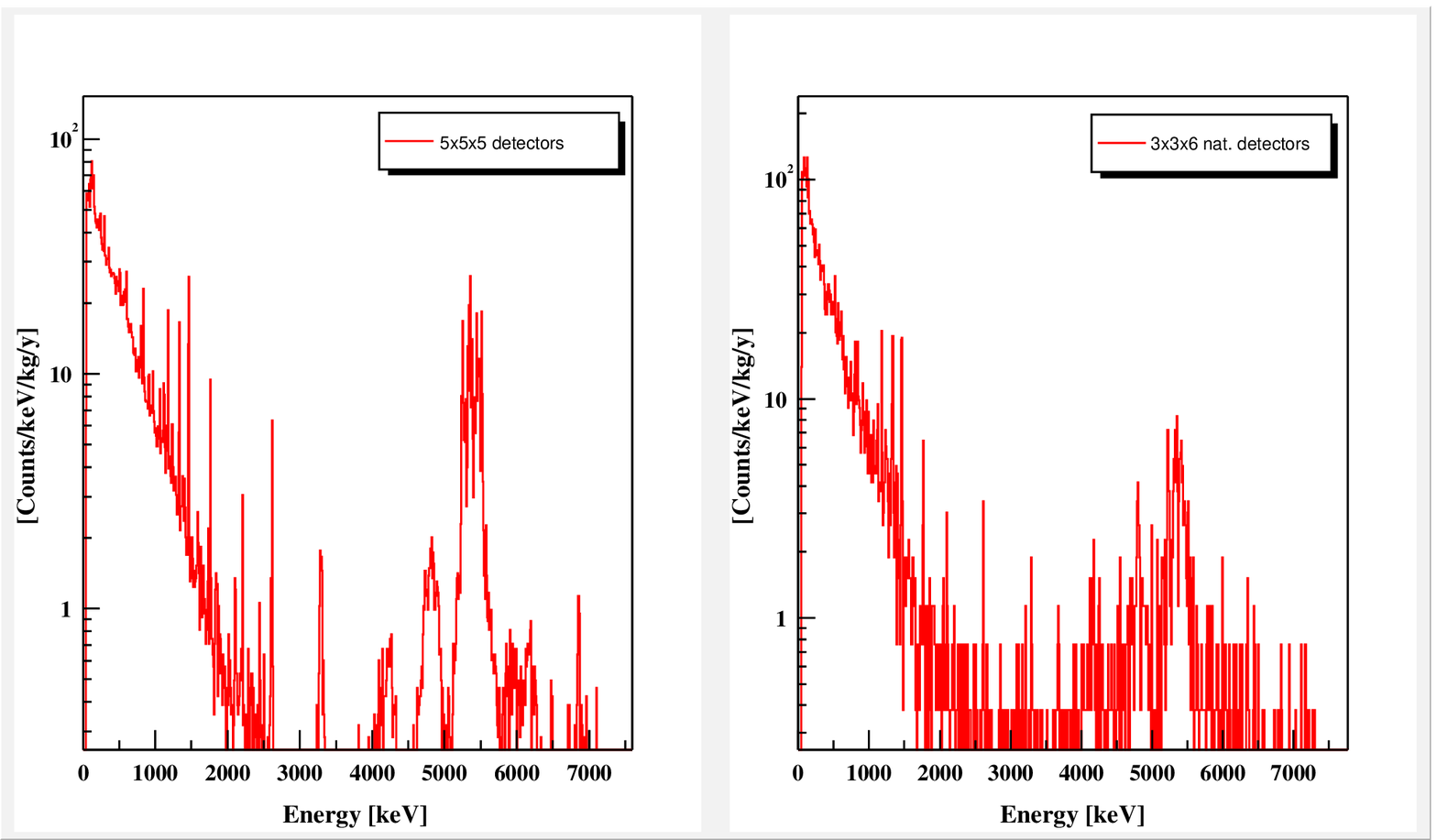}
 \end{center}
 \caption{Summed background spectra from the operating \ciccio and (natural abundance) \magro crystals.}
 \label{fig:fondoB}
\end{figure}

The total collected statistics are therefore of \ca 4.76 kg \per year for \ciccio crystals and \ca 0.53 kg \per year for the \magro ones. The corresponding background spectra are shown in Fig.~\ref{fig:fondoB}. 
The gamma lines due to \cosn, \kq and of the \udt and  \thdt chains are clearly visible. 

These lines due to contamination of the apparatus, are not visible in the spectra of the  single detectors: they appear after summing the 29 detectors, and are a good check of the calibration and stability of the detectors during the background measurement. Also visible are the gamma lines due to Te activation ($^{121}$Te, $^{121m}$Te, $^{123m}$Te, $^{125m}$Te and $^{127m}$Te)  and those due to Cu activation ($^{57}$Co, $^{58}$Co, $^{60}$Co and $^{54}$Mn) by cosmic ray neutrons while above ground.

The FWHM resolution of \ciccio detectors at low energy, as evaluated on the 122 keV gamma line of  $^{57}$Co, is \ca 2.8 keV. The $^{208}$Tl gamma line at 2615 keV - clearly visible in the background sum spectrum - is used to evaluate the energy resolution in the region of double beta decay, the FWHM is  8.7 keV. 

\begin{figure}
 \begin{center}
 \includegraphics[width=0.99\textwidth]{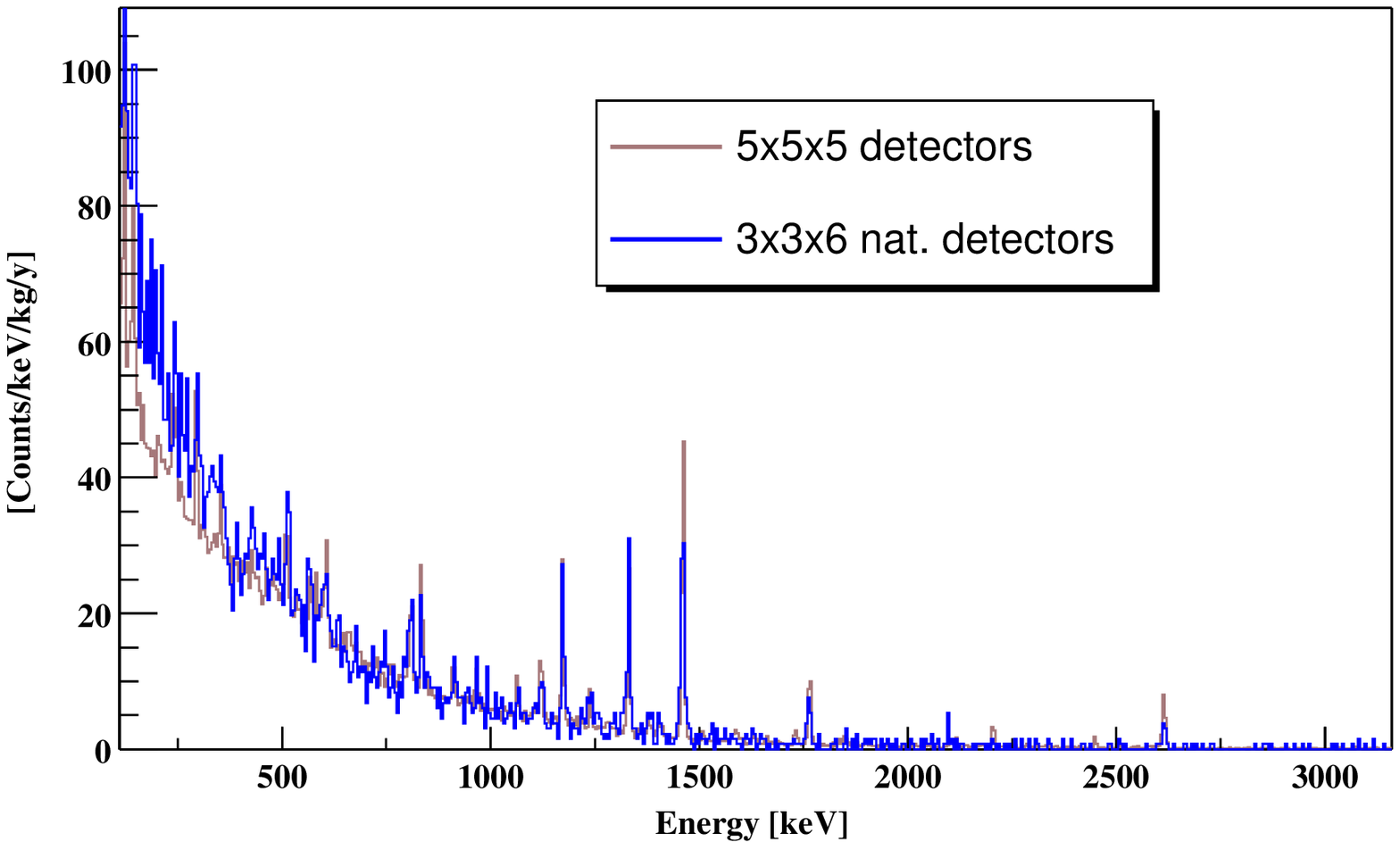}
 \end{center}
 \caption{Comparison between the background of the \ciccio crystals 
 and that of the natural \magro crystals in the gamma region.}
 \label{fig:fondogamma}
\end{figure}

\begin{figure}
 \begin{center}
 \includegraphics[width=0.99\textwidth]{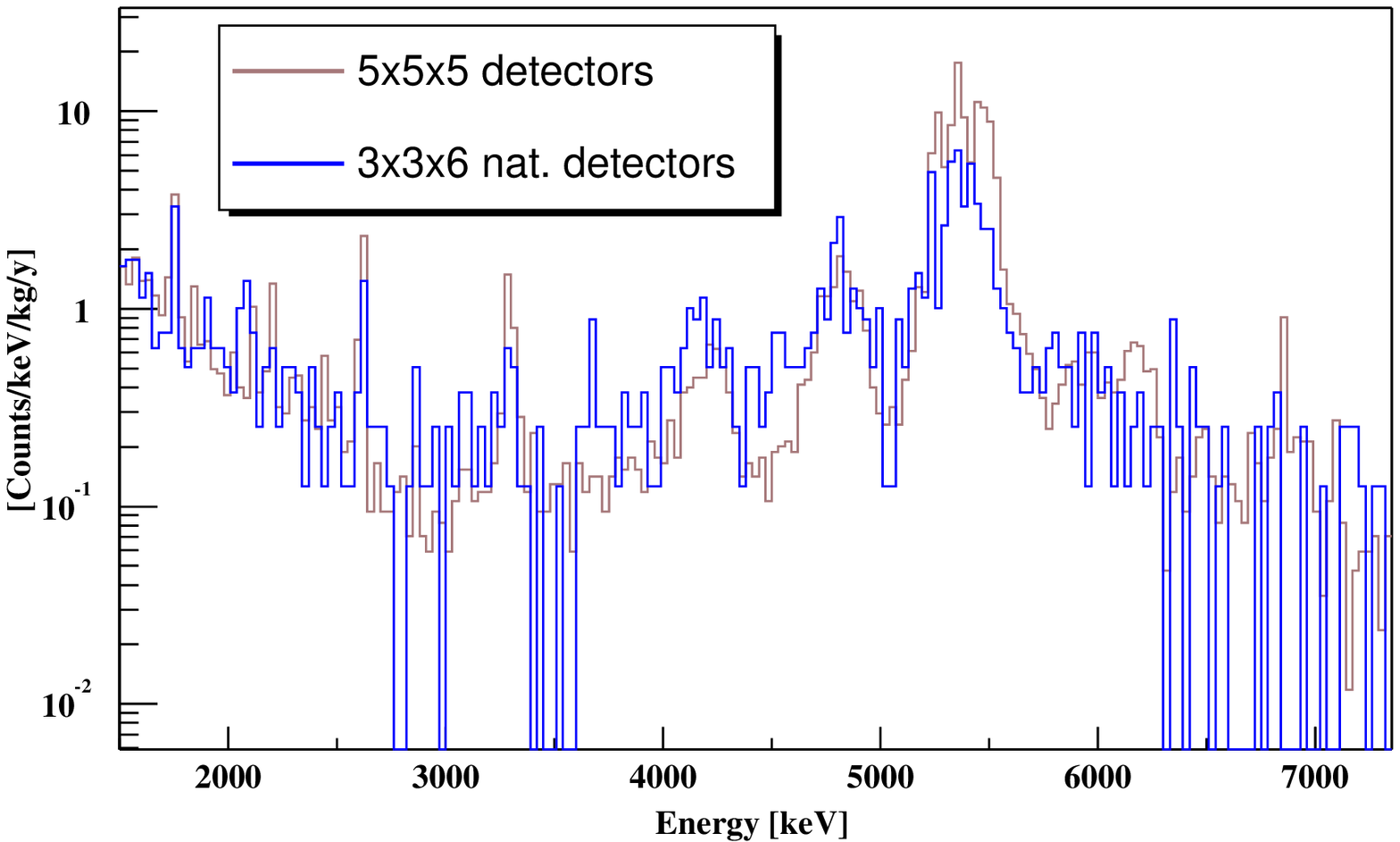}
 \end{center}
 \caption{Comparison between the background of the \ciccio crystals 
 and that of the natural \magro crystals in the alfa region.}
 \label{fig:fondoalfa}
\end{figure}

 The \magro and \ciccio crystal background spectra (sum of all the anticoincidence spectra of the detectors) are compared in Fig.~\ref{fig:fondogamma} and Fig.~\ref{fig:fondoalfa}. 

Here the statistical accuracy is much less, nevertheless the gamma lines, not visible in the single detector
spectra, are clearly visible in the background sum spectrum. The FWHM resolution at low energy, measured on the 122 keV gamma line of $^{57}$Co, is \ca 1.5 keV. The FWHM resolution on the \tld gamma line at 2615 keV is evaluated to be of about 12 keV (this value has a large error due to the poor statistical significance of the peak).

\subsubsection{CUORICINO background analysis}

Because of the low number of counts, any evaluation of the possible background sources necessarily has a large statistical uncertainty. Some preliminary considerations are however possible. They are based on the analysis of the sum spectra of the operating small and large crystals (the count rates of the single detector spectra are too low to consider them separately).
In Fig.~\ref{fig:fondogamma} and Fig.~\ref{fig:fondoalfa} a comparison between the background spectra of the large and small crystals is shown. General spectral shapes and counting rates (when normalised to the mass of the crystals) are quite similar. On the other hand the intensities of the gamma lines do not show a clear behaviour: the \bidq 1764 keV  (the only clearly visible one of the U chain) and the \tld 2615 keV lines (the only clearly visible one of the Th chain) seem to scale with the efficiency of the detectors (according to the results of a Monte Carlo analysis, the ratio of the detection efficiency of \ciccio and \magro crystals is \ca 3). The \kq line at 1460 keV has the same intensity per unit mass on the two kind of detectors. The \coss lines at 1173 and 1332 keV, on the contrary show a higher intensity per unit mass in small crystals. Moreover they undergo a larger reduction by the anticoincidence cuts with respect to other $\gamma$--lines. 

\begin{figure}
 \begin{center}
 \includegraphics[width=0.9\textwidth]{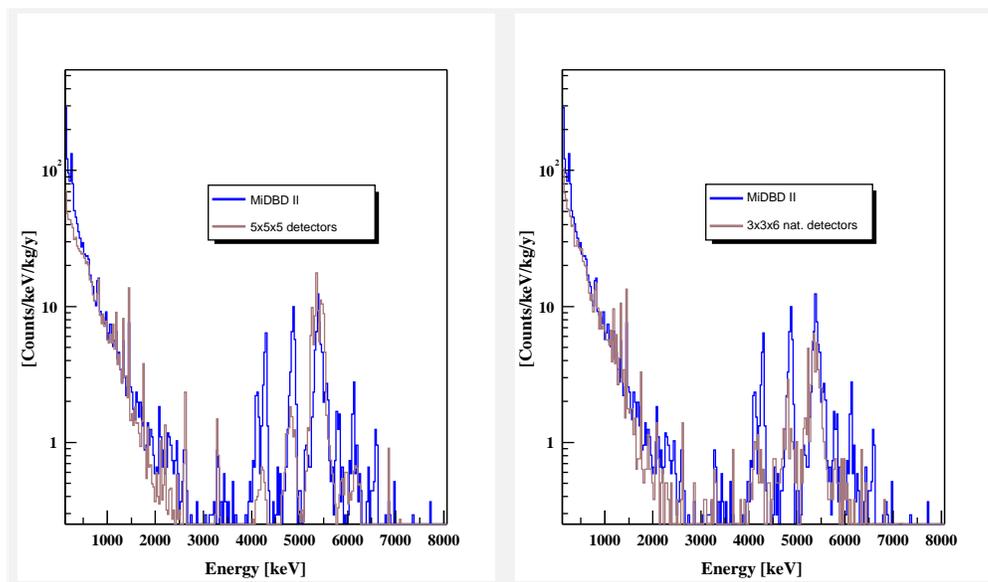}
 \end{center}
 \caption{Comparison between the background spectra of CUORICINO crystals and the background measured in MiDBD-II experiment.}
 \label{fig:cfrQII}
\end{figure}

\begin{figure}
 \begin{center}
 \includegraphics[angle=270,width=0.9\textwidth]{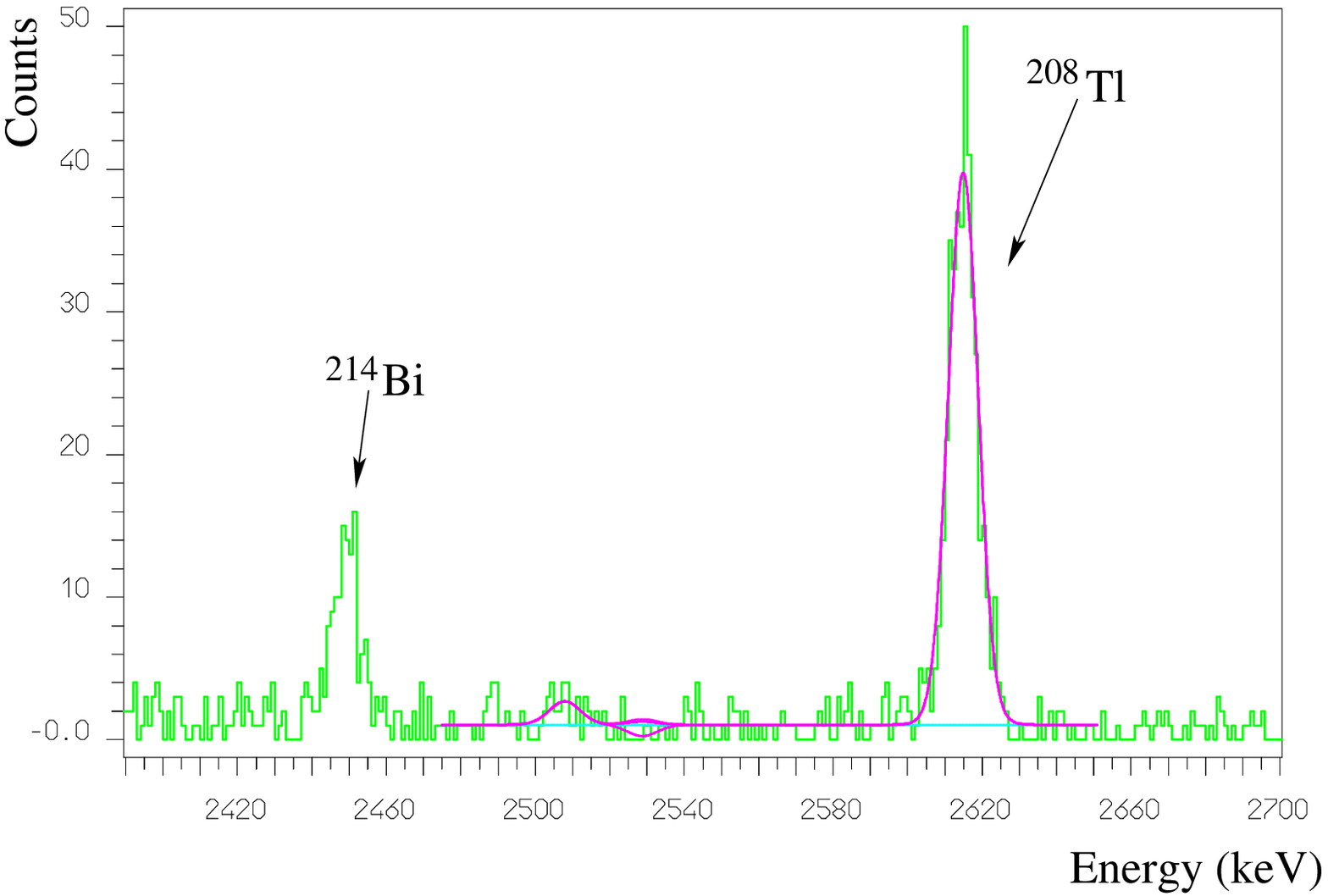}
 \end{center}
 \caption{Summed background spectrum (5.29 kg\per y) from all the operating 
 crystals in the region of neutrinoless double beta decay of $^{130}$Te (Q-value=2528.8 keV).}
 \label{fig:fondoBbb}
\end{figure}

In Fig.~\ref{fig:cfrQII} a comparison between the background spectra of CUORICINO crystals and the background measured in MiDBD-II (20 crystal array second run) is shown. 

In Table ~\ref{tab:crate} the counting rates in different energy regions are reported. It is clearly evident that an improvement was obtained in the 2-3 MeV region (a reduction of \ca 30\% in the counting rates per unit mass either of the small and of the large crystals) despite the reduced internal lead shield used in  CUORICINO (\ca 1.2 cm thickness of roman lead) with respect to MiDBD-II (\ca 3 cm thickness of roman lead). An even better reduction was obtained in the 3-4 MeV region where background is mainly dominated by surface contamination contributions. 

\begin{table} 
\begin{center}
\begin{tabular}{ccccc}
  \hline\hline
  counts/keV/kg/year & 1-2 MeV & 2-3 MeV & 3-4 MeV& 4-5 MeV \\
  \hline
	MiDBD-II 	& 3.21$\pm$0.08  	&  0.61$\pm$0.04	&  0.29$\pm$0.02	& 1.88$\pm$0.06 \\
	3\per3\per6 natural 	& 3.29$\pm$0.11  	&  0.38$\pm$0.04	&  0.24$\pm$0.03	& 0.78$\pm$0.05 \\
	5\per5\per5 		& 3.25$\pm$0.09  	&  0.41$\pm$0.03	&  0.17$\pm$0.02	& 0.55$\pm$0.03 \\	
  \hline\hline
\end{tabular}
\end{center}
\caption{Counting rates per unit mass in MiDBD-II and in CUORICINO.} \label{tab:crate}
\end{table}

With higher statistical accuracy it will be possible to analyze the reasons for this variation of the counting rates and also for the differences between small and large crystals. 
To do that, it is indeed necessary to disentangle the different sources of background by considering the counting rates in smaller regions as that between the two gamma peaks at 2448 and 2615 keV (\udt and \thdt chains respectively) - where we should  have contributions by both degraded $\alpha$'s from surface contaminations and \tld $\gamma$'s from bulk contaminations - and the region just above the \tld 2615 keV line where background sources should be limited to surface contaminations.
Moreover, it should be stressed that the effect of disconnected detectors in reducing the efficiency of the anticoincidence cut (a fundamental tool to study and localize background sources) is not negligible.

\begin{table}[bth]
\begin{center}
\caption{\label{tab:Qino_contrib} Estimate of the relative contributions of the different sources responsible for the background measured in CUORICINO.}
\vspace{2mm}
\begin{tabular}{cccc}
\hline\hline
Source & \tld & \BBz region & 3-4 MeV region\\
\hline
\teod \udt and \thdt surface contamination & - &10\pom 5\% &20\pom 10\% \\
Cu \udt and \thdt surface contamination & \ca 15\% &50\pom 20\% &80\pom 10\% \\
\thdt contamination of cryostat Cu shields & \ca 85\% &30\pom 10\% & -  \\
\hline\hline
\end{tabular}
\end{center}
\end{table}

By using the same background model resulting from the MiDBD analysis, and taking into account that we changed both the detector structure and the internal set-up of the cryostat (the roman lead shield and the cryostat copper radiation shields) we evaluated its consistency for CUORICINO. 
Because of the poor statistical significance of the data so far collected, the analysis has to be considered as preliminary (it was in particular limited to the \ciccio detectors for which the statistical accuracy is better) even if its results already tend to identify the most important sources of the CUORICINO background besides giving a first quantitative guess of their relevance. The percentage result of the measured background in the DBD region, is summarized in table~\ref{tab:Qino_contrib}.
It should be noticed that CUORICINO data seem to indicate a shallower depth for the crystal surface contaminations relative to MiDBD. A better determination of the actual depth and density profile of this contamination will be available when a fully operating CUORICINO and a better statistical accuracy will allow a more significant analysis of the multiple events.

\subsubsection{Analysis of the CUORICINO background above 3 MeV}
\begin{figure}
 \begin{center}
 \includegraphics[width=0.9\textwidth]{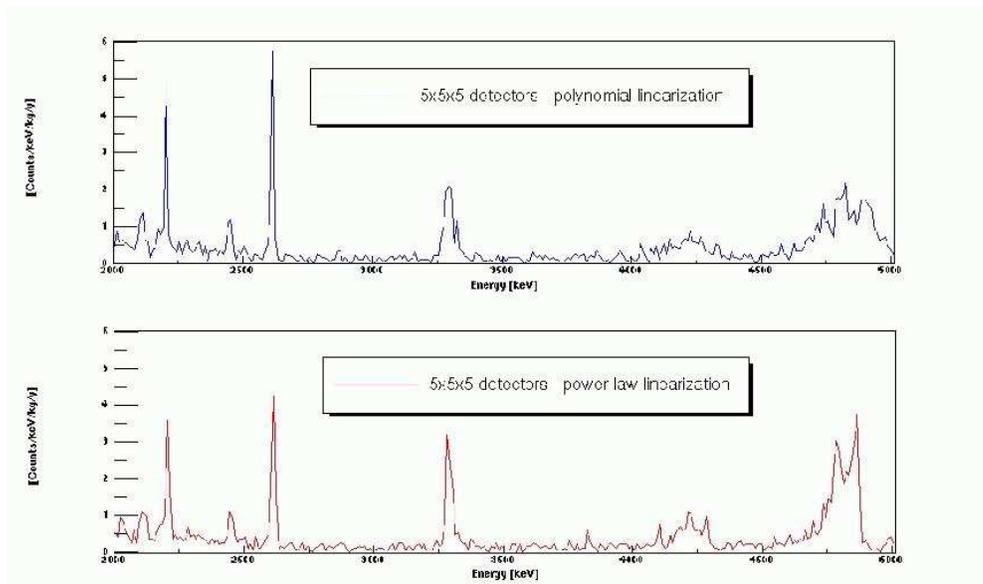}
 \end{center}
 \caption{Comparison between spectra obtained with different linearization methods. The appearence of clear alpha structures with the \emph{power law} metod is evident.}
 \label{fig:cfr_cal}
\end{figure}

 The following analysis tries to identify the background sources responsible of
the CUORICINO counting rate in the region above 3 MeV. 
 The analisys is based on both the coincidence (not used in the previous analyses) and the anticoincidence spectra collected with \ciccio detectors during the first run of CUORICINO and calibrated with a new, more reliable, techinque in the alpha region. 
 The single spectra have been calibrated and linearized in energy with a new 
technique that yields quite better results in the alpha region: both the gamma 
lines of the \thdt source measurements and the alpha line of \podd clearly visible in the background spectra are used, assuming a power law dependence for the relashionship between pulse amplitude and particle energy. The anticoincidence sum spectrum of \ciccio detectors calibrated in this way is compared in Fig.~\ref{fig:cfr_cal} with the analogous spectrum used in the previous analysis and calibrated only with gama lines (assuming here a polynomial relashionship between pulse amplitude and particle energy and extrapolating this calibration in the alpha region). 

 As already mentioned in previous discussions, alpha and beta particles produced by environmental radioactivity can give important contributions to background above 3 MeV. Because of their low range these particles can only be located either in the crystals or on the surfaces of the materials directly facing the crystals: the copper holder or, less likely, small components placed near the crystals or on the crystal surface such as the NTD thermistor, the Si heater, their gold wires, the glue and other smaller parts. Monte Carlo simulations of the coincidence and anticoincidence sum spectra of \ciccio detectors produced by radioactive contaminations of the above mentioned elements have been obtained using our GEANT4 based code. These spectra have been compared with the measured spectra in order to identify the actual contaminations responsible of the measured background.

\begin{figure}
 \begin{center}
 \includegraphics[width=0.9\textwidth]{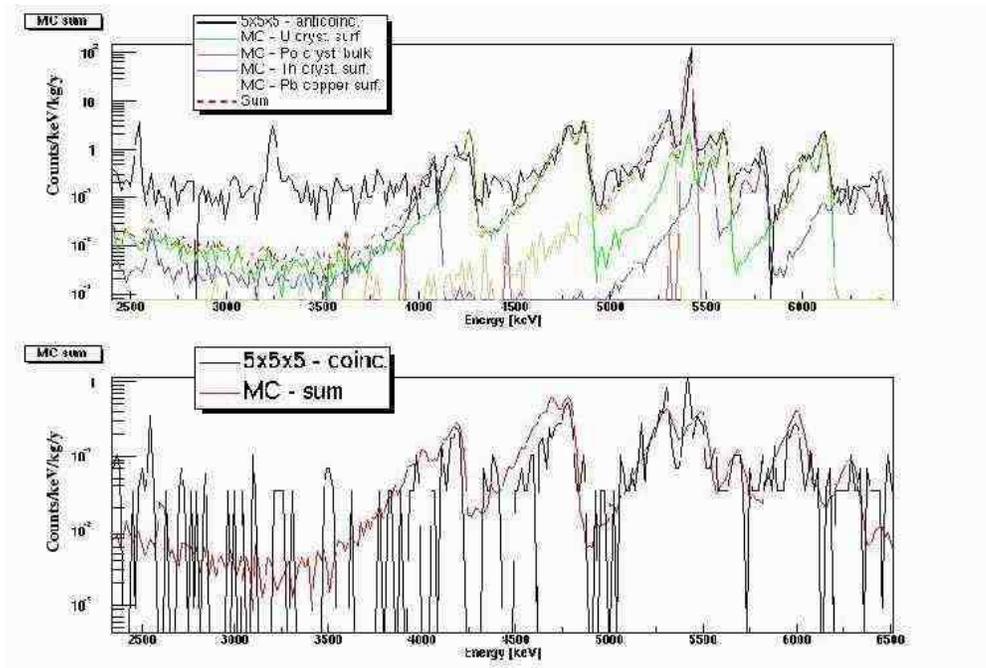}
 \end{center}
 \caption{Comparison between Monte Carlo and CUORICINO anticoincidence (top) and coincidence (bottom) spectra in the case of the \teod crystal surface contaminations ($\lambda$ \ca 1 $\mu$m) specified in the figure.}
 \label{fig:picchi}
\end{figure}
 
\begin{figure}
 \begin{center}
 \includegraphics[width=0.9\textwidth]{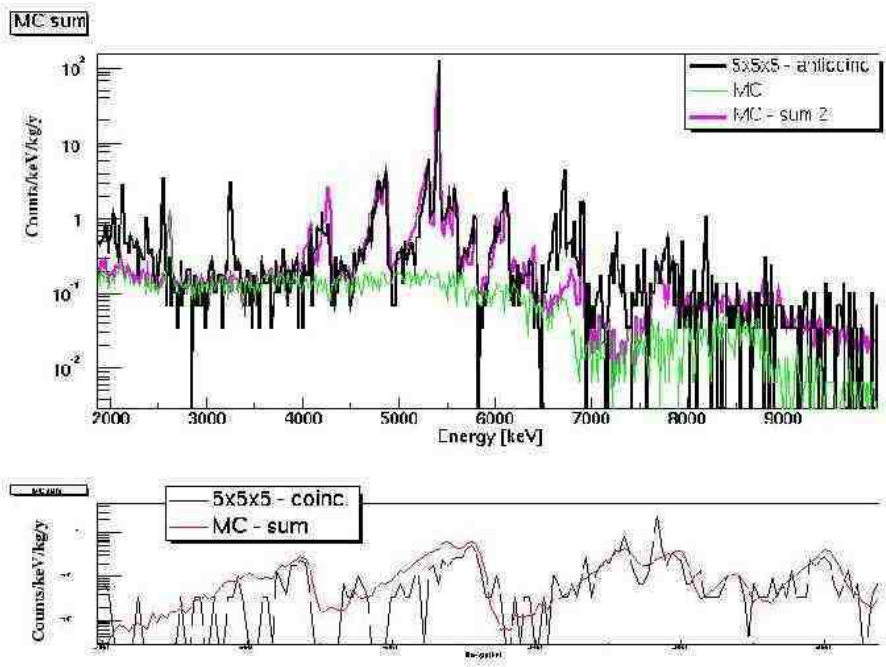}
 \end{center}
 \caption{Comparison between Monte Carlo and CUORICINO anticoincidence (top) and coincidence (bottom) spectra when a Th surface contamination of the detector copper holder (green line) is added (pink line) to the crystal conatminations considered in fig.~\ref{fig:picchi}.}
 \label{fig:picchi_continuo}
\end{figure}

 As a first step, bulk and surface contaminations of the crystals have been considered. Both these contaminations give rise to peaks (centered at the transition energy of the decay) in the anticoincidence spectrum, but in the case of bulk contaminations these peaks are gaussian and symmetric while in the case of surface contaminations the peaks show a low energy tail. On the contrary, the only possible sources contributing to the coincidence spectrum are the surface contaminations of the crystals. The general shape and the structures appearing in both background spectra strongly depend on the depth and density profile assumed for the contamination (an exponentially decreasing function characterized by a \emph{depth} $\lambda$: $\rho$(x) = Ae$^{-\lambda x}$ yielded the best results).  
 The results of the comparison between Monte Carlo simulations and measured spectra allow to prove that the alpha peaks of the anticoincidence spectrum of Cuoricino are due to a surface (and not bulk) \thdt and \udt contamination of the crystals with the only exception of the \podd line which has to be attributed to a bulk contamination of the crystals.
 The shape of the peaks in the anticoincidence spectrum and the shape of the coincidence spectrum is fully accounted for when a surface contamination \emph{depth} of the order of 1 $\mu$m or less is assumed.
 As shown in Fig.~\ref{fig:picchi} the identified crystal contaminations yield a satisfactory explanation of the coincidence spectrum. Some contribution to the continuous part of the anticoincidence spectrum seems however still missing.
 In order to explain this continuum, a further source, not contributing to the coincidence spectrum, has to be considered. Such a source can be a \thdt or \udt contamination on the copper surface, as shown in Fig.~\ref{fig:picchi_continuo}. The \emph{depth} of the contamination should be of the order of \ca 5 $\mu$m (a deeper contamination would produce too high gamma peaks while a thinner contamination would give rise to structure in the anticoincidence spectrum) with a total activity in the first 1.5 $\mu$m of the order of 10$^{-9}$ g/g either in \udt or \thdt or both. 
 
\vskip 2cm 
This analysis shows the importance of continuing the CUORICINO background measurement in order to improve the statistical accuracy of the above described results. These results, on the other hand, show that with a substantial reduction of the copper and \teod crystal surface contamination levels the sensitivity goal for CUORE (i.e. a background counting rate of the order of 0.01 counts/keV/kg/y or better) can be reached.

\subsubsection{Preliminary measurements of the Copper surface contaminations}
According to the suggestions of the LNGS Scientific Committee we have investigated various possibilities to identify and measure the surface contamination of the detector Copper holders which is expected to represent the main contribution to the CUORE background level in the neutrinoless double beta decay region. We would like to remind that the bulk contamination of Uranium and Thorium in copper has been found to be less than a part per trillion, while the analysis of the MIBETA and CUORICINO data indicates a surface contamination by two to three orders of magnitude larger. Degraded  $\alpha$ and $\beta$ particles from this source have  been found to represent a substantial contribution of the background in the region of neutrinoless double beta decay. Two main measurement procedures were considered:
\begin{itemize}
\item{dissolution of different superficial layers of Copper in ultra pure acid and analysys of the resulting  solution with Inductive Coupled Plasma Mass Spectroscopy (ICPMS)} 
\item{laser ablation of a number of distributed spots  on the Copper surface followed by ICPMS analysis of the dissolved material. We did that in Canada, but the sensitivity seem still low (~ 10-8 g/g)}
\end{itemize}
Other approaches like Total Reflection X-Ray Fluorescence (TRXRF) and Secondary Ion Mass Spectroscopy (SIMS) are also going to be considered, even if  they allow to investigate only the very superficial layer of the Copper samples.

The most powerful method was found by us to be a High Resolution  (HR) ICPMS with which preliminary results were obtained in collaboration by the Gran Sasso Laboratory (M.Balata and S.Nisi) and the Mass Spectroscopy group of the European Center of Ispra (P.Tricherini). The procedure has been the following: samples of Copper were attacked with ultrapure nitric acid  with different attacking times in order to extract the material from different thickness. The corresponding solutions were then analyzed in ISPRA   with a HR-ICPMS device characterized by a sensitivity up to 10-15 g/g of U and Th in the solution. The sensitivity in g/g of the dissolved Copper was lower by about three orders of magnitude due to the small copper concentration in the solution. The preliminary results of a  measurement carried out on a sample of the same Copper used for the CUORICINO detector holders is shown in Fig. 2.
 It can be seen that the average concentration of Uranium and Thorium strongly decreases when considering copper layers of 1.48 and 7.5 micron respectively (only a limited number of dissolved layers of different thickness could be analyzed in the preliminary measurement). This trend of surface contamination as a function of depth is in excellent agreement with that evaluated from our CUORICINO background model (comparison between  Monte Carlo simulations and actual background spectra). 

\subsubsection{CUORICINO vs. CUORE}
 While being a self-consistent experiment, CUORICINO finds its place in the CUORE project as a test facility intended to verify the technical feasibility of CUORE. Indeed the good results obtained as far as the detector performances are concerned prove that the realization of CUORE could be more straightforward than expected. There is still some  $R\& D$ that is required to improve detector reproducibility in order to reduce the spread in the main parameters that characterise detector performances (as the pulse height, the energy resolution  and the detector signal rise and decay times). 
Nevertheless the good results obtained in CUORICINO prove that the increase in the number of bolometers and in the total mass of the array do not substantially affect the quality of the experiment: the MiDBD experiment with 3 times fewer detectors, and a mass 6 times lower than CUORICINO, had a similar energy resolution in the background sum spectrum.
For what concerns background evaluation, as stated above CUORICINO is an excellent tool to predict the expected background for CUORE, when the different geometrical structure of the two setups is taken into account. 
Indeed CUORE will have a tightly packed structure where the operation of the detectors in anticoincidence will allow a strong reduction of background. Moreover the lead shield designed for CUORE will be optimized in order to practically cancel the background coming from outside. Such an optimization was not possible in CUORICINO which had to be housed in an already existing cryostat characterized by a limited space. Indeed to mount CUORICINO in the same cryostat that was used for MiDBD, the internal Roman lead shield thickness was necessarily reduced thus increasing the background due to $\gamma$ rays from the refrigerator materials. This will obviously be not a problem for CUORE. The CUORICINO background results, although preliminary, are very promising. They demonstrate both the effectiveness of the procedures applied during its preparation and our knowledge of its main background contributions. The detailed study of CUORICINO background (i.e. with better statistical accuracy) will give  more insight into background sources.\\

\clearpage
\section{CUORE project} \label{sec:CUORE}

\begin{figure}[h]
 \begin{center}
 \includegraphics[width=0.55\textwidth]{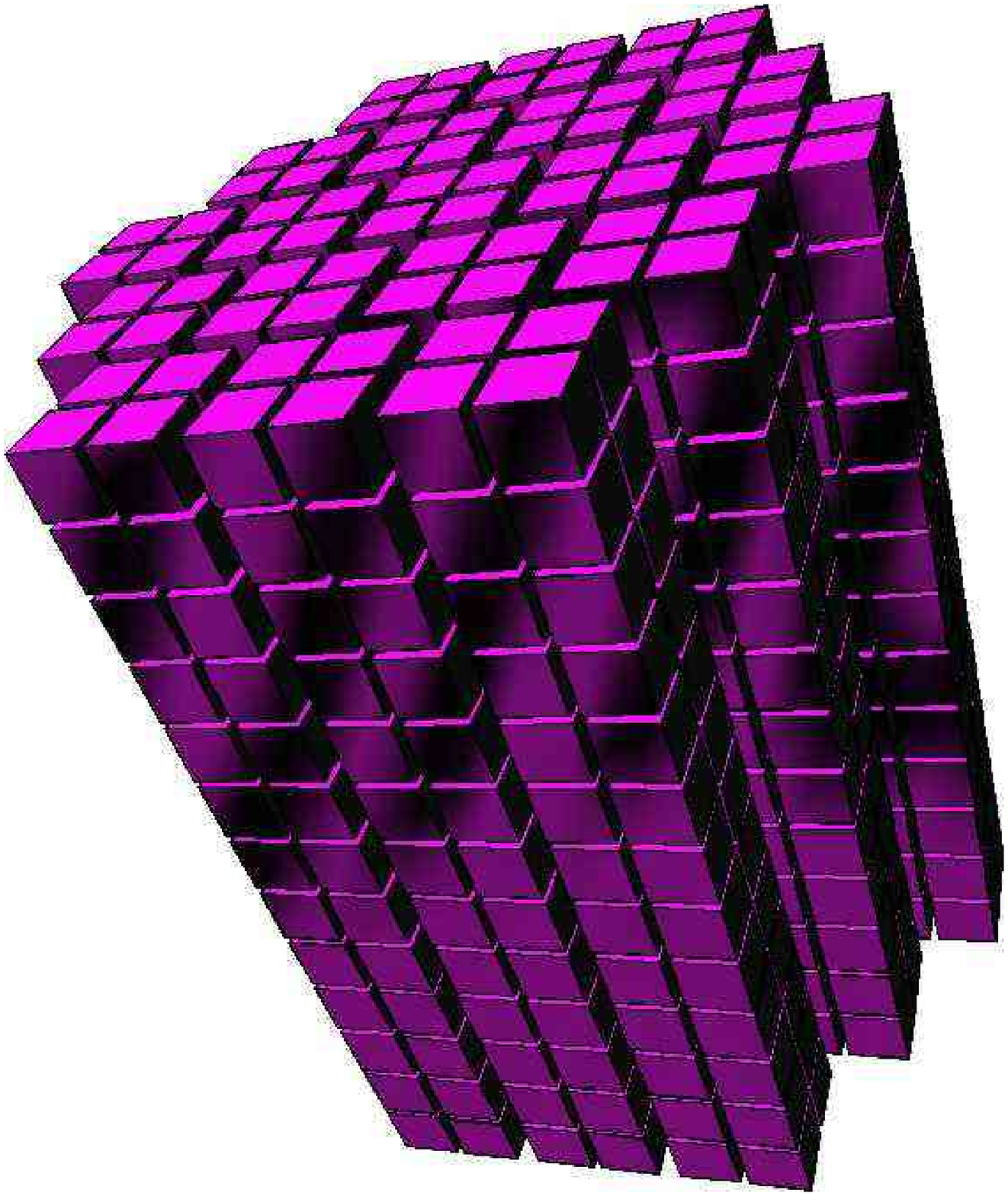}
 \includegraphics[width=0.3\textwidth]{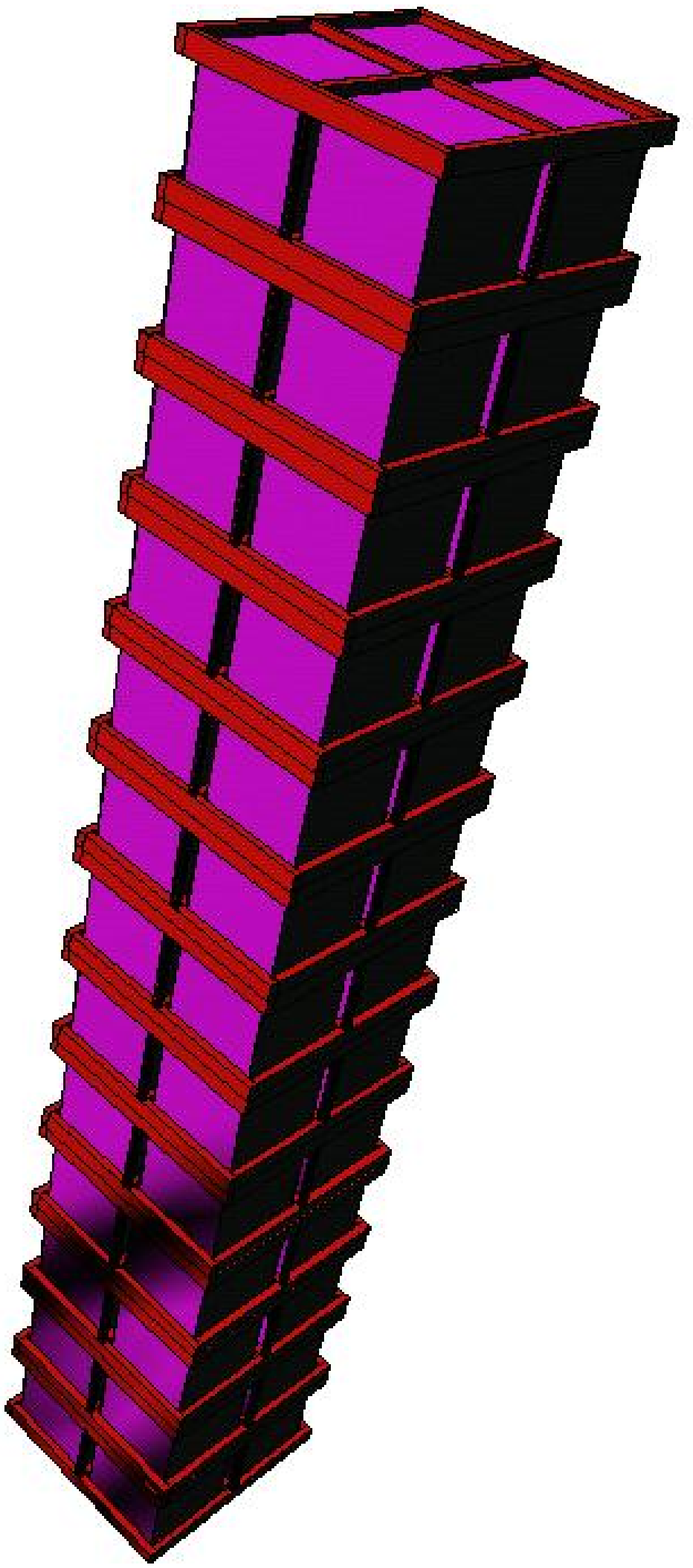}
 \end{center}
 \caption{The CUORE detector (left), one of the 19 towers (right).}
 \label{fig:cuore_cyl}
\end{figure}

The CUORE detector will consist of an array of 988 \teod bolometers arranged in a cylindrical configuration of 19 towers of 52 crystals each (Fig.~\ref{fig:cuore_cyl}). The principle of operation of these bolometers is now well understood (sec.~\ref{sec:bolometers}) and since \teod is a dielectric and diamagnetic material, according to the Debye Law (\ref{eq:Debye}), the heat capacity of a single crystal at sufficiently low temperature can be so low that also a small energy release in the crystal can result in a measurable temperature rise. 
This temperature change can be recorded with thermal sensors and in particular using Neutron Transmutation Doped (NTD) germanium thermistors. These devices were developed and produced at the Lawrence Berkeley National Laboratory (LBNL) and UC Berkeley Department of Material Science \cite{haller84}. They have been made uniquely uniform in their response and sensitivity by neutron exposure control with neutron absorbing foils accompanying the germanium in the reactor \cite{norman00}.

The \teod crystals are produced by the Shanghai Institute for Ceramics (SICCAS),
China and they will be the source of 750 g \teod crystals for CUORE \cite{alessandrello02}.

A single CUORE detector consists of a \ciccio single crystal of \teod that acts both as a detector and source. The detectors will be supported in a copper frame in the 19 tower configuration shown in Fig.~\ref{fig:cuore_cyl}.
The frame, and dilution refrigerator mixing chamber to which it is thermally connected, forms the heat sink, while the PTFE (Polytetrafluoroethylene or TEFLON) stand-offs provide the thermal impedance which delays the re-cooling of the bolometers. The bolometers operate at \ca~10 mK.

As described above (sec.~\ref{sec:CUORICINO}), the excellent performance of the \ciccio crystals to be used in CUORE has been already proved by CUORICINO~\cite{alessandrello02,alessandrello00} which corresponds to a tower of  length and number of floors slightly larger than in CUORE. We adopted this solution to make use of the full experimental volume available  in the dilution refrigerator previously used for the MiDBD experiment.

The CUORE detector then will be the final version of a three step development comprising: MiDBD, CUORICINO, and finally CUORE. This, plus the fact that CUORE requires no isotopic enrichment, (the isotopic abundance of  \tect is 33.8\%) puts CUORE ahead of other options of truly next generation \BBz experiments. The technology, though novel, is developed and to a large degree proven (sec.~\ref{sec:bolometers} and \ref{sec:MiCUORICINO}).

\subsection{Single CUORE detector} \label{sec:single}

A single CUORE detector is a \ciccio single crystal of \teod grown with ultrapure \teod powders and polished on the surfaces. Crystals of \teod have a tetragonal structure, and are grown along the (001) axis. The two axes normal to this axis are crystallographically equivalent, a fact relevant to their use in the search for solar axions discussed previously (sec:\ref{sec:axions}). The surface hardness is not the same for all sides, which complicates the crystal polishing. We have shown that repeated thermal cycling does not damage the crystals, as in the cases of crystals of other tellurium compounds, or those of tellurium metal.
The Debye temperature of the \teod crystals was specially measured for the CUORE project as 232 K \cite{barucci01}. This differs from the previous value of 272 K in the literature \cite{white00}. The specific heat of the \teod crystals was found to follow the Debye law down to 60 mK; the heat capacity of the 750 g crystals is  2.3\per 10$^{-9}$ J/K  extrapolated down to 10 mK.

The NTD thermistors are attached by epoxy to the crystal and are operated in the Variable Range Hopping (VRH) conduction regime with a Coulomb gap~\cite{giuliani93,mott69,shklovskii71}. One of the most significant parameters characterizing the thermal response is the sensitivity A, already introduced in section \ref{sec:bolometers} and defined in equation (\ref{eq:sens}).
For our NTD thermistors this parameter ranges between 7 and 10. The resistance behaviour  follows the relation:
\begin{equation}\label{eq:rdit}
    R = R_{0} \exp (T_{0}/T)^{\gamma}; \hskip1cm \gamma = 1/2.
\end{equation}
(see the analogous behavior for the resistivity in equation (\ref{eq:resistivity})).

The VRH regime occurs in Ge when it is ''doped'' close to the Metal to Insulator Transition, which is \ca~6\per 10$^{16}$ atoms / cm$^{3}$. This is achieved by thermal neutron radiation in a nuclear reactor (see section\ref{sec:ntd}). The sensitivity parameter A, depends on the neutron irradiation dose. Therefore, each thermistor must be characterized at operating temperatures, as described in Sec.~\ref{subsec:thermistors} and, for Si thermistors, in Ref.\cite{alessandrello99}.

\begin{figure}
 \begin{center}
 \includegraphics[width=0.85\textwidth]{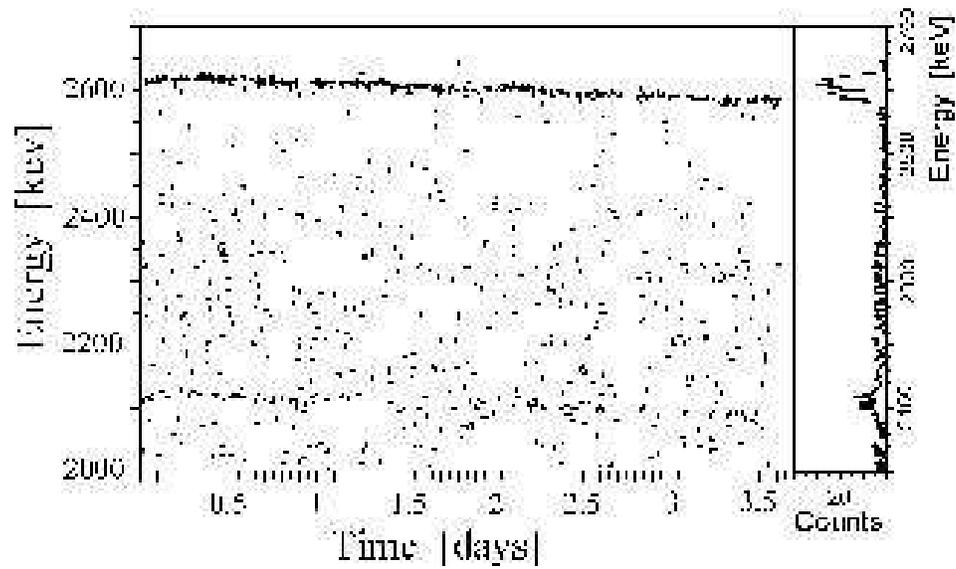}
 \vskip 2cm
 \includegraphics[width=0.85\textwidth]{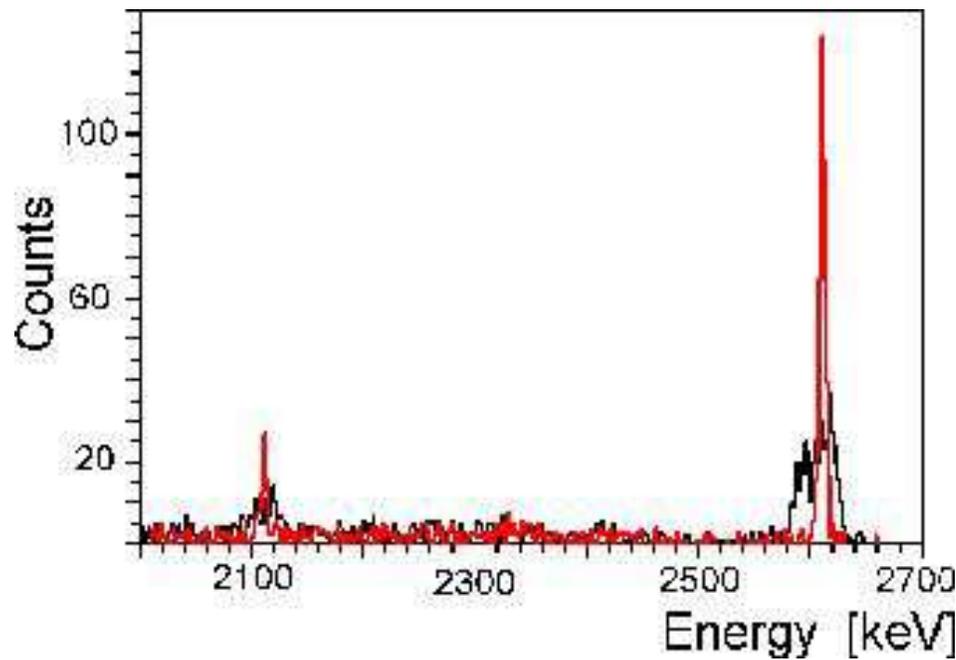}
 \end{center}
 \caption{Effects of the system thermal instabilities on the worsening of the detector energy resolution (above). Comparison between calibration spectra before (black curve: $\Delta E_{2615}^{FWHM}$ = 30~keV) and after (red curve: $\Delta E_{2615}^{FWHM}$ = 6~keV) the stabilization procedure.}
 \label{fig:stab}
\end{figure}

It is very important to optimize the neutron irradiation exposure and to make the exposures as uniform as possible. It is not possible to evaluate the thermistor material directly from the reactor because of the long half life of $^{71}$Ge  (11.43 days). A delay of several months is required to see if the Ge needs more exposure. To circumvent this difficulty, the Ge material is accompanied by foils of metal with long-lived $(n,\gamma)$ radioactive product nuclides. Accordingly, the neutron exposure of the Ge can be determined accurately, and uniformity of exposure is achieved. This technique was developed recently by the Lawrence Berkeley National Laboratory group of the CUORE Collaboration \cite{norman00}. Following the neutron exposure and radioactive decay period, the NTD germanium is first heat treated to repair the crystal structure then cut into 3\per 3\per 1 mm$^{3}$ strips.
The thermistors are glued to the \teod crystal by 9 spots of Araldit rapid, Ciba Geigy (now Novartis) epoxy, of 0.4 to 0.7 mm deposited on the crystal surface by an array of pins. The height of each spot is 50 $\mu$m. This procedure was found to be reasonably reliable and reproducible in the MiDBD experiment \cite{alessandrello00a}. The heat conductance of the epoxy spots was measured in Milan and the phenomenological relation was found to be \ca 2.6 \per 10$^{-4}$ (T[K])$^{3}$ watts per degree kelvin per spot.  The stabilization of the response of bolometers is crucial because of the unavoidable small variations in the temperature of the heat bath that change the detector gain (and consequently deteriorates the energy resolution; fig.~\ref{fig:stab}).

This problem is successfully addressed by means of a Joule heater glued on to each crystal. The heater is used to inject a uniform energy in the crystal and the thermal gain is monitored and corrected off-line (see Sec.~\ref{sec:analysis}).
The heaters are Si chips with a heavily doped meander structure with a constant resistance between 50 to  100 k$\Omega$. They are manufactured by the ITC - IRST company in Trento, Italy.
Electrical connections are made with two 50~$\mu$m diameter gold wires, ball bonded to metalized surfaces on the thermistor. The gold wires are crimped into a copper tube, which is inserted into a larger one forming the electrical connection, and avoiding low temperature solder which contains \pbdd and traces of other radioisotopes. The larger copper tube, \ca 14~mm  long and  2~mm  in diameter, is glued to the copper frame that supports the crystals. This tube is thermally connected to the frame but electrically insulated.
The mounting of the \teod crystals is crucial to detector performance, and must fulfill a number of sometimes contradictory criteria:
\begin{enumerate}
\item{the crystals must be rigidly secured to the frame to prevent power dissipation by friction caused by unavoidable vibrations, that can prevent the crystal from reaching the required temperature and can produce low frequency noise;}

\item{the thermal conductance to the heat sink (copper frame) must be low enough to delay the re-cooling of the crystal, following a heat pulse, such that the pulse decay time (re-cooling time) is much longer than the rise time;}

\item{however, the heat conductance must be high enough to guarantee efficient cooling;}

\item{the frame must compensate for the differences in thermal expansion coefficients of the various materials used;}

\item{and finally, only materials selected for low radioactivity can be used.}
\end{enumerate}
For CUORE, only two materials will be used, copper and PTFE; they can both be obtained with very low levels of radioactivity.
Copper has a thermal conductivity and specific heat high enough to be an ideal heat bath, and has excellent mechanical properties; it machines well, and has excellent tensil, tortional and compressional strengths.

PTFE is used between the copper frame and the crystals. It has low heat conductance and low heat leak \cite{pobell92}. It compensates for the differences between coefficients of thermal expansion of copper and of \teod. The heat conductance between the crystal and heat sink was measured in a similar configuration, but with a contact area larger by a factor of 2 \cite{alessandrello97}.

The measurements of the temperature dependence of specific heats and thermal conductivities is crucially important for developing a model that explains detector performance and that can be used to improve thedetector design (sec.\ref{sec:bolometers}).

\subsection{The NTD thermistors}\label{sec:ntd}

The key to the success of CUORE is the production of large numbers of nearly identical neutron-transmutation-doped (NTD) Ge thermistors. These are made by neutron transmutation doping of ultra-pure Ge in a nuclear reactor to obtain the proper characteristics of resistance and A, the variation of resistance with temperature. Melt-doped Ge crystals cannot achieve the necessary uniformity due to a variety of dopant segregation effects.  The only technique available for producing such uniform doping is NTD.  In typical applications, the neutron absorption probability for a 3 mm thick wafer of Ge is small, on the order of 3~\%, leading to a very homogenous, uniform absorption process.  The most important aspect of this process is that $^{70}$Ge transmutes into Ga, an acceptor, and  $^{74}$Ge transmutes into As, a donor, the primary active dopants in NTD Ge.  In this process, one places the Ge in a nuclear reactor where the following reactions take place:
\begin{eqnarray}\label{eq:ntdacc1}
^{70}Ge\;(21\%) + n & \to & ^{71}Ge\;  (\sigma_T = 3.43\pm0.17 b,\;\sigma_R = 1.5 b )\nonumber \\
^{71}Ge	 & \to & ^{71}Ga\;  (t_{1/2} =11.4 day) \hskip  1cm 		Acceptor
\end{eqnarray}
\begin{eqnarray}\label{eq:ntdacc2}
^{74}Ge\;(36\%) + n & \to & ^{75}Ge\;  (\sigma_T = 0.51\pm0.08 b,\;\sigma_R = 1.0\pm 0.2 b )\nonumber \\
^{75}Ge	 & \to & ^{75}As\;  (t_{1/2} =83 min) \hskip  1cm 		Donor
\end{eqnarray}
\begin{eqnarray}\label{eq:ntdacc3}
^{76}Ge\;(7.4\%) + n & \to & ^{77}Ge\;  (\sigma_T = 0.16\pm0.014 b,\;\sigma_R = 2.0\pm 0.35 b )\nonumber \\
^{77}Ge	 & \to & ^{77}Se\;  (t_{1/2} =38.8 hr) \hskip  1cm 		Double Donor
\end{eqnarray}

where $\sigma_T$ and $\sigma_R$ refer to the thermal and epithermal neutron capture cross sections, respectively.  

Since the doping level of the Ge needs to be on the order of $1\times10^{17}$ atoms/cm$^3$, a very high flux reactor, such as that at the University of Missouri or the Massachusetts Institute of Technology, is necessary to do the doping in a reasonable time. Even more important is the stability of the neutron flux and the energy distribution as measured by the Cd ratio.  A large Cd ratio (>200) is desirable.  In addition, a minimal fast (>5 MeV) flux is necessary to eliminate contaminants such as $^3$H, $^{65}$Zn, and $^{68}$Ge produced by fast neutrons. These radioactive contaminants degrade the detector operation because of internal radioactive decay and therefore increased background.

In order to determine both the thermal ($\Phi_T$) and epithermal ($\Phi_R$) neutron fluences that the Ge wafers are exposed to during the reactor irradiations, we developed a method using multiple metallic monitors that are irradiated with every set of Ge samples. For each monitor foil, the  amount of radioactivity produced depends on the sum ($\sigma_T\Phi_T+\sigma_R\Phi_R$). Each monitor provides one equation in two unknowns, the solution to which can be depicted as a line on the graph of thermal flux vs. epithermal flux.  For multiple elements, multiple lines can be drawn on this same graph, and they will have a common intersection at the solution of values of thermal and epithermal flux that satisfy all conditions.  Essentially, this is a graphical solution to several equations in two unknowns.  Elements with significantly different ratios of $\sigma_T/\sigma_R$ provide significantly different slopes to the lines and a better determination of the intersection.  Fe and Zr are a good combination since they have significantly different $\sigma_T/\sigma_R$ and long half-lives.

\begin{figure}
 \begin{center}
 \includegraphics[width=1.\textwidth]{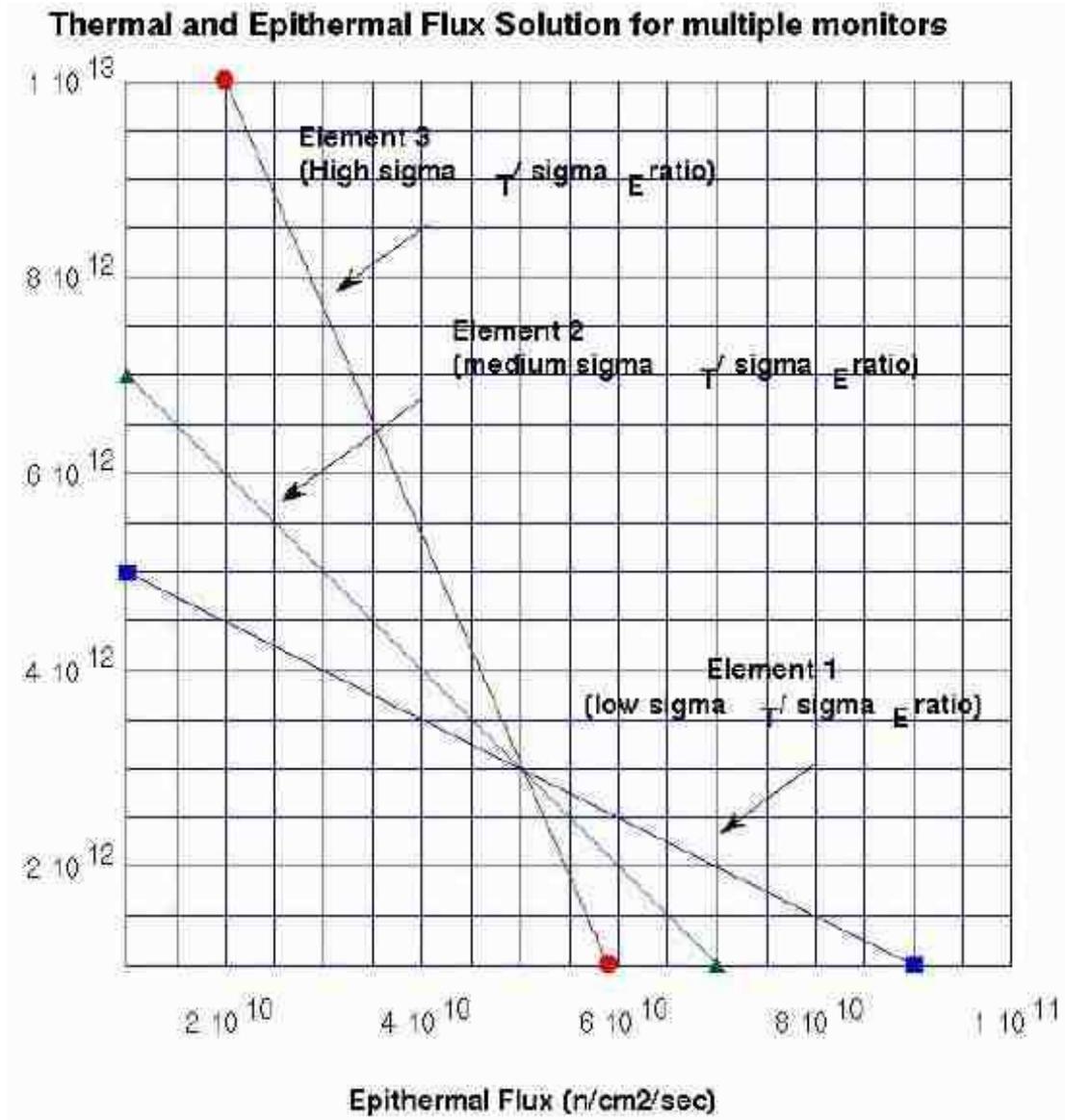}
 \end{center}
 \caption{Graph depicting the solution for thermal and epithermal flux for a set of three monitors, each with different ratios of thermal and epithermal cross-sections.}
 \label{fig:ntd1}
\end{figure}

\begin{figure}
 \begin{center}
 \includegraphics[width=0.9\textwidth]{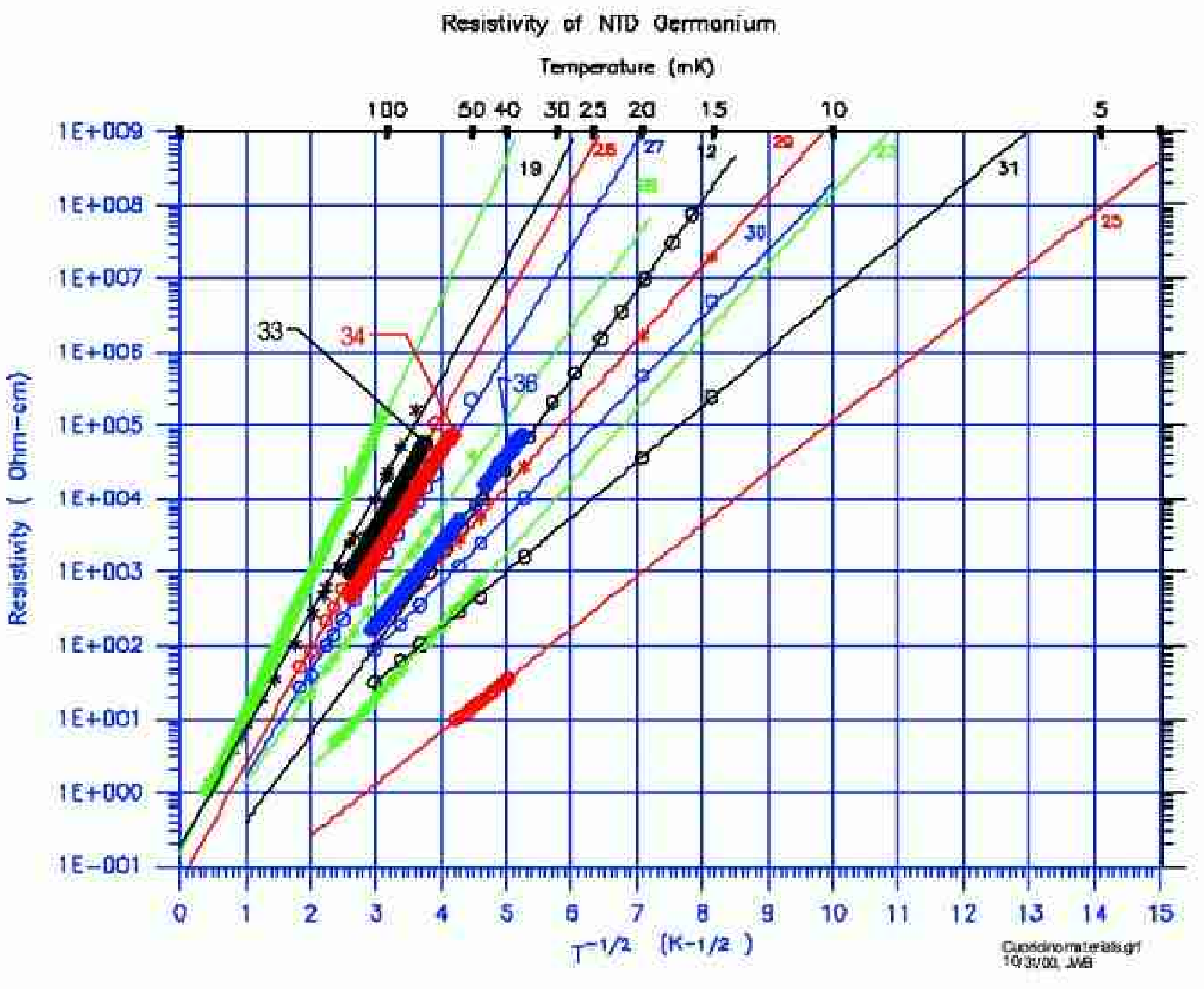}
 \end{center}
 \caption{Resistivity of various batches of NTD Germanium as a function of 1/$\sqrt{T}$.}
 \label{fig:ntd2}
\end{figure}

Once the thermal and epithermal fluences have been determined, the concentration of dopants produced in the germanium can be calculated.  The significant quantity involved in thermistor performance is the net dopant concentration which is equal to the concentration of Ga atoms minus the concentration of As atoms minus twice the concentration of Se atoms.   Unfortunately, to measure the thermal performance of the thermistor, one needs to wait for the decay of the activation product $^{71}$Ge (11.4 day), which requires approximately one year.  
Following this one year of decay, some of the NTD Ge material was made into thermistors and sent to the University of Florence, for testing at low temperatures.  Samples were cooled to 30 mK and resistance vs. temperature measurements were made over approximately the 300-30 mK range.  These curves were compared tos similar curves from other batches of NTD Ge as shown in  figure \ref{fig:ntd2}.

\begin{figure}
 \begin{center}
 \includegraphics[width=1\textwidth]{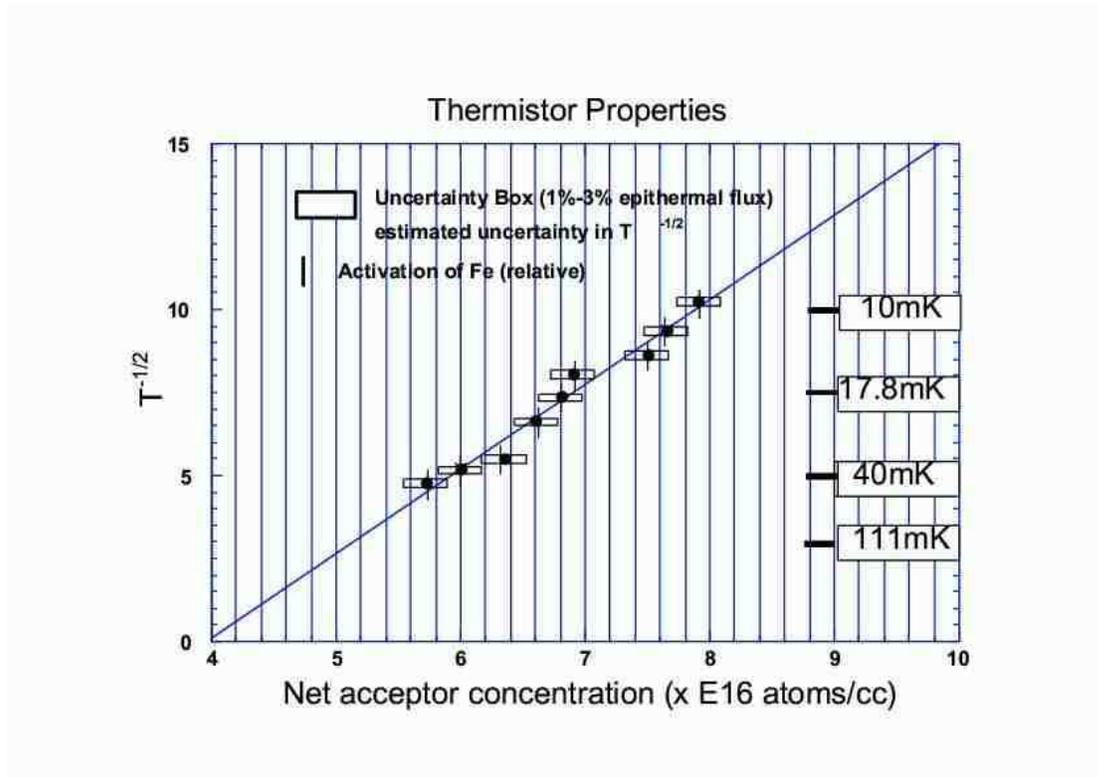}
 \end{center}
 \caption{Thermistor performance as a function of the calculated net acceptor concentration.  The first six points (from the left) involved only one irradiation (and show a relatively large spread in deviations from the line) while the last three points are following the second irradiation and have a very small spread from the line.}
 \label{fig:ntd3}
\end{figure}

For the various batches we ran (NTD-33-38) the data can be combined to show the variation of the temperature performance vs the calculated net acceptor concentrations.  These results are illustrated in Figure \ref{fig:ntd3}.  The three points on the right side of the data set were the results of two irradiations, and the middle of these three points has properties nearly the same at those of NTD-31, currently being used in CUORICINO.

The problem of producing large numbers of identical thermistors is quite complicated.  These thermistors, nearly identical to NTD-31, were made with an iterative process whereby the second doping was calculated following a measurement of the first.  This is a very inefficient process for producing large numbers of thermistors.  Fortunately, the same method used to make the thermistors can be used to analyze the results.  Neutron capture on $^{71}$Ga produces $^{72}$Ga (14 hour half life) and $^{75}$As activates to $^{76}$As (26.3 hour half life).  Both these isotopes emit gamma rays of high enough energy and intensity to make precise measurements of both Ga and As doping levels.  Selenium cannot be measured by Instrumental Neutron Activation Analysis (INAA) because the product nuclide, $^{77}$Se is stable. Similarly, Secondary Ion Mass Spectroscopy (SIMS) and other mass spectroscopic techniques can be used to determine the dopant concentrations on a much shorter time scale.  Two, or even three, irradiations will be required to produce thermistors with properties close enough to be used in the large array of CUORE.  These techniques will be further refined in order to allow faster and more uniform production of the thousands of thermistors required for CUORE.

The plan to make these irradiations more predictable is to use the NTD facility at the MIT reactor with multiple irradiations on a single large batch of Ge wafers.  Each wafer will have its own monitor, and each will be separately analyzed to determine the appropriate subsequent irradiation to bring them to the right temperature characteristics.

Previous measurements  carried out in the Gran Sasso Laboratory~\cite{alessandrello92} have shown that the residual acrtivity of the NTD thermistors becomes fully tolerable in an experiment with thermal detectors already a few months after irradiation.

\subsection{Modular structure of the CUORE detector}\label{sec:modular}

As stated earlier, the CUORE array comprises 988 \teod bolometers, grouped in 247 modules of 4 bolometers each.
These are arranged in 19 towers of 52 crystals each, or a stack 13 modules high. The towers are assembled in a cylindrical structure as shown in Fig.~\ref{fig:cuore_cyl}.

\begin{figure}
    \begin{minipage}[c]{1\textwidth}%
      \centering
      \includegraphics[width=0.5\textwidth]{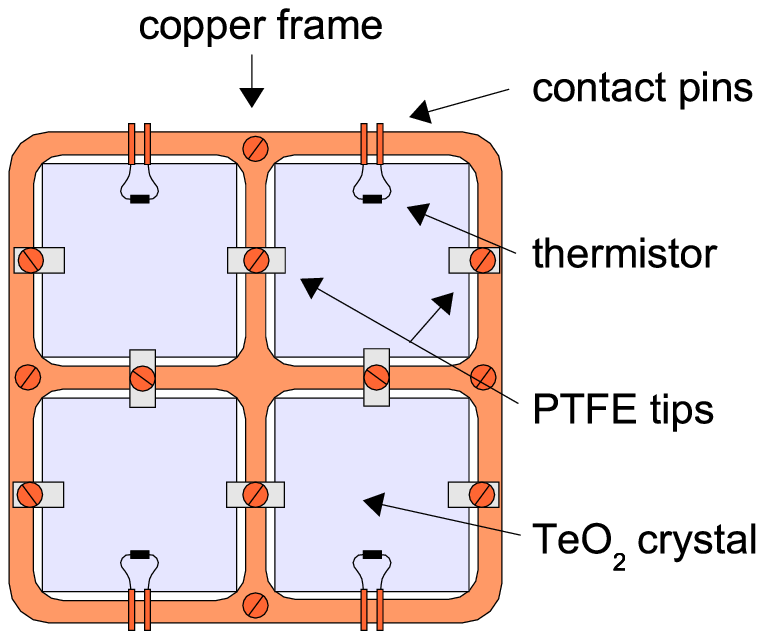}%
      \includegraphics[width=0.5\textwidth]{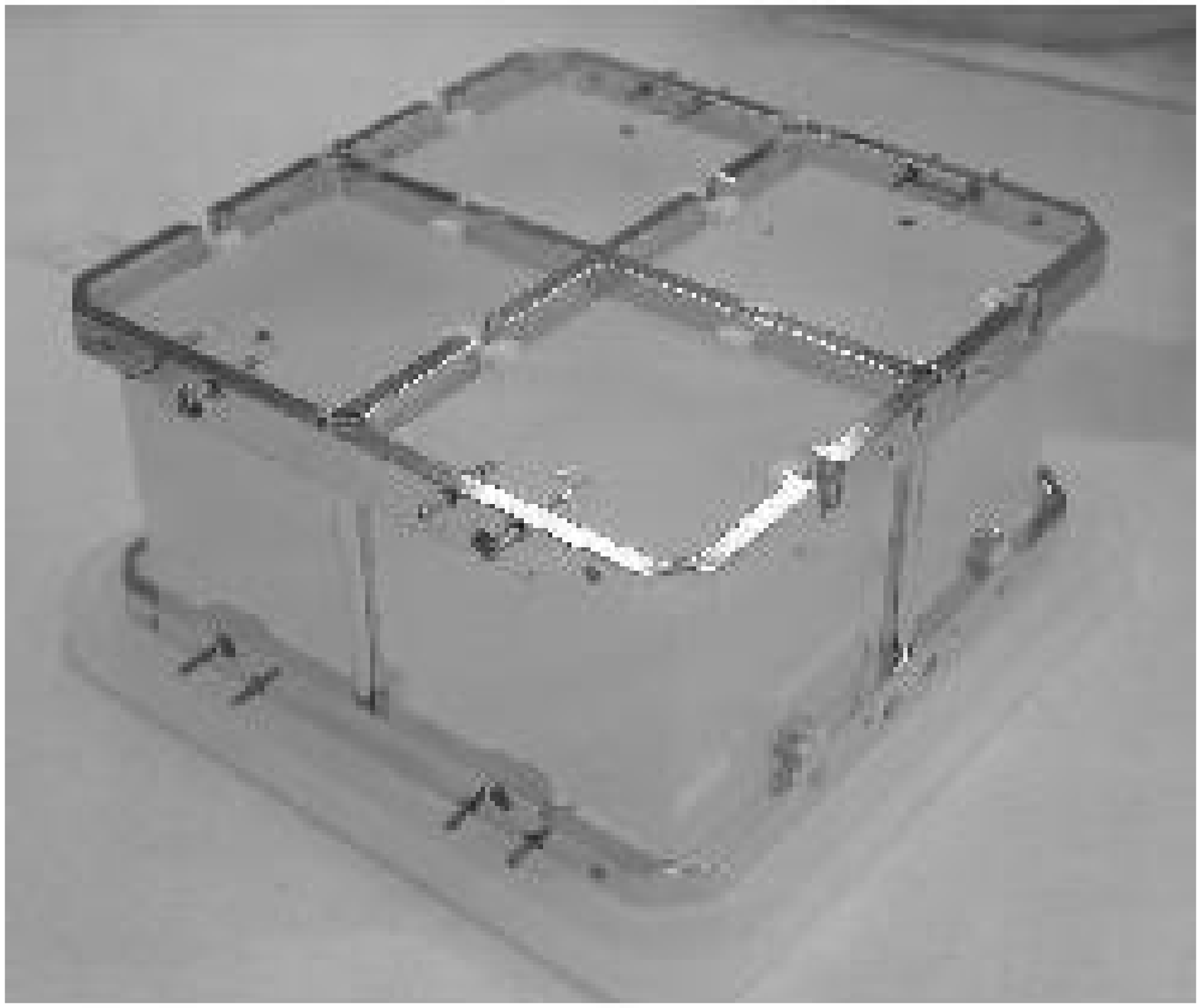}%
    \end{minipage}%
 \caption{A four detector module.}
\label{piano}
\end{figure}

The structure of each four-detector module is shown in Fig.~\ref{piano}. The four crystals are held between two copper frames joined by copper columns. PTFE pieces are inserted between the copper and \teod, as a heat impedance and to clamp the crystals.
There is a 6~mm gap between crystals with no material between them.

\begin{figure}
 \begin{center}
 \includegraphics[width=0.8\textwidth]{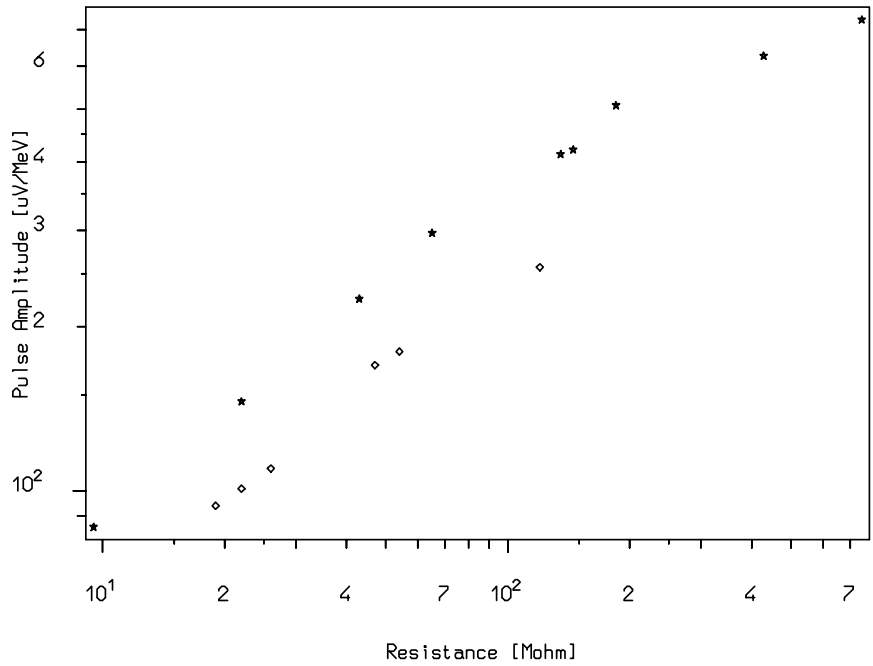}
 \end{center}
 \caption{Improvement in the Detector Merit Curve by increasing the thermistor
 size: 3\per3\per1 mm$^{3}$ thermistors (stars) vs. 3\per1.5\per0.4 mm$^{3}$ (diamonds).}
 \label{DMC-2}
\end{figure}

\begin{figure}
 \begin{center}
 \includegraphics[width=0.8\textwidth]{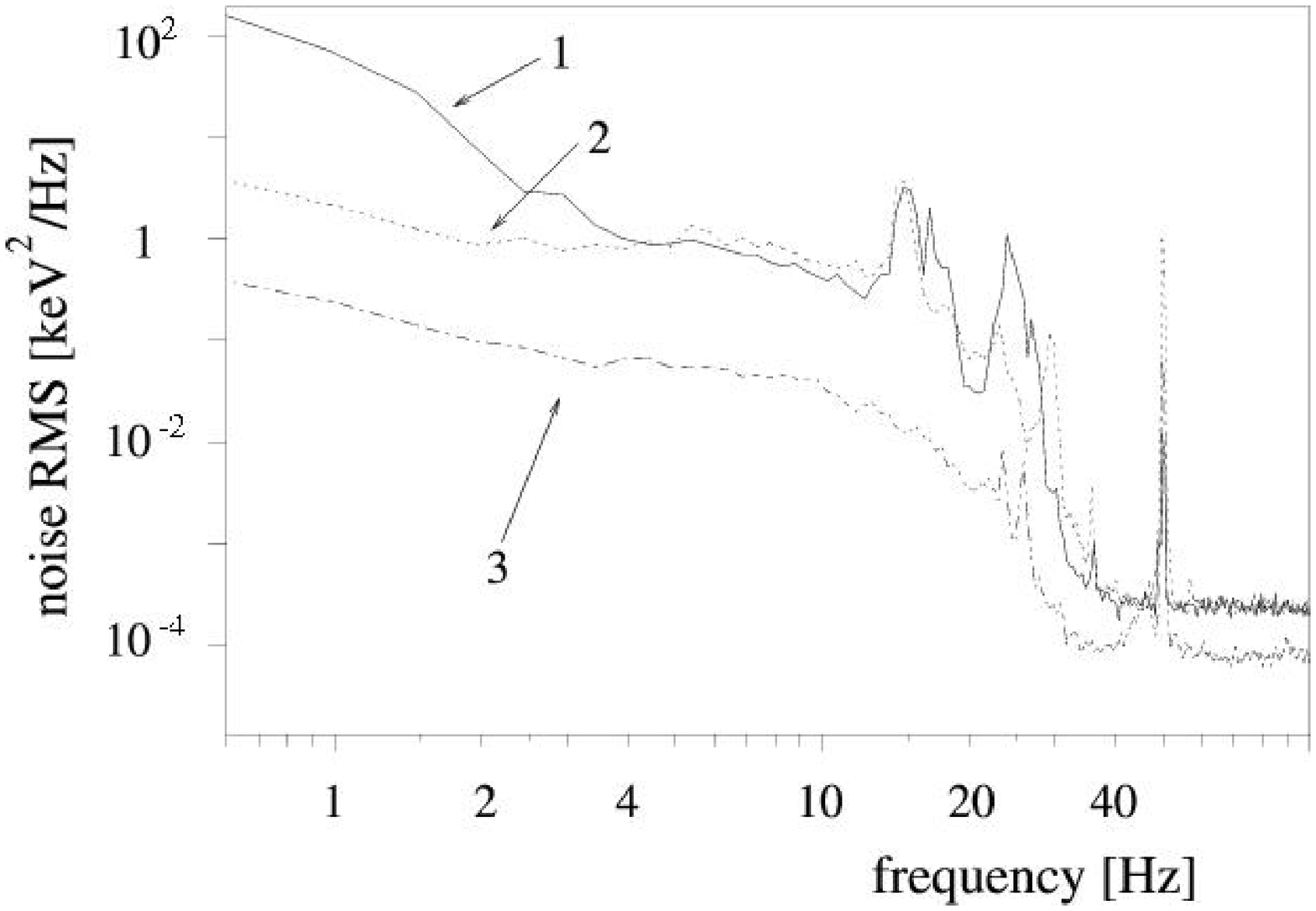}
 \end{center}
 \caption{Noise Power Spectra without damping suspension (1), with
 damping suspension only (2) and with dumping suspension + cold electronics (3).}
 \label{noise}
\end{figure}

The four detectors are mechanically coupled; some of the PTFE blocks and springs act simultaneously on two crystals. Tests of this and a similar structure, but with \magro crystals, clearly demonstrate that the CUORE technology is viable. The signal to noise ratio was improved by replacing the 3\per1.5\per0.4 mm$^{3}$  Ge thermistor by one with dimensions 3\per3\per1 mm$^{3} $, which significantly improved the DMC shown in Fig.~\ref{DMC-2}, and by utilizing cold electronics Fig.~\ref{noise}. A further noise reduction was finally obtained by suspending the detector by means of a stainless steel spring.

\begin{figure}
 \begin{center}
 \includegraphics[width=0.8\textwidth]{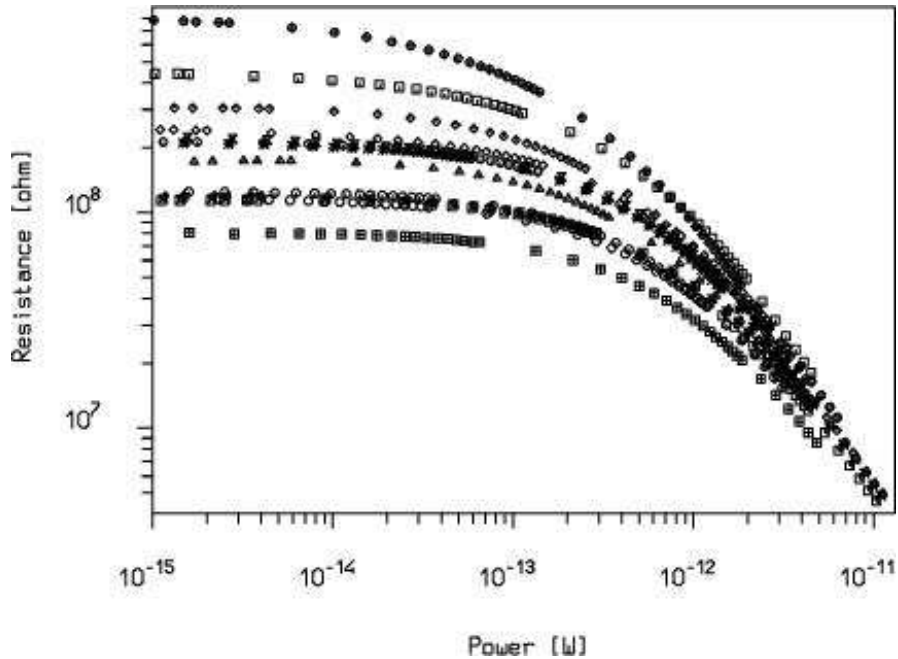}
 \end{center}
 \caption{Fifteen thermistor load curves for the 20 crystal array.}
 \label{load_curve}
\end{figure}

Reproducibility was tested on the MiDBD array, which had a sufficient number of detectors (20) operating simultaneously. The reproducibility is satisfactory as shown in the load curves in Fig.~\ref{load_curve}. Four-detector modules of 50\per50\per50 mm$^{3}$ crystals were also successfully tested, with pulse amplitudes spanning an interval from 50 to 150 $\mu$V/MeV at \ca 100~M$\Omega $ operation point, in agreement with the thermal model. The full width at half maximum FWHM is \ca 1~keV for low energy gamma peaks, and \ca 5 to 10~keV  at 2.6 MeV.
A stack of thirteen, 4-detector modules, will be connected by two vertical copper bars. The details of the wiring of the towers is still under development; however, the current plan is for 50 $\mu$m diameter twisted pairs of nylon coated constantane wire running along the vertical copper bars. On the top of each tower there would be 52 twisted pairs for the thermistors and four for the heaters. Two more twisted pairs will be required for diagnostic thermometers, one on top and one on the bottom of the tower.

The CUORICINO detector presently taking data at LNGS consists of one of the 52 detector towers with the exception of the two planes containing the \magro crystals discussed earlier. CUORE is an array of 19 towers, each of which will be suspended independently from a large cylindrical copper plate, thermally connected to the mixing chamber of the dilution refrigerator (DR).
In this manner, tests of CUORICINO will be tests of the CUORE array. The array will be suspended from the mixing chamber by a vertical spring to decouple the detector from the vibrations of the dilution refrigerator. Tests were performed with piezo-electric accelerometers and vibrators to optimize the suspension (see Fig.~\ref{noise}). As a result a pendulum mechanical suspension was designed for the entire CUORE array.

The heavy shielding of CUORE will present non-trivial problems in the energy calibration with radioactive sources. The presently considered option is to use radioactive metal wires encapsulated in PTFE tubes. The tubes would be placed vertically between the towers, and free to slide along a fixed tube to the appropriate calibration point. A vacuum tight sliding seal will allow the part of the tube containing wire to be inserted and extracted from the cryostat.

\subsection{Location}
\begin{figure}
 \begin{center}
 \includegraphics[width=0.95\textwidth]{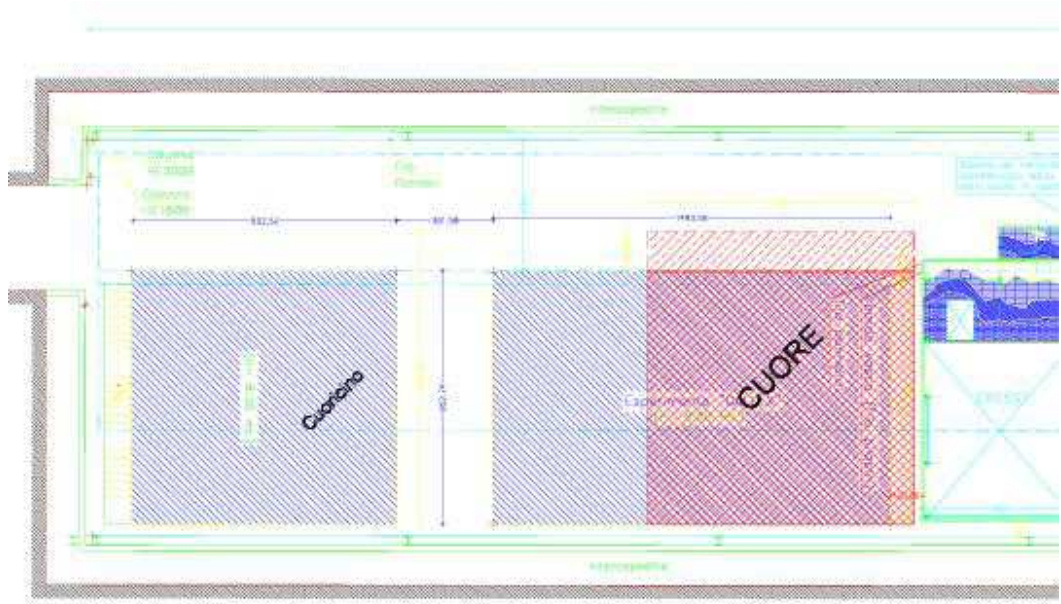}
 \includegraphics[width=0.95\textwidth]{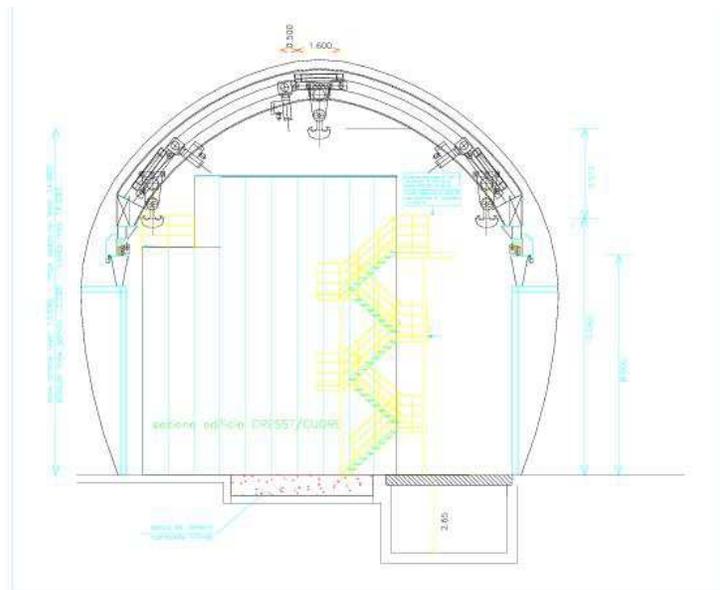}
 \end{center}
 \caption{Final location  of CUORE in hall A of the underground Laboratori Nazionali del Gran Sasso.}
 \label{fig:qlngs}
\end{figure}

CUORE will be located in the underground hall A of Laboratori Nazionali del Gran Sasso (LNGS, L'Aquila - Italy) at a depth of 3400 m.w.e. where the muon flux is reduced to \ca 3\per10$^{-8}$ $\mu$/cm$^{-2}$/s and the neutron flux to \ca 10$^{-6}$ n/cm$^{-2}$/s. After the approval of CUORE by the Gran Sasso Scientific Committee its final location just near the CRESST installation has been decided as shown in  Fig.~\ref{fig:qlngs}

\clearpage
\section{Cryogenic system} \label{sec:cryogenic}

The CUORE bolometers will operate at temperatures between 7 and 10~mK. This will require an extremely powerful dilution refrigerator (DR). At these temperatures, the cooling power of a DR varies approximately as T$^{2}$. Estimates were made of the parasitic power the detector and DR would receive from: heat transfer of the residual helium gas in the inner vacuum chamber (IVC), power radiated from the 50 $mK$ shield facing the detector, and from vibrational energy (microphonic noise). The estimated value is \ca 1~$\mu$ W  at 7 mK, using reasonable values for the residual gas pressure and achievable surface quality for radiation transfer. The resulting estimate for the radiation contribution was negligible. CUORE will utilize a similar system to that of the Nautilus Collaboration that cools a 2-ton gravitational antenna. That system experienced a parasitic power of 10$\mu$W  from unknown sources.

\begin{figure}
 \begin{center}
 \includegraphics[width=0.8\textwidth]{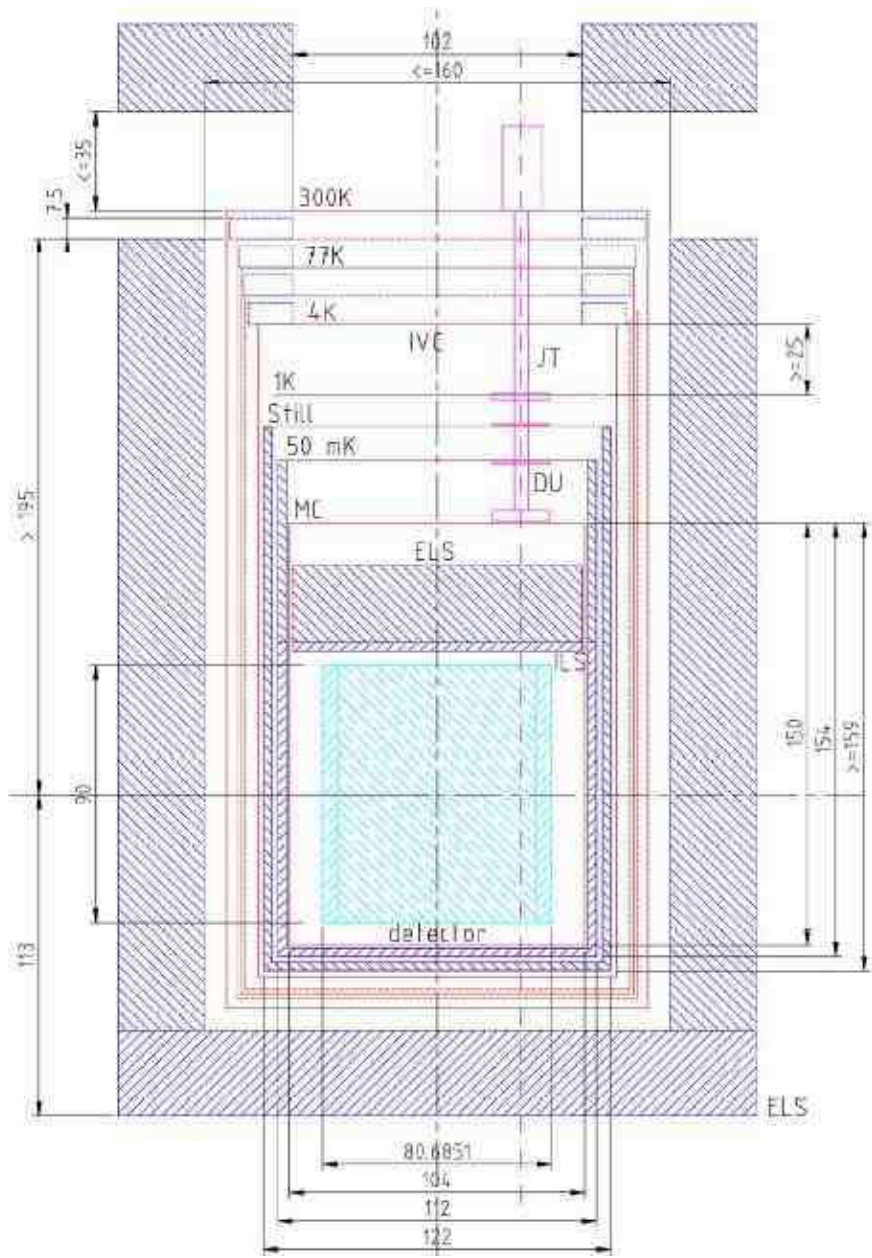}
 \end{center}
 \caption{CUORE cryostat and shielding.}
 \label{fig:dewar}
\end{figure}

The CUORE detector will be cooled by a $He^3/He^4$ refrigerator with a cooling power of 3~mW  at  120~mK. Refrigerators with the required characteristics are technically feasible. One example is the DRS-3000 DR model constructed by the Kamerling Onnes Laboratory in Leiden. 
The unit is shown in Fig.~\ref{fig:dewar}, inserted in the dewar. 
One important design feature is the 50~mm diameter clear access to the mixing chamber to allow a rod, suspended from an external structure, to suspend the detector array to minimize vibrations from direct connection to the mixing chamber. The temperature of the rod will be controlled in stages along its length by flexible thermal contacts to avoid vibrations.

The dewar housing the DR will have a jacket of liquid nitrogen (LN) to avoid the need of superinsulation which is not free of radioactivity. The system is designed with several tubes passing through the LN bath, the liquid He bath, and the IVC, to allow refilling with cryogenic liquids, and for sufficient feed through for electrical connections. 
Liquefiers will provide constant liquid levels for the two baths over long periods of operation. The He that evaporates from the main bath and from the 1K-pot will be recovered in a buffer tank and re-liquefied with a helium liquefier with a capacity of several tens of l/day. The gas will be cooled to \ca 70 K, then to 10~K, by a two stage Gifford-MacMahon cycle. This will be followed by a Joule-Thompson cycle that liquefies it and injects it back into the main bath.
The nitrogen liquefier, on the other hand, will require only a single stage Gifford-MacMahon cycle to cool a cold finger above the LN bath. It will condense the evaporated N$_{2}$ and let it fall back into the bath.

This complex cryogenic system will require a constant monitoring system. It must be capable of surveying the operating conditions to allow prompt intervention to correct malfunctions rapidly. In addition, the operating conditions of the entire cryogenic system must be recorded for future correlation with instabilities in the experimental data (section \ref{sec:slow_control}).

Besides housing the CUORE detector, the dewar will be able to support an effective system of roman lead shields maintained at the low temperature close to the detectors (Fig.\ref{fig:dewar} and section \ref{sec:shields}).

\subsection {Development of a Pulse-Tube-assisted cryostat}
\label{subsec:PT} 
\begin{figure}
 \begin{center}
 \includegraphics[width=0.8\textwidth]{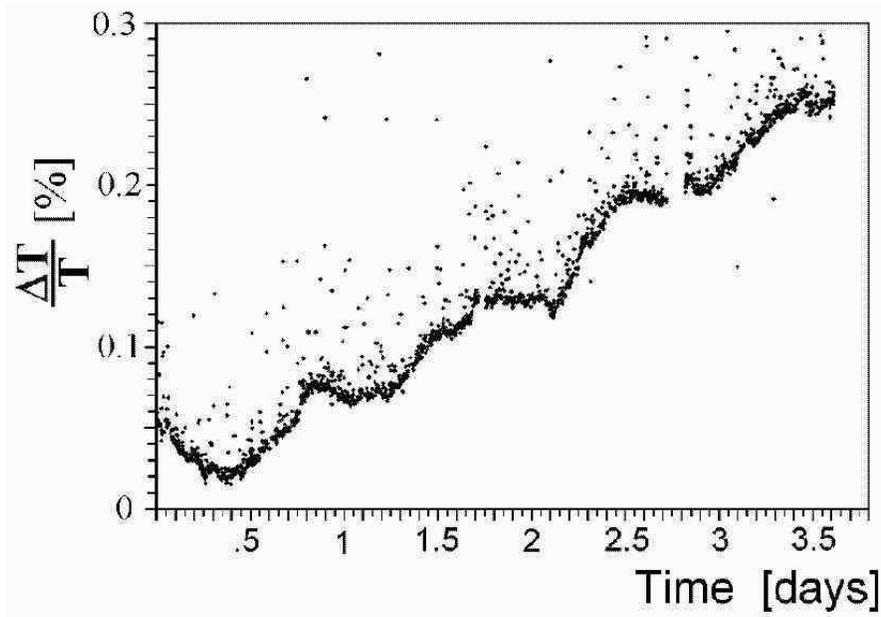}
 \end{center}
 \caption{Temperature instability as observed in a typical MiDBD measurement.}
 \label{fig:T_inst}
\end{figure}

The experience collected by the proponents in running CUORICINO, and by the Milano group in running the previous 20~detector arrays, shows that a constant (or very slowing varying, of the order of 1~\%/day) level of liquid helium in the main bath of the cryostat helps in achieving a steady response of the detectors. Any variation of the main bath level translates into small changes of the mixture flow and therefore of the refrigerator cooling power. These changes are harmful for the stability of the bolometer output and could spoil the effectiveness of the correction procedure (usually referred as ``stabilization''). Detectors need to be operated in a DC-coupling mode because their baseline level reflects the temperature changes and is used for the detector response \emph{stabilization} (Fig.~\ref{fig:T_inst} and section \ref{sec:analysis}). A fundamental point is that the stabilization procedure is effective only if the detector operation point (and therefore the baseline level) is kept inside a narrow environment of the initial value. This cannot be achieved in presence of relatively large or fast changes of the liquid helium level.

The basic design described above for CUORE, is based on the successful operation of CUORICINO and MiDBD in which the helium level was kept almost steady by means of a helium re-liquefier. This system is however based on a twenty-year-old technology and has exhibited many weak points during the multi-annual operation of the bolometric arrays, such as unpredictable variations of the liquefaction efficiency and microphonic noise introduced by the cryogenic heads of the GM cycle. We believe that these problems could be even more serious in the CUORE cryogenic system, due to the much larger boil-off of the CUORE cryostat (see section \ref{sec:cryogenic}). For this reason we will also consider the opportunity to rely on a traditional liquefier which liquefies externally the helium gas coming from the main bath and the 1~K~pot through recovery lines. In this case, an active stabilization of the holder temperature should compensate for the base temperature variations induced by the relatively fast decrease of the liquid level and the traumatic changes induced by the frequent main bath refills.

However, we will study (at the R\&D level) an alternative cryogenic configuration which could allow us to keep the 4~K stage temperature constant for the whole duration of the experiment. This solution is based on a new commercial device provided by the modern cryogenic technology: a Pulse Tube (PT) cooler. PT is a novel type of mechanical cooling system exhibiting extremely low vibration performance not reached by conventional Gifford-McMahon or Stirling coolers. PT coolers consist of a compressor, a rotary valve and a cold head. Their advantages are:
\begin{itemize}
\item {low maintenance, due to the absence of moving displacers and cold seals;} 
\item {low noise, due to the lack of moving parts in the cold head, which assures a very quiet system;} \item {low cost, due to its simplicity, which allows the PT to be manufactured at less cost than common cryo-refrigeration systems.}
\end{itemize}

Commercial two stage PT's provide two cold points, typically at 80~K and 4~K, with a cooling power of the order of 0.5~W at 4~K. In this framework, a liquid-helium-free medium-power dilution refrigerator was installed in September 2003 at the Insubria University. This refrigerator does not use cryogenic liquids at all: a PT cools the 80~K and the 4~K shields, while the $^3$He-$^4$He mixture is condensed by a Joule-Thompson expansion after thermalization at the 4~K stage. This refrigerator will be a precious test-bench for this new technique, in terms of reliability and noise level. In the event of convincing performance of this system, we will design a refrigerator for CUORE endowed with an array of PT's with the aim to stop or reduce the main bath boil-off or, as an extreme solution, to get rid completely of cryogenic fluids. This solution would solve much of the stability problem at the root.

\subsection{Shielding requirements}\label{sec:shields}

Double-beta decay experiments, as well as dark matter searches and searches for rare events in general, require deep underground locations and ultra low radioactive environments.
This latter requirement is usually accomplished by placing severe constraints on the selection of the materials used and by the realization of proper shielding surrounding the detectors. In the case that complicated structures like dilution refrigerators are used, the problem of their shielding and construction using only low radioactive contamination materials is complex. 
Moreover, since in MiDBD and CUORICINO a non negligible contribution to the background rate in the \BBz region was recognized as originating from environmental radioactivity, and radioactive contaminations of the cryostat structure (sections~\ref{sec:MiCUORICINO} and \ref{sec:simulation}), the design of effective shields (specially those directly surrounding the detector) is crucial to reach the sensitivity goal of CUORE.

Part of the bulk lead shielding will be placed inside of the cryostat, and part outside (Fig.~\ref{fig:dewar}). This will shield the detector from environmental radioactivity and from radioactive contaminations of the dewar structure, also reducing the total amount of lead required. A $ 4\pi $ layer of ultra-low background lead will constitute 3~cm thick walls of the cubic structure of the array. This layer will be Roman lead whose \pbdd activity was measured to be less than  4~mBq/kg \cite{alessandrello98a}.
As for CUORICINO, the dilution refrigerator will be constructed from materials specially selected for low levels of radioactivity (section~\ref{sec:simulation}). Nevertheless, these levels might be higher than can be tolerated by the sensitivity requirements. The main goal of the inner lead shield will be therefore of cancelling their dangerous contributions to the CUORE background level. 

The top of the detector array will be protected by two Pb layers of \ca 1~m diameter and 10~cm thickness each, with a 10~cm diameter central bore to accommodate the copper cold finger that supports the detector and the narrow neck of two radiation shields of the refrigerator that are at temperatures of 50~mK and 600~mK. The layer close to the detector will be made of high quality lead with an activity of 16~Bq/kg of \pbddn. The upper layer will be made of modern lead with an activity of \pbdd of 150~Bq/kg.  Another layer of lead, 17~cm thick, and with a dimaeter of \ca 40~cm will be placed directly on the top face of the detector.
It will be constructed from low activity lead of 16~Bq/kg of \pbddn.
This configuration is designed so that the minimum path to the detector from the IVC and the dilution unit is 20~cm of lead. Finally, outside the dewar, there will be two 10~cm thicknesses of lead, 16~Bq/kg of \pbdd for the inner layer, and 200~Bq/kg for the outer layer.
The lead shield will be surrounded with a 10~cm thick box of borated polyethylene that will also function as an hermetically sealed enclosure to exclude radon. It will be flushed constantly with dry nitrogen.

The entire dewar, detector, and shield will be enclosed in a Faraday cage to exclude electromagnetic disturbances that also constitute a source of background, albeit at low energies, important to dark matter and solar axion searches.
The addition of a muon veto surrounding the entire structure will be also considered.

\clearpage
\section{Required laboratory} \label{sec:hut}

The entire CUORE setup will be installed underground inside a proper building which will have to guarantee room for all the setup parts (e.g. cryogenics, electronics, shieldings), for a controlled area to be used during assembling procedures and for all normal monitoring activities.

The hut in which the CUORE experiment will be housed will have therefore to satisfy the following main requirements:
\begin{itemize}
\item {the cryostat and the detector have to be sustained by means of two independent structures, in order to avoid any possible transmission of mechanical vibrations to the bolometers from the cryogenic facility.}

\item {the dilution refrigerator has to be kept at a suitable height from the floor to give the possibility of an easy dismounting of the thermal and radioactivity shields when necessary.}

\item {the area where the cryostat and the front-end electronics are located has to be shielded from electromagnetic interferences by means of a Faraday cage.}

\item {the area where the detector will be assembled must be a
clean room in order to avoid radioactive contamination of the
array in the mounting phase.}
\end{itemize}

\begin{figure}
 \begin{center}
\ifx\pdfoutput\undefined 
 \includegraphics[width=0.9\textwidth]{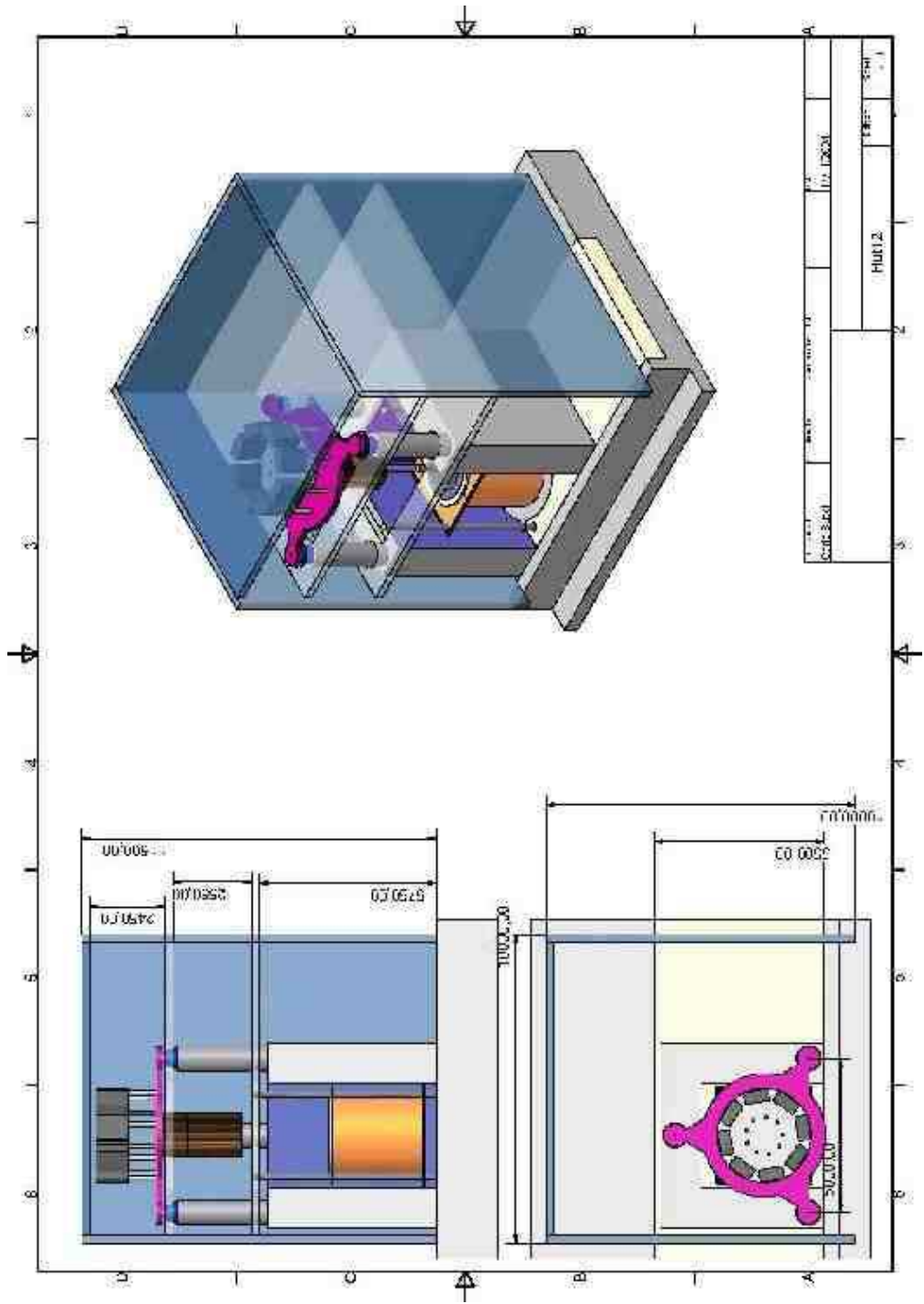}
\else
 \includegraphics[width=1.4\textwidth,angle=90]{Pictures/Qhut}
\fi
 \end{center}
 \caption{Scheme of the CUORE setup installation.}
 \label{fig:CUORE_cryo}
\end{figure}

Following these requirements we made a schematic design of the hut which will be divided in three levels as shown in Fig.~\ref{fig:CUORE_cryo}. At the ground floor a platform of reinforced concrete, with a central aperture of about 3 m diameter, will be placed. In this aperture a lifting platform, able to raise at least 50~tons, will be realized with four couples screw-nut, driven by four motors simultaneously controlled. The radioactivity shielding will be placed on the lifting platform; in this way it will be easy to pull down the shieldings whenever is necessary to open the cryostat. On the ground level, the pumps of the cryostat and the compressors for the liquefiers (or for the pulse tubes) will also be placed.

The dilution refrigerator will be suspended at about 3.5~m from the ground in order to guarantee an easy dismounting of the thermal shields (Fig.~\ref{fig:CUORE_cryo}).
The concrete platform at the ground floor will be the base for two sets of three pillars each: the first set will go up to the floor of the second level and will support a plate to which the cryostat top flange will be anchored; the second set will go up to the ceiling of the second level and will bear the support
structure of the detector. An appropriate design for the mechanical suspension for the CUORE array is under study. We have however seen that it is relatively easy to realize a suitable device since the only unavoidable requirement is the achievement of a low natural vibration frequency of the system (below 0.5 Hz).

The first level will consist of two separated sections: a clean room (at least class 100), where all the CUORE detector parts will be assembled and stored, and a control room  for the cryostat.

The second level will also be separated in two sections. The first one,  containing the top of the cryostat, the support structure and the front-end electronics, will be surrounded by a Faraday cage. The second will be used for the data acquisition system.

\subsection {Cooling water requirements}\label{subsec:cooling} 
Depending on the cryostat design, the thermal power required to cool the pumps and compressors can vary between 40 and 60~kW. The water cooling circuit of the LNGS underground laboratories could in principle easily accomplish to our needs but our experience in past years tells us that this circuit has to reach better reliability and purity of the water. Therefore we will install a water chiller of adequate cooling power that in the future can work as a backup solution to cover possible failures of the laboratory circuit.

\subsection {Power requirements}\label{subsec:power}

In order to guarantee a continuous operation of the experiment, all the equipment has to be powered by an Uninterrupted Power Supply (UPS) that should have a power in the range 70--100~kVA, depending on the setup design. Only the water chiller can be powered by a normal power supply, provided that water circulation pumps are under UPS and the water circuit has a buffer big enough to resist to power interruptions for the same amount of time the UPS does.

\subsection {Safety}\label{subsec:safety}
The CUORE experiment doesn't use dangerous materials and the only safety requirements concern therefore the use of cryogenic liquids. Furthermore the amount of liquid nitrogen and liquid helium contained in the cryostat will be of the order of 100 liters and then, also in case of an abrupt evaporation the amount of gas released to air is limited to few hundreds of cubic meters.
Of course, in the case of a cryostat using pulse tubes (instead of cryogenic liquids) this issue would also disappear.

\subsection {Space requirement}\label{subsec:space} 
The space required to allocate CUORE is 10\per10 m$^{2}$ and we consider it sufficient for the installation of th experiment. Moreover a few square meters outside the hut are required in order to accomodate the water chiller (it should be placed in a bypass outside the main hall not too far from the hut) and the ventilation system of the clean room (close to the
hut).

\clearpage
\section{Electronics} \label{sec:electronics}

Although CUORICINO has many fewer channels with respect CUORE, the design of its electronics has been developed with the aim of fully satisfying the requirements (electrical specification, space occupation, cost, etc.) of CUORE. Therefore, the description of the front-end set-up we are proposing for CUORE is very close to the actual front-end of CUORICINO. Very few modifications are needed to upgrade the set-up of CUORICINO to that of CUORE.

\subsection{General Description}

With the CUORICINO experiment two front-end readouts are being investigated:  a configuration consisting of 38 channels completely operated at room temperature (``warm electronics'': WFE) and a system of 24 channels characterized by an additional first differential unity gain buffer stage at low temperature (``cold electronics'': CFE).

The connection between the thermistors and the first stage of the electronics is via a twisted pair of 1~m (CFE) and 5~m (WFE) in length. The main reason for these two prototypes is to assess the effect of the microphonic noise of the connecting wires on the detector energy resolution at low energy, in order to achieve the maximum possible sensitivity at threshold. Since other factors dominate the detector resolution at higher energies, an equivalent performance is expected there for the two set-ups.

Regardless of the configuration considered, the design of the electronic system must satisfy many specifications. Among them, the four major issues can be summarized as follows:
\begin{enumerate}
\item{It should be characterized by a small noise contribution, in the frequency band of the detector signal (\ca one to few tens Hz), generated by the biasing circuitry and the preamplifier.}
\item{It should accommodate the broad spread of manufacturing characteristics that is typical of the bolometer/thermistor modules. In CUORICINO, for example, the optimum bias current of some modules is as low as 150~pA, while is as high as 300~pA for others. As a consequence the signal amplitude exhibits a corresponding spread.}
\item{It should enable in situ measurement of the bolometer DC characteristics.}
\item{The design should incorporate a high level of remote programmability in order to avoid mechanical interventions in the vicinity of the bolometers during the measurements.}
\end{enumerate}

In the following subsections we shall give a brief discussion of the circuitries able to address these issues and presently used for the CUORICINO readout. An extended description of the CUORICINO electronics can be found in reference~\cite{arnaboldi02}.

Among some of the important tasks addressed by the readout electronics which are not reported here, a particular attention was given to the setup power supply and the shielding against the electromagnetic interferences. A very precise, multi-output, pulse generator instrument was finally developed to allow a continuous calibration of the energy conversion gain of the whole array (\emph{stabilization}).

\subsection{Thermistor biasing and detector wiring}

\begin{figure}
 \begin{center}
 \includegraphics[width=0.95\textwidth]{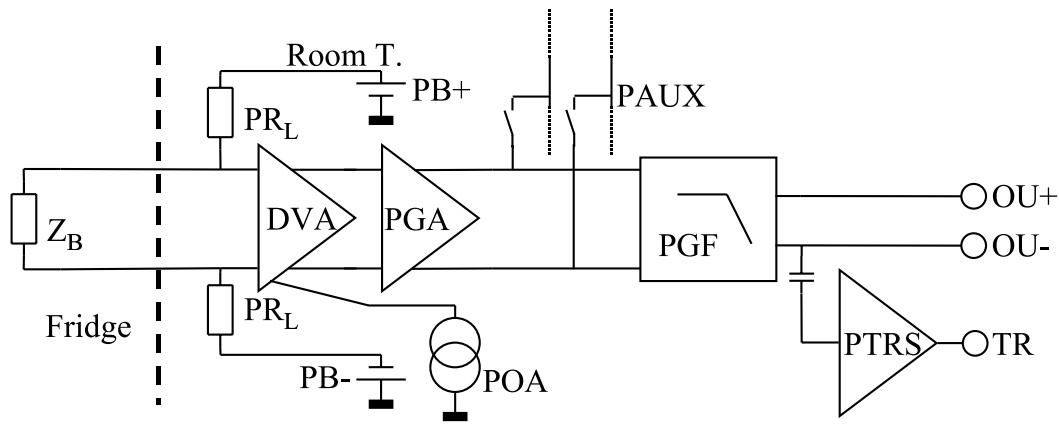}
 \end{center}
 \caption{Schematic diagram of the front-end electronic circuit operated at room temperature (WFE).}
\label{WormElectronics}
\end{figure}

\begin{figure}
 \begin{center}
 \includegraphics[width=0.95\textwidth]{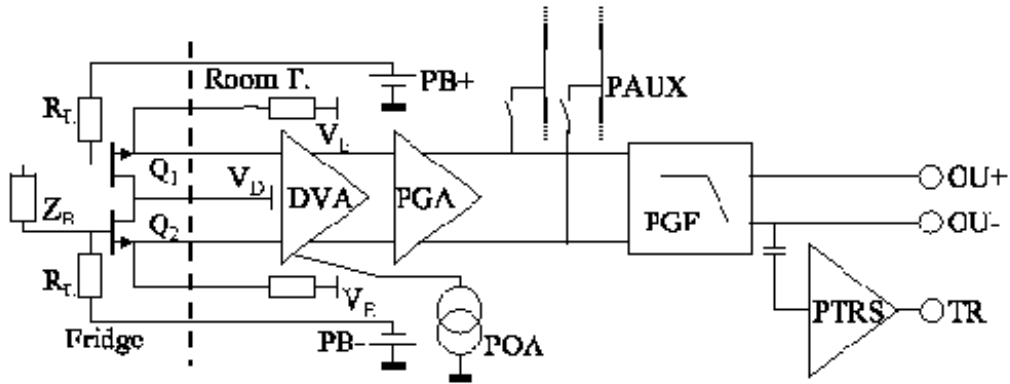}
 \end{center}
 \caption{Schematic diagram of the front-end having the first stage, the differential buffer $Q_1, Q_2$, at cold.}
\label{ColdElectronics}
\end{figure}

In Fig.~\ref{WormElectronics} and Fig.~\ref{ColdElectronics} are shown the two schematic diagrams of the readout solutions for CUORICINO. As can be easily seen, they differ by the presence of the cold buffer stage in one of them (Q$_1$ and Q$_2$ in Fig.~\ref{ColdElectronics}). From left to right of both figures we find the detector, its biasing network, the first stage of amplification (warm/cold) and so on. Any block characterized by a capital P as the first letter of its label is equipped with at least one remotely programmable functionality.
Thermistor biasing is accomplished by applying a programmable voltage across a series combination of the thermistor and two load resistors.

For the set-up of Fig.~\ref{WormElectronics} the load resistors, $PR_L$, are also at room temperature.
The analysis and minimization of the contribution to the low frequency (LF) noise from large resistors under bias conditions, is the subject of an independent report \cite{arnaboldi02a}.
Under computer control, the biasing resistors can assume values in the range 11--54~\gohm. This facilitates the measurement of the bolometer's I-V characteristics at large bias.
The impedance at the input of the electronics is sufficiently high that parasitic conductance in the circuit board is an issue that demands great attention.
Precautions that we took to suppress parasitic conductance were:
\begin{itemize}
\item{segregation of the two high impedance detector nodes by a circuit board layer;}
\item{strategic placement of guard traces around the solder junctions of high-impedance component leads;}
\item{maintenance of a minimum spacing of \ca 5~mm between high-impedance traces;}
\item{placement of slots at critical locations on the circuit board to avoid conductance through minute amounts of soldering residue that sometimes escapes the standard cleaning procedures.}
\end{itemize}
Altogether, these solutions yielded excellent results.

In the set-up of Fig.~\ref{ColdElectronics} the pair of load resistors R$_L$ is also put close to the buffer location, at low temperature. The rest of the biasing set-up is unchanged, but the R$_L$ value is 54 \gohm \ and cannot be switched to 11 \gohm.

The application of the bias voltage to the load resistors is accomplished by the programmable bias attenuator (PB+, PB-), an especially designed Digital to Analog Converter (DAC) characterized by an optically isolated digital input. It can generate the voltages PB+ and PB- of both Fig.~\ref{WormElectronics} and Fig.~\ref{ColdElectronics}, symmetrically with respect to ground (PB-=-PB+).
The programmable bias attenuator can also provide the additional feature of inverting the polarity of the bias voltage.
This feature is very useful, since by reversing the bias voltage during DC characterization, it is possible to account for the thermoelectric effect operating in the leads that connect the front-end electronics to the thermistors \cite{alessandrello97a}. To simplify measurements and improve accuracy, an additional bus PAUX on Fig.~\ref{WormElectronics} and Fig.~\ref{ColdElectronics} can be connected to any selected channel to perform the voltage reading with a very precise multimeter, independently of the acquisition system. 
The biasing network is equipped with a heavy low pass filter that reduces to a negligible level whatever noise originates from the power system or from the network itself.

In CUORICINO, the wiring of all the electrical elements at low temperature has been realized by means of shielded twisted cables (superconducting coaxial cables with a CuNi and 100 $\mu$m diameter NbTi twisted pair). Five stages of thermalization cut the thermal conductance of the cables themselves. The five stages are located at 4.2~K, 1.2~K, 120~mK, 50~mK and 5~mK plates and are composed of sapphire windows (10\per20\per1 mm$^3$) that are sputtered with gold on the bottom, and have 12 1~mm wide tracks (gold-platinum) on the top, where the twisted cables are soldered. 

We are studying the possibility of upgrading this set-up with a less expensive solution. First, the sapphire windows are now going to be replaced with flexible Capton windows with thinner and shorter Cu tracks. A simple pressure on their center against the cold plate allows an effective thermalization of these windows. 
The next step will be the realization of very thin Cu or CuNi tracks sputtered on a flexible Capton substrate. This new approach seems to be feasible thanks to the small radioactive background measured for Capton.

\subsection{Preamplifier and Cold buffer stage}

The design of the preamplifier (DVA in Fig.~\ref{WormElectronics} and Fig.~\ref{ColdElectronics}) satisfies four main requirements:
\begin{enumerate}
\item{Because the typical source impedance and signal characteristics of a bolometer, the voltage amplifier is the conventional approach \cite{mather82}. Its performance is not superior to a charge or current configuration \cite{fascilla01}, but it allows floating inputs which make biasing naturally compatible with DC coupling.}
\item{The long leads (in the WFE option) connecting the preamplifiers with the bolometers have the drawback of acting as antennas for microphonic noise. We addressed this problem by making the preamplifier input differential. Compared with a single-sided input the amplifier noise power is doubled, but the troublesome (common mode) microphonic noise originating in the leads can be effectively suppressed. Moreover, the differential configuration suppresses interchannel
crosstalk.}
\item{The signal bandwidth of our bolometers extends from 1~Hz to few tens Hz. We then require that the preamplifier noise, both series and parallel, be minimized in this frequency interval. The CUORICINO preamplifier is characterized by a voltage noise floor of 3 nV/$\sqrt Hz$, that becomes 4 nV/$\sqrt Hz$ at 1 Hz. Its noise current is 4 fA/$\sqrt Hz$.}
\item{Monitoring the baseline of each bolometer, and correlating the  baseline shift with the shifts in the overall system gain, requires a high thermal stability of the preamplifier. A special circuit has been developed that allows the drift to be maintainedbelow 0.2 $\mu$ V/C.}
\end{enumerate}

With the aim of minimizing the microphonic noise due to the length of the leads, we have realized the set-up of Fig.~\ref{ColdElectronics}. It's main feature is that the first buffer stage of amplification operated at low temperature. The box housing the preamplifiers is in fact anchored to the 4.2~K plate of the dilution refrigerator, next to the detectors. The buffer stage is then warmed in order to reach the optimum temperature for Silicon, around 110~K.
Each cold stage is a differential buffer (Q$_1$ and Q$_2$) composed of two Silicon JFETs that work in the source follower configuration \cite{arnaboldi:tbs}. In this way the length of the leads that connect the detector to this additional buffer stage is reduced to about 1~m, while the remaining wires towards the external stage at room temperature are driven by the low output impedance of the differential buffer. The cold buffer stage features a negligible current noise, a floor noise of about 2 nV/$\sqrt Hz$ and a noise of about 10 nV/$\sqrt Hz$ at 1~Hz. Each cold buffer, together with the load resistors, is located on boards that accommodate 6 channels. The boards are fabricated on a substrate made of PTFE 25\% reinforced fiberglass. 
This solution achieves two goals. First, the small thermal conductance of the PTFE minimizes the injected thermal power needed to increase the JFET temperature from 4.2~K to the optimum 110~K. Second, the radioactivity level of PTFE is lower by a factor of about two with respect to pure fiberglass. In addition, the printed circuit boards have been realized with a thickness of only 0.5~mm thus halving their mass.
The boards are suspended by means of (low thermal conductivity) nylon wires inside thermally shielded copper boxes which can contain two of them. 
Although a bit more complex than WFE, the CFE option has the advantage of offering an effective noise reduction (see Fig.~\ref{noise}).

For both configurations, the input offset is the sum of the contributions from the asymmetry of the JFETs and the bias voltage of the NTD thermistor.
The JFET asymmetry may contribute up to $\pm$ 20~mV to the offset. During normal operation,  thermistor biasing contributes no more than 20~mV, but for load curves acquisition the bias voltage must be adjustable to both polarities and to a magnitude as high as $|50mV|$.
The offset corrector circuit (POA in Fig.~\ref{WormElectronics} and Fig.~\ref{ColdElectronics}) is designed to compensate input offsets from -80~mV to +80~mV, under the control of the digital section.
The POA was implemented with a circuit solution that minimizes any noise contribution, especially at LF~\cite{arnaboldi02}. Once fired by the computer, the POA adjusts the output offset of the selected readout channel to a default value of 1~V. By request, this value can be changed remotely to an arbitrary level.

An AC (DC) test signal can be injected directly to the DVA inputs in order to test the front-end of any selected channel and the load resistors of the WFE channels. During this operation, a programmable system of relays disconnects the detector (or the cold stage) from the DVA. 

\subsection{Programmable gain amplifier}

The outputs of the preamplifier drive the differential inputs of a programmable gain amplifier, (PGA in Fig.~\ref{WormElectronics} and Fig.~\ref{ColdElectronics}).
The amplifier output to the preamplifier input ratio (voltage gain) can take values in the range 220--5000~V/V.
The differential outputs travel on an analog bus to the filter circuit PGF, which is followed by analog to digital conversion of the acquisition system.

\subsection{Antialiasing filter and analog trigger}

In CUORICINO, optimal filtering algorithms are applied to the acquired detector signals (sec.~\ref{sec:analysis}).
To ensure adequate frequency response we have implemented an antialiasing filter (PGF in Fig.~\ref{WormElectronics} and Fig.~\ref{ColdElectronics}) at the downstream end of the analog signal processing.
This is an active six-pole Bessel low-pass filter, which yields a roll-off of 120 $dB/decade$.
In order to adapte the useful bandwidth to each bolometer module, the cutoff frequency can be adjusted from the computer.

The input of the trigger circuit (PTRS in Fig.~\ref{WormElectronics} and Fig.~\ref{ColdElectronics}) is AC coupled to the output of the antialiasing filter. The core of PTRS is an analog amplifier. 
The gain of PTRS is adjustable from the computer.
The AC coupling produces a negative excursion of the trigger circuit input that, if uncompensated, would result in a shift of the baseline.
A baseline restoring section \cite{liguori99} has been added to PTRS to suppresses this effect.
Pulses from a bolometer may exhibit a trailing undershoot because of the coupling of the inductive part of the bolometer's dynamic impedance to the parasitic capacitance in shunt with it.
After differentiation by the AC coupling, the undershoot of large pulses has the potential to re-trigger.
An RC network \cite{arnaboldi:afp} within PTRS cancels the effect of the undershoot and thus suppresses re-triggering. Its time constant has been made adjustable from the computer.
The final element in the trigger circuit is a two-pole Bessel low-pass filter.
The cutoff of this filter tracks the cutoff of the antialiasing filter PGF,  but is higher in frequency by 4~Hz.
A single trigger-Bessel module card accommodates three channels.

\subsection{Calibrating Pulse Generator System}

Very long measurements need an accurate energy calibration. A radioactive source cannot be left or put (on request) close to our detectors, because of its unavoidable contributions to the dead time and background level. Our solution is based on the use of both a radioactive source positioned close to the array (typically once a month, just outside the cryostat) and an always-present very stable electrical calibrating system (heater), which inject signals at a faster rate. The radioactive source fixes the energy scale of the detector while the heater effectively corrects gain variations (sec.~\ref{sec:analysis}).

The electrical calibrating system is composed of a small heater resistor glued on each crystal. Across any heater resistor a small voltage pulse is developed which dissipates a small power, that generates a signal very similar to that of an impinging particle.

The accuracy of our calibrating system depends on the stability of the heater resistor, whose characteristics are described in \cite{alessandrello98s}, and to the electrical pulse generator. To this aim we have developed and are using a very stable pulse generator having the following features:
\begin{enumerate}
\item{It is modular and provides 4 independent outputs for each card used.} 
\item{Each output generates square pulses that can be completely programmable in amplitude and width, and can span a range corresponding to pulse amplitudes equivalent to a few tens of keV up to more than 20 MeV. }
\item{Each pulse occurrence can be driven by an external input (e.g. DAQ), in order to allow for a complete tagging.}
\item{The stability of the pulses is within 2~ppm in any parameter.}
\end{enumerate}

A complete detailed description of this device is given in \cite{arnaboldi03b}.

\subsection{Voltage Supply system}

The voltage supply system is composed of two main stages. The first stage fulfils a pre-adjustment of \pom 13~V by means of several 75~W switching regulators. Special filtering sections are used to suppress any high frequency spike. The second stage outputs provide the desired voltages in the ranges -12--+12~V for the output stage of the front-end and in the range  -10--+10~V for the cold stages and the preamplifiers. The \pom 10~V output is used also as the reference for the offset correcting circuits. For this reason a very stable (a few ppm/$^{o}$C drift) and low noise (1.3 $\mu$ Vpp in the 0.1 Hz to 10 Hz frequency range, at a rated current of 1.5~A circuit) was designed and built~\cite{pessina99}. A single second stage voltage supply supports 24 complete channels.

\subsection{Digital control}

A digital control system (DCS) enables a computer assisted regulation of the most significant analog front-end parameters: the thermistor bias current, the input offset correction, the gain, the bandwidth, the application of test signals to the inputs, the calibrating pulse injection to the detectors, etc.
During data acquisition the DCS is generally idle.
In this state even the DCS clock shuts down, thus eliminating the possibility that DCS activity will contribute to noise on the analog outputs.

To minimize elecromagnetic interferences (EMI) and ground loops, the communication between the DAQ (sec.~\ref{sec:readout}) and the DCS is made with an asynchronous, serial, bi-directional optical fiber link bus.

\subsection{EMI Shielding}

We foresee, as is partially done now in CUORICINO, surrounding the entire cryostat and the front-end electronics with a Faraday cage able to shield also low frequency magnetic fields. For this purpose the internal walls of the cage will be covered with a sheet of 0.1--0.2~mm thickness of SKUDOTECH, a very high permeability metal~\cite{skudotech:tm}. A sound absorbing material will also cover the internal walls. The EMI shield will be split in two sections.
A first Faraday cage will surround the cryostat and its lead/polyethylene shields. The very front-end, fed by the detector connections (output signals and input bias), will be organized in 12 towers (each composed of 6 standard 19 inch-racks) and connected to the second stage located in a smaller Faraday cage sharing one of its walls with the main Faraday cage. The small Faraday cage will house the antialiasing filter, the trigger circuit and the acquisition system optically connected via TCP/IP to the outside world. The choice for a set-up split in two separate rooms allows a distribution of the power dissipation over a large space and an effective reduction of the digital circuits EMI.


\subsection{Noise considerations}

The above described analog WFE produces a noise of the order of 50~nV (FWHM) referred to its input, over the 10~Hz bandwidth of the signal.
The 54~\gohm\ load resistors produce 80~nV in the same bandwidth so that the combined noise is \ca 100~nV (FWHM).
On the other hand, for the CFE option, due to the presence of the cold buffer stage, the series power noise is increased by a factor of about two, while parallel noise is decreased by a factor of at least three, mainly due to the use of the cold load resistors. In this way the overall noise contribution becomes about 90~nV (FWHM).
In both cases the noise is considerably below the 1.0~$\mu$V (FWHM) level observed for one of our best performing bolometers. We believe therefore that the analog front end is more than adequate for the CUORE needs.

The two front-end solutions have the same level of noise, but the presence of the cold buffer stage, which reduces the high impedance path, should further improve the common mode noise due to the vibration of the leads.
Unfortunately, there is another source of microphonics which is not a common mode: the mechanical friction between the detector and its mounting. If this source would result in a noise level much greater than that of the connecting leads, the use of a cold buffer stage would be unnecessary. This would allow a reduction of the connecting leads by a factor of two and would avoid thermal power injection inside the dilution refrigerator. It should be noticed that the reduction of the parasitic capacitance obtained by shortening the high impedance path is not a concern  in the case of our very slow signals.

\clearpage
\section{Data acquisition system (DAQ)}\label{sec:readout}

In this section we describe the architecture of the readout and DAQ system for CUORE. 
In particular we report the main requirements that are driving the
system design and describe the general architecture of the system oulining a few open options. We also give the block diagram of the main readout board and finally describe the main DAQ system and its software layout. 

\subsection{Requirements}\label{sec:readout_req}

As described in the previous sections, the signals coming from each crystal are amplified and shaped by a preamplifier that generates pulses well described by the following typical parameters:

\begin{itemize}
\item Voltage signals with maximum peak amplitudes as high as 10 V; the signals can be either differential or unipolar; due to the large energy range, the peak amplitude can also be as low as few mV.
\item Rise times of the order of 30 ms, falling times up to an order of magnitude slower.
\item Total duration up to several seconds.
\end{itemize}

The DAQ system must be designed to sample and read out the signals of about 1000 channels\footnote{As an option, it has been proposed to add a second sensor to each crystal. This would of course double the number of channels, but would have a negligible impact on the architecture proposed here, except of course an almost linear increase of the total cost of the system.} of this type with virtually zero dead time, independent readout among channels, self-triggering capability, and, last but not least, optimized cost.

In order not to spoil the energy resolution over the whole range (from 5 keV up to 15 MeV), real 16 bit sampling is required. In addition, the pulse must be sampled for 5 s, in order to get a good measurement of the tail and of the background level. A typical sampling rate of 1 kHz seems adequate, but it is highly recommended to allow for a programmable sampling speed around this value. With the aforementioned values, the typical event size for one
channel is 5000x2 = 10 kbytes.

It is also very important to measure the background level (baseline), both before and after the end of the pulse, therefore it is required to have a suitable internal circular memory and a logic capable of recovering the pre-trigger signal conditions.

The choice for real 16 bit sampling poses severe requirements on the noise level acceptable at the channel input. Assuming a maximum range of 10 V, the noise level must be below 150 $\mu$V r.m.s.

Being that CUORE is designed to search for very rare events like \BB decays or dark matter induced nuclear recoils, the requirements for the DAQ system for data throughtput are not severe at all. The maximum foreseen data rate is
given by the calibration procedure, that might produce at most few events per minute per channel. As described in section \ref{sec:readout_daq}, this rate does not pose any special requirement on the DAQ system, even assuming
a complete sampling of the channels without zero suppression.

\subsection{General architecture}\label{sec:readout_arch}

The CUORE DAQ system must be able to digitize, acquire, analyze online, store and display the pulses generated by 1000 \teod bolometers.

Each channel must be completely independent from the others and the readout must introduce zero dead time.

In order to minimize the cost and allow for a maximum flexibility in the on-line analysis and triggering, the system proposed here is mostly based on the large and cheap computing resources that are now commercially available, reducing the complexity, and therefore the cost, of the readout boards as much as possible.
\begin{figure}
\begin{center}
\includegraphics[width=0.9\textwidth]{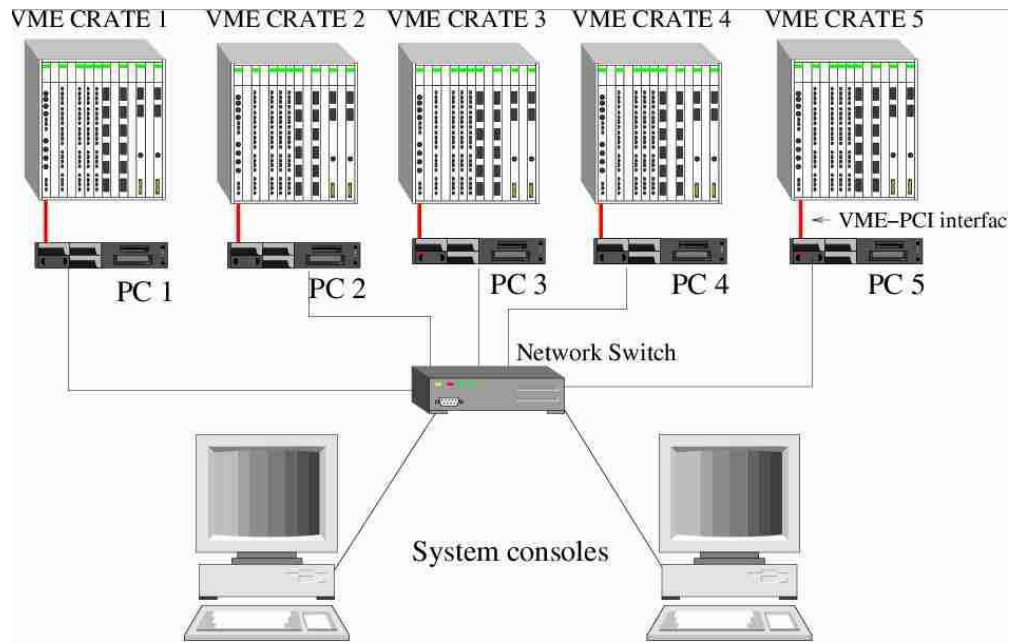}
\end{center}
\caption{Block diagram of the proposed DAQ system. The digitizing boards are housed in 5 VME crates, each controlled and read by a dedicated Linux PC that uses a VME-PCI interface to access the VME bus. All computers are connected via a network (normal 100 Mbit ethernet or fast optical link if required) to a set of consoles that provide graphical user interface and event display. A custom designed software runs on all computers.}
\label{fig:block_diagram_main}
\end{figure}

The general structure of the system is shown in Fig.~\ref{fig:block_diagram_main}. It is composed of a set of personal computers, a network infrastructure, a PCI based VME interface and 5 or more VME crates with custom designed digitizing boards.

We propose to design and build a simple digitizing VME board with full 16 bit dynamic range (possibly 24), large internal circular buffers, optional self-triggering capability with a programmable threshold and digitizing speed programmable in the range from 0.01 KHz up to 10 KHz.

Each board can probably handle 12 channels, to allow reading 240 channels per VME crate (5 VME crates\footnote{Should the two sensor options be adopted, the number of required crates would be 10; additionally, we might succeed in designing a 16 channel digitizing board that would reduce the number of VME crates to 4 or 8. The general structure of the system would not be changed at all.} would then be required with 1000 channels).

In this structure the digital boards are just pure digitizers, that continuously convert the input analogue signal into a stream of digital words; when a trigger occurs, either generated by the remote front end electronics or 
internally generated by the discriminator, the words stored in the internal buffers are copied to another memory that can be accessed by the main DAQ software via VME bus. The structure of the board guarantees that the channels remain independent and fulfils the zero dead-time requirement.

Each board issues an interrupt signal on the VME bus whenever a trigger occurs; in this way the main software may work either in interrupt or in polling mode.

The large and cheap computing power available nowadays yields a large flexibility in the possible on-line filtering and processing of the data. 
Full on-line  analysis is possible with this structure, allowing a fast and efficient monitoring of the experimental conditions.

\begin{figure}[thb]
\begin{center}
\includegraphics[width=0.9\textwidth]{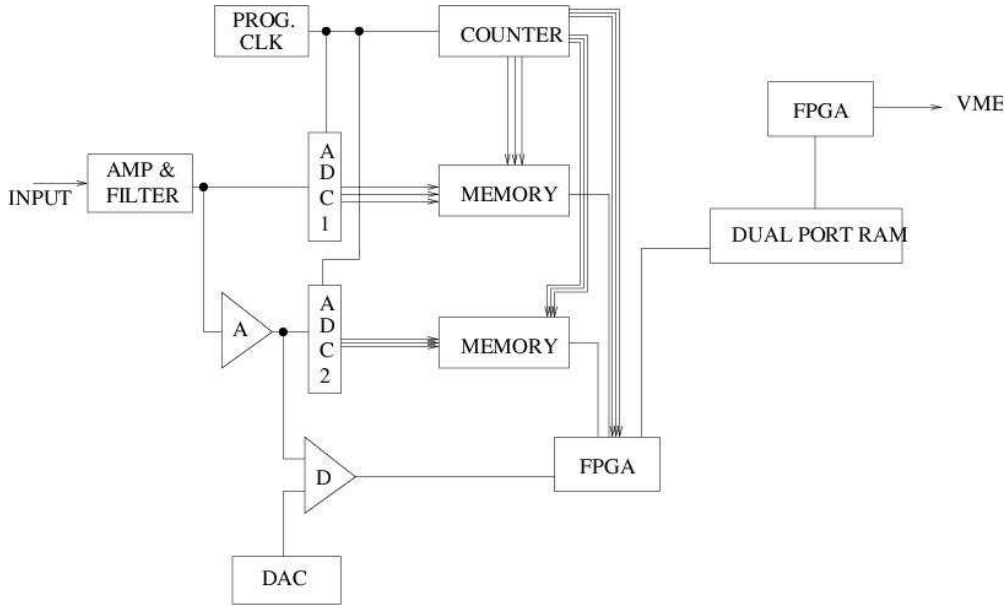}
\end{center}
\caption{Base design of the digitizing board. In this "minimal" version the board simply has the function to convert the input signal into a stream of words that are transferred to the DAQ computers via VME bus.}
\label{fig:board_block_diagram_1}
\end{figure}

The next two sections give a somewhat more detailed description of the digital board and of the computing and software architecture.

\subsection{Readout board}\label{sec:readout_board}

We are considering two possible block diagrams for the digitizing board. The first is the one sketched in the previous section and shown in Fig. \ref{fig:board_block_diagram_1}.

In this design, the digitizing board is very simple and most of the job is left to the software that runs on the DAQ computers; in fact, the board is a pure free-running digitizer, with the minimum amount of logic needed to recover the data from a few hundred ms before the trigger until several seconds after the trigger. The trigger can be either external, i.e. generated by the front end electronics and sent to the board via a NIM or TTL signal, or can be internally generated with a simple discriminator.
To achieve the required resolution (real 16 bits), we will probably put two 12 bit ADCs in parallel, with a precise amplifier in front of the least significant ADC. In this way 24 bits will in fact be available. The amplifier gain will be chosen to leave a large overlap between the two ADCs, allowing a precise intercalibration.

\begin{figure}[bht]
\begin{center}
\includegraphics[width=0.6\textwidth,angle=-90]{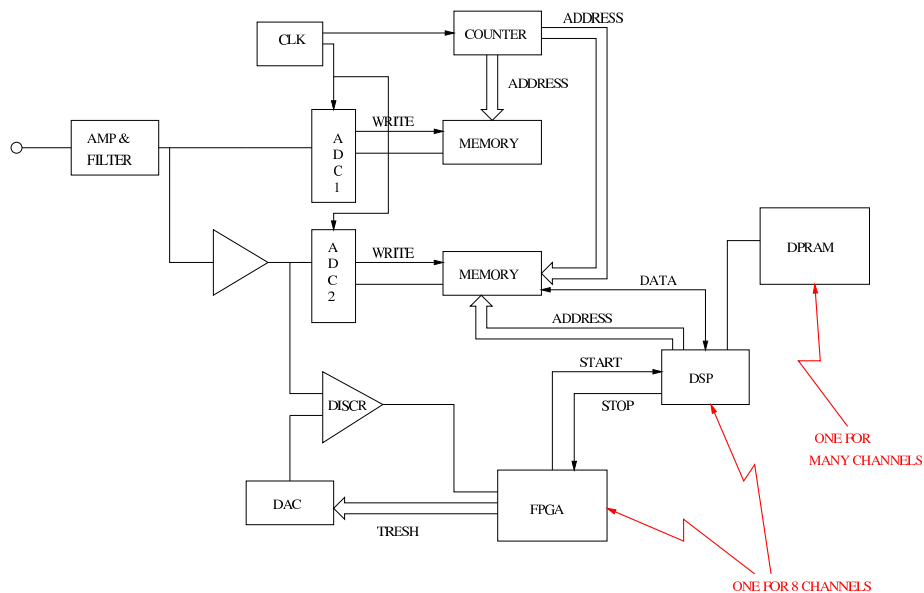}
\end{center}
\caption{Optional enhanced design of the digitizing board. In this version the board is equipped with a DSP that can perform additional computing functions like event selection or digital filtering.}
\label{fig:board_block_diagram_2}
\end{figure}

When a trigger occurs, the FPGA logic will transfer the data (a programmable time window around the trigger signal) to a Dual Port Ram that can be accessed via VME bus. A VME interrupt will be also generated when this transfer is completed, allowing the DAQ computer to start read-out. It will be possible to use this board either in interrupt mode or in polling mode. Block transfer D32 will be supported too, allowing very fast transfer. The Dual Port Ram will be large enough to keep several minutes of data at normal trigger rate during calibration.

We are also considering a second approach, a little more complex from the point of view of hardware design, that includes a DSP in the digitizing board (see Fig. \ref{fig:board_block_diagram_2}). With this solution, the DSP could be used to perform a first level data filtering and implement ``smart'' trigger conditions in order to reduce the data rate. In our opinion, with the data rate foreseen for CUORE, it will be possible to perform all this higher level functions using the computing power of the DAQ system, allowing for more flexibility and reducing the cost of the system. Therefore we prefer the scheme shown in Fig.  \ref{fig:board_block_diagram_1} as our baseline  design and leave the second design as an option. 

\subsection{Computing system and DAQ layout}\label{sec:readout_daq}

The system shown in Fig.~\ref{fig:block_diagram_main}, with the baseline option for the digitizing board shown in Fig.~\ref{fig:board_block_diagram_1}, requires a significant amount of computing power because most of the  functions are delegated to the software. 

As a baseline design, we propose that each VME crate will house 20 boards, each of them capable of digitizing 12 channels. Therefore we need 5 VME crates to read 1000 channels. Each VME crate will be connected to a PC via a
PCI-VME interface. These 5 PCs will be connected via network to another set of PCs that will perform event building, online analysis, graphical display, system control and monitor.

The network infrastructure will be rather simple, with a single switch star-connected to all PCs. Normal 100 Mbit ethernet is probably more than enough for CUORE but it would be easy to implement this structure using 1 Gbit optical link without any architectural modifications. We think, therefore, that this very standard network
structure is adequate.

We propose to use Linux based PCs for all computers and to use root as a software environment. The software running on each computer will be custom designed while the consoles and the graphical interfaces will be root based. 
A possible WEB interface to the system is foreseen, though network security problem will probably limit its use.

\subsection{Slow control}\label{sec:slow_control}

During a measurement, all the parameters of the apparatus must be monitored and registered. This is particularly required in long running time experiments where the acquired data can be ``contaminated'' with problems related to experimental system conditions. In particular in the case of cryogenic detectors many inconveniences can arise from the cryogenic apparatus. The experience gained with MiDBD and CUORICINO detectors  showed us the fundamental role played by the temperature instabilities and by the vibrations induced by the malfunctioning of the cryostat and liquefier.
In order to correlate these frequently unavoidable problems it is very important to continuously check the main parameters of the cryogenic system (e.g. pressures, mixture flows, cryogenic liquid levels and others). The cryogenic apparatus must be equipped with transducers able to transpose these data in a voltage value to be acquired by the DAQ during the experiment. 
This \emph{environmental data} acquisition system can be implemented directly in the DAQ or be designed independently and correlated off-line with the experimental data by means of a time synchronization between the two systems. The choice of the best approach can definitely be taken later on; the control system and DAQ designs have in fact little intereference and can evolve independently.
At the moment we are not planning an active automatic control of the apparatus. This could be designed later, when a better definition of the cryogenic and detector apparatuses will exist. On the other hand, we would like to implement a human remote control(e.g. WEB interface) in order to maintain the apparatus in the best measurement condition without the necessity of the constant presence of a shifter inside the underground laboratory.

\clearpage
\section{Construction procedures}\label{sec:PROC}
The only possibility for the CUORE collaboration to satisfy the sensitivity requirements will be to use severe procedures for the construction of the experimental setup. According to the experience gained in CUORICINO, this will imply rigorous material selection and construction procedures. 

\subsection{Material Selection}\label{sec:MATE}

One of the most critical points of the CUORE experiment will be the selection of all the materials used in the construction of the detector. Since the selection criteria must result in the minimum possible contribution to the background level, it is obvious that this will be accomplished only if the starting quality of the materials will be sufficiently good from the point of view of radio-purity. 
We can divide the materials needed for CUORE in three main categories: i) materials for crystal growth; ii) materials for the detector structure and iii) materials for the shielding structures. Material selection is important also for other parts of the experimental setup (e.g. the cryostat) but in principle these parts can be heavily shielded so the required limits on their radioactive contaminations are less stringent (section~\ref{sec:shields}).

\subsubsection{Materials for crystal growth}

\teod is a molecule composed of only two atomic species: Te and O. In practice if we want to produce a good stoichiometric material, the oxidation of the Te metal must be realized in few well defined steps. In each step the selection of the compound used will be very critical.
The first material to be chosen is obviously Te. The analyzed metallic Te for the CUORICINO production showed limits below the sensitivity of the HPGe detectors installed in LNGS (apart from the well known contamination in \pbdd). 
\teod powders were also measured at the LNGS low activity underground laboratory. Contamination values are reported in table~\ref{tab:materials}. 
The presence of \podd can be ascribed either to a genuine \podd contamination or to \pbdd. Since Pb contaminants should be limited in the CUORE experiment, we will try to investigate this point in detail. Two acids are usually used for metallic Te oxidation: Nitric and Hydrocloric. The ratio used in this preparation is: 1~kg of Te, 1.3~kg of HCl and 2.5~kg of HNO$_3$\cite{siccas:pc}. So we need to check with the same accuracy the radio-purity of the acids and of the Te metal. For CUORICINO we have measured the two acids involved in the crystal production using ICPMS with quite good results: HCl showed a contamination of 1.3~ppt of \udt and <1~ppt of \thdtn, while HNO$_3$ showed 13.1~ppt of \udt and 112.3~ppt of \thdtn. One of the main problems in CUORE  will be the systematic control of these values during all the crystal production operations.
There is another very critical point at this stage: a large quantity of water (few liters per kg) is used to ``clean'' the \teod powder just after their production and before the growth procedure. This operation can be of course very dangerous because of the possible presence of radioactive contaminations in the water that can be transferred to the powder. The impact of these contaminations could explain the higher contamination observed in the \teod powders with respect to the starting Te metal.
The growing procedure usually reduces the contamination levels of the final crystal with respect to the bare material. However it is usually difficult to quantify the actual reduction in the produced crystals. On the other hand, an increase of the number of growing cycles - although undoubtely useful - would be very expensive.
The last critical point in the preparation of the crystals will be the surface treatment made in order to assure a good contact between crystals and semiconductor chips (thermistors and silicon resistances). The data obtained with MiDBD showed that a relevant contribution to the background in the double beta decay region could be identified with the presence of surface contaminations on the crystals. In the  production of the early \magro \teod crystals, this consideration was strongly supported by the measurement of the lapping powders used by the factory. In CUORICINO we carried out the final lapping procedures using a specially selected Al$_2$O$_3$ powder. This showed a radioactive contamination few orders of magnitude lower with respect to the previous ones (tab.~\ref{tab:materials}). However, because of the large spread in the grains dimension of this polishing powder, only roughly smooth surfaces could be obtained. Alternative approaches are in principle possible (e.g. surface acid attack, ion beam sputtering, etc.); the careful test of these methodologies and the development of powerful dignostic techniques will be considered in the early CUORE R\&D phase (see section \ref{sec:SURF}).

\subsubsection{Materials for the detector structure}
The experience gained in MiDBD and CUORICINO indicates that all the structure of the CUORE detector should be mainly made of Cu. Only minor components will be built with different materials. In the case of Cu, we have indications that very low limits for the bulk contaminations of the U and Th chains can be reached in specially selected productions. We have measured a number of samples for CUORICINO and found that the Oxygen Free High-Conductivity Cu is definitely better than the standard electrolytic Cu. Moreover some specific productions, made by our colleagues in Germany~\cite{heusser:pc}, have shown very low limits in radioactive contaminations, at levels sufficient for CUORE sensitivity requirements (see section \ref{sec:simulation}). We must plan and control the production of Cu for CUORE directly with the foundry in order to check the contamination of the material immediately. Moreover the Cu must be stored underground immediately after production to avoid cosmic ray activation. This poses severe conditions on the physical size of the produced Cu in order to minimize the time needed by the mechanical factories for the production of the final pieces of CUORE.

Another important material in the detector structure is PTFE. The spacers holding the detector in the right position are made with this material and therefore the radioactive contaminations of PTFE can be very dangerous for the background level of the detector. For organic materials there are two different problems: first it is necessary to have very low levels of intrinsic radioactivity, second it is crucial to control the exposure of these materials to air radioactive pollution (typically Radon). In general the radioactivity of organic compounds like PTFE, Nylon etc can be very low, as demonstrated by the Borexino collaboration~\cite{borexino98}. Concerning the diffusion of Radon it is necessary to maintain the selected materials in a very clean environment and, to minimize the possible radioactive contamination present in the volume next to the surface, the external part must be mechanically removed before the realization of the spacers.

\begin{table}
\caption{Radioactive contamination measurements for CUORICINO materials. All values are in mBq/kg. Al$_2$O$_3$-1 is the original lapping powder used at SICCAS while Al$_2$O$_3$-2 is a certified 0.7 $\mu$m Sumitomo powder. }\label{tab:materials}
\begin{center}
\begin{tabular}{lccccccc}
\hline\hline
& $^{238}$U   & $^{226}$Ra  & $^{214}$Bi  & $^{232}$Th  & $^{228}$Ac  & $^{208}$Tl  & $^{40}$K\\
\hline
Au wires &  <0.99 & <1.58 & <.34 & <0.33 & <.29 & <.37 & 1.3$\pm$0.2 \\
Al$_2$O$_3$-1    &  12.1$\pm$1.8 & 24.8$\pm$3.7 & 8.2$\pm$1.2 & 4.85$\pm$0.7 & 6.4$\pm$0.9 & 3.2$\pm$0.5 & 10.2$\pm$1.5 \\
Al$_2$O$_3$-2 &  <60.1 & 21.9$\pm$3.3 & <1.8 & <2.4 & <2.1 & <2.0 & <16 \\
Cu screws &  <18.4 & <11.0 & <7.3 & <13.6 & <14.0 & <14.0 & 18.3 \\
\teod     &  11.5$\pm$1.7 & 30.5$\pm$4.6 & 1.7$\pm$0.2 & 3.4$\pm$0.5 & 2.4$\pm$0.36 & 1.7$\pm$0.2 & 6.6$\pm$0.9 \\
tin   &  42.3$\pm$6.1 & 45.1$\pm$6.2 & 39.0$\pm$5.8 & 18.4$\pm$2.7 & 18.4$\pm$2.7 & 18.4$\pm$2.7 & <65 \\ 
\hline\hline
\end{tabular}
\end{center}
\end{table} 
 
The other materials involved in the detector construction are characterized by very small masses. Also in this case however, a good radiopurity level is required in order to avoid any possible dangerous contribution to the background level. In that sense, the selection of the epoxy glue (connection between chips and crystals), of the gold bonding wires (electrical connections), of the electrical pins, etc. must be carefully controlled. The indications obtained from our previous tests, show very good levels for epoxies while are non conclusive for gold wires and pins. All these components will be measured again to improve their contamination limits; alternative approaches (e.g. neutron activation) will also be considered.

The last critical part of the detector assembly is represented by the wires used for the electrical connection between the detector and first thermalization stage on the inner lead shield (they are not shielded). It is very difficult, at the moment, to define their actual contribution to the background level since we still lack a defined design of the final electrical connection. As a matter of fact we will choose the best among a number of different wire types (constantan, superconductor, etc) and will then design the wiring system using only previously selected radio-clean materials. 

It should be stressed that all the materials must match two different requirements: they must be characterized by low radioactive background and, at the same time, ensure a good performance in a cryogenic environment. Of course, is not obvious that to satisfy these very different conditions unavoidable difficulties might not arise in correctly defining the material for the specific application.

\subsubsection{Shielding materials}

The radioactive shields will be made of lead. To minimize the contribution from the lead itself, we need to build different concentric shields made with lead characterized by different contents of \pbdd. It should be stressed that the presence of this nuclide can contribute to the background spectra with the bremssthrahlug photons produced by the beta decays of the $^{210}$Bi. 
The internal lead shield, to be placed as close as possible to the detector at cryogenic temperatures, will consist of ancient roman lead that contains a level of \pbdd lower than 4 mBq/kg~\cite{alessandrello93}( section \ref{sec:shields}). The intermediate lead shield can be realized with commercial lead with a \pbdd contamination of the order of few Bq/kg. The external part will be constructed with standard commercial lead that normally has contaminations of the order of few tens of Bq/kg in \pbdd. All the above requirements were already satisfied by the lead used in CUORICINO.

\subsubsection{Other components of the experimental setup}

There are some parts of the cryogenic environment that need careful design, not only from the point of view of the cryostat performance, but also for their possible contributions to the radioactive background. In particular, although the upper part of the cryostat will be shielded with respect to the detector volume with a heavy layer of lead, its most massive components should be constructed with materials characterized by an acceptable level of impurities. We can indicate a rough limit for these materials and reduce at a minimum their required amount. 
Of course, the requirements connected to producing a well performing cryogenic apparatus must also be taken into account. Some possible materials that satisfy both requirements have already been found during the last few years for CUORICINO. This job will continue in the future with the systematic selection task needed for the  specific application to CUORE.

\subsection{Construction environment}\label{sec:ENVIR}

The \teod crystals will be fabricated at the Institute for Ceramics in Shanghai (SICCAS), China, from low-radioactivity ore. Our prior experience with similar crystals fabricated for MiDBD and CUORICINO indicates that the final stages of lapping and polishing must be done in a clean, radioactivity-free environment, recognizing that these detectors are the most-sensitive detectors available for traces of radioactive materials. Two possible options can be envisaged for the choice and set-up of this environment: a) a conservative option, which follows strictly the CUORICINO experience; b) an innovative option, which foresees a substantial use of out-sourcing, therefore requiring the development of new environments or the modification of existing environments presently not suitable for the delicate operations involved in the CUORE assembly. The main points of the two options are outlined below.
\begin{itemize}
\item[(a)] In the conservative option, the crystals will be grown in China, cut in a cubical shape and lapped using alumina powder down to a grain size of about 15~$\mu$m. The control of the residual activity in the lapping pads and in the alumina will be performed (see subsection \ref{sec:MATE}). The lapped  crystals will be delivered by ship in order to limit exposure to the high-energy hadronic component of the cosmic rays and stored underground. 
The final most delicate polishing operation will be accomplished directly in the Gran Sasso laboratory, possibly underground or at least keeping each crystal outside the underground laboratory just for the time required for polishing. A special environment has to be developed for this purpose, consisting of a clean room where several polishers can be installed in order to proceed in parallel with many crystals. A radon-free atmosphere consisting of synthetic air should be realized in the clean room. Presently, in the hall C of the underground Gran Sasso laboratory, there exists a facility which with some modifications, could have the required features: it is the combination of clean rooms developed for the BOREXINO experiment. If available at the time of the CUORE construction, this would be an ideal environment for the polishing procedure. Otherwise, a similar structure must be built. 
After polishing, each crystal will be stored underground in a radon-free container. Once the polishing operation is completed, in the same clean room where it was realized, the thermistors will be attached to the \teod crystals. Finally, the crystals will be inserted in their copper frames and the thermistor wires connected to the read-out pads. Each 4-detector module will be stored independently in a radon-free container and moved to the cryostat area. The final assembly of the modules will be realized directly in the lower level of the cryostat room, that will have clean room features. 
As far as the whole detector will be completed, it will be connected to the dilution unit of the refrigerator and will be ready for the final wire connections.
\item[(b)] In the innovative option, a large fraction of the detector assembly work will be performed in China, from where the crystals will be shipped in the form of completed 4-detector modules in special radon-free containers. A clean room environment (class 1000 or better, with radon-abatement) must be set up either at SICCAS or at some other facility under the supervision of LBNL scientists. These same scientists will develop procedures to be used for the production of the crystals. The final polishing operation will be performed by skilled Chinese personnel, but oversight, either in the form of an employee on site or remote access, must be maintained. Finally, these crystals should be instrumented and packaged into elementary modules, i.e. groups of 4 crystals, frames, thermistors, and heaters before leaving the clean area. Even this operation will be performed by supervised Chinese manpower, after an appropriate training period. Once all the elementary modules have been completed, they will be shipped to the underground Gran Sasso laboratory as quickly as possible to reduce cosmic ray activation. The final assembling will be performed in Gran Sasso as in option (a).
\end{itemize}

An intermediate solution between the two options is possible. The Chinese part of the work in option (b) could be terminated at the final polishing phase, while the delicate operation of thermistor attaching and 4-module realization would be realized in the Gran Sasso laboratory. In this way, no polishing set-up is required in Europe, but the detector assembly operations will be entirely accomplished underground in the Gran Sasso laboratory, where skillful physicists and technicians can more easily check the mounting procedure.

An important point to be considered for the choice between the two options is the total exposure to cosmic rays of the crystals from their growth to their final storage underground. Of course, this exposure will be much shorter in case of option (a). An estimate of the production rate of radio-nuclides generated by nuclear reactions induced by high-energy hadrons in the crystals has been already done using the COSMO program (see section \ref{sec:simulation}). Cosmogenic \coss in \teod crystals could be a limiting contribution for CUORE so that exposure to cosmic rays must be limited as possible. 
Monte Carlo simulations have shown that a total exposure period of 4 months to cosmic rays would give a contribution to the \BBz CUORE background below the sensitivity requirement (see section \ref{sec:simulation}). Moreover, a recent experimental determination of the \coss production cross section by protons on Te (0.63$\pm$0.15 mb for 1.85~GeV p~\cite{norman03}) results in a strong disagreement with COSMO (which yields a much larger one) so that the allowed exposure period could be much longer. An independent activation measurement carried out by exposing a 51.1 g TeO2 crystal to a beam of \ca 10$^{13}$ 24 GeV protons has found a cross section of 1.6 mb instead than 2 mb as predicted by COSMO. Since however the effect of cosmic rays is expected to be dominant in the region of 1-2 GeV a new activation measurement with protons of 1.4 GeV is planned at the ISOLDE facility of CERN. A similar measurement is planned at Berkeley. Their reults will help in the final decision about the option to be chosen.

\subsection{Low background technology transfer to the crystal growth company}
\label{sec:TRANSF}
In either case, many critical operations in terms of radioactivity control must be accomplished in China, while more will be required in the case of option (b). Therefore, it is mandatory that low background concepts will be transferred to SICCAS in order to get satisfactory results as for crystal radiopurity.
\begin{itemize}
\item[--] The operations which precede crystallization involve the use of chemical products. In particular, metallic tellurium must be transformed into TeO$_2$ powder: there are indications that U and Th impurities are added in this phase, that must be carefully checked, both in terms of procedure and of used reagents. In particular, radiopure reagents must be selected by the CUORE collaboration (see section \ref{sec:MATE}) and provided to the Chinese company.
\item[--] There are indications that the Pt crucible used for crystal growth contaminates the crystal with the isotope $^{190}$Pt, which is responsible for an alpha line at 3.25 MeV, very dangerous if degraded in energy. Therefore, new crucibles (possibly in Ir) should be developed in collaboration with SICCAS.
\item [--] The materials to be used for lapping and polishing (see section \ref{sec:MATE}), such as pads and powders, must be carefully checked and provided to SICCAS. This operation is even more critical in case of option (b) of section \ref{sec:ENVIR}. Some experience has been gathered during CUORICINO 
construction, where special pads and powders down to 0.5 $\mu$m grain size were selected (see section \ref{sec:MATE}). This selection must be continued until a complete and satisfactory identification of the polishing materials is accomplished. The materials will be delivered to China and used under the 
supervision of members of the CUORE collaboration.
\item [--] In the event that option (b) of section \ref{sec:ENVIR} is selected, the copper frames, the thermistors, the heaters and all the tools necessary  for the elementary module assembly must be shipped to China in special containers protecting them from exposure to radon and other contaminating 
agents. The Chinese personnel involved in the module assembly will be trained to fabricate and store the modules in the proper conditions in terms of radiopurity.
\end{itemize}

\subsection{CUORE assembly procedure}
\label{sec:ASSEM}

We list below the main steps which precede and constitute the CUORE assembly procedure, indicating the typical time necessary to accomplish each step, evaluated on the basis of the CUORICINO experience and discussions with SICCAS technicians. Under the general assumption that the total assembly procedure needs to last no longer than 12 months, these indications will be used in the subsequent section to estimate the manpower needed, the amount of work that needs to be done ``in parallel'' and to satisfy the CUORE construction schedule.
\begin{itemize}
\item[--] {\sl Crystal production}. Performed entirely in China. Estimated production rate: about 60 crystals/month. Total production time: 18 months. In this phase the crystals are produced with lapped surfaces, down to a \ca 15 $\mu$m roughness.
\item[--] {\sl Crystal polishing}. Performed in China or in the Gran Sasso laboratory. The experience gathered with polishing crystals in Como for the CUORICINO experiment shows that 3 persons can polish 16 crystals/week with two polishers. In order to reach a higher rate of the order of 50 crystals/week, as required in order to remain on schedule with the complete assembly, 3 groups will have to work in parallel.
\item[--] {\sl Thermistor preparation}. Performed in Berkeley (thermistor development and production) and in Como / Florence (bonding and characterization). The thermistors will be all ready for bonding and characterization 3 years after the start-up of the CUORE project. The bonding rate is about 20 thermistors/day. Two persons working in parallel with two bonding machines (already available) will bond 200 thermistors/week. Total bonding time: 5 weeks. The characterization will be performed in parallel on samples bonded in the first week and should not require additional time.
\item[--] {\sl Heater preparation}. Performed in Trento and in Milano. Eighteen months will be necessary to produce 2000 heaters, assuming a reasonable 50\% yield. The bonding time will be 10 weeks at the same conditions mentioned in the previous point. The characterization, that cannot be by sample, will have a rate of 50 heaters/week. Therefore, about 1 year is needed for heater selection. Their production and characterization can take place during crystals growth, because no R\&D is required for them.
\item[--] {\sl Heater and thermistor gluing}. Performed in China or in the Gran Sasso laboratory. The CUORICINO experience shows that 2 persons can glue thermistors and heaters with a rate of 15 crystals/week, using the special tools developed for this aim. 
\item[--] {\sl Frame preparation}. Performed in the Gran Sasso laboratory. Once the frames and the other copper elements for the typical 4-element module are machined and cleaned (see sections \ref{sec:single} and \ref{sec:PROC}), a preparation work preceding the module mounting must be performed, consisting in assembling the Teflon supports, the read-out pads and in wiring locally the frame. It is reasonable to assume that 2 persons can prepare 20 frames (= 1 tower)/week.
\item[--] {\sl Crystal insertion in the frames}. Performed in China or in the Gran Sasso laboratory. The CUORICINO experience shows that 2 persons can insert crystals in 2 planes/day. 
\item[--] {\sl Connection of the gold wires}. Performed in China or in the Gran Sasso laboratory. The CUORICINO experience shows that 1 person can wires in 1 plane/day.
\item[--] {\sl Assembly of the towers}. Performed in the Gran Sasso laboratory. The CUORICINO experience shows that 2 persons working together can form a tower of 10 elementary modules in about one week.
\item[--] {\sl Assembly of the cube}. Performed in the Gran Sasso laboratory. A reasonable assumption is that 2 persons can assemble the 19 towers into the CUORE cylinder in about 1 month, provided that special tools for towers manipulation and movement have been realized previously. Of course, they will need to work without anyone else around the cryostat, and this must be taken into account.
\item[--] {\sl Cryostat wiring}. Performed in a laboratory to be defined and in the Gran Sasso laboratory. In the case of CUORICINO, the preparation of a complete set of wires for the tower was made by 2 persons and took around 1 month. This operation can be made almost completely on the bench, before the construction of the CUORE cube, by several groups in parallel, so as to prepare all the wires in 6 months. An additional time of about 1 week/tower is required to lay the wires in the cryostat. This operation should be made by 2 persons working together. Unfortunately, this work cannot be done in parallel, since no more than 2 people can work on the cryostat simultaneously. Therefore, another 6 months should be considered for laying all the wires in the cryostat.
\item[--] {\sl Final wire connection and check of the contacts}. Performed in the Gran Sasso laboratory. Provided that the wires in the cryostat have been installed and that a clever method has been developed for connecting low thermal conductivity thin wires (i.e. special low radioactivity connectors), this job should be perfomed in about 1 month by 2 persons working simultaneously on the cryostat. Of course, this job is the only one that cannot be done in parallel with any of the other operations.
\end{itemize}

The proponents would like to stress that in the previous points no improvement in terms of simplification of the procedures was considered with respect to the CUORICINO experience. There are several innovations (larger crystals and consequent reduction of the total channels, simplified methods for thermistor-absorber coupling, simplified methods for wiring, etc.) which could speed up the assembly procedure. Some of these innovations are discussed in section \ref{sec:INNO}.

On the other hand, if the method for the abatement of the surface radioactivity presented in section \ref{sec:SURF} should be adopted, the assembly procedure will present an increased complexity and might imply some delay. 

\begin{figure}
\begin{center}
\includegraphics[width=0.65\textwidth]{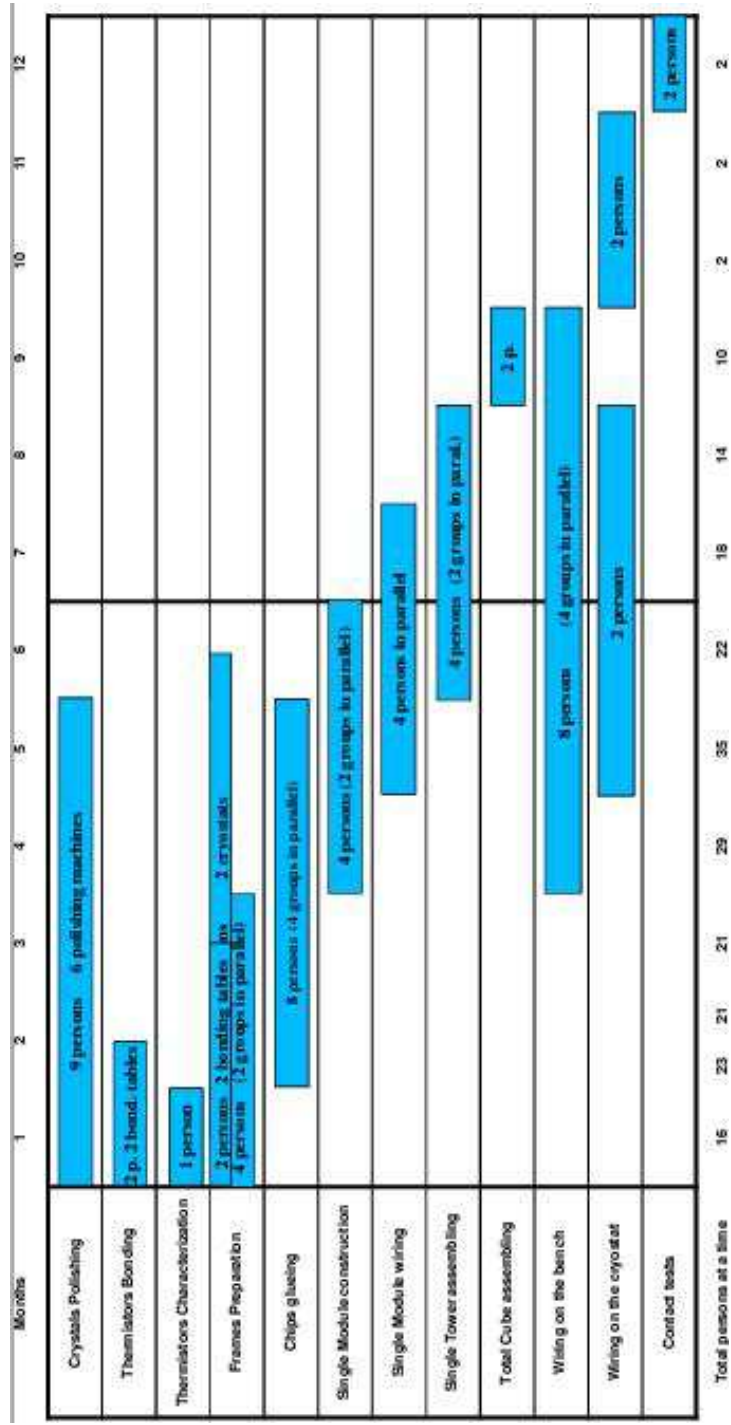}
\end{center}
\caption{Gantt diagram illustrating the CUORE detector fabrication schedule.}
\label{fig:SCHEDULE}
\end{figure}

\subsection{Work breakdown and construction schedule for the CUORE detector}
\label{sec:SCHED}

On the basis of the previous section, it is possible to attempt a work breakdown and a construction schedule for the construction of the CUORE detector. The various items described in section \ref{sec:ASSEM} can be reported synoptically in a Gantt diagram (see fig.~\ref{fig:SCHEDULE}), taking care of the two following points:
\begin{itemize}
\item[1)] The assembly of the CUORE cube must be completed in about one year. Considering that two years are necessary for the growth and preparation of the crystals, this means that CUORE will be cooled down for the first time three years after the growth of the first crystal. It is not feasible to span over  three years the 4-detector module fabrication, along with crystal growth, since this would conflict with the time necessary to produce and clean the copper elements and with the Research and Development program.
\item[2)] In order to concentrate the detector assembly in one year, and taking into account the requirements in terms of time and personnel for the realization of the various items, some of the tasks reported above need to be accomplished by more personnel working in parallel. For example, the operation of {\sl chip gluing} will be performed by three groups of two people each, since only in this way it will be possible to accomplish this task in about three months.
\end{itemize}

Once again, we stress that this construction schedule refers to the CUORE baseline. Innovations (consider for example Nd-based crystals or surface sensitive detectors) would inevitably modify the schedule in a way which is difficult to predict now, and might imply some delay. However, the reported analysis shows that the construction of the CUORE  detector in its conventional form is feasible in one year, but at the price of an extensive use of manpower. Notice that in some phases of the assembly, 35 people must work simultaneously and full time on the detector preparation. Given the size of the CUORE collaboration and the limited availability of technical staff, it is hardly conceivable to realize this formidable task without a substantial use of external resources.

\subsection{General time schedule of the CUORE project}
\label{sec:QSCHED}
\begin{figure}
\begin{center}
\ifx\pdfoutput\undefined 
\includegraphics[width=0.7\textwidth]{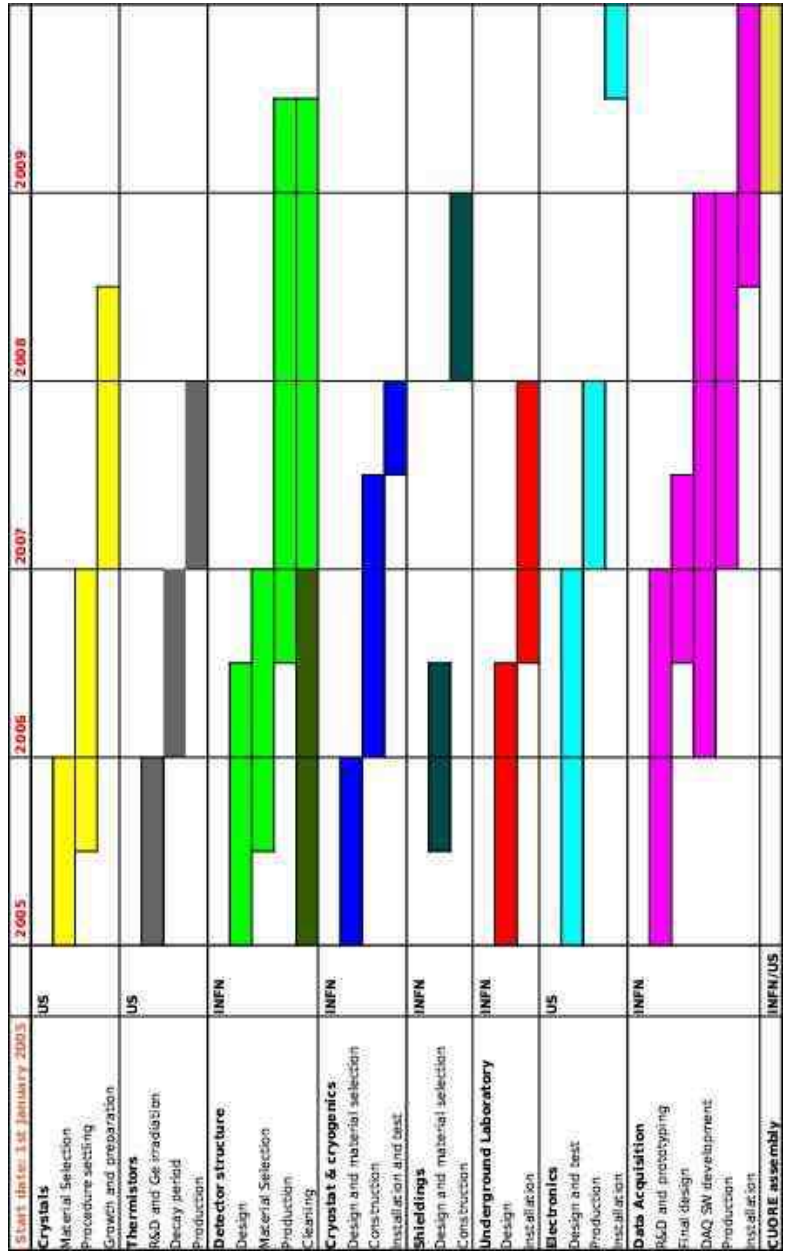}
\else
\includegraphics[width=1.3\textwidth,angle=90]{Pictures/QTime}
\fi
\end{center}
\caption{Gantt diagram illustrating the CUORE project time schedule.}
\label{fig:QSCHEDULE}
\end{figure}

As synoptically reported in the Gantt diagram of Figure~\ref{fig:QSCHEDULE} the general time schedule of the CUORE project requires approximately 5 years after approval. This summarizes the time sequence and duration of all the various tasks described in the previous sections. It should be stressed that the deadlines of many tasks are mandatory since often other jobs  depend on them. However, most of the activities foreseen for the first 18-24 months can run in parallel without much interference. Therefore, most of the required R\&D activities could  proceed during this period without resulting in sizeable delays on the project schedule. This is particularly true (as specified by the dashed filled area in fig.~\ref{fig:QSCHEDULE}) for the improvement in the surface (mainly copper) cleaning procedures.

\clearpage
\section{Off-line analysis} \label{sec:analysis}

The main goal of the off-line analysis is the extraction of the relevant physics informations from the large amount of raw data recorded by the DAQ system. This task includes both the raw detector pulse analysis (e.g. amplitude evaluation, noise rejection, gain instability and linearity correction) aiming at producing reliable n-ple's (a proper number of parameters fully describing each bolometric pulse) and energy spectra, and the following multi-dimensional analysis aiming at obtaining the sought physics results (e.g. \BBz or DM interactions). The identification of the single background sources is also an important goal of the second-level analysis.

\subsection{First-level analysis}
As described in a previous section, when the output voltage of one detector exceeds the trigger threshold,
the acquisition system records a number of converted signal samples. The acquired time window (\ca few sec) must fully contain the pulse development in order to allow an accurate description of its waveform. The existence of a pre-trigger interval just prior to the production of the pulse (``baseline'') guarantees that a small fraction of the number of acquired samples can be used to measure the DC level of the detector (which can be directly correlated to the detector temperature).

The fact that for each trigger an entire waveform is sampled and recorded implies that off-line analysis of a large body of data will require a significant effort. Nevertheless, this effort will be justified because of the useful information that can be extracted from the signal waveform. The following are important goals for the (first-level or \emph{pulse}) analysis (FLA):
\begin{enumerate}
\item{maximization of the signal to noise ratio for the best estimate of the pulse amplitude. This is accomplished by means of the optimum filter (OF) technique \cite{gatti86};}
\item{correction of the effects of system instabilities that change the response function of the detectors (gain ``stabilization'');}
\item{rejection of the spurious triggered pulses by means of pulse shape analysis;}
\item{identification and rejection of radioactive background pulses by means of coincidence analysis.}
\end{enumerate}

The OF technique is frequently used with bolometers to evaluate the amplitude of a signal superimposed on stochastic noise. This algorithm has proven to provide the best estimate of the pulse amplitude under general conditions. Relative to a simple \emph{maximum--minimum} algorithm, this technique allows the evaluation of the signal amplitude with much higher efficiency resulting in an effective improvement of the detector energy resolution. The following information is needed to implement the OF technique: the detector response function (i.e. the shape of the signal in a zero noise condition) and the noise power spectrum. Once these are known, the OF transfer function is easily obtained and used as a digital filter for the acquired pulses. The role of the OF transfer function is to weight the frequency components of the signal in order to suppress
those frequencies that are highly influenced by noise. The amplitude of the pulse is then evaluated using optimally filtered pulses. Effective pulse-shape parameters can be deduced from the comparison of the filtered pulses with the filtered response function (e.g. rms difference after proper time synchronization).

In processing the data off-line, the following parameters are evaluated and recorded to disk for each digitized pulse (\emph{n-ple}):
\begin{enumerate}
\item{the \emph{channel number} i.e., the number of ADC channel that exceeded the trigger threshold;}
\item{the \emph{absolute time} at which the pulse occurred with a precision of 0.1~msec;}
\item{the \emph{OF amplitude} i.e. the amplitude of the optimally filtered signals}
\item{the \emph{baseline}, obtained by averaging a proper number of  points from the pre-trigger interval. Since the detectors are DC coupled, it provides a direct measurement of the detector temperature at the creation of the signal;}
\item{the signal \emph{rise} and signal \emph{decay times};}
\item{the \emph{pulse shape parameters}, obtained by comparing the acquired pulse with the expected response function of the bolometer after OF or adaptive filters. A further powerful technique is based on the use of artificial neural networks (ANN);}
\item{the \emph{pile-up fraction}. Pile-up is usually efficiently rejected by pulse-shape analysis even if this technique can't identify the rejected pile-up events. In order to improve the pile-up rejection and quantitatively evaluate its rate (e.g. for short-time coincidence analysis), the Wiener-filter algorithm is implemented~\cite{press91}.}
\end{enumerate}

The next PLA step is the gain instability correction.
The OF amplitudes are corrected to reduce or cancel the effects of system instabilities responsible for the variation of the ratio between the energy E deposited into a given crystal and the amplitude $\Delta$ V of the
corresponding electrical pulse. According to equation (\ref{eq:maximum}) of our very naive detector model there are three instabilities that can modify the ratio $\Delta$ V/E (where V=V$_b$*G is the output voltage given by the product of the bolometer voltage V$_b$ and the electronics gain G): i) a variation in the electronic gain G, ii) a variation in the bias V$_{Tot}$ and finally iii) a variation in the temperature T$_b$ of the crystal.
The electronic system is designed to guarantee a stability of G and V$_{Tot}$ within 0.1\% . It is however, much more difficult to maintain stability within 0.1\% of the detector temperature on long time scales.
At a temperature of 10~mK this would require maintaining the temperature of 1000 crystals to an accuracy of 10 $\mu$ K for a period of several days.

To overcome this problem, and as already mentioned in previous sections, a silicon resistor glued to each crystal is used as a heater to produce a reference pulse in the detector. It is connected to a high precision programmable pulser that produces a fast voltage pulse every few minutes dissipating the same amount of energy
(E$_{ref}$) into the crystal each time. These voltage pulses mimic pulses produced in the crystal by particle interactions and are used to measure the value of the ratio $\Delta$ V/E. Any variation of the amplitude of the reference pulse is due to variations of the $ \Delta$V/E  ratio.  The OF amplitude of the reference pulse is
therefore used to measure, every few minutes, the actual value of $ \Delta$ V/E while the baseline of the reference pulse provides the contemporary measurement of the value of T. A fit is then used to obtain the values of $ \Delta$ V/E as a function of temperature.
Therefore, in this step of the off-line analysis, the OF amplitude of each pulse is corrected according to the given value of $ \Delta$ V/E(T[t])  for the detector temperature at which the pulse has been generated.
The effectiveness of this technique has been proven in the MiDBD  experiment, where a typical
temperature fluctuation over a day ranged from a few tenths to \ca 100 $\mu $K .
After correction, these fluctuations were reduced to less than  1~$\mu$ K \cite{alessandrello98s} (Fig.~\ref{fig:stab}).

Pulse shape analysis is very useful in rejecting spurious signals produced by microphonics and electronic noise. A confidence level is determined for each pulse shape parameter and for the rise and decay time of each pulse.
Signals falling within these intervals are defined as ``true'' (or physical) pulses, while signals having one or more of their parameters outside of the relevant interval are rejected as noise. The use of more than one pulse shape parameter results in better reliability of the rejection technique.

The linearization of the detector response is critically important for energy calibration. The final step in data processing is the conversion of the OF amplitudes into energy values. The naive bolometer model
previously used assumes linearity; several parameters depend however on the crystal temperature, rendering the corresponding equation non-linear. Accordingly, the relation between $ \Delta $ V and E will be obtained periodically by the use of radioactive calibration sources. The ratio $ \Delta V/E $ will be measured for several gamma lines, and the data will be fit to the model previously described, but taking into consideration the fact that the bolometer resistance and the crystal heat capacity are temperature dependent. This will provide the calibration function of E as a function of $ \Delta$V, that will then be used to convert the OF
amplitudes into energy values.

Finally, the close packed array of the crystals will allow the rejection of events that left energy in more than one single crystal. This will be particularly useful in rejecting a very high energy gamma rays that enter from outside of the array. The small angle Compton scattering in a single crystal can mimic a double beta decay event, a dark matter scattering, or a solar axion. The probability that this photon would escape the rest of the array without a second interaction is small. In addition, background events from radioactivity within the structure of the array will also have a significant probability of depositing energy in more than one crystal. This will also be true for high and intermediate energy neutrons. In the final stage of off-line analysis these coincidence events can be identified from the data which will contain the detector number, signal time, pulse energy, and pile-up parameter.

\subsection{Second-level analysis}
Since the main goal of CUORE is the search for \BBz of \tect, the same will also hold for the second-level analysis (SLA). In CUORICINO, most of the tools of this analysis are powered by the experience gained by the Milano group during the previous experiments on \BBz  of various nuclides. Since many institutions participating in CUORE have a similar experience, we are planning to merge the various analysis techniques of the different groups to apply them to double beta decay processes in order to obtain a single powerful standard analysis procedure. Nevertheless, it will be essential to be able to guarantee a safe cross-check of the results, so that independent analyses from the various groups will be encouraged.

Detection efficiencies for the various transitions will be evaluated by means Monte Carlo simulations specifically designed for the CUORE detector. An updated collection of the most recent calculations for nuclear matrix elements will be kept and updated (according to the various nuclear models). That will be essential to derive reliable results for the effective neutrino mass parameter, the right handed current parameters, or the majoron coupling parameter. 

Physics analysis for Dark Matter and axion searches in CUORE will be completely designed and realized in the framework of CUORICINO. In order to improve the noise rejection near threshold an Artificial Neural Network (ANN) approach will be improved (the possibility of an ANN trigger is under investigation). The effectiveness and reliability of the rejection techniques based on the selection of multi-dimensional confidence intervals (strongly depending on the suitability of the chosen parameters) will be tested in CUORICINO and methods to calibrate the rejection efficiency will be defined. Time stability of threshold parameters will be a major issue. 

\clearpage
\section{Simulation and predicted performances} \label{sec:simulation}

The goal of CUORE is to achieve a background rate in the range 0.001 to 0.01 co\-un\-ts\-/(keV\dot kg\dot y) at the \BBz transition energy of \tect (2528.8 keV). A low counting rate near threshold (that will be of the order of \ca 5--10 keV) is also foreseen and will allow CUORE to produce results for Dark Matter and Axions searches.
In Sec.~\ref{sec:bkgsim} we present a very conservative evaluation of the background attainable with CUORE; this is mainly based on the state of the art of detector design and of radioactive contaminations. This rather pessimistic approach in background evaluation is the only one that presently guarantees a reliable prediction. However CUORE construction will require about five years (sec.~\ref{sec:QSCHED}); there will therefore be time enough for R\& D dedicated to background reduction as is foreseen in most next generations experiments. 

\subsection{Background simulations} \label{sec:bkgsim}
\begin{figure}[h]
 \begin{center}
 \includegraphics[width=0.7\textwidth]{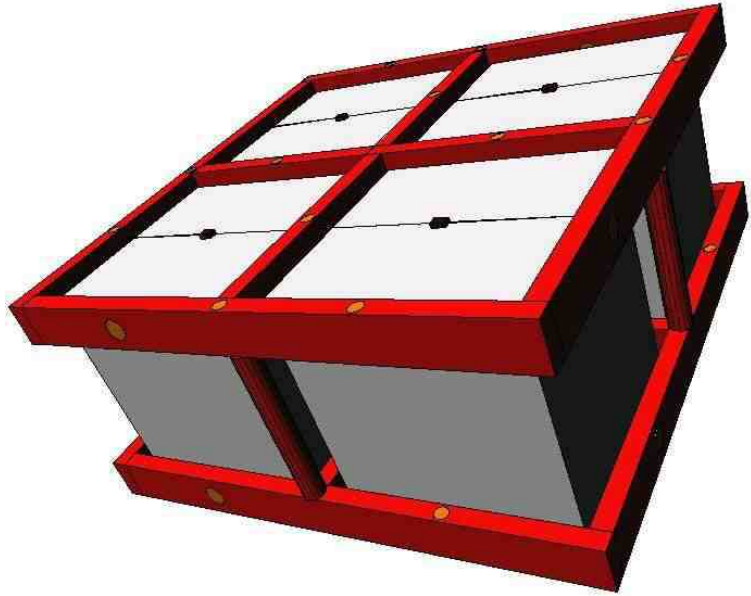}
 \end{center}
 \caption{Details of the CUORE single modules accounted for in the Monte Carlo simulations}
 \label{fig:mcframe}
\end{figure}

Radioactive contamination of individual construction materials, as well as the laboratory environment, were measured and the impact on detector performance determined by Monte Carlo computations.
The code is based on the GEANT-4 package; it models the shields, the cryostat, the detector structure and the detector array.
Even smallest details of the detector apparatus (copper frames, screws, signal wires, NTD thermistors, etc.; Fig. \ref{fig:mcframe}) and of the cryogenic setup were taken into account.

\begin{table}[h]
\caption{Bulk contamination levels (in picograms per gram) used in the simulation for \teod, copper and lead.} \label{tab:contaminazioni}
\begin{center}
\begin{tabular}{cccccc}
  \hline\hline
  Contaminant & $^{232}Th$ & $^{238}U$ & $^{40}K$ & $^{210}Pb$ & $^{60}Co$ \\
  \hline
  \teod & 0.5 & 0.1 & 1 & 10 $\mu$Bq/kg & 0.2 $\mu$Bq/kg \\
  copper & 4 & 2 & 1 & 0 &  10 $\mu$Bq/kg\\
  Roman lead & 2 & 1 & 1 & 4 mBq/kg & 0 \\
  16 Bq/kg lead & 2 & 1 & 1 & 16 Bq/kg & 0 \\
  \hline\hline
\end{tabular}
\end{center}
\end{table}

It includes the propagation of photons, electrons, alpha particles and heavy ions (nuclear recoils from alpha emission) as well as neutrons and muons.
For radioactive chains or radioactive isotopes alpha, beta and gamma/X rays emissions are considered according to their branching ratios. The time structure of the decay chains is taken into account and the transport of nuclear recoils from alpha emissions is included.

The considered background sources are:
\begin{enumerate}
\item{bulk and surface contamination of the construction materials from the  $ ^{238}U $, $ ^{232}Th $ chains and $ ^{40}K $ and $ ^{210}Pb $ isotopes;}
\item{bulk contamination of construction materials due to cosmogenic activation;}
\item{neutron and muon flux in the Gran Sasso Laboratory;}
\item{gamma ray flux from natural radioactivity in the Gran Sasso Laboratory;}
\item{background from the \BBd decay.}
\end{enumerate}

In the next sections these background sources and their contribution to CUORE background are discussed. 
It will be shown that, in a very conservative approach where the CUORE-tower mechanical structure is assumed identical to the structure used in CUORICINO, considering the worst possible condition for bulk contaminations (i.e. all the contamination equal to the present 90\% C.L. measured upper limits (Table~\ref{tab:contaminazioni})) and assuming for surface contamination a reduction of about a factor ten with respect to CUORICINO, the CUORE background will be \ca 0.007~co\-un\-ts\-/(keV\dot kg\dot y) at the \BBz transition and \ca 0.05~co\-un\-ts\-/(keV\dot kg\dot d) near threshold.

\subsection{Bulk contaminations} \label{sec:bulkcont}

The main contribution to background from bulk contaminants comes from the cryostat structure (cryostat radiation shields), the heavy structures close to the detectors (the copper mounting structure of the array, the Roman lead box and the two lead disks on the top of the array) and from the detectors themselves (the \teod crystals). The radioactivity levels used in the computations for these materials are given in Table~\ref{tab:contaminazioni}. These levels are nearly equal to the best upper limits obtained for the radioactive content of these same materials as shown in Table~\ref{tab:contam_mis}.

\begin{table}[b]
\caption{Available 90\% C.L. upper limits for bulk contaminations of \teodn, copper and lead (levels in picograms per gram if not differently indicated).} \label{tab:contam_mis}
\begin{center}
\begin{tabular}{ccccccc}
  \hline\hline
  Contaminant & method & \thdt & \udt & \kq & \pbdd & \coss \\
  \hline
  \teod & bolometric & 0.7 & 0.1 & 1. & 100 $\mu$Bq/kg & 1 $\mu$Bq/kg \\
  Copper~\cite{heusser:pca} & Ge diodes & 5.6 & 2 & 0.3 & - &  10 $\mu$Bq/kg\\
  Roman lead & Ge diodes & 50 & 30 & 2 & 4 mBq/kg & - \\
  Low act. lead~\cite{laubenstein:pc} & Ge diodes & 3.4 & 2.7 & 1.7\pom 0.3 & 23.4\pom 2.4 Bq/kg & 18\pom 1 $\mu$Bq/kg\\
  \hline\hline
\end{tabular}
\end{center}
\end{table}

All the values reported in this table are 90\% upper limits. Indeed in all cases no evidence of the presence of radioactive contaminants was obtained for the material examined. The contamination levels of \teod reported in Table~\ref{tab:contam_mis} were obtained from the data collected in the first run of CUORICINO with the \ciccio crystals. Copper \cite{heusser:pca} and lead \cite{laubenstein:pc} contaminations were determined through low activity Ge spectrometry. These data represent the best evaluation reported in literature for the radioactive contamination of \teodn, copper and lead. 
From a comparison between Table~\ref{tab:contam_mis} and Table~\ref{tab:contaminazioni} it is clear that the contamination levels assumed for the CUORE simulation are perfectely compatible with the presently available upper limits. The only exception is that of \coss in \teod crystals, whose presence as a cosmogenic contamination of the crystals will be discussed later (sec.~\ref{sec:cosmog}).

\begin{figure}[h]
 \begin{center}
 \includegraphics[width=1\textwidth]{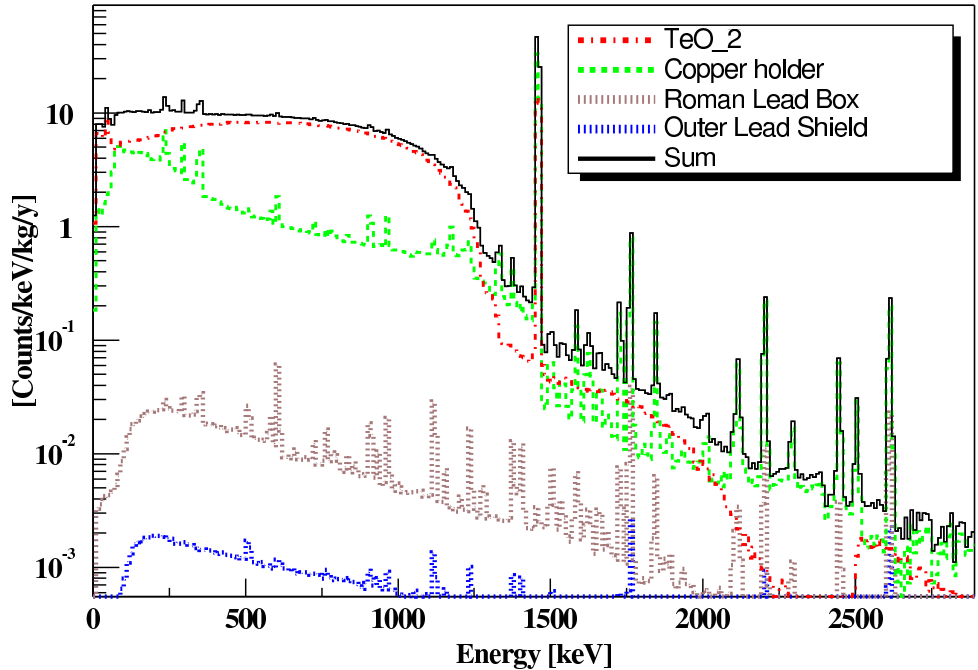}
 \end{center}
 \caption{Simulated spectra for bulk contaminations of the \teod crystals, the Copper  structure, the Roman Lead shield and the outer Lead shield. Each spectrum is obtained by summing the  simulated anticoincidence spectra of all the CUORE detectors.}
 \label{fig:MCbulk1}
\end{figure}

\begin{figure}
 \begin{center}
 \includegraphics[angle=90,width=0.8\textwidth]{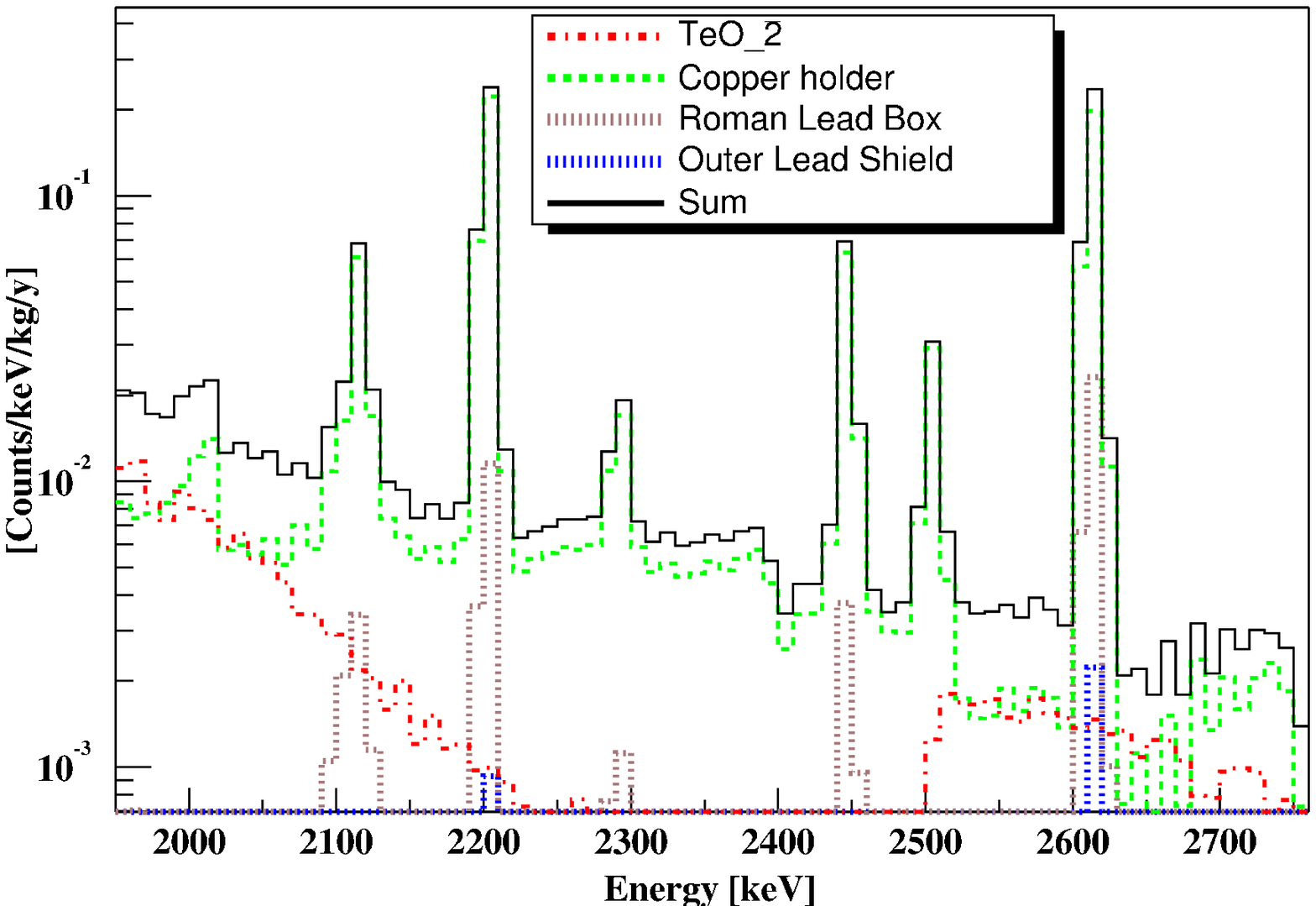}
 \end{center}
 \caption{Double beta decay region of the simulated spectra for bulk contaminations of the \teod crystals, the Copper structure, the Roman Lead shield and the outer Lead shield. Each spectrum is obtained by summing the  simulated anticoincidence spectra of all the CUORE detectors.}
 \label{fig:MCbulk2}
\end{figure}

Concerning small (mass) components of the CUORE detector, which we have not considered so far in the simulation (e.g. NTD Ge thermistors, Si heaters, glue layers, pins, wires and soldering material,etc.), a very careful selection according to their contamination is planned for the future. In some case we already have upper limits on their contamination that make their contribution to the CUORE background negligible; in other cases 
further measurements will be required.

\begin{figure}
 \begin{center}
 \includegraphics[angle=90,width=0.8\textwidth]{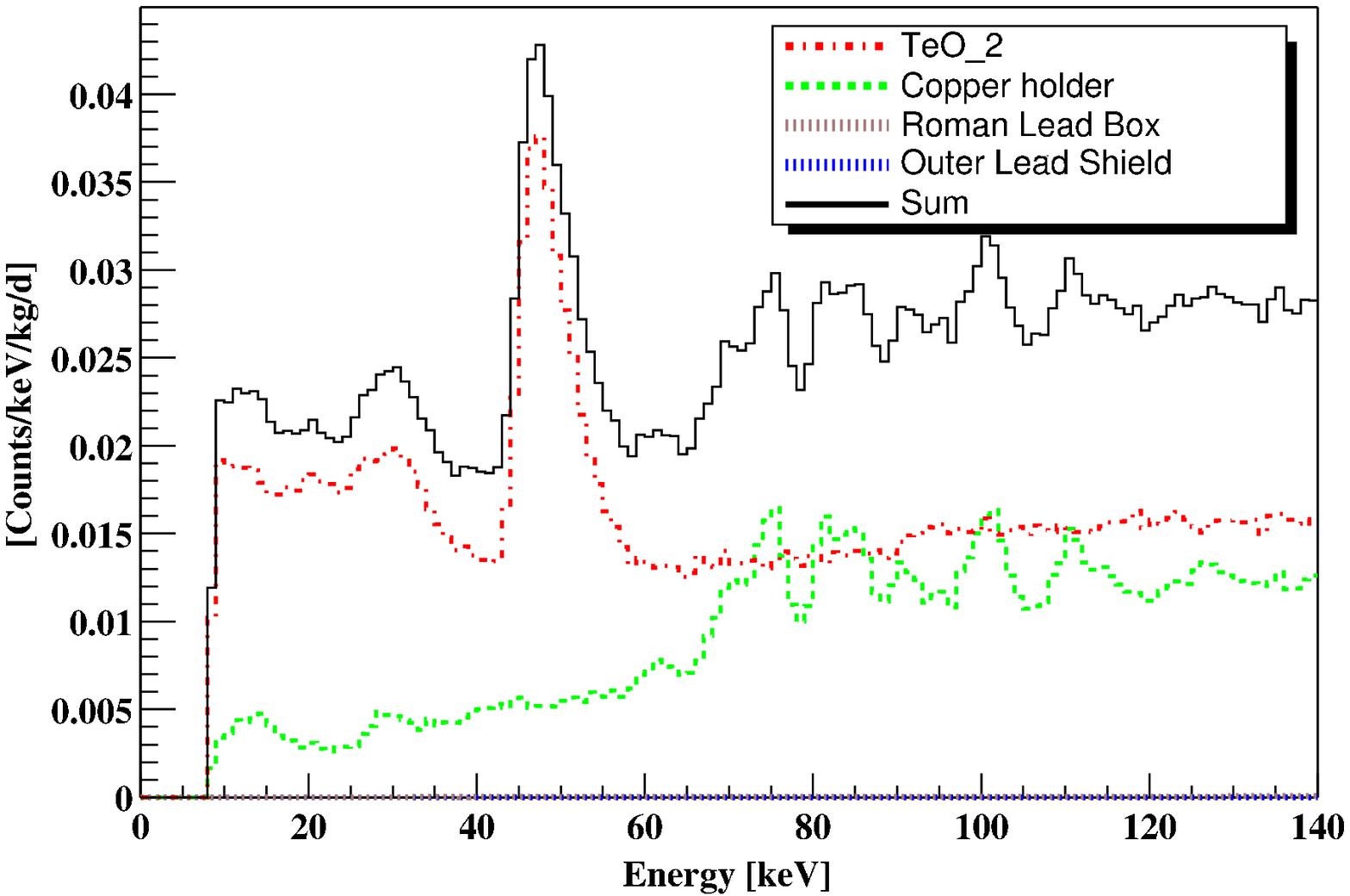}
 \end{center}
 \caption{Dark matter region of the simulated spectra for contaminations of the \teod crystals, the Copper  structure, the Roman Lead shield and the outer Lead shield. Each spectrum is obtained by summing the  simulated spectra of all the CUORE detectors after the anticoincidence cut.}
 \label{fig:MCbulk3}
\end{figure}

The results of the Monte Carlo simulations (see Fig. \ref{fig:MCbulk1}) using the contamination levels discussed here are given in Table~\ref{tab:bulk} for the \BBz decay and the low energy (10-50~keV) regions. The assumed threshold is 10~keV and only values obtained after requiring an anti-coincidence between detectors are indicated.
These values have to be considered as upper limits on the possible contribution of bulk contaminations to the CUORE background; they prove that with the already available materials, background levels lower than \ca 4\per 10$^{-3}$~counts/keV/kg/y in the \BBz decay  region (see Fig .\ref{fig:MCbulk2}) and \ca 3\per 10$^{-2}$~counts/keV/kg/d in the low energy (10-50~keV) region (see Fig.\ref{fig:MCbulk3}) are assured.

\begin{table}
\begin{center}
\begin{tabular}{ccccc}
  \hline\hline
  Simulated & \teod & Cu & Pb & TOTAL \\
  element & crystals & structure & shields &  \\
  \hline \BBz decay region & & & & \\
  counts/keV/kg/y & 1.6\per 10$^{-3}$ & 1.5\per 10$^{-3}$  & 7.0\per 10$^{-4}$ & 3.8\per 10$^{-3}$\\
  \hline dark matter region & & & &  \\
  counts/keV/kg/d & 2.3\per 10$^{-2}$ & 9.6\per 10$^{-4}$  & 5.0\per 10$^{-5}$  & 2.4\per 10$^{-2}$\\
  \hline\hline
\end{tabular}
\end{center}
\caption{Computed background in the \BBz decay and in the low energy regions for bulk contaminations in the different elements, the Cu structure accounts for the detector mounting structure and the 50~mK shield.} \label{tab:bulk}
\end{table}

\subsection{Surface contaminations} \label{sec:surfcont}

Surface contaminations contribute to the background only when they are localized on the crystals or on the copper mounting structure directly facing them. As learned in MiDBD and CUORICINO (sec.~\ref{sec:MiCUORICINO}), the presence of even a very low level of radioactive impurities on the surfaces of the detectors can produce a non negligible contribution to the \BBz background level. Unfortunately this kind of contamination, that is important mainly for detectors without a surface dead layer (as is the case of bolometers), is poorly studied. Lacking any data coming from direct measurements of the typical impurity levels present on \teod and copper surfaces, we rely on the results obtained in MiDBD and CUORICINO. 

These indicate that both a surface contamination of the crystals and of the copper surface of the mounting structure is present. In the case of \teod, we know that the contamination is mainly \udt and that its presence is strictly connected to the kind of surface treatment undergone by the crystals. In the CUORICINO \ciccio crystals this contamination produces a background counting rate in the \BBz region of about (4\pom 3)\per 10$^{-2}$~counts/keV/kg/y (table~\ref{tab:Qino_contrib}). 
Concerning the copper surface contamination, we still miss a clear indication of its origin and identity (i.e. we are not yet able to distinguih between contributions generated by different sources), but we have an evaluation of the \BBz  background counting that can be ascribed to them (whatever they are, \udt or \thdtn): \ca 0.1\pom 0.05~counts/keV/kg/y.

According to a Monte Carlo simulation of the CUORE detector, based on the CUORICINO contamination levels, we obtained a background counting rate in the \BBz region (after the anticoincidence cut) of about 1.6\per 10$^{-2}$~counts/keV/kg/y and 5.8\per 10$^{-2}$~counts/keV/kg/y for the \teod crystals and the copper structure respectively~\ref{tab:CUORE-sup}.
The goal of CUORE is to reduce the surface contribution by a factor at least 20 with respect to this evaluation, obtaining a background coming from surfaces of about 3\per 10$^{-3}$~counts/keV/kg/y.
Improvements of a factor of \ca 1.5--2 are expected simply by possible reductions of the copper mounting structure dimensions (the Monte Carlo simulation we used so far refers to an identical mechanical structure for the CUORE and CUORICINO single detector modules). 
A reduction by a factor of at least ten of the copper (\teod) surface contamination is therefore the first milestone of CUORE. 

\begin{figure}
 \begin{center}
 \includegraphics[angle=90,width=0.8\textwidth]{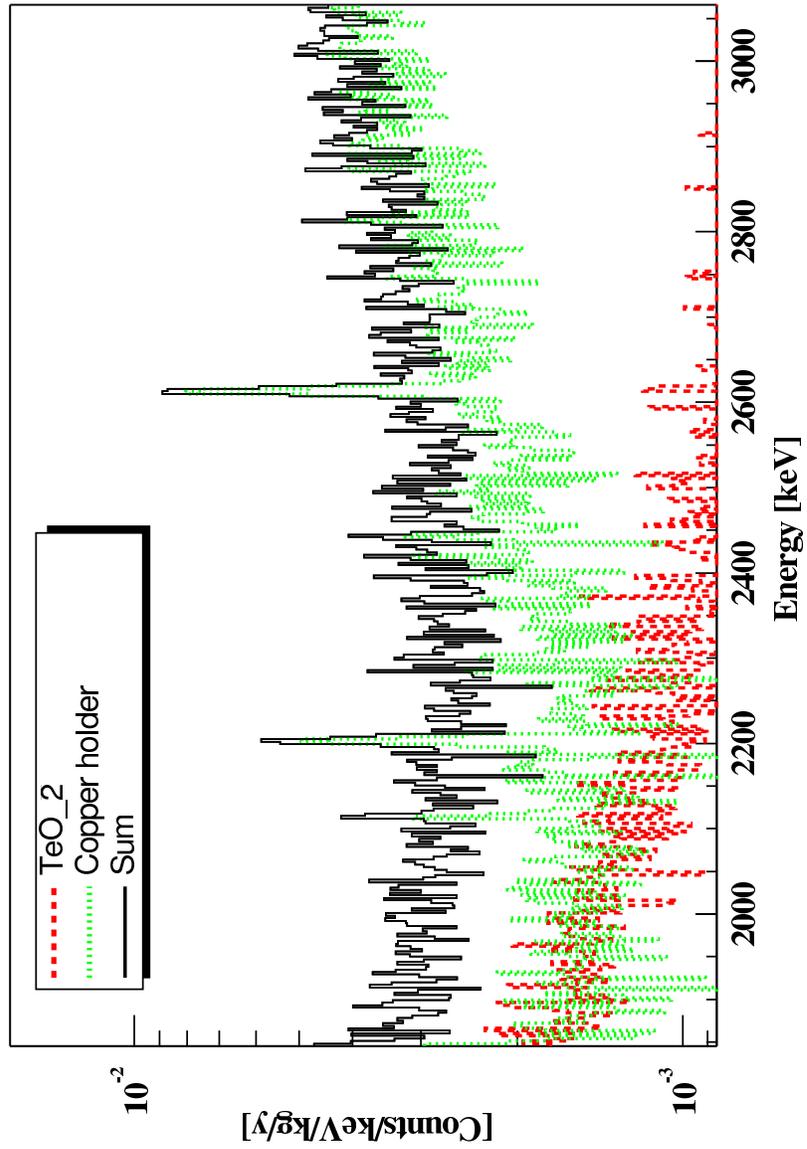}
 \end{center}
 \caption{\BBz region of the simulated spectra for the surface contaminations of the \teod crystals and of the Copper  structure. Each spectrum is obtained by summing the simulated spectra of all the CUORE detectors aftre the anticoincidence cut.}
 \label{fig:MCsup2}
\end{figure}

\begin{figure}
 \begin{center}
 \includegraphics[angle=90,width=0.8\textwidth]{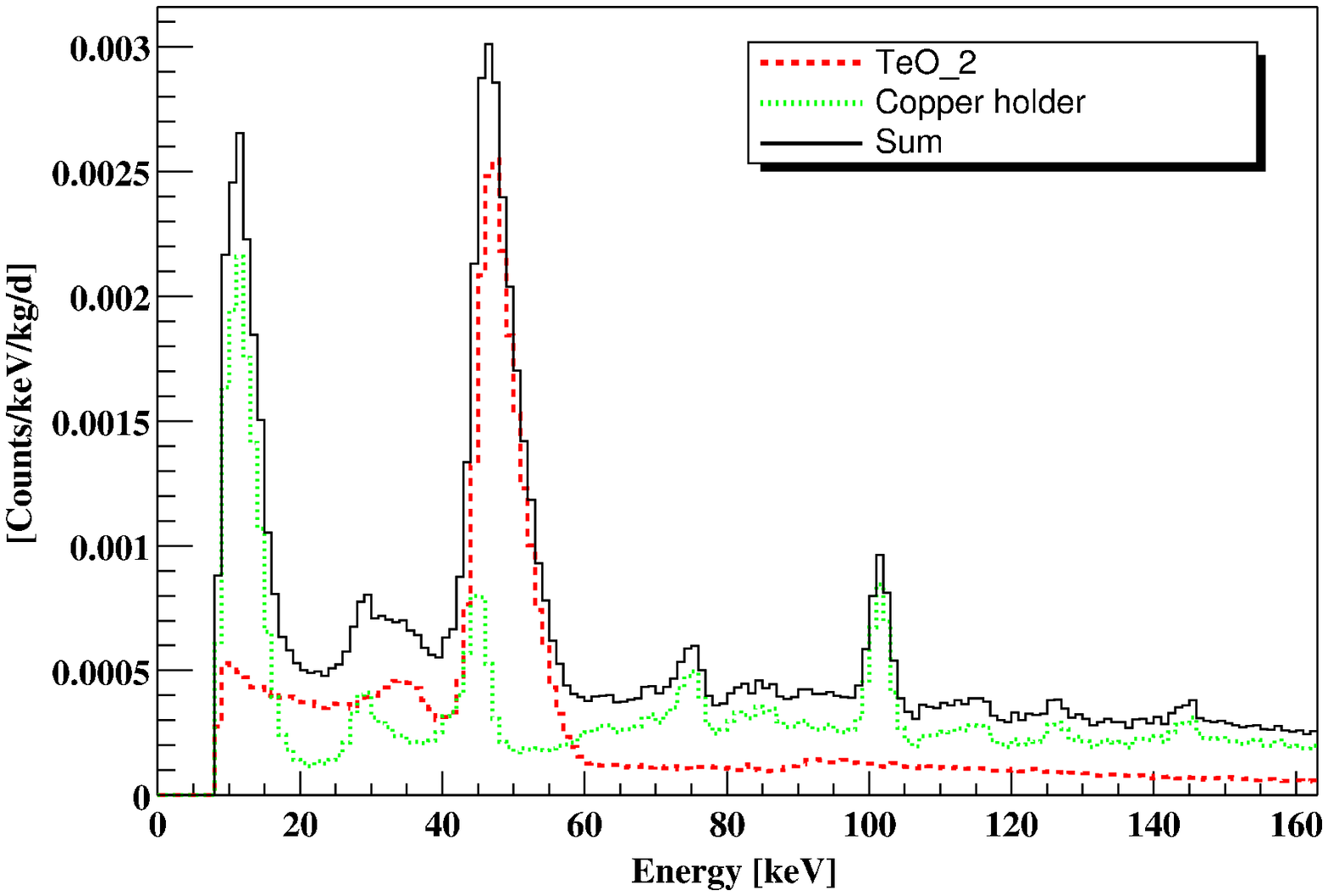}
 \end{center}
 \caption{Dark matter region of the simulated spectra for the surface contaminations of \teod crystals and of the Copper structure. Each spectrum is obtained by summing the simulated anticoincidence spectra of all the CUORE detectors.}
 \label{fig:MCsup3}
\end{figure}

The expected shape of the background produced by surface contaminations in CUORE (after a reduction by a factor 20 of both the crystal and copper surface contaminations but assuming the same mechanical structure of CUORICINO) is given in Figures \ref{fig:MCsup2} and \ref{fig:MCsup3}. The corresponding background contributions in the two regions of interest are summarized in Table~\ref{tab:surfbkg}. 

As discussed in Sec.~\ref{sec:MATE}, the MiDBD crystal surfaces were contaminated, probably during the polishing procedure, due to the use of highly contaminated powders. Polishing powders with a much lower radioactive content are however commercially available \cite{sumitomo:tm} and have already been used for CUORICINO (tab.~\ref{tab:materials}) obtaining a clear improvement in the surface contamination. Complete over all control of the surface treatments undergone by the crystals after growth, and the use of radiopure substances should guarantee an even better result than that obtained in CUORICINO.

A similar situation holds for copper. In MiDBD and in CUORICINO the copper surfaces were treated with an etching procedure \cite{palmieri96} optimized in order to reduce impurities on surfaces before the sputtering process.

This procedure significantly improved the surface quality of copper and reduced its surface contamination. However it was not optimized from the point of view of background. The use of low contaminated liquids (water and inorganic acids are available with \udt and \thdt contamination levels lower than 0.1~pg/g) in a low background environment will allow a sizeable improvement of the surface contamination of copper. 

This (i.e. similar to CUORICINO but with much more care to prevent radioactive contamination) and other improved cleaning procedures (sec.~\ref{sec:SURF}) are under study and will represent a consistent part of the CUORE R\&D activities during the first two years after approval.
Fast diagnostic methods to measure the surface contamination levels of Copper and determine the effectiveness of the adopted surface cleaning procedures are also under investigation. 
Preliminary results indicate that high resolution ICPMS could satisfy our requirements \cite{trincherini:pc}. Sensitivities achievable with this technique for \udt and \thdt are in the range of 10$^{-(3-2)}$~ppt of solution under investigation. Since the maximum solution concentration can be 2\per 10$^{-3}$, the expected sensitivity on the solute contamination (Copper) would be in the range 1--10~ppt of \udt and \thdtn, i.e. enough for our requirements. From the above discussion in fact, the surface contamination measured in CUORICINO (assuming a 1~$\mu$m copper layer) is of the order of \ca 1--10~ppb of \udt and \thdtn. Of course radio-purities comparable with the instrumental sensitivity are required for liquids (mainly acids and water) to be used in the solution preparation (i.e. copper surface removal). Liquids  satisfying this requirement are however already commonly available. 

\begin{table}
\begin{center}
\begin{tabular}{ccccc}
\multicolumn{5}{l}{Cylindrical structure}\\
\hline\hline
   Simulated        &   \thdt                  &   \udt	                &   \coss                &    Sum  \\
	\hline
   \teod   & 212$\pm$23.6     &    14.5$\pm$5.11 &    1200$\pm$19.5  &    1420$\pm$31.0   \\
   Cu Box & 393$\pm$46.3     &    84.8$\pm$28.3 &    -  &    4784$\pm$54.3   \\
   Cu Bars  & 29.7$\pm$4.08     &    27.1$\pm$5.12 &    -  &    56.8$\pm$6.55    \\
   Cu Frames& 163$\pm$21.5     &    233$\pm$33.7 &    -  &    397$\pm$39.9   \\
   Cu 50mK Shield& 31.9$\pm$2.21     &    - 		&    -  &    31.9$\pm$2.21    \\
   Roman Pb Shield& 604$\pm$24.4     &    - 		&    -  &      604$\pm$24.4  \\
    Sum   & 1430$\pm$61.5     &    360$\pm$44.6 &    1200$\pm$19.5  &    2390$\pm$74.5    \\
\hline\hline
\end{tabular}
\caption{Computed background (after the anti-coincidence cut) in the \BBz energy region for bulk contaminations in the different elements for cylindrical structure in units of 10$^{-6}$ c/keV/kg/y.}\label{tab:surfbkg}
\end{center}
\end{table}

\begin{table}
\begin{center}
\begin{tabular}{ccc}
\hline\hline
Element      &  Contamination	       &  Contribution to the\BBz region \\
 	     &    Bq/cm$^2$	       &	  c/keV/kg/y		       \\
\hline
\teod & 1.9E-08 $\pm$ 9.8E-09   &     1.6E-02 $\pm$ 8.4E-03	 \\
Copper       & 4.9E-08 $\pm$ 1.9E-08   &     5.8E-02 $\pm$ 6.6E-03	 \\
Total        & 3.4E-08 $\pm$ 1.3E-08   &     7.4E-02 $\pm$ 1.1E-02	 \\

\hline\hline
\end{tabular}
\caption{Estimated upper contribution to the CUORE \BBz region from surface contaminations obtained by using the surface contamination levels evaluated for CUORICINO and assuming an exponential density profile with $\lambda$=1~$\mu$m for \teod crystals (\udt) and $\lambda$=5~$\mu$m for Copper (\udt and  \thdt).}\label{tab:CUORE-sup}
\end{center}
\end{table}

\subsection{Cosmogenic contribution} \label{sec:cosmog}
 
Cosmogenic activation is produced by cosmic rays when the crystals are above ground during fabrication and shipping of the crystals from the factory to the underground laboratory. To determine the activation we used COSMO, a code based on computed cross sections, to estimate the type and amount of radionuclei produced by cosmic rays on \teod \cite{cosmo:desc}. The radionuclei produced by the activation of tellurium are mostly tellurium isotopes (A=121,123,125,127) as well as $^{124}$Sb, $^{125}$Sb, $^{60}$Co and tritium, these last three being of more concern because of their long half-life ($^{125}$Sb:~beta decay of 2.7 years, end-point energy of 767 keV, $^{60}$Co:~beta decay of 5.27 years end-point energy 2823 keV and $^{3}$H:~beta decay of 12.3 years, end-point energy of 18 keV). 

\begin{figure}
 \begin{center}
 \includegraphics[angle=90,width=0.8\textwidth]{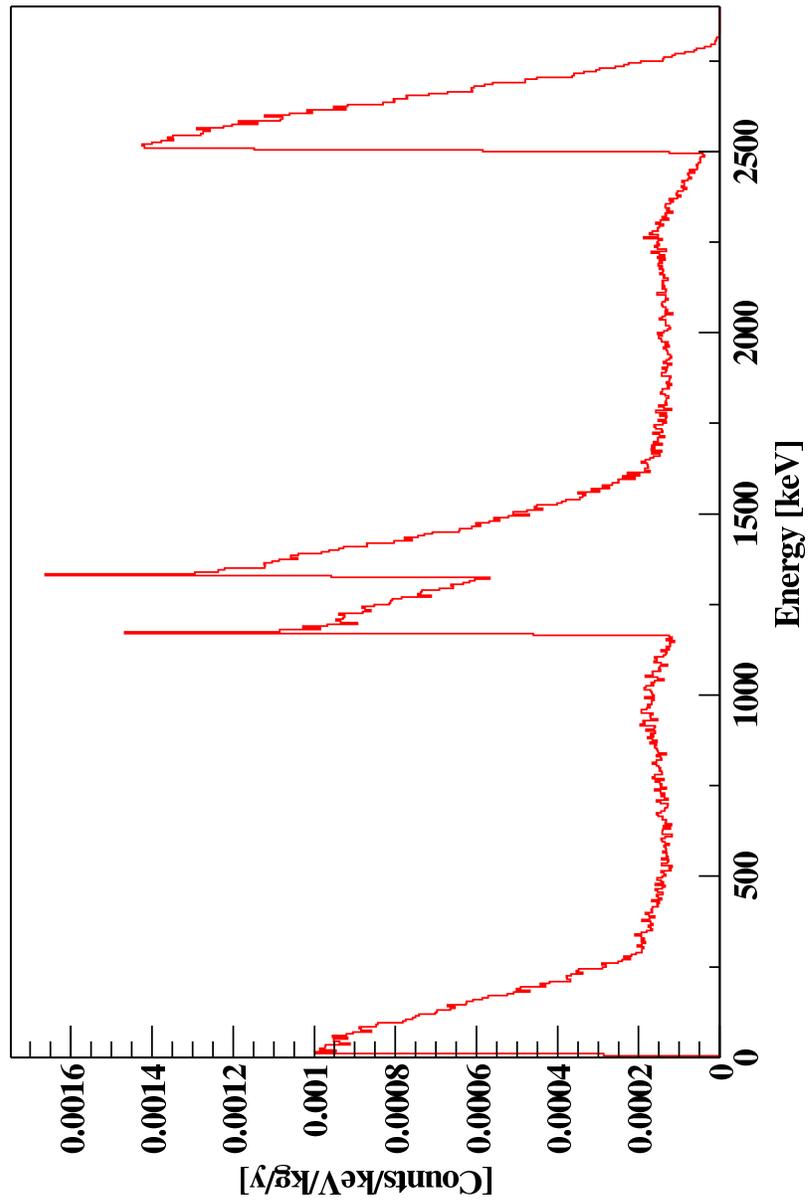}
 \end{center}
 \caption{Simulated spectrum for the assumed $^{60}$Co bulk contaminations of \teod crystals. The spectrum is obtained by summing the
 simulated anticoincidence spectra of all the CUORE detectors.}
 \label{fig:MCbulkCo}
\end{figure}

We have studied, toghether with SICCAS who produces the crystals, and on the basis of our present knowledge of the cosmic rays production rates, a possible time schedule for the crystal growth and shipping to Gran Sasso that can guarantee the required low level of $^{60}$Co (see Fig .\ref{fig:MCbulkCo}). 
In the case of CUORE, the control on crystal production will be severe.
A single 750 g crystal can be grown in about two months, while to grow the 988 CUORE detectors will require 18 months. 
Once grown the crystals will be shipped to Italy and stored underground, therefore their  total exposition to cosmic rays could be limited to about 4 months.  
The total induced activities remaining after 2 years underground have been estimated \cite{cosmo:desc} and the consequent contribution to the detector counting rate was deduced by a MC simulation. The radionuclei that contribute to background throught their $\beta^-$ decay are:
\begin{itemize}
\item{the long living $^{60}$Co isotope in the \BBz region, with an activity of \ca 0.2~$\mu$Bq/kg, while  a minor contribution is due to the isotopes $^{110m}$Ag and $^{124}$Sb whose activity is 4 times lower and fast decreasing with time;}
\item{in the dark matter regions the long lived nuclei of $^{3}$H and $^{125}$Sb with an activity of \ca 7~$\mu$Bq/kg for the former and of \ca 15~$\mu$Bq/kg for the latter.}
\end{itemize}
The influence of $^{60}$Co in CUORE background was already considered in the evaluation of the contributions due to bulk contaminations of the crystals, while the contribution of $^{3}$H and $^{125}$Sb to the dark matter region is completely negligible if compared to the intrinsic background from all other sources. 

As already mentioned in section~\ref{sec:ENVIR}, a recent experimental determination of the \coss production cross section on Te by 1.85~GeV protons has been obtained at LNBL~\cite{norman03}. The measured value (0.63$\pm$0.15 mb) is in strong disagreement with the value used by the COSMO program. A similar disagreement, even if at a lowerextent, has been found on the basis of an activation measurement carried out at CERN at an energy of 24 GeV, which is however less important for cosmic ray activation. The contribution to CUORE \BBz background by cosmogenic \coss discussed above, could be therefore overestimated and the allowed exposure period for \teod crystals could be consequently much longer. Further measurements of the \coss production cross section are therefore of crucial importance and we plan to have, in the near future, other direct measurements by irradiating Tellurium samples with proton beams of suitable energies.

\subsection{Underground neutron, $\mu$ and $\gamma$ interactions}

As mentioned earlier, neither contributions from underground cosmic muons nor neutrons have been taken into account in detail in the estimation of the background. 
However, the following simplified arguments will serve to have an approximate idea of their contribution.  The depth of the LNGS (3500 m.w.e) reduces the muon flux down to \ca 2\per 10$^{-8}$~cm$^{-2}$s$^{-1}$, but a further effective reduction could be obtained with the use of an efficient (99.9\%) active veto for muons traversing the CUORE setup in order to tag possible events associated with them. The muon-induced contribution to the background is therefore expected to be negligible. 

On the other hand, the heavy shieldings surrounding the CUORE detector will substantially reduce the event rate due to environmental radiation of various origin (neutrons and photons), environmental radioactivity (natural decay chains U/Th, $^{210}$Pb, $^{40}$K, \dots), as well as muon interactions in the surroundings rock or in the shielding itself. 
 
Neutrons may constitute a worrisome background for the dark matter experiment because for appropriate neutron energies (few MeV) they can produce nuclear recoils ($\lesssim $100 keV) in the detector target nuclei which would mimic WIMP interactions. Simple kinematics implies that in the case of tellurium, neutrons of 1(5) MeV could elastically scatter off tellurium nuclei producing recoils of energies up to 31 (154) keV. In general, one considers neutrons of two origins: from radioactivity in the surroundings or muon-induced. Depending on the 
overburden of the underground site (i.e., depending on the muon flux), muon-induced neutrons are produced, at lesser or greater rate, both inside and outside the shielding. They are moderated (according to their energies) by the polyethylene/lead shield (when produced outside) or tagged by the muon veto coincidence (when produced within the passive shielding). 

In the case of external neutrons (from the rocks, from fission processes or from (n,$\alpha$) reactions, as well as neutrons originated by muons in the walls of the underground site), the environmental neutron flux has been measured in LNGS. The result is of \ca 1\per 10$^{-6}$ cm$^{-2}$s$^{-1}$ for the thermal component, \ca 2\per 10$^{-6}$ cm$^{-2}$s$^{-1}$ for the epithermal and \ca 2\per 10$^{-7}$ cm$^{-2}$s$^{-1}$ for energies over 2.5 MeV. 
They are fairly well moderated by the polyethylene and eventually absorbed or captured.  

We have carried out a Monte Carlo simulation of the propagation of neutrons through the 10 cm thick borated polyethylene shield of CUORE. The result is that the neutron  induced event rate on the entire energy range (from threhsold to 10 MeV) is much lower than the contribution due the bulk contamination of crystals.  
Neutrons produced by muon interaction inside the shielding materials are very scarce and they can be efficiently tagged by the muon veto.
We have estimated that for the LNGS muon flux (2.5\per 10$^{-8}$ $\mu$/(cm$^{2}\,$s)), muons would produce in the CUORE shielding of polyethylene (10 cm) and lead (20 cm)),  about \ca 0.04 neutrons/day in the polyethylene shield and \ca  25 neutrons/day in the lead shell.
So, independent of the mechanism used to reject or tag the events associated to neutrons, their rather small number is expected to play a secondary role in the total background compared with other main sources of background. 

A preliminary evaluation of the influence of the environmental  $\gamma$  background in Gran Sasso resulted in a negligible contribution for the  \BBz region and a contribution similar to that of bulk contaminations for the dark matter region.

A more complete and detailed study of the background rates from external sources for CUORE is underway and will be used for the optimization of shieldings and muon veto.

\subsection{Two neutrinos double beta decay background}

Using the present upper limits for the \tect \BBd~half-life, the unavoidable background produced by the \BBd decay in the dark matter region is lower than 10$^{-4}$~co\-un\-ts\-/(keV\dot kg\dot y) and is completely negligible in the  \BBz region. This is true because of the relatively good energy resolution of the \teod bolometers.

\clearpage
\section {Innovations in the CUORE structure}\label{sec:INNO}

The CUORE structure described up to now is based on the results obtained in CUORICINO and on the experience gathered with the test-runs performed in the hall C cryostat on the \ciccio crystals. We have been very conservative in the design of every CUORE element. However, we foresee the requirement for a significant activity of Research and Development during the next few years, in parallel (but with lower priority) with the realization of the CUORE baseline structure outlined so far. In fact, there are indications that several aspects of the CUORE detectors can be improved substantially with respect to the present performance (for instance, single element reproducibility) and some of them {\em need} to be improved dramatically (mostly, the background level due to surface contamination that could spoil CUORE, and a large part of its scientific value, if not controlled adequately).

In this section, we outline the R\&D work planned for CUORE and discuss the possible upgrading of the present structure. We would like to stress that all this R\&D activity can be carried out in Hall C without any interruption of the presently running  CUORICINO experiment.

\subsection {Innovation in the single module}\label{subsec:INNO_SINGLE}

\subsubsection {Optimization of the thermistor sensitivity}
As described in section \ref{sec:single}, our current NTD thermistors (belonging to series \#~31) have a T$_0$ value of about 3.0~K. This parameter has never been optimized and was obtained only in the attempt to get a tolerable resistance value between 10 and 15~mK (not higher than a few hundreds of M$\Omega$). In the two versions of the MiDBD arrays, and in the CUORICINO experiment, we had however to heat the refrigerator base temperature up in order to achieve these resistance values. The naturally achieved base temperature would have lead to too high resistances, of the order of few G$\Omega$. There are two reasons why lower T$_0$ (and therefore lower resistivities at equal temperatures) would probably imply better detector performance, in spite of the lower thermistor sensitivity:
\begin{itemize}
\item {it would be possible to operate detectors either at lower temperatures (therefore with higher signals) or at the same temperature as now, but with lower resistances (therefore with lower noise);}
\item {it is well known that lower T$_0$ thermistors have better electron-phonon coupling (see for instance \cite{alessandrello00b}), favoring therefore a more efficient heat transmission to the thermistor electrons with consequent higher pulse amplitudes.}
\end{itemize}
An R\&D program, lasting about one year, will therefore define the optimum T$_0$ for CUORE thermistors. Several thermistor species with  T$_0$'s ranging from~2 through 3~K will be produced by the Berkeley group and test detectors will be produced in order to select the best doping level in time for the production of the final thermistor batch.

\subsubsection {Innovative thermal couplings in detector construction}
A set of glues will be selected to improve the coupling between NTD~Ge~thermistors and energy absorbing crystals. Methods will be studied to deposit controlled amounts of glue in a reproducible way, in the form of spots and/or a uniform layer. Measurement of the low temperature thermal coupling provided by the various glues will be performed, in order to define the most conductive coupling. Of course, all these cryogenic tests must be accompanied by a careful analysis of glue radioactivity, so as not to introduce dangerous radio-impurities. Special NTD Ge thermistors with ``pedestals'' and ``legs'' will be developed in order to reduce the surface touching the crystals. This reduction will allow the use of a glue layer at the interface in spite of the large difference in thermal contraction coefficients between glues and crystalline solids.

In addition, the proponents intend to perform tests in a thin film laboratory on eutectic coupling between NTD~Ge~thermistors and energy absorbing crystal. The most promising couplings will be realized in small scale prototype detectors and compared with the standard 9-araldyte-spot coupling, mentioned in section \ref{sec:model}.

Innovative methods to hold and couple the absorbing crystals to the thermal bath will be studied, such as special TEFLON supports, nylon threads, simple gluing etc., with and without additional calibrated thermal links. Prototype small scale detectors will be studied.

Finally, the proponents will perform tests of real-size CUORE modules in LNGS hall C dilution refrigerator with the innovative couplings determined in the prototype small scale detectors. If new couplings, and a related procedures for their series realization, will be defined within three years since the start of the CUORE project, they will be introduced in the final CUORE structure.

\subsubsection {Modification of the basic crystal size}
There is clear experimental evidence that a crystal mass increased by about a factor two should not spoil basic detector performance. At present, the energy resolution at 2.6~MeV is higher by a significant factor (typically a factor \ca 5) than that expected from the detector noise. Therefore, an increase of the detector mass by a factor two should affect the intrinsic detector resolution because of the consequent reduction of the signal amplitude, but should influence only marginally the energy resolution in the Double Beta Decay region. Of course, this argument does not apply to the search for Dark Matter and Solar Axions, where threshold, and therefore signal-to-noise ratio at low energies, is the crucial parameter. However, in a more Double-Beta-Decay-oriented version of the CUORE project, an increase of the crystal size (with equal total mass) could bring some advantages, mainly of practical and economical nature (fewer electronic channels, shorter and easier assembly operations). The proponents have in mind in particular elementary modules based on 6\per6\per6 cm$^3$ crystals, whose mass is larger by a factor 1.73 than present crystals' one. A possible configuration consists of 16 towers arranged in a 4\per4~matrix. Each tower will be a stack of 9~modules. As in the present configuration, each elementary module will consist of 4~crystals held in common frames. The total mass would be very similar to the baseline CUORE structure (namely, 756~kg, assuming the same density that provides 750 kg for the present structure), and an almost cubic shape for the array would be preserved. Another approach consists of giving up the 4-detector arrangement of the single module (which might not be testable in the present refrigerators because of its size) and to move to stacks of single crystals. Even in this case, the favored arrangement is a 8\per8\per9~(height) three-dimensional crystal matrix. In both solutions, the total number of channels would be 576 instead of 1000.

Negotiations for the order and purchase of 6\per6\per6~\teod crystals are in progress. 
Two such larger crystals  have been already ordered and should be tested in Gran Sasso in early 2004.

A final decision on the crystal size should be taken not later than 18 months after CUORE start: this should leave sufficient time for the series production of the crystals. That implies a hard R\&D work on 6\per6\per6~\teod crystals in the first year.

\subsection {Possible insertion  in CUORE of crystals containing compounds of other nuclear candidates for \BBz}
\label{subsec:NEOD}
As pointed out before many compounds containing \BBz candidate nuclei could be inserted in form of single or multiple crystals in the structure of CUORE. Some of these compounds which look as particularly favourable from the thermal and nuclear point of view are listed in Table~\ref{tab:DBDnuclei}. 

\begin{table}
\begin{center}
\caption{Most favorable \BBz candidate nuclei.}
\begin{tabular}{lcc}
\hline\hline
Compound & Isotopic abundance (\%)& Transition energy (keV)\\
\hline
$^{48}$CaF$_2$ & 0.0187 & 4272 \\
$^{76}$Ge & 7.44 & 2038.7\\
$^{100}$MoPbO$_4$ &  9.63 & 3034 \\
$^{116}$CdWO$_4$ & 7.49  & 2804 \\
$^{130}$TeO$_2$ & 34 & 2529  \\
$^{150}$NdF$_3$ & 4.64 & 3368  \\
\hline\hline
\end{tabular}
\label{tab:DBDnuclei}
\end{center}
\end{table}

All these compounds have been  tested as absorbers of thermal detectors. These tests were successful, but some difficulty was found for NdF$_3$ and NdGaO$_3$, due to the paramagnetic properties of Neodimium.
A particularly good candidate would be a crystal of $^{48}$CaF$_2$. In fact a ``scintillating bolometer'' where light and heat were simultaneously detected was already operated~\cite{alessandrello98}. Operation of these detectors in coincidence mode between heat and light pulses would be very effective in the suppression of the \BBz region background due to degraded alpha particles. Unfortunately the very low isotopic abundance of $^{48}$Ca  makes this experiment fiscally impossible, unless new processes for enrichment of this isotope be discovered. Other scintillating bolometers like those with CdWO$_4$ or MoPbO$_4$ can obviously be envisaged. We would  like to specify here that also \teod crystals have been found to emit light~\cite{demarcillac03a}. 
Their yield is however very low (about 50 eV per MeV) which make them inefficient  for the operation of a scintillating bolometer. This approach would be in any case much less effective than the active shield with Ge or Si bolometers, which will be considered later. This shield would in fact suppress not only the surface activity due to alpha particles, but also the surface activity due to beta decays.

A dedicated R\%D work is foreseen for the Neodimium case, since $^{150}$Nd is probably the best candidate to search for \BBz decay \cite{elliott02}, thanks to the high transition probability due both to favorable nuclear matrix elements and to the high transition energy (3.37~MeV). Unfortunately, $^{150}$Nd natural isotopic abundance is low (5.6~\%) and no dielectric diamagnetic compound of Nd is known to exist. The first problem can probably be solved thanks to innovative techniques of isotopic enrichment, such as ICR (Ion Cyclotron Resonance) method \cite{dolgolenko00}, while the second one poses a challenge for the bolometric technique. The main difficulty, as already mentioned, is due to the magnetic properties of Nd: all the Nd compounds compatible with growing large crystals present a magnetic ordering implying inevitably high specific heat at low temperatures. However, the chance to realize a working Nd-based bolometer relies on our ignorance. The thermal models which attempt to explain how a large-mass, phonon-mediated bolometer works are not completely convincing. In particular, there are indications that a non-thermal component of the signal, due to ``high'' energy phonons, may be important (see section \ref{sec:bolometers}, last paragraph). In this case, the specific heat could be a scarcely relevant parameter: the right question would be rather how long a non-thermal phonon lives in a crystal with a magnetic ordering and which chance it has to transmit its energy to the NTD~thermistor. We have decided to follow a strictly experimental approach and to plan bolometric tests on commercially available crystals containing Nd. We have performed tests on existing NdF$_3$ (a paramagnetic material with huge Shotckey peaks at about 70~mK) and NdGaO$_3$ (which presents a more promising antiferromagnetic ordering below \ca 1~K) crystals in the hall C dilution refrigerator. In particular tests carried out with two cubic crystals of NdGaO$_3$  of  1.3 and 3 cm side have not allowed to reach, after weeks of cooling, temperature lower than \ca 40 mK even if  with different time profiles. These temperatures would not be sufficient to produce bolometers with acceptable energy resolution.

In about two years from the start of the CUORE project a final answer on the feasibility of a Nd-based bolometer must be given. In the event of a positive response, a part of the CUORE array (probably not more than \ca 10~\% in terms of mass) could consist of Nd-based crystals. The Nd~section of the array could be even more sensitive to Majorana neutrino mass than the Te~section, in spite of the lower mass. In this case, CUORE would be a formidable two-fold experiment, capable to cross check within itself a positive result in the search for \BBz~decay.

\subsection {Development of surface sensitive \teod elements}
\label{subsec:SURF}
Perhaps, the most important objective of the CUORE R\&D program consists of developing methods to control and to reduce at a negligible level the radioactive background coming from surface events. This background is presently observed in the large mass bolometers of the CUORICINO experiment and was observed in the previous 20~detector arrays. 
Monte Carlo simulations (see section \ref{sec:surfcont}) show that the dominant background contribution in the \BBz region of interest (around 2.5~MeV) is originated by passive elements which surround the detectors (in particular, the copper supports for the crystals, which work also as a 5~mK thermal bath for the detectors). This background is to be ascribed to surface contaminations in Th and U, which emit energy-degraded alpha particles able to deposit their energy at the detector surfaces, populating the 2000-4000 keV spectral region. The presently designed CUORE configuration would not allow a substantial improvement of this background (expected to be of the order of 0.1~counts/(keV kg y)) unless sizeable improvements in the reduction of the surface contributions would be obtained. 

\begin{figure}[h]
\begin{center}\leavevmode
\includegraphics[width=0.9\linewidth]{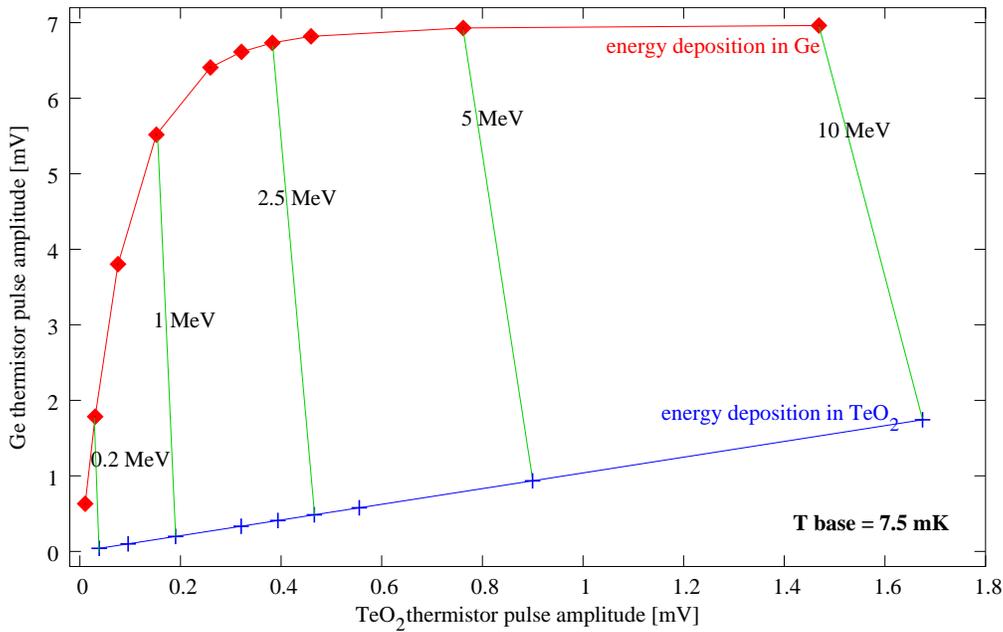}
\caption{Scatter plot of the pulse amplitudes from \teod thermistor versus the pulse amplitudes from shield thermistor.}
\label{fig:SP_SURF}
\end{center}
\end{figure}

The objective of a specific R\&D program has to be therefore the realization of elements whose surface background is reduced down to 0.001--0.001~counts/(keV kg y) in order to satisfy the CUORE sensitivity requirement. This result can be achieved in three ways:
\begin{itemize}
\item {to improve dramatically the quality of the surface treatment of copper (this option is not discussed here, see section \ref{sec:SURF}, and should be pursued in parallel with the two following ones;}
\item {to realize single CUORICINO and CUORE elements with a radically innovative structure, in which the copper surface facing the detectors is reduced by a factor at least 100-1000;}
\item {to realize bolometers able to distinguish an event originated at the surface from one originated in the bulk.}
\end{itemize}

As far as the second item is concerned, the R\&D program consists of identifying materials which can replace copper in forming the main part of the detector holder (best candidates are plastics, such as TEFLON, nylon or acrylic materials, achievable with very high radiopurity). Detector holders based on these new materials have to be developed, first for reduced-scale detectors, and secondly, after selection of the most promising solutions, for modules with the final size foreseen for the CUORE experiment. 

As far as the third item is concerned, the basic idea consists in the realization of active shields in the form of thin, large-area, ultrapure, Ge~or Si~bolometers, by which the \teod crystals will be surrounded so as to get rid of possible events originating in the germanium (which could release part of their energy in the main detector) with the anti-coincidence technique. A powerful method would be to thermally couple these active shields to the main \teod detector. The~Ge or Si~auxiliary bolometers would be attached at the main crystal providing almost complete coverage without a dedicated holder. In this way, a composite bolometer will be realized with multiple read-out, capable to distinguish the origin of the event (active shields or \teod crystal) by means of the comparison among pulses coming from the different elements. A degraded alpha coming from outside would release its energy in the shield, originating a thermal pulse that, seen by the shield thermistor, would be much higher than a pulse corresponding to the same energy released directly in the main \teod crystal. In fig. \ref{fig:SP_SURF} it is possible to appreciate the efficiency of the separation method. A scatter plot is shown, where the amplitude of the pulses collected with the \teod thermistor is plotted against the amplitude of the corresponding pulses collected with the shield thermistor. Two clearly separated curves identify the events generated in the shield with respect to the events generated in the main crystal. The pulses have been simulated using a multi-stage thermal model of the composite bolometer.
In order to reduce the readout channels, the six thermistors of the shields could be connected in series or in parallel. In this way, the total number of channels would double. We believe that this is a tolerable complication. The size of the auxiliary bolometers would be 50\per 50\per 0.3~mm$^3$ or 60\per 60\per 0.3~mm$^3$: therefore, the main crystal plus the shields would form a cube with a side only 0.6~mm larger than the main crystal alone, preserving the general CUORE structure and rendering the assembling procedure substantially unchanged. This proposed solution has the potential to control the problem of the surface radioactivity relying on a technological improvement of the detector rather than on a better cleanness of the employed materials.

\begin{figure}[h]
\begin{center}\leavevmode
\includegraphics[width=0.9\linewidth]{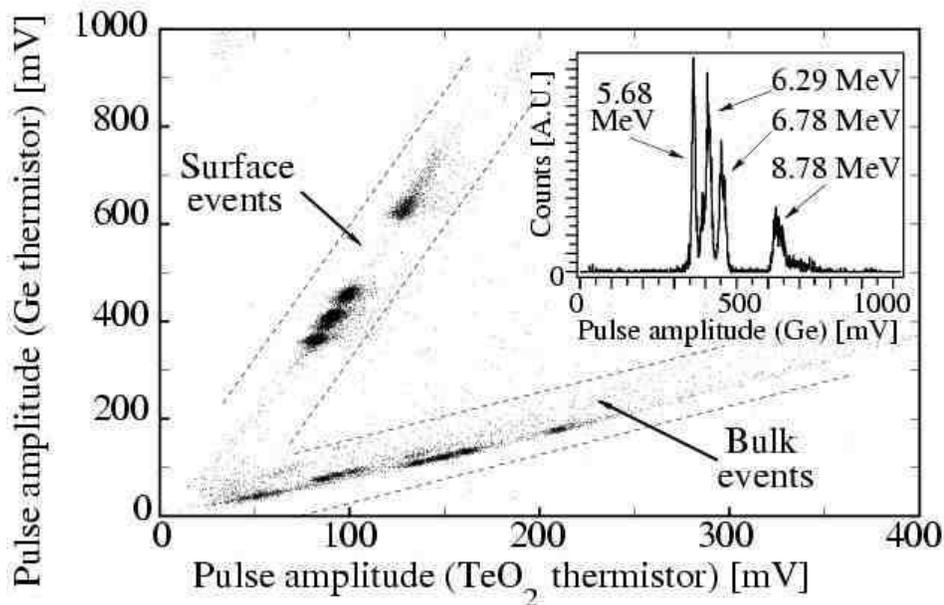}
\caption{Scatter plot of the pulse amplitudes from \teod thermistor versus the pulse amplitudes from shield thermistor obtained with a 12~g \teod surface sensitive detector.}
\label{fig:sp_12}
\end{center}
\end{figure}

\begin{figure}[h]
\begin{center}\leavevmode
\includegraphics[width=0.9\linewidth]{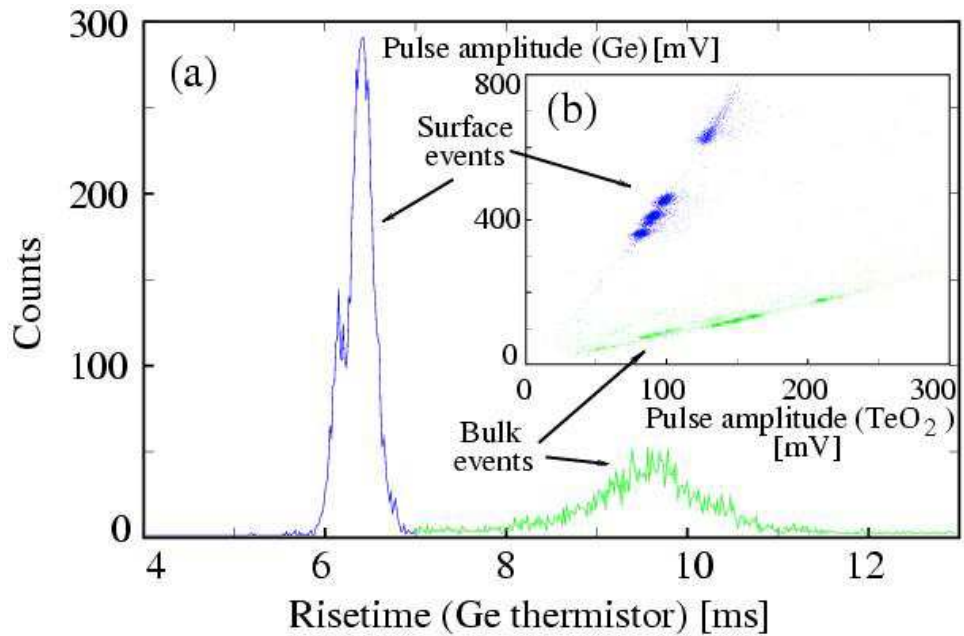}
\caption{Rise time distribution of the pulses from the Ge layer thermistor of a 12~g \teod surface sensitive detector.}
\label{fig:sprt_12}
\end{center}
\end{figure}

In order to test the exposed principles, several small scale prototype detectors were realized. In all of them the main absorber consisted of a \teod single crystal, with masses ranging from 12~g (2\per2\per0.5~cm$^3$) to 48~g (2\per2\per2~cm$^3$).  A Ge or Si single crystal with a 1.5\per1.5~cm$^2$ surface and 500~$\mu$m thickness was glued at a 2\per2~cm$^2$ face by means of four epoxy spots (1 mm diameter and 50~$\mu$m thick). Neutron Transmutation Doped (NTD) Ge thermistors were glued at the main absorbers and at the auxiliary Ge or Si bolometers with six epoxy spots (0.5 mm diameter and 50 mm thickness). The NTD chip size was 3\per1.5\per1~mm$^3$ in all cases. The resistance-temperature behavior of the chips is parameterized as usual by the formula $R(T) = R_0 e^{(T_0 / T )^{1/2}}$. The values of R$_0$ and T$_0$ were respectively 2.65 Ohm and 7.8 K. The coupling of the main absorber to the heat bath was realized by means PTFE blocks, as in the CUORICINO modules. The Ge or Si active layers were exposed to $\alpha$ particles. The source was obtained by a shallow implant of  $^{224}$Ra nuclides onto a piece of copper tape, facing the Ge or Si layer. $^{224}$Ra is an a emitter with a half life of 3.66 d in equilibrium with its $\alpha$ and $\beta$ emitting daughters. The main $\alpha$ lines are at 5.68, 6.29, 6.78 and 8.78 MeV. Two weak lines sum up at about 6.06 MeV. 
The detectors were cooled down in a low power dilution refrigerator located in the Cryogenic Laboratory of the Insubria University (Como, Italy). The detector thermistors were DC biased through a voltage supply and a 20 GOhm  room-temperature load resistance. Typical operation temperatures were T \ca 25 mK, corresponding to a thermistor resistance R \ca 100 MOhm. The voltage across the thermistor was typically \ca 25 mV. 
We report here about the results obtained with a 12 g detector with a Ge active layer. Similar results were obtained with the other devices (no remarkable difference was observed between Ge and Si shields). The energy absorber was a \teod crystal with a rectangular shape (2\per2\per0.5 cm$^3$ dimension). The fast pulses coming from the active Ge  layer had a rise time of \ca 6.5 ms and a decay time of \ca 40 ms, while the pulses coming from the main absorber were remarkably slower (rise time \ca 9.7 ms, decay time \ca 70 ms). The scatter plot obtained is shown in fig.~\ref{fig:sp_12}. 
The two bands, corresponding to surface and bulk events, are well separated and the 4 main $\alpha$ lines of the source are clearly appreciable. A powerful separation between the two event types can be achieved also by rise time selection. In fig.~\ref{fig:sprt_12}(a) the rise time distribution is shown for pulses from the layer thermistor. The fast surface events are cleanly separated by the slow bulk events. A selection of the ``fast'' events identifies clearly the surface event band, as shown in fig. ~\ref{fig:sprt_12}(b).
The rise time analysis shows that the scatter plot is redundant to select surface events. It might be convenient then to get rid also of the main absorber thermistor, reducing the read-out channel to one as in the conventional case. On the other hand, a detailed analysis will be performed to check if pulse shape discrimination can allow to identify surface events using signals only from the main absorber thermistor. In this case, the Ge or Si layers would operate as signal-shape modifiers and would not require additional thermistors, leading to a substantial simplification of the device.
This experiment demonstrates dramatically the power of the technique we have developed to identify a superficial impact point in low temperature calorimeters.

\subsection {Development of scintillating \teod bolometers}
\label{subsec:SCINT}
Measuring at the same time the light and heat produced in the detector is an attempt already exploited in the quest for the background reduction~\cite{shutt93,bobin97}. 
In the DBD, the electrons of the decay will produce at the same time phonons and light signals (in the region of MeV), while the alpha particles from residual radioactive decays are expected to release a negligible signal in light. In the case of WIMP interactions, the signal is produced by nuclear recoil, then without scintillation, contrary to the main background, given by beta and gamma radioactive decays. In the latter case however, the expected signals are in the region of 10 KeV. 
Light signals in bolometric detectors can be detected by a second bolometer, thermally decoupled form the main crystal.
An R\&D program is developed in the framework of the CUORE collaboration to produce such a hybrid detector based on \teodn. A scintillation signal in a \teod bolometer operated at 20 mK~was previously reported~\cite{demarcillac03} and has been confirmed by our preliminary measurements. The measured light yield reported was quite low (0.67\% that of BGO) but sufficient to distinguish alpha particles and gammas at energies in the MeV range. 
The scintillation in pure \teod crystals may be due to exciton luminescence or to the presence of a defect, for example V-O (oxygen vacancy with one trapped electron). Anyway, whatever the nature of this scintillation could be, it is clear that a larger light yield will be welcome for an improved discrimination power on particles. This can be done only by doping the \teod crystal with an activator. 
The crystal growth of doped \teod crystals is not easy~\cite{lukasiewicz92}. This is why the choice of the activator ion to enter in the \teod lattice (substitutional position is preferred) has to be carefully studied keeping in mind two main criteria: luminescence properties of the free ion and host lattice possibility to accept the dopant. In our case, a supplementary constraint is imposed by the fact that the dopant for obvious reasons, has to have stable isotopes only. On the other hand, it is to be noted that in this application practically no restriction is to be imposed on the emission spectrum and/or on the decay time.
With respect to the host lattice possibility to accept the dopant, several parameters have to be taken into consideration: ionic radius, oxidation number, coordination number, type of outer atomic shell, electronegativity and chemical hardness of the ion. 
A first campaign of crystal growth has already started in collaboration with the Research Institute for Solid State Physics and Optics of the Hungarian Academy of Sciences
using four different dopant activators. We expect to get preliminary results at the end of 2004, in order to select the most promising dopant for further attempts.\\

According to option (a) of section~\ref{sec:ENVIR}, the R\&D program about reduction of surface radioactivity must be completed at most within two years from the start of CUORE, in order to have time to transfer the new detector concepts to the CUORE mass production.

\clearpage
\section {Cleaning of CUORE structure: crystals, frames and ancillaries}\label{sec:SURF}

A peculiar property of thermal detectors is that they are active over the entire volume and therefore strongly subject to radioactive surface contaminations. This contribution dominates the CUORICINO \BBz background (sec.~\ref{sec:CUORICINO}) and could be a dominant contribution also in CUORE (sec.~\ref{sec:simulation}) The material surface cleaning problem is therefore crucial for CUORE success. Notable improvements in the background reduction were in fact obtained by applying standard surface cleaning techniques to the detector elements (e.g. \teod crystal lapping) and to the detector mounting structure (copper box and frames, teflon stand-ups, contact pins, etc.). In a sense, MiDBD and CUORICINO represent a successfull R\&D for CUORE material surface cleaning. In order to reach a background level lower than 0.01 c/(keV~kg~y) however, further improvements are required. In particular, \emph{all} possible surface cleaning techniques must be considered in order to choose the most appropriate. Moreover, since cleaning techniques are deeply selective, effective contamination analysis methods must also be considered/developed. A series of up-to-date cleaning methods are summarized in the following. Each of them shows positive and negative features that must be considered and compared in order to reach a final decision before the CUORE detector mounting starts. 
As discussed in sections \ref{sec:CUORICINO} and \ref{sec:simulation}, a factor of ten reduction in the copper surface contaminations would be enough to reach the 0.01~counts/(kev\dot kg\dot y) \BBz region background level in CUORE. The R\&D program about surface cleaning (which must include also the development of powerful diagnostic methods) must be completed at most within two years from the start of CUORE (sec.~\ref{sec:QSCHED}). 

\subsection{Surface cleaning techniques}
 A surface contaminant is defined as any material deposited on a substrate that interferes with any process involving the surface layer up to a certain depth or that affects surface properties or surface stability in an undesirable way. Surface contaminants can cover the whole surface (e.g. oxide layers or adsorbed hydrocarbon layers), can be limited to restricted areas (e.g. particles or fingerprints) or can be dispersed into an epidermal layer for a certain depth (e.g. inclusions of foreign material or diffused impurities). Recontamination (for example by adsorbed vapors from the atmosphere) means contamination picked up after the cleaning process in the external processing environment and
before any further processing begins.

The variability of the contamination is what mainly affects cleaning reproducibility. Therefore, to fail the control contamination means not to control cleaning, and it will be expected that this will be the main problem for the cleaning operation of the whole CUORE structure. Cleaning is the reduction of surface contamination to an acceptable level in order to have no significant amounts of undesirable material. But what are the CUORE requirements to get an adequately clean surface? In the absence of a clear answer to this question, the strategy suggested by the proponents is to transfer to CUORE the know-how developed at INFN Legnaro National Laboratory (LNL) in the topic of surface treatments for Superconducting Cavities either made of bulk Niobium or Niobium Sputtered Copper. In addition to that, Padua University and the LNL of INFN are currently holding a Master in ``Surface Treatments to innovative mechanical technologies for the Industry''. This experience involves both industries and the material science section of Padua University, and represents the best humus for facing and solving ``CUORE problems''.

Our goal is twofold: we would like to clean the CUORE copper structure and to process all the \teod Crystals. In the following, we will list all the possible cleaning processes that could be significant for our purpose. Then, on the basis of the exposed advantages and disadvantages, we will select among them the best processes for cleaning the CUORE structure. In this way we will propose two different surface treatments, the former for the copper structure, the latter for the \teod crystals.

The scope of reviewing all the possible treatments advisable for CUORE is however not only limited to the selection of the best cleaning steps. It is also a warning in getting deep in any issue where cleaning is concerned, in order to have the highest possible control of contamination.

Not knowing if the surface is contaminated, indeed, does not mean that the surface is clean. Simple operations, normally
considered of little importance can be detrimental. For instance, it would be absolutely normal to degrease pieces, before cleaning, by means of Alkaline cleaners. However it is absolutely mandatory to know that alkaline cleaners generally contain dissolved silicates, carbonates, borates, citrates and even molybdate. Residuals of these particles can remain attached in micro-valleys of a microrough surface.

\paragraph{Abrasive cleaning\\}
The removal of gross contamination by abrasive cleaning includes the use of:
\begin{itemize}
\item{Abrasive surfaces: sandpaper, emery paper, steel wool, scotch-brite, scouring pads, etc.}
\item{Abrasive powders in a paste or fluid carrier: SiC, Al$_2$O$_3$, diamond, precipitated calcium carbonate (CaCO$_3$), talc, kaolin, wheat starch. Cerium Oxide (widely used in glass lenses industrial cleaning was avoided for CUORICINO/CUORE \teod crystals because of its high level of radioactive impurities).}
\item{Impacting particles entrained in a high velocity gas or liquid stream: vapor honing, glass bead blasting, liquid
honing, grit blasting, sand blasting.}
\item{Abrasives combined with an etchant to provide chemical-mechanical abrasion and polishing.}
\end{itemize}

 Abrasive particles can be used wet or dry and with various particle sizes. Particles can be accelerated in a high velocity
water stream or in a gas stream by using a pressurized siphon system. 
In addition to removing gross contamination, the following results can be obtained:
\begin{itemize}
\item{Abrasive embedding into the surface even up to hundreds of microns}
\item{Roughening of the surface}
\item{Introduction of microcracks into the surface of brittle materials}
\item{Production of unacceptable distortion of metallic surface due to the compressive stresses introduced by a king of ``shot peening''}
\end{itemize}

Blasting done with plastic media is less damaging than with harder grits; however it can leave organic contamination on the surface). There is however a clean alternative to all that: the cleanest mechanical polishing is \emph{ice micro-droplets} blasting in a high pressure nitrogen stream. This system has been conceived for the cleaning of superconducting cavities, but it is only in the prototype stage and any application to CUORE should start ab initio.

\paragraph{Chemical etching\\}
Chemical etching removes surface material along with strongly adherent contaminants.
Advisable procedures before chemical etching are:
\begin{itemize}
\item{Alkaline pre-cleaning, in order to get uniform wetting and etching}
\item{Pickling, i.e. the removal of the large amounts of oxides formed during fabrication process.}
\item{Bright dip, which is nothing else than a Mild Pickling}
\end{itemize}

 It must be kept into account that acid cleaning of metals can have the detrimental effect of introducing hydrogen into the
surface. If hydrogen embrittlement is a concern, the part should be high-temperature vacuum baked after etching. Sometimes chemical etching does not remove some constituents from a surface and leaves a smut that must be removed by further etching. Whenever other elements in addition to Copper frames should be treated, a chemical etching using Etchant Vapor instead of a Liquid Phase could be also considered.

\begin{figure}
 \begin{center}
 \includegraphics[width=0.85\textwidth]{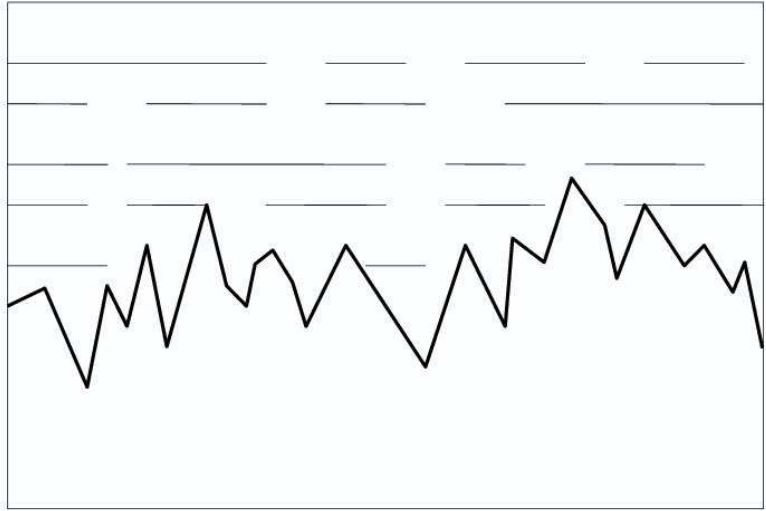}
 \end{center}
 \caption{Zoom of the micro-roughness of a metal surface immersed in an electrolytic bath}
 \label{fig:roughness}
\end{figure}

\paragraph{Electrochemical polishing\\}
In an electrolysis cell, the surface of an electrical conductor can be removed by making it the anode of an electrolytic
cell and ``de-plating'' the material. This is \emph{electro-etching} or \emph{electrocleaning} or \emph{electrodissolution}. The higher the current density, the more roughened the surface (fig.~\ref{fig:roughness}). Oxygen is released at the anode: it will react oxidizing contaminants on the surface at low anodic potentials.
 
Electropolishing is a much smoother process, where the working current density is carefully chosen. In case of Cu electropolished in a H$_3$PO$_4$ solution, in a first approximation, the leveling action is due to the protection of the smooth areas by a deposited phosphate material, and to the erosion of the exposed peaks.

Electropolishing leaves a phosphate film on the surface, which can be removed either by leaving pieces in the solution without any applied voltage or by an HCl rinse.
 
\begin{figure}
 \begin{center}
 \includegraphics[width=0.9\textwidth]{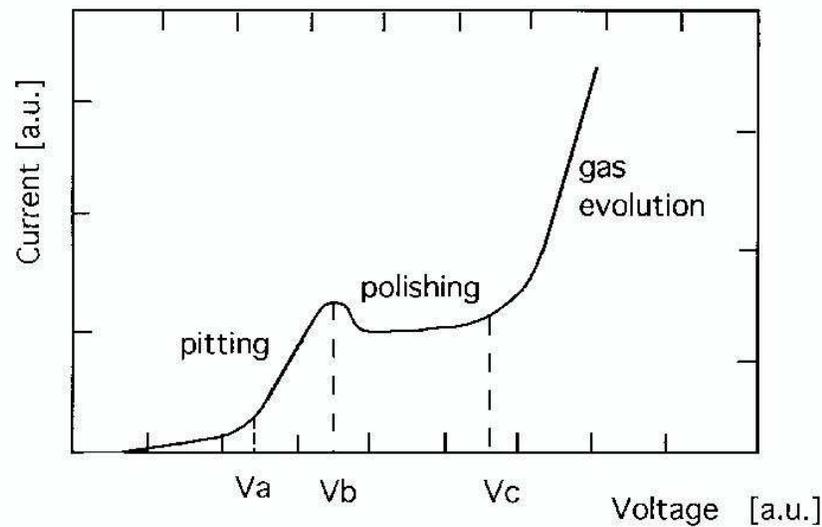}
 \end{center}
 \caption{Typical I-V characteristic for electropolishing}
 \label{fig:electropolishing}
\end{figure}

A typical I-V characteristic for the electropolishing of Copper in Ortophosphoric acid solution in the case of planar and parallel face electrodes and with negligible edge effects  is shown in fig.~\ref{fig:electropolishing}.
 The following behaviour is observed at the voltages corresponding to different sections of the curve:
\begin{itemize}
\item{Over section Va and Vb the current increases as a linear function of the voltage. Material dissolution happens with a too  low rate. The process is accompanied by the evolution of Oxygen  bubbles that promote local pittingof the surface of the piece by sticking to the anode.}
\item{Polishing effect is observed between Vb and Vc; the grain structure is brought into relief, as long as the process takes place. However roughness levelling and brilliant surface are obtained at voltages close to V$_c$. Here at the end of the plateau there is the minimum of the evolution of Oxygen bubbles. Even a minimum amount of bubbles can represent a limitation to the roughness smoothing action. Migrating toward the top they produce undesired vertical traces depending on the solution agitation.}
\item{At a higher potential, the gas evolution becomes stronger and the surface erosion is accompanied by pitting. A better surface quality is obtainable at voltages well above Vc, since Oxygen bubbles have not enough time to stick to the surface.}
\end{itemize}
 
In the framework of more than 18 years experience in the field of superconducting cavities, the LNL of INFN have well-established experience in Copper electropolishing. At LNL, indeed, a method for controlling the electropolishing process has been developed.
We try to minimize parasitic problems (such as gas evolution) by searching the best working point of the I-V characteristic and by following it during the whole process.
The method consists in a computer assisted monitoring of the I-V characteristic. The process is driven by voltage. An automatic program displays the numeric derivative of I versus V. The working point is chosen as the minimum of such a derivative, i.e. the minimum of bath differential conductance that corresponds to the point of maximum resistance of the viscous layer.

\begin{figure}
 \begin{center}
 \includegraphics[width=0.95\textwidth]{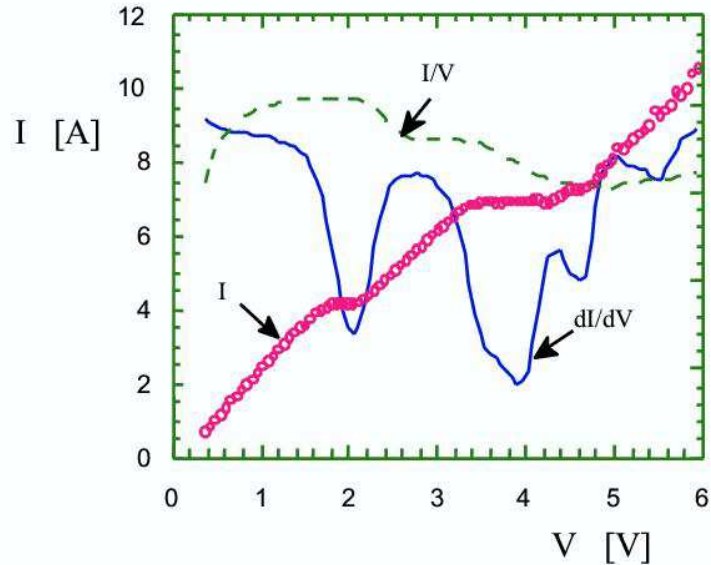}
 \end{center}
 \caption{I-V characteristic curve for a standard electropolishing process (circles). The differential conductivity curve dI/dV (continuous line) and the ratio I/V (shaded line) are displayed versus voltage. We interpret the first minimum in differential conductance as due to edge effects and to the not uniform distance between electrodes}
 \label{fig:ivelectro}
\end{figure}

In figure~\ref{fig:ivelectro}, we display the differential conductance compared to the ratio I-V and together with the I-V
characteristic. Some literature approximates the minimum of dI/dV with the minimum of I/V. Already from the figure below, it is visible that the two minima differ quite substantially in voltage.
In any case that of an electrolytic cell is a non-linear circuit, hence the solution conductivity is a differential quantity while, the ratio I/V has no any physical sense in such a case.
 
Our method allows the determination automatically of the best working point, when electropolishing. Its great advantage is that, by computer controlling, the operator does not need to know what is the metal being cleaned, and what acid is being used.

\paragraph{Electroless electrolytic cleaning\\}
 This method exploits the difference in electromotive potentials to remove material from one surface and deposit it on another. Such displacement-type electrolytic eleaning processes are commonly known for Silver. When immersing Silver in an undiluted solution of pure household ammonia contained in an Al or Mg tray, the Silver surface is cleaned as the Al or Mg is oxidized.

\paragraph{Deburring\\}
Deburring is the removal of the rough edges produced in cutting or shearing. Deburring is performed by: abrasion, chemical etching, laser vaporization and ``flash deburring'' (which uses a thermal pulse from an exploding gas-Oxygen mixture to heat and vaporize the thin metal protrusions).

\paragraph{Solvent cleaning\\}
Solvents dissolve the contaminant taking it into solution. It is possible to use: i) Polar solvents, such as water and water-alcohol mixtures, used to dissolve ionic salts which are polar contaminates; ii) Non-polar solvents, such as chlorinated hydrocarbon solvents, used to remove non-polar contaminates such as oil. A mixture of solvents can dissolve both polar and non-polar contaminates. A solvent's effectiveness is determined by the solubility parameter for specific contaminants (i.e. the saturation maximum amount of a specific contaminant dissolvable in a specific amount of the solvent). Of course the solubility parameter will be temperature dependent.

Solvent cleaning leaves a surface layer of residue which must be removed. This removal can involve: i) water rinsing; ii) an
elevated temperature; iii) other solvents that displace the surface layer. For example, both for cristals and copper frames; a solvent wipe-clean cleaning sequence might be Trichloroethylene-Acetone-Methanol-Isopropanol. Good polar solvents for a variety of polar contaminants (such as ionic salts) are Water and Water-Alcohol mixtures (typically in a ratio of
1:1). The addition of alcohol lowers water surface energy allowing it to penetrate into ``hideouts'' to remove hidden contamination.

Water and water-alcohol mixtures are not good solvents for non-polar contaminants such as oils.

\paragraph{Chlorinated and chlorofluorocarbon solvents\\}
Chlorinated hydrocarbon solvents, such as Trichloroethylene, are preferred to hydrocarbon solvents for their lower flammability, but there is concern with their toxicity and carcinogenic properties.
Chlorofluorocarbons are not carcinogenic and can be used in large quantities. They are, however, less effective than chlorinates solvents in dissolving contaminants.

1,1,1-trichloroethane (TCA-CHCCl$_3$, or methyl chloroform) is a common chlorinated solvent widely used in vapor degreasers. TCA has high Permissive Exposure Level (PEL), is classed as non-volatile, has low toxicity, but it has a high Ozone Depleting Potential (ODP) rating. Hence, in order to meet regulation standards, Vapors must be contained in enclosed vapor systems.

A possible alternative to TCA is Methyl-ene chloride (MEC-CH$_2$Cl$_2$). It has a rather low PEL, but it has a boiling point of 39.8°C. This makes it more difficult to handle but it has the advantage of being applicable for vapor degreasing temperature-sensitive materials. It must be kept in mind that MEC is a very aggressive solvent; it can damage plastics and rubbers.

Perchloroethylene (PCE or PERC - Cl$_2$C:CCl$_2$) is particularly suitable for being applied to Copper frames after machining. It has a high boiling point (121.1°C), is useful for dissolving heavy greases, and it can contain a large amount of water without degrading its solvency power.

Solvents can be mixed to give synergistic cleaning actions: a typical blend used for metal vapor degreasing is the Azeotrope
mixture of Freon TF with methylene chloride or with ethanol.  When using solvents however great attention must be payed to
the performed operation. Some solvents can react with the surface being cleaned. For example, Chlorinated solvents can react with water to form HCl which can react with many metals, particularly Al, Mg, Be, Zn , but also Cu and Te to form inorganic salts.

\paragraph{Liquid CO$_2$ as a solvent\\}
Liquid CO$_2$ has low surface tension and low viscosity. LCO$_2$ at 20°C, which is below the Supercritical Point, is a well-known good solvent for cleaning metals

\paragraph{Supercritical fluids\\}
If a gas, such as CO$_2$, is compressed to its ``critical pressure'', it liquefies to become a ``critical fluid''. If it is
also heated above its ``critical temperature'', it becomes a supercritical fluid. Critical fluids and supercritical fluids are good solvents for many medium-molecular-weight, non-polar or slightly polar organics. Solubility parameter for which CO$_2$ can vary from 0 in the gas phase to 10 under high pressure supercritical conditions (SCF-C0$_2$- critical point 31°C, 74 bar pressure). Values of 6-8 are typical.

Supercritical C0$_2$ fluid has the following advantages:
\begin{itemize}
\item{it is stable}
\item{has low toxicity}
\item{minimal cost}
\item{is a solvent for many organic materials}
\end{itemize}

\paragraph{Semi-aqueous cleaners\\}
Semi-aqueous clenears refer to solutions of natural or synthetic organic solvents, used in conjunction with water in some
part of cleaning cycle. Water immiscible semi-aqueous cleaners include: terpenes; high-molecular-weight esters; petroleum
hydrocarbons; glycol esters.
Terpenes are natural hydrocarbons such as: the d-limonene; the alpha- and beta-pinenes, derived from citrus and pine oils.
They may be as effective as the CFCs in many instances, but have a greater tendency to leave residues and suffer from the fact that they are slow drying.

Water miscible semi-aqueous cleaners are low-molecular- weight alcohols; ketones; esters; and organic amines. Acetone
(CH$_3$COCH$_3$) removes heavy oils quite effectively but tends to leave a residue, that should be removed by a methanol rinse.

\paragraph{Saponifiers, soaps, and detergents\\}
Alkaline saponifiers (generally silicate and phosphate-based) convert organic fats to water soluble soaps. Alkaline
cleaners are generally used hot. Mild alkaline cleaners (pH of 8-10) often have dissolved silicates, carbonates, borates, and citrates and should be used to clean alkaline-sensitive materials, such as Al and Mg. Strong alkaline cleaners, caustics, (pH = 12) may have water, sodium silicate, sodium molybdate and sodium fluoroborate.

Sodium silicate developes electrostatic forces that displace the contaminants while depositing a glassy film that prevents
recontamination. The glassy material is generally removed in the DI water rinse. After using alkaline cleaners, the surface should be followed by an acid dip prior to the water rinse to remove alkali salts since alkali salts adhere strongly to surfaces and are difficult to remove by ``water rinsing''. Clean oxide surfaces, in particular \teod monocrystals strongly adsorb hydrocarbons, and detergents or solvents normally do not completely remove the hydrocarbons; alkaline or oxidative cleaners must be used to remove the remaining hydrocarbons.

Detergent cleaning is a comparatively mild cleaning technique. In detergent cleaning, the detergent surrounds contaminants, emulsifying them into suspension, without actually dissolving the material. This emulsifying action is assisted by wetting agents and surfactants which loosen the contaminants from the surface. The most common detergents are soaps which are the
water-soluble reaction product of a fatty acid ester and an alkali (usually sodium hydroxide). In solutions, pH adjusters are used to aid in the cleaning action. Generally it is found that basic solutions clean better than acidic solutions if chemical etching is not involved. The pH of the cleaning solution is often adjusted to be basic, using ammonia or ammonium hydroxide.

Chelating agents (sequestering agents) keep into solution the normally in-soluble phosphates. Glass cleaning solutions use
chelating agents such as ethylene diamine tetraacetic acid (EDTA) and citric acid with salts containing hydroxyl and amine
substitutes.

\paragraph{Ultrasonic cleaning\\}
Low frequency ultrasonic cleaning relies on the jetting action of collapsing cavitation bubbles in contact with a surface
to provide a high pressure jet of fluid against the surface. Ultrasonic cleaning is often a good way to remove loosely adhering particles after a grinding or abrasive procedure and can be used with solvents to remove adsorbed contaminants. Ultrasonic jetting is good for removal of large particles but less efficient as the particle size decreases into the submicron range. The cavitation bubbles are formed by the tension portion of an ultrasonic wave in a fluid media and grow with time. The size that can be attained depends inversely on the frequency and the surface tension of the fluid. High frequencies (>60 kHz) give smaller bubbles and a higher bubble density. The ultrasonic wave is produced by magnetostrictive or electrostrictive transducers which can be attached to the fluid-containing tank walls or immersed in the fluid in the form of a probe that can concentrate the ultrasonic energy into a small area. Typically the transducers operate at 18-120 kHz, at an energy density of about 100 watts/gal of fluid. The ultrasonic cleaner size can be from 5 gallons for a small cleaner, up to very large systems using many transducers. The size of cavitation bubbles in the fluid depends on the vapor pressure, surface energy, and temperature of the fluid. For example, pure water at 60°C and 40 kHz has a maximum cavitation bubble size of about 100 microns if a surfactant is present, the bubble size is smaller due to the lowered surface energy. The jet pressure from the collapsing bubble can be as high as 300 psi. The cavitation jetting is more energetic for cooler media and when there are no gases in the bubble to hinder its collapse.

Ultrasonic cleaning must be used with care on \teod crystals, since the jetting action can produce high pressures that cause
erosion and introduce fractures in the surface.

\paragraph{Megasonic cleaning\\}
High frequency (>400 kHz) ultrasonic cleaning does not cause cavitation. Instead, the action consists of a train of wave fronts that sweep across a smooth surface producing disruption of the viscous boundary layers on the substrate surface by viscous drag. The resulting pressure is less than 50 psi and does not hurt fragile surfaces. High frequency transducers can be focused to restrict the area of impact and allow lateral fluid flow from the area of concentration.

Megasonic cleaning utilizes high frequency (850-900 kHz) transducers to produce non-cavitating pressure waves. The megasonic agitation system is applicable to smooth surfaces, particularly for removing particles, but does not work well on configured surfaces where the surface is shadowed from the pressure wave.

\paragraph{Wipe-clean\\}
In some cases the surface can not be immersed in a fluid and must be cleaned by wiping. Wiping with a fluid should be done with a moist lint-free cloth or sponge which has no extractables when in contact with the wiping-fluid. The wiping motion should be a rolling motion such that contamination that is picked-up does not come into contact with the surface as wiping proceeds.

\paragraph{Strippable coatings\\}
Particles can be removed from surfaces by covering the surface with a liquid polymer, allowing it to solidify, then mechanically stripping (peeling) the polymer from the surface. This technique is used by the optics industry to remove particles from mirror surfaces and protect surfaces from abrasion during assembly, and in silicon technology to remove particulates from silicon wafers. There are many types of ``strip coats,'' each coating leaves different residues on stripping and have differing corrosion compatibility with surfaces. Hydrocarbon residues left by strippable coatings can be removed by oxidation techniques. A strippable coating is not necessarily organic, but it can be also a sputtered getter film, for example a Niobium, Hafnium or Titanium film sputtered onto copper for Oxygen purification. The film is then removed by chemical etching. 

\paragraph{Rinsing\\}
After any wet cleaning process the surface should be thoroughly rinsed in pure or ultrapure liquid, usually water, before allowing to dry. This avoids leaving residues on the surface. A common rinsing technique is to use successive rinses (cascading rinsing) in pure or ultrapure water until the rinse water retains a high resistivity (e.g., >12 megohm). A problem can be the ``dragout'' of one fluid with the part which then contaminates the subsequent fluid tank. For the beginning rinse, a sheeting agent can be added that lowers the surface tension of the water and aids in flowing
the rinse water off the surface. After rinsing, the surface should be dried as quickly as possible since the residual water film on the surface will cause particles to stick to the surface and on drying the particles will adhere very tenaciously.

\paragraph{Drying and outgassing\\}
Displacement drying uses anhydrous fluids, such as isopropanol, anhydrous ethyl alcohol denatured with acetone or methanol, or a commercial drying agent. One of the best drying techniques is a ``vapor dry'' where the cold surface is immersed in the vapor above a heated anhydrous alcohol sump. The cold surface condenses the alcohol vapor which flows off into the sump taking water and particulates with it. When the surface becomes hot, condensation ceases and the hot surface, when withdrawn dries rapidly.
Surfaces can be blown dry using a low velocity dry gas stream. When blowing, a nozzle with a 0.2 micron or smaller
particulate filter should be used in the nozzle. In addition, when drying insulator surfaces, the gas should be ionized to prevent charge build-up on the surface. The gas can be ionized with an electronic (corona), or a laser ionizer. But be aware that Electronic ionizers can arc and produce particulates.
A high velocity jet of gas can be shaped to blow-off a moving surface. The jet should impact the on-coming wet surface at
about a 30° angle. Hot gas drying or evaporative drying, uses the recirculation of hot dry, filtered air over the surface to promote evaporation. This drying technique has the problem of ``water spotting'' if the fluid is not ultrapure.

Volatile materials from the bulk of the material are removed by outgassing. Obviously TeO$_2$ monocrystals will not outgas
appreciably, while Outgassing is especially important for Copper.
Since diffusion is required, the time to outgas a material may be very lengthy if the diffusion rate is slow and/or the diffusion distance is long. Generally metals primarily outgas hydrogen, particularly that taken up during acid cleaning or
electropolishing.

After drying, the cleaned parts should be stored and transported in a manner that does not unduly recontaminate the
parts.

\paragraph{Reactive cleaning\\}
Reactive cleaning uses liquids, gases, vapors, or plasmas to react with the contaminant to form a volatile or soluble reaction product. Reactive cleaning liquids are often oxidizing solutions. Many acid-based systems can be used as oxidants. Nitric acid can also be used as the oxidizing agent. Nitric acid with hydrofluoric acid is used to oxidize/etch surfaces. Hydrogen peroxide (H$_2$O$_2$) is a good oxidizing solution for cleaning glass.[64] Often boiling 30 unstabilized H$_2$0$_2$ is used. Hydrogen peroxide is often stabilized, which reduces the release of free oxygen. Unstabilized H$_2$O$_2$ must
be stored in a refrigerator to slow decomposition. Hydrogen peroxide is sometimes used with ammonium ``hydroxide'', to increase the complexing of surface contaminants. However the decomposition rate of the H$_2$O$_2$ is greatly increased by combination with ammonium hydroxide.

Oxidative cleaning can be performed using chlorine-containing chemicals. For example, a water slurry of sodium
dichloroisocyanurate (i.e. swimming pool chlorine) which has 63\% available chlorine, can be used to scrub an oxide surface to remove hydrocarbon contamination. This combines mechanical scrubbing with oxidation and improves the cleaning action.

Anodic oxidation in an electrolysis cell can be used to clean surfaces. The contaminated layer can be removed by anodically oxidizing the surface in an electrolytic cell. The oxide layer must be subsequently removed.

Gaseous oxidation cleaning can be used on surfaces where surface oxidation is not a problem. Oxidation is usually
accomplished using oxygen, chlorine, fluorine, ozone, or Nitric oxide which creates volatile reaction products such as CO and CO$_2$.
Reactive gas cleaning may use a reaction with a gas at high temperature to form a volatile material. High temperature air fire is an excellent way to clean surfaces that are not degraded by high temperature.

\paragraph{Ozone cleaning\\}
The use of oxidation by Ozone created by ultraviolet radiation (UV/Ozone cleaning) at atmospheric pressure and low
temperature has greatly simplified the production, storage and maintenance of hydro-carbon-free surfaces. Historically the UV/0, cleaning process was developed because of the need to clean very delicate quartz oscillators. Any attempt to clean the quartz using physical contact caused breakage. The UV/O3, cleaning technique provided a non-contacting way to clean the delicate quartz plates.
The short wavelength radiation causes bond scission in the hydrocarbon contaminants and generates ozone which reacts with
carbon to form volatile CO and CO$_2$.
Typical exposure times for UV/03 cleaning are from a few minutes to remove a few monolayers of hydrocarbon contamination to
hours or days or weeks for storage of cleaned surfaces. The UV/03 cleaning technique has the advantage that it can be used as a dry, in-line cleaning technique at atmospheric pressure. The UV/03 cleaning technique is also useful for cleaning holes (vias) in surfaces.

\paragraph{Hydrogen cleaning\\}
 High temperature hydrogen or forming gas (90\% N$_2$:10\% H$_2$), can be used to remove hydrocarbon contamination from a surface by hydrogenating the material and making it more volatile. Hydrogen reduction of oxide layers can be used to clean surfaces in a furnace environment. However, Hydrogen cleaning can also change the surface chemistry. For example, hydrogen firing of \teod could produce metallic Tellurium surface by reducing the Oxyde to metal on the surface.

\paragraph{Reactive plasma cleaning and etching\\}
Reactive plasma cleaning uses a reactive species in the plasma to react with the surface to form a volatile species which
leaves the surface at much lower temperatures than those used for reactive gas cleaning. The additional requirement on reactive plasma cleaning is that it does not leave a residue. Oxygen or hydrogen are often used for plasma cleaning while fluorine (from SF$_6$, CF$_4$ , CHF$_3$, C$_2$F$_6$ C$_3$F$_8$, or SF$_6$) and chlorine (from HCl, CCl$_4$, or BCl$_3$) are used for plasma etching. The reactive plasma cleaning/etching technique is typically specific and can be used to selectively remove the oxide from the surface and then have a low etch rate for the substrate material.

Oxygen (or air) plasmas are very effective in removing hydro-carbons and absorbed water vapor from surfaces. A typical plasma cleaner has plasma generated by an rf discharge and the surfaces to be cleaned are in a ``remote'' or ``downstream'' location and not in the plasma generation region. The surface attains a potential (sheath potential) that is negative with respect to the plasma, and ions are accelerated from the plasma to the surface. For the case of a ``cold plasma'' which has low energy particles, this sheath potential will only be a few volts. When the plasma particles are more energetic or the electrons are accelerated to the surface, the sheath potential can be tens of volts. In addition to being bombarded by ions, the surface in contact with the plasma will be bombarded by ``activated species'', excited species, thermal species, and high energy photons (UV and, under some conditions, soft X-rays). Ions and excited species will release their energies of ionization or excitation when they impinge on the surface.

Plasma etchers and strippers typically use more aggressive reactant gases such as chlorine or fluorine and are constructed to withstand corrosion and pump the particulates that are often formed in the etching and stripping process.

When plasma is etching copper, a copper chloride (CuCl$_2$) residue is left on the surface which can be volatilized by heating to above 200 C. Often mixtures of gases are used for etching and cleaning. Oxygen is often added to the fluorine plasma to promote the formation of atomic fluorine and to oxidize the surface and thus increase the etch rate. One of the most common gas mixtures used to etch is 96 CF4. Helium is often added to dilute the mixture and to increase the thermal conductivity of the plasma thus reducing the temperature rise of the substrate during etching.

\paragraph{Plasma cleaning\\}
 Plasmas can be used to clean surfaces in the deposition system in the same manner as they are used to condition vacuum
surfaces. In some PVD deposition systems which are not normally used with a plasma, a ``glow bar'' or ``glow plate'' is used as the cathodic or anodic electrode of a DC discharge to create the plasma. The larger the area of the surface, the better the plasma distribution in the system. Plasma cleaning can be done using an inert gas plasma or can use a plasma containing a reactive gaseous species to form a volatile reaction product from the interaction of the gaseous species and the surface species.

``Ion scrubbing'' of a surface occurs when a surface, which is in contact with an inert gas plasma, develops a wall sheath and is bombarded by inert gas ions accelerated across this wall sheath.
Generally the ion energy is too low to cause surface damage or physical sputtering, but is effective in the desorption of
adsorbed surface contaminants such as water.

Reactive plasma cleaning/etching can be done in the deposition system in much the same way as was described in the
``external'' plasma cleaning. The surface in contact with a plasma containing reactive species develops a negative potential with respect to the plasma (self-bias). Ions, along with neutrals and ``activated'' species, of the reactive species bombard the surface producing volatile reaction product either with contaminants (cleaning) or the substrate material (etching). The most common reactive gas used is oxygen or air.

A major concern in any plasma process is to obtain a uniform plasma over the surface. Some plasma-generation configurations are more amenable to uniformity than are others. The magnetron configuration is one where plasma uniformity is difficult to
obtain except in certain applications such as passing the substrate through the plasma of a planar magnetron. For CUORE
magnetron configurations using moving electromagnetic fields could be conceived in order to obtain uniform etching without substrate damage.

The reactive etching/cleaning processes produce volatile species which may be deposited in other parts of the system where
there are different plasma conditions. This may have a detrimental effect on the gas handling/ pumping system and can be a source of particulates in the etching system, that must be periodically reconditioned.

\paragraph{Sputter cleaning\\}
Sputter cleaning uses physical sputtering, not chemical reaction, to remove some of the surface layer, which includes the
contaminates. Sputter cleaning is a kind of universal etch since conceptually everything can be removed by the sputtering process.
However certain types of surface contamination, such as particles and inclusions of inorganics compounds, are very difficult to remove by sputtering because of their shape. This cleaning process can be easily integrated into the deposition process so as to allow no time for recontamination between the cleaning and the deposition process as is done in ion plating. Sputter cleaning however could produce detrimental surface damage on \teod crystals, unless very low power and low density plasmas are ignited. For TeO$_2$, indeed Helium plasmas are foreseen.

During sputter cleaning, the bombarding gas may become incorporated into the surface and subsequently released on
heating. Sputtering from a plasma environment has the disadvantage that gaseous contamination in the plasma becomes activated and can react with the surface being cleaned; also sputtered species can be returned to the surface by scattering (redeposition) and contaminate surface species can be recoil-implanted into the surface. Sometimes this makes sputter cleaning difficult. Ion milling, where ion beam sputtering is used to remove surface material, can be done in a vacuum environment where the sputtered species are not redeposited on the substrate surface and gaseous contamination is rapidly pumped away.

\paragraph{Ion beam cleaning\\}
The use of ion and neutral beams allow the cleaning/etching of a surface in a good vacuum environment. Energetic ion beams of reactive species can be used to clean/etch surfaces and the process is called Reactive Ion Beam Etching. Beams of uncharged radicals of reactive species (H, Cl, F) can be used to clean surfaces in vacuum. The use of energetic inert gas ion beams to bombard a surface concurrently with a molecular beam of the etchant gas (Ion Beam Assisted Etching-IBAE) shows enhanced etching over either the inert ion bombardment (sputtering) or the molecular beam alone.

\subsection{Conclusions}

 From the analysis of the possibilities listed above, we learn that two different cleaning procedures must be conceived for
Copper and for Crystals.

For cleaning the Copper structure delivered after fabrication, indeed, we propose the following procedure:
\begin{itemize}
\item{Mechanical finishing}
\item{Degreasing}
\item{Tumbling or vibratory finishing}
\item{Rinsing}
\item{Bright dipping}
\item{Rinsing}
\item{Electropolishing}
\item{Rinsing}
\item{Ultrasound rinsing}
\item{Chemical etching}
\item{Copper passivation}
\item{Rinsing}
\item{Ultrasound rinsing}
\item{Drying}
\item{Vacuum degassing}
\item{Sputter cleaning}
\item{Storing in controlled atmosphere environment}
\end{itemize}

On the other side for cleaning \teod crystals delivered after fabrication, indeed, we propose the following procedure:
\begin{itemize}
\item{Alcohol Rinsing}
\item{Drying}
\item{Helium gas low pressure and low power Sputter cleaning}
\item{Ion beam cleaning}
\item{Storing in controlled atmosphere environment}
\end{itemize}

Tests of both procedures will be carried out in the next feature to assess their effectiveness.

\clearpage
\section{Enrichment option}\label{sec:enrich}
CUORE is the only proposed next generation \BBz decay experiment that plans the use of non-enriched material. Indeed, the high isotopic abundance of \tect allows very high sensitivities even using the much cheaper natural abundant material. This obviously implies that the construction of a larger array would not only be technically possible but also fiscally acceptable. On the other hand the high natural abundance of \tect allows a much simpler and less expensive enrichment procedure than that for other double beta decay candidates. Moreover \teod crystals made with \tect isotopically enriched material have been already tested during the MiDBD experiment with results consistent with those obtained for natural crystals, both from the point of view of detector performance and detector background. 
As discussed in section~\ref{sec:Qprospects}, a 95\% enrichment in \tect of CUORE with a background of 0.001 counts/(keV\dot kg\dot y) would imply a 5 year sensitivity  F$_D$ of 1.9\per 10$^{27}$ y implying a sensitivity on the Majorana mass parameter in the range 8-45 meV.

We are presently considering a "partial enrichment" option for CUORE which would leave unaltered the CUORE structure shown in Fig.\ref{fig:cuore_cyl}. Four central columns of CUORE would contain 9 inner planes made by 95\% enriched \teod  crystals with the upper and lower two planes made by natural ones. We would have therefore in total 144 enriched and 844 natural crystals, respectively. The sensitivity of CUORE on the Majorana neutrino mass would be almost doubled if the background of the enriched crystal would be the same as that of the natural  ones. In addition comparison between the spectra from the enriched and natural crystals would  be precious to understand the origin of the background especially if a peak indicative of neutrinoless double beta decay would be found. Our plan would be to produce in the USA and test at LNGS three \ciccio crystals enriched in \tect plus three natural ones, within two year from the beginning of the project to ensure that the thermal and background properties would be the same as those of CUORICINO. If this would not be the case the option of CUORE entirely made by natural crystals will be adopted.

\clearpage
\section {Structure of the collaboration}\label{sec:collaboration}

The CUORE collaboration involves physicists from 14 active groups in Europe and United States, plus about 50 technicians and a number of undergraduate students. Five Laboratories of international relevance participate to it: the Lawrence Berkeley National Laboratory, the Lawrence Livermore National Laboratory, the  Laboratori Nazionali del Gran Sasso, the  Laboratori Nazionali di Legnaro and the Kamerling Omnes Cryogenic Laboratory in Leiden. 
All the specific services of these Laboratories will participate to the various technical problems involved by CUORE: cryogenics, chemical, construction, material properties, electronics, low radioactivity measurement etc. In addition a collaboration has been initiated with the Department of Structural Engineering  of the Politecnico of Milan, one of the leaders in Europe, for the design and construction simulation of the CUORE structure. 
With respect to the other experiments on the same subject, a peculiarity of CUORE is the multidisciplinary aspect required by its construction and operation. For this reason we would like to mention the ``historical'' background of the various participating group and their familiarity and internationally acknowledged skill  in the specific fields of the project.  We will also mention how they will collaborate on the various tasks.

\begin{enumerate}
\item{The Berkeley group, in addition to its well known experience in low energy nuclear physics, has an outstanding skill in the fabrication and operation of  NTD (Neutron Trasmutation Doped) thermistors obtained by controlled irradiation in two USA reactors. Since 15 years they have established a close collaboration with the Milano group in the development of thermal detectors.}
\item{The Milan, South Carolina and Zaragoza groups have largely contributed in the field of double beta decay experiment since more than 30 years. The South Carolina group is concentrating on the issue of other candidate \BBz bolometer materials.}
\item{The Leiden group is one of the leaders in the world arena in the construction and operation of dilution refrigerators and more in general in the field of Cryogenics}
\item{the Berkeley, Livermore, Milan (with Como) , South Carolina and Zaragoza groups are working since tens of years and often in collaborarion in the study and reduction of radioactive background}
\item{the Legnaro group  is a leading  team in Europe in the field of surface treatment and cleaning}
\item{the Florence group has since tens of years a deep knowledge in the material properties at low and very low temperature, and is presently collaborating with Como and Milano in the characterization of a various samples of NTD thermistors produced in Berkeley}
\item{The Genoa group was and is involved since many years in the field of data acquisition}
\item{The Milan group is the only one in the world  performing searches on double beta decay with the cryogenic technique since fifteen years}
\end{enumerate}
All these groups are since the last year collaborating to the development and operation of CUORICINO, which is a first step towards the realization of CUORE. 

\clearpage
\section{Cost statement}
\label{sec:cost}
In this section we give an estimate of the costs of the basic CUORE design described in this proposal with possible sources of these funds.

\begin{table}[h]
\caption{Breakdown of the total CUORE cost estimates (kEuro)}\label{CUORE:cost}
\begin{center}
\begin{tabular}{lcccccccccc}
  \hline\hline
Start date:	&	&2005	&2006	&2007	&2008	&2009	&Total	&US	&INFN\\
1st January 2005&&&&&&&&\\
  \hline
Crystals			&US	&50	&50	&2800	&1400	&	&4300	&4300	&\\
Thermistors			&US	&100	&	&500	&	&	&600	&600	&\\
Detector structure		&INFN	&170	&150	&150	&100	&100	&670	&	&670\\
Cleaning			&INFN	&50	&350	&200	&50	&50	&700	&	&700\\
Cryostat \& cryogenics		&INFN	&100	&1000	&100	&	&	&1200	&	&1200\\
Shieldings			&INFN	&50	&50	&	&500	&	&600	&	&600\\
Underground Laboratory		&INFN	&50	&300	&500	&	&	&850	&	&850\\
Electronics			&US	&40	&40	&480	&	&30	&590	&590	&\\
Data Acquisition		&INFN	&10	&30	&155	&155	&170	&520	&	&520\\
CUORE Installation		&INFN/US&	&	&50	&50	&900	&1000	&500	&500\\
Contingency			&INFN/US&300	&700	&700	&600	&600	&2900	&1450	&1450\\
\hline									
Total cost			&&920	&2670	&5635	&2855	&1850	&13930	&7440	&6490\\
  \hline\hline
\end{tabular}
\end{center}
\end{table}

\paragraph{\teod crystals\\}
The final cost of the 988 \ciccio \teod crystals, as agreed some time ago with SICCAS if of the order of 4,300 Euro per crystal (comprehensive of Te metal, chemical reduction to \teod, crystal growth and preliminary lapping). This would amount to a total of about 4.3 million Euro. 

\paragraph{Shields\\}
According to the current quotation of normal and low activity lead and the  basic design described in section \ref{sec:shields}, the total cost would amount to 500,000 Euro. An additional cost of 100,000 Euro would be required for the external neutron shield.

\paragraph{Electronics\\}
As described in section \ref{sec:electronics}, the front-end electronics for CUORICINO was designed taking into account CUORE requirements. Few modifications will be therefore required and the cost of each channel (from thermistor bias circuit to DAQ input) can be precisely estimated: 590 Euro. This cost would include also a 10\% excess of spare channels.

\paragraph{Dilution refrigerator and cryogenic equipment\\}
This section would include the dilution refrigerator system, the CUORE cryostat and the liquid helium and nitrogen systems. A rough estimate amounts to one million Euro.

\paragraph{Underground laboratory\\}
As described in section \ref{sec:hut} the underground building will house (besides the CUORE cryogenics, shields, electronics and DAQ) a clean room for detector assembling and a Faraday cage for front-end electronics. The total cost of these structures would amount to 850,000 Euro (200,000 for the Faraday cage, 400,000 for the clean room, 150,000 for the laboratory and 100,000 for the UPS's).

\paragraph{Detector structure and machining\\}
The cost of the detector structure materials will be dominated by the required amount of copper (5 tons) whose cost can be estimated around 50,000 Euro. PTFE and other materials would cost 10,000 Euro. The main conribution would come however from detector structure machining whose cost is estimated of the order of 500,000 Euro.

\paragraph{DAQ system\\}
At present, only a very rough estimate of the cost of the DAQ system is possible. In order to satisfy CUORE requirements in fact, a new design is required with respect to CUORICINO (sec.\ref{sec:readout}). A rough estimate of the total cost amounts to 400 Euro/ch. Additional 50,000 Euro will be required for the slow control system.

\paragraph{Surface cleaning facilities and procedures\\}
As previously discussed (sec.\ref{sec:SURF}), surface cleaning for CUORE requires an intensive R\&D phase. A final cost for the whole item will be possible only after we'll get a final decision. A rough estimate is however possible by extrapolating our CUORICINO experience. In particular, excluding the cost of cleaning materials (which could be extremely variable in dependence of the purity requirements) we can estimate a total budget of 700,000 Euro, including both R\&D (mainly at INFN LNL) during the first 1-2 years and facilities/procedures costs (at LNGS).

\paragraph{NTD thermistors\\}
In order to produce 2000 NTD Ge thermistors for CUORE, we estimate a cost of approximately 600,000 Eu which will be used to pay for the purchase of the Ge  material, the reactor irradiations, the determination of the doping levels achieved, and the processing of the NTD material into thermistors.  We also estimate the time required to do this production to be on the order of three years.  

\clearpage
\section{Conclusions}

It should be abundantly clear from the foregoing discussion that the most important remaining issue in neutrino physics is the determination of the neutrino mass scale. Neutrino oscillation experiments have clearly demonstrated that:
\begin{enumerate}
\item{neutrinos have mass and that their mass eigenstates mix;}
\item{the mixing matrix elements have values that imply that next generation \BBz decay experiments should be able to determine the scale of neutrino mass in the case of Majorana neutrinos.}
\end{enumerate}
The main advantage of CUORE is that it will utilize \teod crystals of natural isotopic abundance Te even if an enriched option could also be considered at an affordable cost. CUORE will offer also the unique possibility to search for \BBz of isotopes other than Te. On the other hand, the bolometric technique has yielded energy resolutions approaching those of intrinsic Ge detectors and energy resolution is crucially important for an actual observation of \BBz and a measurement of \amnu.  Important results on background reduction have already been obtained in CUORICINO which represents also a valid test of the CUORE project feasibility . With a little R\&D CUORE will reach a 5--year sensitivity of \ca 30 meV on \amnun, which could decrease to 15 meV if the full CUORE R\&D program will be successful.

We would like to add that CUORE is the only second generation double beta experiment based on a pilot perfectly operating prototype setup (CUORICINO) which consists  essentially in one of the CUORE  towers,  thus ensuring the complete feasibility of its larger "brother". 

The present and future results of CUORICINO will obviously be precious to anticipate the performance of CUORE. Last but not leat, CUORE is the only experiment with an already approved and fixed underground location. 

\newpage
During the preparation of this Proposal to which he validly
contributed, Angel Morales passed away. We lost an outstanding physicist,
a devoted teacher and a dear friend.

\end{document}